\documentclass[
superscriptaddress,
 amsmath,amssymb,
 aps,
onecolumn,
nofootinbib
]{revtex4-2}
\usepackage{rotating}
\usepackage{graphicx,comment}

\usepackage{booktabs}
\usepackage{bbold}
\usepackage{dcolumn}
\usepackage{bm}
\usepackage{hyperref}
\usepackage{accents} 
\usepackage[dvipsnames]{xcolor}
\usepackage{tabularx}
\usepackage{caption}
\usepackage{wrapfig}
\usepackage{booktabs}
\usepackage{subcaption}
\usepackage{feynmp-auto}
\usepackage{makecell}
\usepackage{comment}
\usepackage{amsmath}
\usepackage{amssymb}
\usepackage{orcidlink}
\usepackage{multirow}
\usepackage{array}

\newcommand{\sv}{\langle \sigma v \rangle}
\newcommand{\Neff}{N_{\rm eff}}
\newcommand{\Tr}{\mathrm{Tr}}

\usepackage{calc}
\newlength{\panelw}\newlength{\panelwi}\newlength{\panelwii}\newlength{\panelwiii}
\newlength{\paneldenom}\newlength{\panelgap}
\newcommand{\panelrow}[3]{%
  \settowidth{\panelwi}{\includegraphics{{#1}.pdf}}%
  \settowidth{\panelwii}{\includegraphics{{#2}.pdf}}%
  \settowidth{\panelwiii}{\includegraphics{{#3}.pdf}}%
  \ifdim\panelwii=0pt\relax\GenericError{}{panelrow: cannot measure #2}{}{}\fi
  \setlength{\paneldenom}{\panelwi+\panelwii+\panelwiii}%
  \setlength{\panelw}{\textwidth-\panelgap}%
  \setlength{\panelw}{\panelw*\ratio{\panelwii}{\paneldenom}}%
  \setlength{\panelwi}{\panelw*\ratio{\panelwi}{\panelwii}}%
  \setlength{\panelwiii}{\panelw*\ratio{\panelwiii}{\panelwii}}%
  \includegraphics[width=\panelwi]{{#1}.pdf}\kern\panelgap 
  \includegraphics[width=\panelw]{{#2}.pdf}%
  \includegraphics[width=\panelwiii]{{#3}.pdf}%
}
\newlength{\relicrefw}\newlength{\relicw}
\newcommand{\relicref}{figures/figures_relic_source/heavyScalar_relic_merged__relic_source.pdf}
\newcommand{\relicwide}[1]{\includegraphics[width=\textwidth]{#1}}
\newcommand{\relicsingle}[1]{%
  \settowidth{\relicrefw}{\includegraphics{\relicref}}%
  \settowidth{\relicw}{\includegraphics{#1}}%
  \ifdim\relicrefw=0pt\relax\GenericError{}{relicsingle: cannot measure \relicref}{}{}\fi
  \setlength{\relicw}{\textwidth*\ratio{\relicw}{\relicrefw}}%
  \includegraphics[width=\relicw]{#1}}

\hypersetup{
  colorlinks   = true, 
  urlcolor     = Maroon, 
  linkcolor    = PineGreen, 
  citecolor   = Maroon 
}

\renewcommand{\arraystretch}{0.85}
\begin{document}


\title{How effective are effective constraints on Sub-GeV dark matter?}

\author{Neal P. Avis Kozar \orcidlink{0000-0003-4875-6782}}%
\email{neal.aviskozar@queensu.ca}
\affiliation{Department of Physics, Engineering Physics, and Astronomy, Queen’s University, Kingston, Ontario, K7L 3N6, Canada}
\affiliation{Arthur B. McDonald Canadian Astroparticle Physics Research Institute, Kingston ON K7L 3N6, Canada}

\author{Christopher Cappiello\orcidlink{0000-0002-7466-9634}}
\email{cappiello@wustl.edu}
\affiliation{Department of Physics and McDonnell Center for the Space Sciences,
Washington University, St. Louis, MO 63130, USA}

\author{Melissa D. Diamond\orcidlink{0000-0003-1221-9475}}%
\email{melissa.d.diamond@mcgill.ca}
\affiliation{Department of Physics, McGill University, Montreal, Quebec, H3A 0G4, Canada}
\affiliation{Arthur B. McDonald Canadian Astroparticle Physics Research Institute, Kingston ON K7L 3N6, Canada}

 \author{Aaron C. Vincent \orcidlink{0000-0003-3872-0743}}%
\email{aaron.vincent@queensu.ca}
\affiliation{Department of Physics, Engineering Physics, and Astronomy, Queen’s University, Kingston, Ontario, K7L 3N6, Canada}
\affiliation{Arthur B. McDonald Canadian Astroparticle Physics Research Institute, Kingston ON K7L 3N6, Canada}

\begin{abstract}
Dark matter that couples directly to nuclei will inevitably pick up effective interactions with mesons, photons and electrons. While these interactions naively lead to more constraints, we show here that they in fact open a new, large parameter space of viable thermally-produced sub-GeV dark matter models. These effective interactions can arise at the loop level, or via mixing of pseudoscalar or axial vector mediators with the neutral mesons. This dramatically alters the phenomenology of purportedly nucleophilic dark matter models, resulting in a new thermal dark matter production channel via photon and meson-sourced freeze-in, and leading to many new constraints and observables for these models.   We calculate the dark matter-photon interaction cross sections and the new meson decay rates that emerge from these nucleophilic models, and make them available to the reader to apply to specific models of their choosing. We derive novel constraints on the dark matter-nucleon cross section in the keV--GeV mass range.  When we apply these new constraints to models with specific choices of mediator mass and ratios between the quark couplings, we find that the limits from effective interactions constrain vast regions of sub-GeV dark matter parameter space. However, if the reheat temperature is low enough, free parameter scans of the $m_{\chi}-\sigma_{\chi N}$ plane reveal large regions of viable sub-GeV parameter space and useful benchmarks for experimental searches. 
\end{abstract}

\maketitle

\tableofcontents
\section{Introduction}

Limits on dark matter (DM) interactions with Standard Model (SM) particles are often presented in the plane of DM-nucleon elastic scattering cross section $\sigma_{\chi N}$ versus DM mass $m_\chi$. Direct detection limits typically require an assumption about how $\sigma_{\chi N}$ relates to the DM-nucleus scattering cross section, of which the most common assumptions are spin-dependent (SD) and spin-independent (SI) interactions, encoding whether the scattering amplitude depends on the total spin $J$ of the target nucleus, or simply the total mass number, respectively. In fact, these are only two of many effective operators that can describe DM-nucleus interactions in the limit of a heavy mediator~\cite{Fan:2010gt,Fitzpatrick:2012ix}.
This lingua franca allows us to superimpose limits (and sometimes hints) from a variety of different experiments, showcasing slow but steady progress as experimental searches dig ever deeper into parameter space. However, the composite nature of nucleons means that these quantities, e.g. $\sigma_{\chi N}^\mathrm{SI/SD}$, necessarily stem from a more fundamental theory with couplings to quarks or gluons, which in turn implies an inevitably richer phenomenology. Indeed, at higher orders in perturbation theory, many couplings with quarks give rise to an effective coupling to photons thanks to loop diagrams of virtual quarks---or at low energies, loops of pions and other light mesons and mixing with light mesons. Once a particle can interact with photons or light mesons, it becomes difficult indeed to hide. 

In Ref. \cite{Diamond:2023fsm}, we showed that loop corrections from chiral effective field theory (ChEFT) lead to strong constraints on WIMP-like DM interacting via a vector mediator, for DM masses below a few GeV, where direct detection experiments lose sensitivity, but electron-recoil and electromagnetic probes can overcome the perturbative suppression. Ref. \cite{Cox:2024rew} performed similar calculations at the EFT operator level (i.e. assuming a mediator mass above the cutoff $\Lambda$ of the theory), showing that meson decay and Big Bang nucleosynthesis (BBN) place severe constraints on DM-nucleon interactions when the DM interacts via a heavy scalar mediator. The same authors further showed \cite{Cox:2025toz} that for a variety of heavy mediators and quark couplings an effective coupling to photons can spoil the relic abundance in generic models of light DM, taking out swaths of low-mass parameter space unless couplings to quarks are finely tuned. 

Of course, these were not the first works to consider effective couplings leading to stronger constraints than tree-level processes. Millicharged DM can arise through loop processes \cite{Holdom:1985ag,Prinz:1998ua,Davidson:2000hf,Pospelov:2008zw,McDermott:2010pa,BaBar:2014zli,Izaguirre:2015eya,Mahdawi:2018euy,Munoz:2018pzp,Emken:2019tni,Caputo:2021eaa,Chiles:2021gxk,Kahn:2021ttr,Gorghetto:2022sue,Romanenko:2023ytm}, and loop-level processes often arise in higher-energy indirect or collider DM searches \cite{Bern:1997ng,Bergstrom:1997fh,Ullio:1997ke,Boudjema:2005hb,Fox:2008kb,Kopp:2009et,Bringmann:2012ez,Bell:2014tta,Bramante:2015una,Fermi-LAT:2016afa,Kumar:2016gxq,Braaten:2017dwq,Bringmann:2018lay,HESS:2018cbt,FileviezPerez:2019rcj,HAWC:2019jvm,Arina:2020udz,Rinchiuso:2020skh,Arina:2021gfn,Foster:2022nva,Kopp:2014tsa,Mohan:2019zrk}. 

In this paper, we expand the scope of Ref.~\cite{Diamond:2023fsm} to scalar, pseudoscalar, and axial vector couplings between the mediator and quarks.  
By including the mediator explicitly, we are thus able to probe both heavy and light mediators, and the transition between the two cases. We additionally allow for parity-violating interactions where the mediator has a scalar coupling to the fermionic DM and a pseudoscalar coupling to the quarks and vice versa. 
As we will show below, the scalar quark couplings can introduce new interactions with photons at the loop level, while pseudoscalar and axial quark couplings interact with photons via direct mixing with the mesons. 

These additional effective couplings have profound effects. The first is production of DM in the early universe via freeze-in from photon \cite{Cox:2024oeb} or meson annihilation, but also depletion from annihilation into these new channels. If the new mediator or the DM are light, the number of effective relativistic degrees of freedom will also change. New observables subject to strong constraints---such as exotic meson decays, annihilation to both gamma rays in the Milky Way and ionizing radiation in the intergalactic medium, scattering with photons, and the effects of self-interactions---also emerge. Light mediators are also bounded by fifth force constraints and stellar cooling limits. 

Plotted in the $m_\chi-\sigma_{\chi N}$ plane, we find that for fixed mediator masses and coupling ratios, the effective interactions with photons and mesons detailed in this work strongly constrain the sub-GeV landscape, largely from overproduction in the Early Universe from processes such as $\gamma \gamma \rightarrow \chi \chi$. This is in keeping with the main conclusions of Refs. \cite{Diamond:2023fsm,Cox:2024rew,Cox:2025toz}. But it is not the whole story: once a few parameters are left free to vary, we will show that it is easy to find large, unexplored swaths of the  $m_\chi-\sigma_{\chi N}$ plane that lead to the observed DM relic abundance $\Omega_\chi h^2 = 0.12$, for all models that we consider here. 

This leads to an exciting prospect: new benchmark regions of thermally-produced, sub-GeV, quark-coupled DM, parts of which can be probed with laboratory experiments.

We begin in Sec. \ref{sec:theory} by describing how we calculated the effective DM-photon interaction cross sections and the dark mediator-induced meson decays.  We present the resulting cross sections and decay rates in Tables \ref{tab:scalar}, \ref{tab:psuedo+scalar}, \ref{tab:pseudoscalar}, \ref{tab:pseudoscalar+scalar}, and \ref{tab:axialvector}.  In Sec. \ref{sec:limts} we lay out the various observables that are impacted by the effective couplings derived in Sec. \ref{sec:theory}, and present the resulting constraints and open parameter space in Sec. \ref{sec:results}.   We conclude in Sec. \ref{sec:conclusion}.

\section{Effective Interactions}
\label{sec:theory}
In this work, we focus on Dirac fermionic DM $\chi$ which interacts with the SM via a scalar or vector mediator that has a scalar, pseudoscalar or axial vector coupling to quarks.  The vector coupling case was explored in Ref. \cite{Diamond:2023fsm}, and at high mediator masses in \cite{Cox:2025toz}, where it was shown to lead to significant loop level DM-electron interactions.   At quark level, the relevant interaction terms are

\begin{equation}
\label{eqn:baseLagrangian}
     \begin{split}
      \mathcal{L} \supset&\text{ }   \phi(g_\chi  \chi\bar{\chi}-ig_{5\chi}\chi\gamma^5\bar{\chi})+\lambda_{\chi}\chi\gamma_{\mu}\gamma_5\bar{\chi}A'^{\mu}
    +\frac{1}{2}m_{\phi}^2\phi^2+\frac{1}{2}m^2_{A'} A'_{\mu}A'^{\mu}
    +\sum_q \left( \lambda_qq\gamma_{\mu}\gamma_5 \bar{q}A'^{\mu} +\phi (-i\tilde{g}_q q\gamma^5\bar{q}+g_q  q \bar{q})\right),  
     \end{split}
\end{equation}
where the sum is over all quark flavors $q$.  We will only focus on the up, down and strange quarks as these are the ones most relevant to interactions with nucleons. While the scalar $\phi$ can have both scalar and pseudoscalar couplings to quarks, we will explore each of these cases individually as their phenomenology is rather different.  We do not include a $\lambda_{\chi}\chi\gamma_{\mu}\bar{\chi}A'^{\mu}$ term here because all effective DM-photon interactions and all meson decays mediated by this vertex and the $\lambda_qq\gamma_{\mu}\gamma_5 \bar{q}A'^{\mu}$ go to zero at lowest loop order.  The case of a vector coupling to quarks through $\lambda_qq\gamma_{\mu} \bar{q}A'^{\mu}$ was explored in Ref.~\cite{Diamond:2023fsm}.

In this work, we will examine the phenomenology of DM which interacts with quarks via several different coupling routes.  We will use the notation ($C_{\chi}\otimes C_q$) to denote which model we are discussing, $C_{\chi}$ reflects the coupling between the $\chi$ and the mediator, with $S$ for a scalar type coupling of the form $\phi \chi\bar{\chi}$, $P$ for a pseudoscalar type coupling of the form $\phi\chi\gamma^5\bar{\chi}$, and $A$ for an axial vector type coupling of the form $\chi\gamma_{\mu}\gamma_5\bar{\chi}A'^{\mu}$.  $C_q$ describes the type of coupling between the mediators and the quarks and follows the same naming convention as $C_{\chi}$. We summarize the models and model labels in Table \ref{tab:models}.
\begin{table}[]
    \centering
    \hspace*{-1 cm}
    \renewcommand{\arraystretch}{1.5}
    \begin{tabular}{|c|c|c|c|c|c|c|}
    
    \hline
       \parbox[t][0.57 cm][c]{1.75cm}{\textbf{Model Label}}  & \parbox[t][0.57 cm][c]{1.75cm}{\textbf{Mediator}} & \parbox[t][0.57 cm][c]{1 cm}{\textbf{Spin}} & \parbox[t][0.65cm][c]{2.5cm}{\textbf{Mediator Coupling to} $\mathbf{\chi}$  }& \parbox[t][1.2cm][t]{2.5cm}{\textbf{Mediator Coupling to Quarks}}&\parbox[t][0.65cm][c]{2.5cm}{\textbf{Nonrelativistic EFT Operator}}&\parbox[t][1.2cm][t]{2cm}{\textbf{Velocity Dependence for} $\mathbf{\sigma_{\chi\bar{\chi}\rightarrow\gamma\gamma}}$}\\
       \hline
         $S\otimes S$& $\phi$&0 & $g_{\chi}\phi\chi\bar{\chi}$&$\sum_qg_{q}\phi q\bar{q}$&$\mathcal O_1\propto \mathbb{1}$&p-wave\\
         \hline
          $P\otimes S$& $\phi$&0 & $-ig_{5\chi}\phi\chi\gamma^5\bar{\chi}$&$\sum_qg_{q}\phi q\bar{q}$&$\mathcal O_{11}\propto i\hat{S}_{\chi}\cdot\frac{\hat{q}}{m_N}$&s-wave\\
         \hline
         $P\otimes P$& $\phi$&0 & $-i g_{5\chi}\phi\chi\gamma^5\bar{\chi}$&$\sum_q-i\tilde{g}_{q}\phi q\gamma^5\bar{q}$&$\mathcal O_6\propto\left(\hat{S}_{\chi}\cdot\frac{\hat{q}}{m_N}\right)\left(\hat{S}_{N}\cdot\frac{\hat{q}}{m_N}\right)$&s-wave\\
         \hline
         $S\otimes P$& $\phi$&0 & $g_{\chi}\phi\chi\bar{\chi}$&$\sum_q-i\tilde{g}_{q}\phi q\gamma^5\bar{q}$&$\mathcal O_{10}\propto i\hat{S}_{N}\cdot\frac{\hat{q}}{m_N}$&p-wave\\
         \hline
         $A\otimes A$& $A_{\mu}'$&1 &$ \lambda_{\chi}\chi\gamma_{\mu}\gamma_5\bar{\chi}A'^{\mu}$&$\sum_q  \lambda_qq\gamma_{\mu}\gamma_5 \bar{q}A'^{\mu}$&$\mathcal O_4\propto \hat{S}_{\chi}\cdot\hat{S}_N$&s-wave\\
         \hline
    \end{tabular}
    \caption{Summary of the dark matter models covered in this paper. The sixth column shows the nonrelativistic effective field theory operator that would correspond with the heavy mediator limit of each model.}
    \label{tab:models}
\end{table}
At energies below the QCD scale, the DM-quark couplings shown in Eq.\,\eqref{eqn:baseLagrangian} manifest as effective couplings between DM and nuclei~\cite{Fan:2010gt,Fitzpatrick:2012ix,Dent:2015zpa} and between DM and light mesons (pions, kaons, and $\eta$ particles).  The DM coupling to light mesons will itself produce new interactions between DM and photons as well as new meson decay channels, both of which can be easier to observe in some circumstances than DM-nucleus scattering.  

The effective DM-meson interactions can be found by introducing the above external mediator terms to the ChEFT Lagrangian, and expanding to leading order.  The relevant terms of the ChEFT Lagrangian are:
\begin{equation}
\label{eqn:ChEFT Lagrangian}
\begin{split}
    \mathcal{L}_{ChEFT}\supset & \frac{F^2}{4}\left(\text{Tr}\left(D_{\mu}U D^{\mu}U^{\dagger}\right) + \text{Tr}\left(\zeta^{\dagger}U+\zeta U^{\dagger}\right)\right)\\
    & +L_4\text{Tr}\left(D_{\mu}U D^{\mu}U^{\dagger}\right)\text{Tr}\left(\zeta^{\dagger}U+\zeta U^{\dagger}\right)+L_5\text{Tr}\left(D_{\mu}U D^{\mu}U^{\dagger}\left(\zeta^{\dagger}U+\zeta U^{\dagger}\right)\right)\\
    &+L_6\text{Tr}\left(U^{\dagger}\zeta+\zeta^{\dagger}U\right)^2+L_7\text{Tr}\left(U^{\dagger}\zeta-\zeta^{\dagger}U\right)^2~,
\end{split}
\end{equation}
where Tr() denotes the trace, and $U= e^{\frac{i}{F}\bf{\pi}}$ contains the mesons:
\begin{equation}
    \pi= 
    \begin{pmatrix}
    \pi^0+\frac{\eta_8}{\sqrt{3}}+\frac{\sqrt{2}}{\sqrt{3}}\eta_0&\sqrt{2}\pi^+ &\sqrt{2}K^+ \\ \sqrt{2}\pi^-&-\pi^0+\frac{\eta_8}{\sqrt{3}}+\frac{\sqrt{2}}{\sqrt{3}}\eta_0 &\sqrt{2}K^0\\
    \sqrt{2}K^-&\sqrt{2}\bar{K}^0 & -\frac{2}{\sqrt{3}}\eta_8+\frac{\sqrt{2}}{\sqrt{3}}\eta_0
    \end{pmatrix}\,.
\end{equation}\,
We note that the $\eta_0$ is not usually included with the octet of mesons presented above, however it will play an important role in interactions involving the pseudoscalar and axial vector couplings to quarks, so we show it here.  For these calculations we treat the $\eta\simeq \eta_8$ and $\eta'\simeq\eta_0$ as the mixing between each pair is small.  $F = 92.4$ MeV \cite{holstein_1990} is the pion decay constant.  Interactions with external scalars and pseudoscalars are captured in the $\zeta$ term where $\zeta=2B_0 (\mathcal{M}+\textbf{g}_q\phi+i\tilde{\textbf{g}}_q\phi)$, $\textbf{g}_q$ = diag($g_u,g_d,g_s$), $\tilde{\textbf{g}}_q$ = diag($\tilde{g}_u,
\tilde{g}_d,\tilde{g}_s$), $B_0 = \frac{m_{\pi^+}^2}{m_u+m_d}\simeq2.8$ GeV where  $m_{\pi^+}$ is the charged pion mass, and $m_q$ represent the quark masses. $\mathcal{M}=$diag$(m_u,m_d,m_s)$ contains the quark masses.  Contributions from the axial coupling of the vector mediator are captured in the derivative terms
\begin{equation}
\begin{split}
    &D_{\mu}U =\partial_{\mu}U-i\textbf{A'}_{\mu} U-iU\textbf{A'}_{\mu}\,,\\
    &D_{\mu}U^{\dagger}=\partial_{\mu}U^{\dagger}+iU^{\dagger}\textbf{A'}_{\mu}^{\dagger} +i\textbf{A'}_{\mu}^{\dagger} U^{\dagger}\,,\\
\end{split}
\end{equation}
where the matrix
$\textbf{A'}_{\mu}=  A'_{\mu}\text{diag}(\lambda_u,\lambda_d,\lambda_s)$, contains the axial couplings to the quarks. The $L_4$ and $L_5$ terms in \eqref{eqn:ChEFT Lagrangian} are only relevant for the axial vector interactions.  The $L_6$, and $L_7$ terms are only relevant for pseudoscalar couplings to quarks.  We take their values from Ref.~\cite{Bijnens_2014}.
\begin{equation}
\label{eq:Ls}
      L_4 = 1.87\times 10^{-3},\quad L_5 = 1.22\times10^{-3},\quad L_6=1.46\times10^{-3},\quad L_7=-0.39\times10^{-3}
\end{equation}

Putting the above equations together and expanding $\mathcal{L}_{\text{ChEFT}}$ to lowest contributing order in $F$ gives the following relevant interaction terms between the DM mediators and the light mesons:
\begin{align}
\label{eq:expandedChEFTL}
    \mathcal{L}_{\text{ChEFT}}\supset &-B_0(g_u+g_d)\phi\pi^+\pi^- -B_0 (g_u+g_s)\phi K^+K^- \nonumber -\frac{B_0
    (g_u+g_d)}{2}\phi\pi^{0}\pi^0\\
    & +m_{\pi^0\phi}^2\phi_b\pi^0_b
    + m_{\eta\phi}^2\phi_b\eta_b +m_{\eta'\phi}^2\phi_b\eta'_b\\
    &-m_{\pi^0A'}A'^{\mu}\partial_{\mu}\pi^0_b
    -m_{\eta A'}A'^{\mu}\partial_{\mu}\eta_b-m_{\eta'A'}A'^{\mu}\partial_{\mu}\eta'_b. \nonumber
\end{align} 
where
\begin{equation}
\label{eq:pseudoaxialmixings}
\begin{split}
m_{\pi^0\phi}^2&=(\tilde{g}_u -\tilde{g}_d)\left(B_0 F+\frac{32B_0^2L_6}{F}\sum_i^{u,s,d}m_i\right)+\frac{32B_0^2L_7}{F}(\tilde{g}_u+\tilde{g}_d+\tilde{g}_s)(m_u-m_d),\\
m_{\eta\phi}^2&=\frac{\sqrt{3}(\tilde{g}_u +\tilde{g}_d-2\tilde{g}_s)}{3}\left(B_0 F+\frac{32B_0^2L_6}{F}\sum_i^{u,s,d}m_i\right)+\frac{32\sqrt{3}B_0^2L_7}{3F}(\tilde{g}_u+\tilde{g}_d+\tilde{g}_s)(m_u+m_d-2m_s),\\
m^2_{\eta'\phi}&=\frac{\sqrt{6}(\tilde{g}_u+\tilde{g}_d+\tilde{g}_s)}{3}\left(B_{0}F+(L_6+L_7)\frac{32B_0^2}{   F}\sum_i^{u,s,d}m_i\right),\\
    m_{\pi^0A'}&= (\lambda_u-\lambda_d)\left(F+\frac{16 B_0 L_4}{F}\sum_i^{u,s,d}m_i\right)+\frac{16 B_0 L_5(\lambda_um_u-\lambda_dm_d)}{F},\\
    m_{\eta A'} &= \frac{\sqrt{3} }{3}\left((\lambda_u+\lambda_d-2\lambda_s)\left(F+\frac{16 B_0 L_4}{F}\sum_i^{u,s,d}m_i\right)+\frac{16 B_0 L_5(\lambda_um_u+\lambda_d m_d-2\lambda_sm_s)}{F}\right),\\
    m_{\eta^{\prime} A'} &= \frac{\sqrt{6} }{3}\left((\lambda_u+\lambda_d+\lambda_s)\left(F+\frac{16 B_0 L_4}{F}\sum_i^{u,s,d}m_i\right)+\frac{16 B_0 L_5(\lambda_um_u+\lambda_dm_d+\lambda_sm_s)}{F}\right).
\end{split}
\end{equation} We note some key features in Eq.~\eqref{eq:expandedChEFTL} that show how the mediators interact with the mesons and impact our calculations below.  The first three terms in Eq.~\eqref{eq:expandedChEFTL} arise from the scalar coupling between the $\phi$ and the quarks.  Of these, the couplings to charged pions and kaons will lead the $\phi$ to interact with photons via loops. The coupling to $\pi^0$ contributes to $\phi$ decay but only negligibly to interactions with photons.  The terms in the second line result from the pseudoscalar coupling between $\phi$ and quarks. They mix the $\phi$ with the neutral mesons. This causes the neutral mesons to pick up a coupling to DM from the $\phi$, and the $\phi$ to pick up a coupling to photons from the neutral mesons (they couple to photons via the chiral anomaly).  This allows the mesons to decay into $\chi\bar{\chi}$ and allows both the mesons and $\phi$ to mediate interactions between the $\chi$ and photons.  The terms in the bottom row capture $A'_{\mu}$ mixing with the light neutral mesons.  These mixing terms will cause the light neutral mesons to  pick up a coupling to the $\chi$ from the $A_{\mu}'$.  As in the pseudoscalar coupling case, this will allow the light neutral mesons to both decay into $\chi\bar{\chi}$ and to mediate $\chi$-photon interactions.  All of the terms marked with a subscript $b$ (``bare'') are for fields which will mix with another field and not ultimately act as the propagating or observed state in calculations.

DM-quark couplings can also cause the $K^+$ to decay to $\pi^+\chi\bar{\chi}$, $\pi^+\phi$, or  $\pi^+A'$.  These decays arise due to terms in the chiral Lagrangian that describe weak interactions between the light mesons.  The relevant terms are
\begin{equation}
    \mathcal{L}\supset \frac{i G_FF^4 } {\sqrt{2}}V_{us}V_{ud}^*\left(g_8\Tr(\lambda (U^{\dagger}D_{\mu}U)(U^{\dagger}D^{\mu}U))+g_8^s(U^{\dagger}D_{\mu}U)_{3,2}Tr(U^{\dagger}D_{\mu}U)\right),
\end{equation}
where $\lambda=(\lambda_6-i\lambda_7)/2$, $\lambda_{6,7}$ are the Gell-Mann matrices, $G_F = 1.17\times 10^{-5}$ GeV$^{-2}$, $V_{us}^* = 0.222$, $V_{ud} = 0.974$, $g_8=3.62$ and $g_8^s=-0.38 g_8$ \cite{Gerard:2005yk}.  The (3,2) subscript denotes the 3,2 term in the matrix.  Expanding this out, the relevant terms for $K^+$ decay are
\begin{equation}
\label{eq:KLagrangian}
    \begin{split}
        \mathcal{L}\supset&\frac{ G_FF^2 } {\sqrt{2}}V_{us}V_{ud}^*\Bigg[-2g_8(\partial_{\mu}\pi^-)(\partial^{\mu}K^+)+\frac{ig_8}{F}\partial_{\mu}K^+(\pi^0\partial^{\mu}\pi^--\pi^-\partial^{\mu}\pi^0)\\
        &+\frac{ig_8\sqrt{3}}{3F}(-3\eta\partial_{\mu}K^+\partial^{\mu}\pi^-+2K^+\partial_{\mu}\eta\partial^{\mu}\pi^-+\pi^-\partial_{\mu}K^+\partial^{\mu}\eta)\\
        &+\frac{i\sqrt{6}}{3F}\partial^{\mu}\eta'_b(\partial_{\mu}K^+\pi^--\partial_{\mu}\pi^-K^+)(2g_8+3g_8^s)\\
        &+2iA'^{\mu}\left(-\partial_{\mu}K^+\pi^-(g_8(\lambda_s+\lambda_u)+g_8^s(\lambda_u+\lambda_s+\lambda_d))+\partial_{\mu}\pi^-K^+(g_8(\lambda_d+\lambda_u)+g_8^s(\lambda_u+\lambda_s+\lambda_d))\right)\Bigg].
    \end{split}
\end{equation}
The first term in the brackets above produces a small kinetic mixing between the $\pi^+$ and the $K^+$.  This combined with the coupling to charged pions and kaons in Eq. \eqref{eq:expandedChEFTL} resulting from a scalar coupling to quarks leads to a small three-point interaction between the $\phi$, $\pi^-$ and $K^+$ that allows $K^+$ to decay into $\pi^+\chi\bar{\chi}$  or $\pi^+\phi$.  The remaining terms in Eq. \eqref{eq:KLagrangian} are for three-point interactions involving the $K^+$, the $\pi^-$ and either the neutral mesons ($\pi^0$, $\eta$ and $\eta'$) or the external vector mediator $A'^{\mu}$.  Mixing between the neutral mesons and the $\phi$, when it has a pseudoscalar coupling to quarks, lets all four particles contribute to $K^+\rightarrow\pi^+\chi\bar{\chi}$, and allows the decay $K^+\rightarrow\pi^+\phi$. The $A'$ terms in the bottom row of Eq. \eqref{eq:KLagrangian} all contribute to $K^+$ decay to $\pi^+ A'$.  The meson mixing with $A'$ means all four particles contribute to the decay rate $K^+\rightarrow\pi^+\chi\bar{\chi}$.  Calculations of these decay rates can be found in detail in Appendix \ref{Ap:Kdecay} and the resulting decay rates can be found in Tables \ref{tab:scalar}, \ref{tab:psuedo+scalar}, \ref{tab:pseudoscalar}, \ref{tab:pseudoscalar+scalar}, and \ref{tab:axialvector}. 

In the sections below, we calculate the cross sections and decay rates that result from these effective interactions.  The results are also summarized in Tables \ref{tab:scalar}, \ref{tab:psuedo+scalar}, \ref{tab:pseudoscalar}, \ref{tab:pseudoscalar+scalar}, and \ref{tab:axialvector}.  In these tables we display the differential cross sections as calculated, and the integrated cross section in two opposite limits, $m_{\phi}^2 \ll |t|$ (an ultralight mediator), and $m_{\phi}^2 \gg |t|$ (a heavy mediator), as is commonly done in direct detection literature. For both the light and heavy mediator limits, we present the scattering cross sections in the non-relativistic limit where $|t| \ll m_{\chi}^2, m_N^2$, as this is the limit most relevant for both scattering experiments and cosmological constraints.  If a limit is not shown then either the full relativistic cross section is needed to obtain limits worked out in this paper, or taking a large or small mediator mass does not simplify the cross section.

For $\chi\gamma\rightarrow\chi\gamma$ scattering, we present the momentum transfer cross section, $\sigma_{MT}$ 
\begin{equation}
    \sigma_{MT} \equiv \int \frac{d\sigma}{d\Omega} (1 - \cos(\theta))d\phi d\cos(\theta),
\end{equation}
as this is what has been constrained by, e.g., observations of Milky Way satellites in Ref.~\cite{Crumrine:2024sdn}. 

One advantage of using the momentum-transfer cross section is that the factor of $(1 - \cos(\theta))$ can keep the momentum-transfer cross section finite in certain cases where the total cross section diverges. 
However, if  $\frac{d\sigma}{d\Omega}\propto\frac{1}{(1-\cos(\theta))^n}$ where $n>1$,  then the momentum transfer cross section also diverges. In these cases, for the sake of comparison with direct detection limits, we follow the approach of scattering experiments and use an effective cross section, $\bar{\sigma}$ instead of the total cross section:
\begin{equation}
    \bar{\sigma} \equiv 4\pi\frac{d\sigma}{d\Omega}(t=-q_0^2).
\end{equation}
We take $q_0=  m_{\chi} v_0$ and $v_0 = 220$ km/s.  Effective cross sections will be marked with $\bar{()}$ in the tables.

Next we turn to the computation of observables in the case of scalar couplings to quarks, then pseudoscalar couplings, then axial vectors.

\subsection{Scalar Quark Couplings - $S\otimes S$ and $P\otimes S$ Cases}

The following sections focus on DM which interacts with quarks via a scalar mediator with a $\sum_qg_q\phi q\bar{q}$ coupling to quarks. All of the scattering, production, and annihilation cross sections as well as the decay rates calculated in this section can be found in the summary Tables \ref{tab:scalar} for the $S\otimes S$ case and \ref{tab:psuedo+scalar}  for the $P\otimes S$ case.  The details of these calculations can be found in the sections below.
\subsubsection{DM-Nucleon Scattering}

 The $\sum_qg_q\phi q\bar{q}$ term in Eq. \eqref{eqn:baseLagrangian}  results in an effective coupling between the $\phi$ and nucleons of the form 
\begin{equation}
    \mathcal{L}\supset h_N \phi N \bar{N},
\end{equation}
where
\begin{equation}
\label{eq:hp} 
    h_N =m_N\left(\frac{g_{u} f_{Nu}}{m_{u}}+\frac{g_{d} f_{Nd}}{m_{d}}+\frac{g_{s} f_{Ns}}{m_{s}}\right)+\frac{2}{27}m_N\left(1-\left(f_{Nu}+f_{Nd}+f_{Ns}\right)\right)\left(\frac{g_{c}}{m_{c}}+\frac{g_{b}}{m_{b}}+\frac{g_{t}}{m_{t}}\right).
\end{equation}
$m_N$ is the nucleon mass, $m_{q={u,d,s,c,b,t}}$ are the quark masses, $g_q$ are the scalar couplings to quarks, and the $f_{Nq}$ represent the matrix elements $\langle N|m_{q} \bar q q|N \rangle$, with numerical values \cite{Dent:2015zpa}:
\begin{equation}
\begin{split}
&f_{pu}=  0.02, \quad f_{pd} =  0.026,\quad f_{ps} = 0.118. \\
&f_{nu}=  0.014, \quad f_{nd} =  0.036,\quad f_{ns} = 0.118.
\end{split}
\end{equation} with (p/n) denoting a (proton/neutron) coupling.

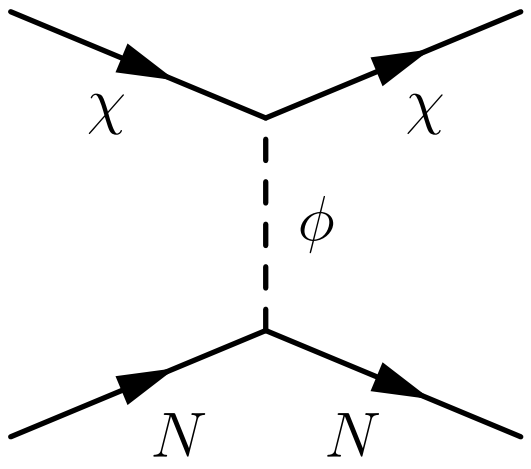
\begin{figure}
    \centering
    \begin{subfigure}{0.3\textwidth}
    \begin{fmffile}{fi-diagram}
 \begin{fmfgraph*}(120,80)
   \fmfleft{i1,i2}
   \fmfright{o1,o2}
   \fmf{fermion,label={\large$N$},label.side=right}{i1,v1}
   \fmf{fermion,label={\large$N$}}{v1,o1}
   \fmf{fermion,label={\large$\chi$}}{i2,v2}
   \fmf{fermion,label={\large$\chi$}}{v2,o2}
   \fmf{dashes,label={\large{$\phi$}}}{v1,v2}
 \end{fmfgraph*}
\end{fmffile}
\caption{\label{fig:bscalar:fi}}
 \end{subfigure}

    \caption{Feynman diagram relevant to $\chi$-nucleon scattering for the $S\otimes S$ and $P\otimes S$ models.}
    \label{fig:bscalar}
\end{figure}

Using Eqs.~\eqref{eqn:baseLagrangian} and \eqref{eq:hp}  we can calculate the tree-level scattering cross section between the $\chi$ and the nucleon as mediated by the $\phi$ for the $S\otimes S$ and $P\otimes S$ models.  The nucleon scattering cross section for both models corresponds to the Feynman diagram in Fig.~\ref{fig:bscalar:fi} and can be found to be 
\begin{equation}
    \left(\frac{d \sigma_{\chi N \rightarrow \chi N}}{d \Omega}\right)^{S\otimes S} = \frac{g_\chi^2 h_N^2}{64 \pi^2} \frac{\left(4m_\chi^2-t\right) \left(4m_N^2-t\right)}{s (m_\phi^2-t)^2},
\end{equation}

\begin{equation}
     \left(\frac{d\sigma_{\chi N\rightarrow\chi N}}{d\Omega}\right)^{P\otimes S}=\frac{-g_{5\chi}^2h_N^2}{64\pi^2}\frac{t(4m_N^2-t)}{s(m_\phi^2-t)^2}.
\end{equation}
The heavy- and light-mediator limits are given in Tables \ref{tab:scalar} and \ref{tab:psuedo+scalar} for the two models. 
\begin{table}[]
   \caption{\textbf{Observable scattering, annihilation and production cross sections and decay rates resulting from DM $\chi$ interacting with quarks via a scalar mediator $\phi$ with $g_{\chi}\phi\chi\bar{\chi}$ and $\sum_qg_q\phi q\bar{q}$ coupling terms}.  $g_{\chi}$, $m_{\chi}$, $m_{\phi}$ and the $g_q$ are free parameters of the theory. $g_{\chi}$ sets the $\chi$ coupling to $\phi$.  $m_{\chi}$ and  $m_{\phi}$ are the $\chi$ and $\phi$ masses respectively.  The differential cross sections are presented in their exact form, calculated to tree level in the nucleon scattering case and the decay case, and calculated at the one-loop level in the photon cases.  $s$ and $t$ represent the Mandelstam variables, with values given in Table~\ref{tab:mandelstam}.  We show the full integrated cross sections in the light-  and heavy-mediator limit where appropriate.  We present the photon annihilation cross section without taking any limits because the overproduction bounds that rely on it require integration over all energies.  $h_N$ is defined in Eq.~\eqref{eq:hp}, $\mathcal{D}$ is defined in Eq.~\eqref{eqn:D-couple}, and $c_{\pi^+K^+\phi}$ is defined in Eq.~\eqref{eqn:gpiKphi}.  These derived constants are all combinations of $B_0$, $F$, $m_{\pi^+}$, $ m_{K^+}$, $m_{\phi}$, $m_{\chi}$ $g_{\chi}$, and the $g_q$'s.  $\mathcal{F}_{S\otimes S}(m_{\chi},m_{\phi})$ is a phase space integral for the three body decay of the $K^+$, defined in Eq.~\eqref{eqn:phasespaces}.  The $\phi$ decay rate shown below only includes terms with decay products with combined masses below $\sim$ 600 MeV, as the decay rate of mediators heavier than this are not relevant to the calculations done in this paper.  Additional decay channels do exist for heavier $\phi$'s.}
    \label{tab:scalar}
\renewcommand{\arraystretch}{1.5}
    \centering
    \hspace*{-1.5 cm} 
    \begin{tabular}{|c|c|c|c|}
    \hline
\multicolumn{4}{|c|}{\textbf{S$\otimes$S}} \\
\hline\rule{0pt}{0pt}
       \parbox[t][0.2cm][c]{2cm}{\textbf{Interaction}} & \parbox[t][0.2cm][c]{3cm}{ \textbf{$\frac{d\sigma}{d\Omega}$ or $\Gamma$ }}& \parbox[t][][t]{3cm}{ $\mathbf{\sigma}$ \textbf{ in Light Mediator Limit}}& \parbox[t][][t]{4cm}{ $\mathbf{\sigma}$ \textbf{ in Heavy Mediator Limit}} \\[12pt] 
        
        \hline
          \rule{0pt}{25pt}$\chi N\rightarrow\chi N $ & $\displaystyle \frac{g_\chi^2 h_N^2}{64 \pi^2} \frac{\left(4m_\chi^2-t\right) \left(4m_N^2-t\right)}{s (m_\phi^2-t)^2
          }$ &$\displaystyle \overline{\frac{g_{\chi}^2h_N^2\mu_{\chi N}^2}{\pi(q_0^2 + m_{\phi}^2)^2}}$
          &$\displaystyle\frac{g_\chi^2 h_N^2}{ \pi}\frac{m_{\chi}^2m_{N}^2}{(m_N+m_{\chi})^2m_{\phi}^4}$
          \\[15pt] \hline 
        \rule{0pt}{25pt} $\chi \gamma\rightarrow\chi \gamma $&$\displaystyle\frac{g_\chi^2 \alpha^2\mathcal{D}}{36864 \pi^4} \frac{t^2 \left(4m_\chi^2 -t\right)}{s (m_\phi^2-t)^2}$&$\displaystyle\left(\frac{g_\chi^2 \alpha^2\mathcal{D}}{2304 \pi^3}\right)_{MT}$&$\displaystyle\left(\frac{g_\chi^2 \alpha^2\mathcal{D}}{288 \pi^3}\frac{E_{\gamma}^4}{ m_{\phi}^4}\right)_{MT}$
        \\[15pt]
         \hline
             \rule{0pt}{20pt}
            $\gamma\gamma\rightarrow\chi\bar{\chi}$&$\displaystyle\frac{g_\chi^2 \alpha^2\mathcal{D}}{36864 \pi^4} \frac{s \left(s-4m_\chi^2\right)}{\left(\left(s-m_\phi^2\right)^2+\left(m_{\phi}\Gamma_{\phi}^{S\otimes S}\right)^2\right)} \sqrt{\frac{s-4m_\chi^2}{s}}$& \multicolumn{2}{|c|}{$\displaystyle\frac{g_\chi^2 \alpha^2 \mathcal{D}}{576 \pi^3}\frac{ E_{\gamma CM} \left( E_{\gamma CM}^2-m_\chi^2\right)^{3/2}}{\left(\left(4E_{\gamma CM}^2-m_\phi^2\right)^2+\left(m_{\phi}\Gamma_{\phi}^{S\otimes S}\right)^2\right)}$ }
             \\[20pt]
         \hline
         
\rule{0pt}{20pt}$\chi\bar{\chi}\rightarrow\gamma\gamma$&$\displaystyle\frac{g_\chi^2\alpha^2 \mathcal{D}}{73728 \pi^4} \frac{s \left(s-4m_\chi^2\right)}{\left(\left(s-m_\phi^2\right)^2+\left(m_{\phi}\Gamma_{\phi}^{S\otimes S}\right)^2\right)} \sqrt{\frac{s}{s-4m_\chi^2}}$&$\displaystyle\frac{g_\chi^2 \alpha^2 \mathcal{D}v_{rel}}{36864 \pi^3} $
&$\displaystyle\frac{g_\chi^2 \alpha^2 \mathcal{D}}{2304 \pi^3} \frac{m_{\chi}^4v_{rel}}{m_\phi^4}$
\\[20pt]
\hline
\rule{0pt}{22pt}$K^+\rightarrow\pi^+\phi$&$\displaystyle \frac{c_{\pi^+ K^+\phi}^2\sqrt{(m_{K^+}^2+m_{\pi^+}^2-m_{\phi}^2)^2-4m_{K^+}^2m_{\pi^+}^2}}{16\pi m_{K^+}^3}$&\multicolumn{2}{|c|}{}\\[7pt]\hline
\rule{0pt}{18pt}$K^+\rightarrow\pi^+\chi\bar{\chi}$&$\displaystyle\frac{g_{\chi}^2c_{\pi^+ K^+\phi}^2}{32\pi^3m_{K^+}}\mathcal{F}_{S\otimes S}(m_{\chi},m_{\phi})$&\multicolumn{2}{|c|}{}\\[7pt]\hline
\rule{0pt}{23pt}$\phi\rightarrow$All&\multicolumn{3}{|c|}{$\displaystyle  \left(\frac{\mathcal{D}\alpha^2m_{\phi}^3}{9216\pi^3}\right)_{\gamma\gamma}+\left(\frac{g_{\chi}^2(m_{\phi}^2-4m_{\chi}^2)^{3/2}}{8\pi m_{\phi}^2}\right)_{\chi\bar{\chi}}
+\left(\frac{B
_0^2(g_u+g_d)^2\sqrt{m_{\phi}^2-4m_{\pi^+}^2}}{16\pi m_{\phi}^2}\right)_{\pi^+\pi^-}+\left(\frac{B
_0^2(g_u+g_d)^2\sqrt{m_{\phi}^2-4m_{\pi^0}^2}}{32\pi m_{\phi}^2}\right)_{\pi^0\pi^0}$}\\[15pt]\hline
    \end{tabular}
\end{table}

\begin{table}[h]
    \caption{\textbf{As in Table \ref{tab:scalar}, but for DM interacting with quarks via a scalar mediator with $-ig_{5\chi}\phi\chi\gamma^5\bar{\chi}$ and $\sum_qg_q\phi q\bar{q}$ coupling terms}. 
    $\mathcal{F}_{P\otimes S}(m_{\chi},m_{\phi})$ is defined in Eq.~\eqref{eqn:phasespacePS}.}
    \label{tab:psuedo+scalar}
    \centering
     \hspace*{-1.5 cm}
    \renewcommand{\arraystretch}{1.5}
    \begin{tabular}{|c|c|c|c|}
             \hline
\multicolumn{4}{|c|}{\textbf{P$\otimes$S}} \\
\hline\rule{0pt}{0pt}
       \parbox[t][0.2cm][c]{2cm}{\textbf{Interaction}} & \parbox[t][0.2cm][c]{3cm}{ \textbf{$\frac{d\sigma}{d\Omega}$ or $\Gamma$ }}& \parbox[t][][t]{3cm}{ $\mathbf{\sigma}$ \textbf{ in Light Mediator Limit}}& \parbox[t][][t]{4cm}{ $\mathbf{\sigma}$ \textbf{ in Heavy Mediator Limit}} \\[12pt] 
        
        \hline
         \rule{0pt}{23pt}$\chi N\rightarrow\chi N $&$\displaystyle\frac{-g_{5\chi}^2h_N^2}{64\pi^2}\frac{t(4m_N^2-t)}{s\left. \left(m_\phi^2-t\right)^2\right.}
         $&$\displaystyle\frac{g_{5\chi}^{2}h_N^2}{8\pi}\frac{1}{m_{\chi}^2v_{rel}^2}$&$\displaystyle\frac{g_{5\chi}^2h_N^2}{2\pi}\frac{m_{\chi}^2m_N^4v_{rel}^2}{(m_{\chi}+m_N)^4m_{\phi}^4}$\\[17pt]
         \hline
         
         \rule{0pt}{23pt}$\chi \gamma\rightarrow\chi \gamma $&$\displaystyle\frac{g_{5\chi}^2 \alpha^2\mathcal{D}}{36864 \pi^4} \frac{-t^3}{s \left(m_\phi^2-t\right)^2}$&$\displaystyle\left( \frac{g_{5\chi}^2 \alpha^2\mathcal{D}}{3456 \pi^3} \frac{E_{\gamma}^2}{m_{\chi}^2}\right)_{MT}$&$\displaystyle\left(\frac{g_{5\chi}^2 \alpha^2\mathcal{D}}{360 \pi^3} \frac{ E_{\gamma}^6}{m_{\chi}^2 m_\phi^4}\right)_{MT}$\\[17pt]
         \hline
         \rule{0pt}{23pt}$\gamma\gamma\rightarrow\chi\bar{\chi}$&$\displaystyle\frac{g_{5\chi}^2 \alpha^2\mathcal{D}}{36864 \pi^4} \frac{s^2}{ \left(\left(s-m_\phi^2\right)^2+\left(m_{\phi}\Gamma_{\phi}^{P\otimes S}\right)^2\right)}\sqrt{\frac{s-4m_\chi^2}{s}}$&\multicolumn{2}{|c|}{$\displaystyle\frac{g_{5\chi}^2 \alpha^2\mathcal{D}}{576 \pi^3} \frac{E_{\gamma CM}^3\sqrt{E_{\gamma CM}^2-m_{\chi}^2}}{ \left(\left(4E_{\gamma CM}^2-m_\phi^2\right)^2+\left(m_{\phi}\Gamma_{\phi}^{P\otimes S}\right)^2\right)}$}\\[20pt]
         \hline
         
\rule{0pt}{20pt}$\chi\bar{\chi}\rightarrow\gamma\gamma$&$\displaystyle\frac{g_{5\chi}^2 \alpha^2\mathcal{D}}{73728 \pi^4} \frac{s^2}{ \left(\left(s-m_\phi^2\right)^2+\left(m_{\phi}\Gamma_{\phi}^{P\otimes S}\right)^2\right)}\sqrt{\frac{s}{s-4m_\chi^2}}$&$\displaystyle\frac{g_{5\chi}^2 \alpha^2\mathcal{D}}{2304 \pi^3} \frac{1}{ v_{rel}}$&$\displaystyle\frac{g_{5\chi}^2 \alpha^2\mathcal{D}}{576 \pi^3} \frac{m_{\chi}^4}{ m_\phi^4v_{rel}}$\\[20pt]
\hline

\rule{0pt}{22pt}$K^+\rightarrow\pi^+\phi$&$\displaystyle \frac{c_{\pi^+ K^+\phi}^2\sqrt{(m_{K^+}^2+m_{\pi^+}^2-m_{\phi}^2)^2-4m_{K^+}^2m_{\pi^+}^2}}{16\pi m_{K^+}^3}$&\multicolumn{2}{|c|}{}\\[7pt]\hline
\rule{0pt}{20pt}

$K^+\rightarrow\pi^+\chi\bar{\chi}$&$\displaystyle\frac{g_{5\chi}^2c_{\pi^+ K^+\phi}^2}{32\pi^3m_{K^+}}\mathcal{F}_{P\otimes S}(m_{\chi},m_{\phi})$&\multicolumn{2}{|c|}{}\\[10pt]\hline
\rule{0pt}
{25pt}$\phi\rightarrow$All&\multicolumn{3}{|c|}{$\displaystyle\left(\frac{\mathcal{D}\alpha^2m_{\phi}^3}{9216 \pi^3}\right)_{\gamma\gamma}+\left(\frac{g_{5\chi}^2(m_{\phi}^2-4m_{\chi}^2)^{1/2}}{8\pi}\right)_{\chi\bar{\chi}}+
\left(\frac{B
_0^2(g_u+g_d)^2\sqrt{m_{\phi}^2-4m_{\pi^+}^2}}{16\pi m_{\phi}^2}\right)_{\pi^+\pi^-}+\left(\frac{B
_0^2(g_u+g_d)^2\sqrt{m_{\phi}^2-4m_{\pi^0}^2}}{32\pi m_{\phi}^2}\right)_{\pi^0\pi^0}$}\\[15pt]

\hline
    \end{tabular}
\end{table}

\subsubsection{DM-Photon Interactions}
The same quark couplings that lead the $\phi$ to scatter with nucleons also couple the $\phi$ to charged pions and kaons. These couplings can be read off from Eq.~\eqref{eq:expandedChEFTL}
\begin{align}
    \label{eq:scalarcheft}\mathcal{L}_{\text{ChEFT}}\supset-B_0(g_{u}+g_{d})\phi\pi^+\pi^--B_0 (g_{u}+g_{s})\phi K^+K^-~.
\end{align}

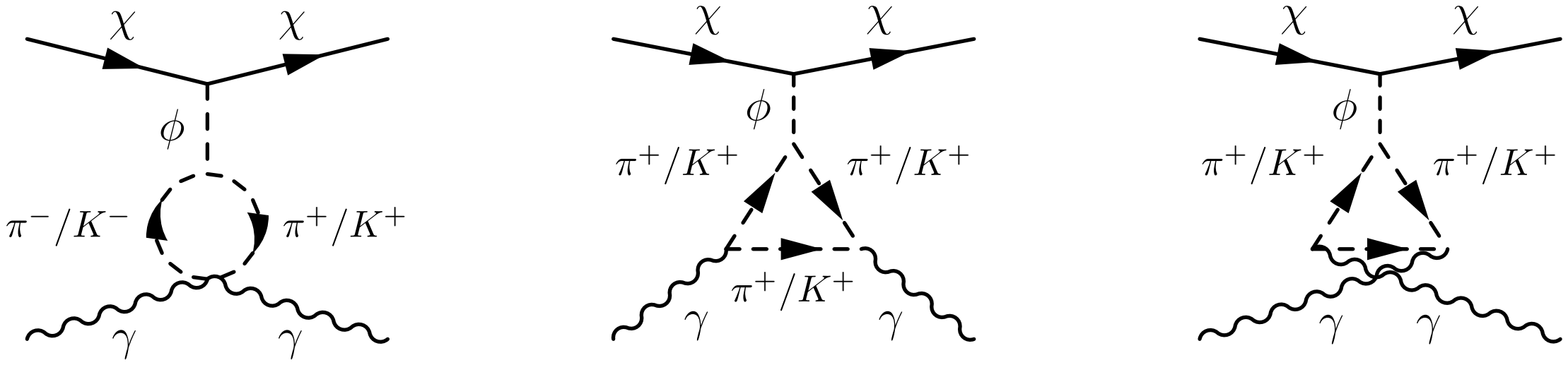
\begin{figure}[h]
    \centering
    \begin{subfigure}{0.3\textwidth}
    \begin{fmffile}{firstloop}
 \begin{fmfgraph*}(120,80)
   \fmfleft{i1,i2}
   \fmfright{o1,o2}
   \fmfforce{(0.5w,0.55h)}{v2}
   \fmfforce{(0.5w,0.2h)}{v3}
   \fmf{fermion,label={\large$\chi$},label.side=left}{i2,v1}
   \fmf{fermion,label={\large$\chi$},label.side=left}{v1,o2}
   \fmf{dashes,label={\large$\phi$}}{v1,v2}
   \fmf{scalar,left,label={\normalsize$\pi^+/K^+$}}{v2,v3}
   \fmf{scalar,label={\normalsize$\pi^-/K^-$},left}{v3,v2}
   \fmf{photon,label={\large{$\gamma$}},label.side=right}{i1,v3}
\fmf{photon,label={\large{$\gamma$}},label.side=left}{o1,v3}
 \end{fmfgraph*}
\end{fmffile}
 \end{subfigure}
 \begin{subfigure}{0.3\textwidth}
    \begin{fmffile}{secondloop}
 \begin{fmfgraph*}(120,80)
   \fmfleft{i1,i2}
   \fmfright{o1,o2}
   \fmfforce{(0.5w,0.65h)}{v2}
   \fmfforce{(0.35w,0.3h)}{v3}
   \fmfforce{(0.65w,0.3h)}{v4}
   \fmf{fermion,label={\large$\chi$},label.side=left}{i2,v1}
   \fmf{fermion,label={\large$\chi$},label.side=left}{v1,o2}
   \fmf{dashes,label={\large$\phi$}}{v1,v2}
   \fmf{scalar,label={\normalsize$\pi^+/K^+$},label.side=left}{v3,v2}
   \fmf{scalar,label={\normalsize$\pi^+/K^+$},label.side=right}{v3,v4}
   \fmf{scalar,label={\normalsize$\pi^+/K^+$},label.side=left}{v2,v4}
   \fmf{photon,label={\large{$\gamma$}},label.side=right}{i1,v3}
\fmf{photon,label={\large{$\gamma$}},label.side=left}{o1,v4}
 \end{fmfgraph*}
\end{fmffile}
 \end{subfigure}
  \begin{subfigure}{0.3\textwidth}
    \begin{fmffile}{thirdloop}
 \begin{fmfgraph*}(120,80)
   \fmfleft{i1,i2}
   \fmfright{o1,o2}
   \fmfforce{(0.5w,0.65h)}{v2}
   \fmfforce{(0.35w,0.3h)}{v3}
   \fmfforce{(0.65w,0.3h)}{v4}
   \fmf{fermion,label={\large$\chi$},label.side=left}{i2,v1}
   \fmf{fermion,label={\large$\chi$},label.side=left}{v1,o2}
   \fmf{dashes,label={\large$\phi$}}{v1,v2}
   \fmf{scalar,label={\normalsize$\pi^+/K^+$},label.side=left}{v3,v2}
   \fmf{scalar}{v3,v4}
   \fmf{scalar,label={\normalsize$\pi^+/K^+$},label.side=left}{v2,v4}
   \fmf{photon,label={\large{$\gamma$}},label.side=right}{i1,v4}
\fmf{photon,label={\large{$\gamma$}},label.side=left}{o1,v3}
 \end{fmfgraph*}
\end{fmffile}
 \end{subfigure}
    \caption{Feynman diagrams relevant for $\chi$-photon scattering for the $S\otimes S$ and $P\otimes S$ models and involving a meson loop.}
    \label{feynmanscalar}
\end{figure}

Through the diagrams shown in Fig. \ref{feynmanscalar}, the $\chi$ can scatter with photons through charged meson loops. 

We calculate the combined amplitude from the three meson loops shown in Figure \ref{feynmanscalar} to obtain an effective coupling for a $\phi\gamma\gamma$ vertex term:
\begin{equation}
     M^{\mu \nu}_{\phi \gamma \gamma}(q_1,q_2) =\frac{-i \alpha}{12 \pi}B_0 \left( \frac{g_u+g_d}{m_{\pi^+}^2}+\frac{g_u+g_s}{m_{K^+}^2}\right)\times\left[ \frac{t g^{\mu\nu}}{2}+  q_2^\mu q_1^\nu \right] \, ,
    \label{eqn:effective-scalar-photon}
\end{equation}
where $\alpha\equiv\frac{e^2}{4\pi}$, and $q_1$ and $q_2$ are the respective momenta of the incoming and outgoing photons.  With this, we can calculate the overall $\chi$-photon scattering amplitude.
\begin{equation}
    \mathcal{M}_{\chi \gamma \rightarrow \chi \gamma} =  \frac{i\bar{\chi}(p_3) (i g_\chi) \chi(p_1)}{t - m_\phi^2} M^{\mu \nu}_{\phi \gamma \gamma}(p_2,p_4) {\epsilon_\nu^4}^* \epsilon_\mu^2 \, ,
\end{equation}
here $p_1$ and $p_3$ refer to the momenta of the incoming and outgoing $\chi$ respectively, and $p_2$ and $p_4$ refer to the momenta of the incoming and outgoing photon.
This gives the differential scattering cross sections 
\begin{equation}
\begin{split}
    \left(\frac{d \sigma_{\chi \gamma \rightarrow \chi \gamma}}{d \Omega}\right)^{S\otimes S}&= \frac{g_\chi^2 \alpha^2\mathcal{D}}{36864 \pi^4} \frac{t^2 \left(4m_\chi^2 -t\right)}{s \left(\left(m_\phi^2-t\right)^2+\left(m_{\phi}\Gamma_{\phi}^{S\otimes S}\right)^2\right)},\\
    \left(\frac{d \sigma_{\chi \gamma \rightarrow \chi \gamma}}{d \Omega}\right)^{P\otimes S}&=\frac{g_{5\chi}^2 \alpha^2\mathcal{D}}{36864 \pi^4} \frac{-t^3}{s \left(\left(m_\phi^2-t\right)^2+\left(m_{\phi}\Gamma_{\phi}^{P\otimes S}\right)^2\right)},
\end{split}
\end{equation}

where 
\begin{equation}
    \mathcal{D} = {\left(\frac{B_0 \left(g_u+g_d\right)}{m_{\pi^+}^2} + \frac{B_0 \left(g_u+g_s\right)}{m_{K^+}^2}\right)}^2 \, .
    \label{eqn:D-couple}
\end{equation}

The annihilation and production cross sections as well as the $K^+$ decay rate into $\pi^+\chi\bar{\chi}$ can be found in Table \ref{tab:scalar} for the $S\otimes S$ case and Table \ref{tab:psuedo+scalar} for the $P\otimes S$ case.\\

 Many of the cross sections depend on the $\phi$ width.  The $\phi$ decay rate is found using the interaction terms in Eq.~\eqref{eqn:baseLagrangian} and Eq.~\eqref{eq:scalarcheft} and shown in Tables \ref{tab:scalar} and \ref{tab:psuedo+scalar}.  The $()_{ii}$ are used to indicate which terms contribute to the decay rate to a specific final state.
\begin{table}[h]
    \centering
    \renewcommand{\arraystretch}{1.3}
    \begin{tabular}{|c|c|}
    \hline
    \textbf{Process} & \textbf{Kinematic variable} \\
    \hline
    \multirow{2}{*}{$\chi N \rightarrow \chi N$}
        & $s = (m_{\chi} + m_N)^2$ \\
        & $t = -2\mu_{\chi N}^2 v_{\rm rel}^2 (1 - \cos\theta)$ \\
    \hline
    \multirow{2}{*}{$\chi \gamma \rightarrow \chi \gamma$}
        & $s = m_{\chi}^2$ \\
        & $t = 2E_{\gamma}^2 (\cos\theta - 1)$ \\
    \hline
    \multirow{2}{*}{$\chi \chi \rightarrow \gamma \gamma$}
        & $s = 4m_{\chi}^2 + m_{\chi}^2 v_{\rm rel}^2$ \\
        & $t$: not used \\
    \hline
    \multirow{2}{*}{$\gamma \gamma \rightarrow \chi \chi$}
        & $s = 4E_{\gamma,\rm CM}^2$ \\
        & $t$: not used \\
    \hline
    \end{tabular}
    \caption{Mandelstam $s$ and $t$ for the scattering and annihilation processes considered. We assume $\chi$--$N$ scattering and $\chi \chi$ annihilation are non-relativistic. For DM--photon scattering we assume $E_{\gamma} \ll m_{\chi}$. The angles $\theta$ are evaluated in the center of momentum frame. }
    \label{tab:mandelstam}
\end{table}

\subsection{Pseudoscalar Quark Couplings - $P\otimes P$ and $S\otimes P$ Cases}
 Next, we consider DM which interacts with quarks via a scalar mediator $\phi$, which couples to quarks via $\sum_q-i\tilde{g}_{q}\phi q\gamma^5\bar{q}$. Here, we compute the DM-nucleon cross section, the DM-photon scattering cross section, annihilation cross sections for DM to photons and vice-versa, and various meson decay rates.   The cross sections and decay rates can be found in the summary Tables \ref{tab:pseudoscalar} for the $P\otimes P$ case and \ref{tab:pseudoscalar+scalar}  for the $S\otimes P$ case.  The details relevant to these calculations can be found in the sections below
\subsubsection{DM-Nucleon Scattering}
A fermion which interacts with quarks via a scalar mediator with a coupling of the form  $\sum_q-i\tilde{g}_{q}\phi q\gamma^5\bar{q}$ will consequently have couplings with nucleons of the form 
\begin{equation}
        \mathcal{L}\supset -i\tilde{h}_N \phi N\gamma^5\bar{N},
    \label{eqn:pseudoscalar-nulceons}
\end{equation}

where \cite{Arina:2014yna} \begin{equation}
\label{eqn:pshp}
    \tilde{h}_{N} = m_N\times\sum_{i=u\,d\,s} \frac{\Delta N_{i}}{m_{i}}\left(\tilde{g}_{i}-\left(\frac{1}{m_{u}}+\frac{1}{m_{d}}+\frac{1}{m_{s}}\right)^{-1}\sum_{j = u \ldots t} \frac{\tilde{g}_{j}}{m_{j}} \right).
\end{equation}
Here $\Delta N_{i}$ represents the quark spin component of the proton $(N=p)$ or neutron $(N=n)$\cite{Liang:2018pis}: 

\begin{equation}
        \Delta p_{u}=\Delta n_{d}= 0.847 , \quad
         \Delta p_{d}= \Delta n_{u}= -0.407, \quad
         \Delta p_{s}=\Delta n_{s} = -0.035 ~.
         \end{equation}
The nucleon scattering cross sections that result from this coupling can be calculated using the Feynman diagram in Fig.~\ref{fig:pseudoscatter:first}. The cross sections both in their full form, as well as in the heavy and light mediator limits, are shown in Table \ref{tab:pseudoscalar} for the $P\otimes P$ case and Table \ref{tab:pseudoscalar+scalar} for the $S\otimes P$ case. 

\subsubsection{DM-Photon Interactions and Meson Decays}
\label{sec:pseudo}
\begin{figure}
    \centering
    \begin{subfigure}{0.3\textwidth}
    \begin{fmffile}{first-diagram}
 \begin{fmfgraph*}(120,80)
   \fmfleft{i1,i2}
   \fmfright{o1,o2}
   \fmf{fermion,label={\large$N$},label.side=right}{i1,v1}
   \fmf{fermion,label={\large$N$}}{v1,o1}
   \fmf{fermion,label={\large$\chi$}}{i2,v2}
   \fmf{fermion,label={\large$\chi$}}{v2,o2}
   \fmf{dashes,label={\large{$\phi$}}}{v1,v2}
 \end{fmfgraph*}

\end{fmffile}
\caption{\label{fig:pseudoscatter:first}}
 \end{subfigure}
    \begin{subfigure}{0.3\textwidth}
    \begin{fmffile}{second-diagram}
 \begin{fmfgraph*}(120,80)
   \fmfleft{i1,i2}
   \fmfright{o1,o2}
   \fmf{photon,label={\large$\gamma$},label.side=right}{i1,v1}
   \fmf{photon,label={\large$\gamma$}}{v1,o1}
   \fmf{fermion,label={\large$\chi$}}{i2,v2}
   \fmf{fermion,label={\large$\chi$}}{v2,o2}
   \fmf{dashes,label={\large{$\phi/\pi_0/\eta/\eta'$}}}{v1,v2}
 \end{fmfgraph*}

\end{fmffile}
\caption{\label{fig:pseudoscatter:second}}
 \end{subfigure}
\begin{subfigure}{0.3\textwidth}
\begin{fmffile}{third-diagram}
 \begin{fmfgraph*}(120,80)
   \fmfleft{i1}
   \fmfright{o1,o2}
   \fmf{fermion,label={\large$\bar{\chi}$}}{o2,v1}
   \fmf{fermion,label={\large$\chi$},label.side=right}{v1,o1}
   \fmf{scalar,label={\large{$\pi_0/\eta/\eta'$}}}{i1,v1}
 \end{fmfgraph*}
\end{fmffile}
    
\caption{\label{fig:pseudoscatter:third}}
\end{subfigure}
    \caption{Feynman diagrams relevant to $\chi$-nucleon scattering (a), $\chi$-photon scattering (b) and meson decay to $\chi\bar{\chi}$ (c) for the $P\otimes P$ and $S\otimes P$ models}
    \label{fig:pseudoscatter}
\end{figure}
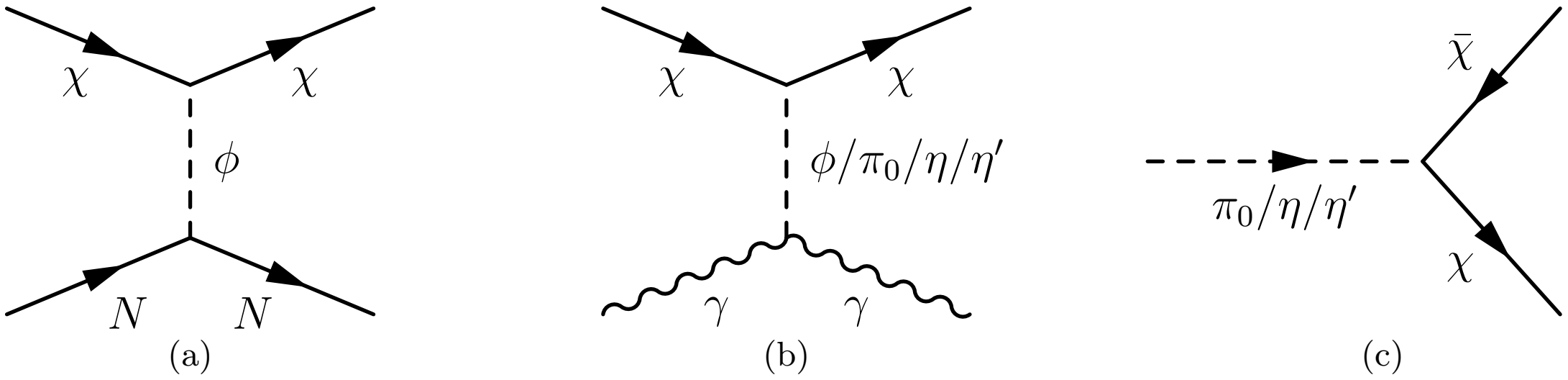
The $\phi$, $\pi^0$, $\eta$, and $\eta'$ coupling terms in Eq.~\eqref{eq:expandedChEFTL}, and shown below, mix the $\phi$ with the light neutral mesons ($\pi^0$, $\eta$, and $\eta'$).  This mixing gives the mesons an effective coupling to $\chi\bar{\chi}$, and gives $\phi$ an effective coupling to photons and several charged mesons, allowing 
all four particles to mediate $K^+$ decay to $\pi^+ \chi\bar{\chi}$ and interactions between $\chi$'s and photons.  These mixings also produce $\pi^0$, $\eta$, and $\eta'$ decays into $\chi\bar{\chi}$ if $m_{\chi}$ is lighter than half the meson mass, and $K^+$ decays to $\pi^+\phi$ when $m_{\phi}<m_{K^+}-m_{\pi^+}$.   In this section we will focus on showing how to determine the effective couplings, as determining the cross sections and decay rates using the Feynman diagrams in Fig \ref{fig:pseudoscatter} follows simply once the rotated couplings are known.  The relevant mixing terms that produce all of these interactions are shown below.

\begin{equation}
\label{mixingap}
    \mathcal{L}\supset m_{\pi^0\phi}^2\phi_{b}\pi^0_b+ m_{\eta \phi}^2\phi_{b}\eta_b+m^2_{\eta'\phi}\phi\eta'_b\,.
\end{equation}
$m_{\phi\pi^0}^2$, $m_{\phi\eta}^2$ and $m_{\phi\eta'}^2$ are defined in Eq.~\eqref{eq:pseudoaxialmixings}.  The effective coupling terms that result from the mixings in Eq.~\eqref{mixingap} are shown in  Eq.~\eqref{expandpseudo}.
The couplings in Eq.~\eqref{mixingap}  complicate the cross section and decay calculations somewhat.  The above mixing terms mean that the $\phi_b$, $\pi^0_b$, $\eta_b$, and $\eta'_b$ do not have well-defined masses in this basis.  In order to compute observables, we rotate into the propagating basis, which removes these mixing terms, giving the new mixed particles well-defined masses.  The $\pi^0$, $\eta$, and $\eta'$ are well-measured particles with experimentally established masses and couplings to the photon.  As such, we treat the terms appearing in Eq. \eqref{eq:expandedChEFTL} and marked with a $b$ as ``bare'' versions of the $\phi$, $\pi^0$, $\eta$, and $\eta'$. The propagating states are linear combinations of these bare particles. Their masses and couplings must coincide with the measured properties of the  $\pi^0$, $\eta$ and $\eta'$.  Thus, the masses of the unmixed mesons are allowed to float freely, and are determined using the observed meson masses, $m_{\phi}$ and the mixing terms shown in  Eq.~\eqref{mixingap}.  We limit ourselves to choices of  $m_{\phi}$ and $\tilde{g}_q$ that produce positive masses for the unmixed mesons and $\phi$.  Negative values for any of these squared masses indicate a theory that is unbounded from below, which should be avoided.

\begin{figure}
    \centering
     \begin{subfigure}{0.47\textwidth}
\includegraphics[width=\linewidth]{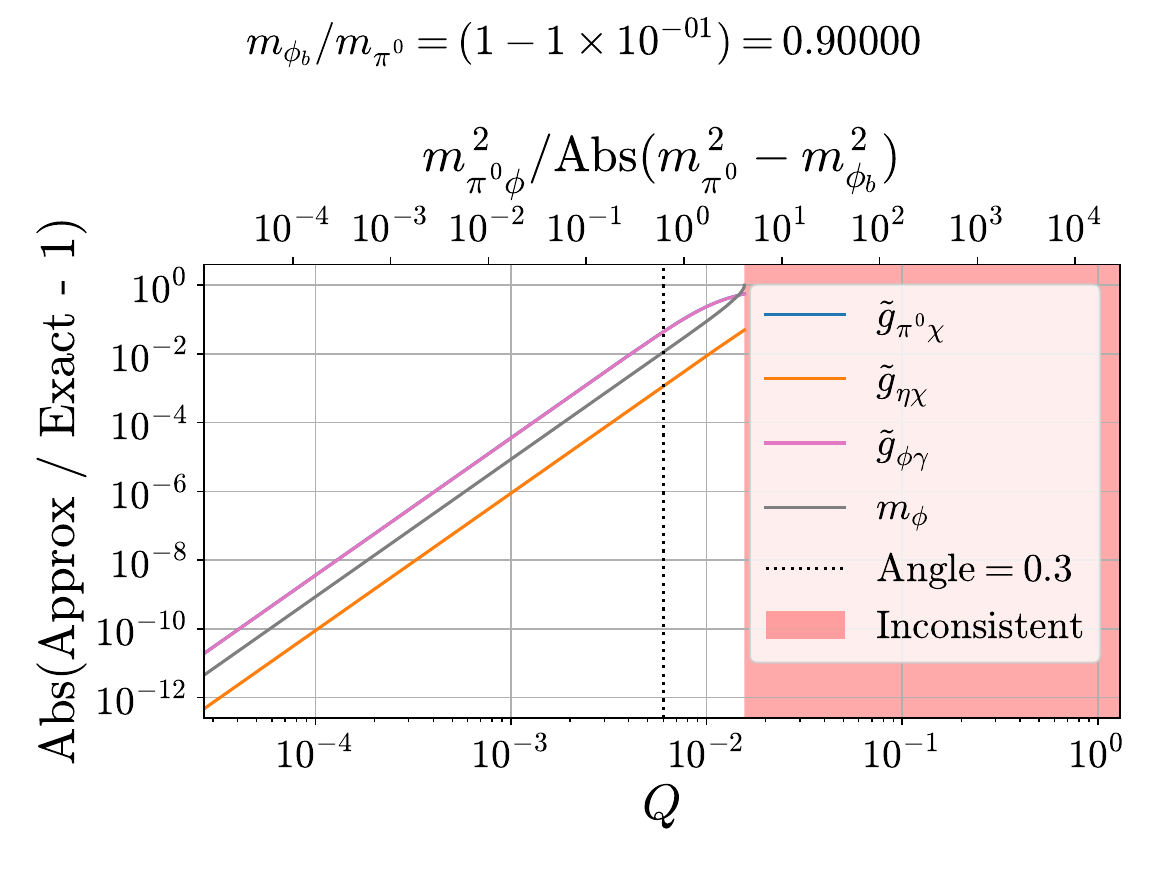}
   
    \end{subfigure}
    \begin{subfigure}{0.47\textwidth}
    \includegraphics[width=\linewidth]{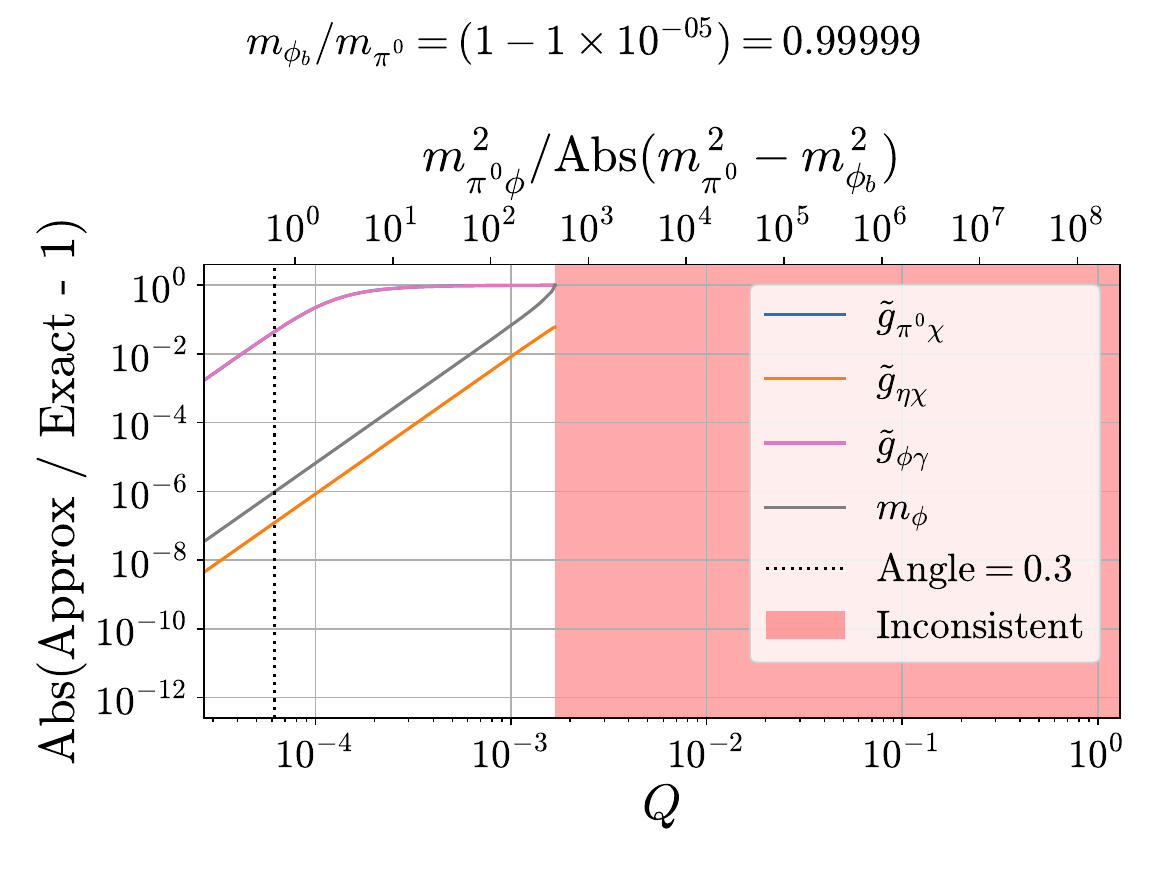}
   
    \end{subfigure}
   \caption{Plots showing the fractional error between the values found using the small angle approximation and the exact calculated values of the meson-$\chi$ couplings,  $m_\phi$, and $\tilde{g}_{\phi\gamma}$ for the $P\otimes P$ and $S\otimes P$ models.  As an example we take the quark couplings to be proportional to their charge, setting  $\tilde{g}_u=2Q$, $\tilde{g}_d=-Q$ and $\tilde{s}_d=-Q$.  The red highlighted region shows where the large mixing causes the DM model to become unstable.  The vertical dotted line marks where the largest rotation angle $=0.3$. The small angle approximation breaks down to the right of this line.  The left plot shows these ratios when $m_{\phi_b} = 0.9 m_{\pi^0}$, while the right plot shows these ratios near a mass resonance, when $m_{\phi_b} = (1-10^{-5}) m_{\pi^0}$. $m_{\pi^0\phi }^2/(m_\phi^2-m_{\pi^0}^2)$ approximates the mixing angle between the $\pi^0$ and the $\phi$ in the small mixing approximation.}
   \label{fig:pseudoapprox}
\end{figure}

In the rotated/propagating basis, the following terms mediate interactions between the $\chi$ and photons.
\begin{equation}
\label{expandpseudo}
\begin{split}
     \mathcal{L}\supset&\phi(-i\tilde{g}_{5\phi\chi}\chi\gamma^5\bar{\chi}+\tilde{g}_{\phi\chi}\chi\bar{\chi})- \frac{1}{2}m_{\phi}^2\phi^2-\frac{\tilde{g}_{\phi\gamma}}{4}F^{\mu\nu}\tilde{F}_{\mu\nu}\phi \\
     &+\pi^0(-i\tilde{g}_{5\pi\chi}\chi\gamma^5\bar{\chi}+\tilde{g}_{\pi^0\chi}\chi\bar{\chi})- \frac{1}{2}m_{\pi^0}^2{\pi^0}^2-\frac{\tilde{g}_{\pi^0\gamma}}{4}F^{\mu\nu}\tilde{F}_{\mu\nu}\pi^0 \\
     &+\eta(-i\tilde{g}_{5\eta\chi}\chi\gamma^5\bar{\chi}+\tilde{g}_{\eta\chi}\chi\bar{\chi})-\frac{1}{2}m_{\eta}^2{\eta}^2-\frac{\tilde{g}_{\eta\gamma}}{4}F^{\mu\nu}\tilde{F}_{\mu\nu}\eta \\
     &+\eta'(-i\tilde{g}_{5\eta'\chi}\chi\gamma^5\bar{\chi}+\tilde{g}_{\eta'\chi}\chi\bar{\chi})- \frac{1}{2}m_{\eta'}^2{\eta'}^2-\frac{\tilde{g}_{\eta'\gamma}}{4}F^{\mu\nu}\tilde{F}_{\mu\nu}\eta' \\    
\end{split}
\end{equation}
where $F^{\mu\nu} = \partial^{\mu}A^{\nu}-\partial^{\nu}A^{\mu}$, $\tilde{F}_{\mu\nu} = \frac{1}{2}\epsilon_{\mu\nu\sigma\rho}{F}^{\sigma \rho}$, $m_{\pi^0}=135$ MeV, $m_{\eta}=548$ MeV, and $m_{\eta'}=958$ MeV.  In general, $m_{\phi_b}\neq m_{\phi}$.  $m_{\phi}$ is the mass of the $\phi$, the particle corresponding to the $\phi_b$ in the propagating basis, and must be calculated numerically using rotation matrices as described in Appendix~\ref{ap:pseudoscalar}. $\tilde{g}_{\pi^0\gamma}$, $\tilde{g}_{\eta\gamma}$, and $\tilde{g}_{\eta'\gamma}$ (shown below) are experimentally measured.  
\begin{equation}
\label{eqn:pseudocouplings}
   \begin{split}
 \tilde{g}_{\pi^0\gamma}&=\frac{e^2}{4\pi^2 F} \\
 \tilde{g}_{\eta\gamma}&=\frac{e^2}{4\pi^2 F_{\eta}}\\
 F_{\eta}&=\sqrt{\frac{\alpha^2 m_{\eta}^3}{64\pi^3\Gamma_{\eta\rightarrow\gamma\gamma}}}\\
 \tilde{g}_{\eta'\gamma}&=\frac{e^2}{4\pi^2 F_{\eta'}}\\
 F_{\eta'}& = \sqrt{\frac{\alpha^2 m_{\eta'}^3}{64\pi^3\Gamma_{\eta'\rightarrow\gamma\gamma}}}\\
\end{split} 
\end{equation}
$\tilde{g}_{\pi^0\chi}$, $\tilde{g}_{\eta\chi}$, $\tilde{g}_{\eta'\chi}$, $\tilde{g}_{\phi\chi}$,  $\tilde{g}_{5\pi^0\chi}$, $\tilde{g}_{5\eta\chi}$, $\tilde{g}_{5\eta'\chi}$, $\tilde{g}_{5\phi\chi}$ and $\tilde{g}_{\phi\gamma}$ emerge as a result of the field rotation.  They depend on both the bare couplings and the rotation matrix applied to the fields to remove the mass mixing terms.  The rotation matrix must be calculated numerically to determine the exact values of the above couplings, as detailed in Appendix \ref{ap:pseudoscalar}.\footnote{A Python script to calculate the rotation matrix, all of the mixing angles within it, and the rotated basis coupling terms can be found at \url{https://github.com/neal-ak/mixing-rotation} with animations sweeping through varying mediator masses. The code plus the data used to create Figs.~\ref{fig:pseudoapprox} and~\ref{fig:axialapprox} can be found in the Zenodo entry~\cite{avis_kozar_2026_24508201}.}

\begin{table}[h]
    \caption{\textbf{As in Table \ref{tab:scalar}, but for DM interacting with quarks via a scalar mediator with $-i{g}_{5\chi}\phi\chi\gamma^5\bar{\chi}$ and $\sum_q-i{g}_{5q}\phi q\gamma^5\bar{q}$ coupling terms}. $\tilde{h}_N$ is defined in Eq.~\eqref{eqn:pshp}, $\tilde{g}_{5j\chi}$  is defined in full in Eq.~\eqref{eq:pschicoup} and in the small angle approximation in Eq.~\eqref{eq:main-pseudocouplingslim}. $\tilde{g}_{j\gamma}$ for the mesons is defined in  Eq.~\eqref{eqn:pseudocouplings}.  $\tilde{g}_{\phi\gamma}$ is defined in full in Eq.~\eqref{eq:phiphoton}, and in the small angle approximation in Eq.~\eqref{eq:main-pseudocouplingslim}.  The meson widths are taken to be $\Gamma_{\pi^0}=7.7\times 10^{-3}$ keV and $\Gamma_{\eta} = 1.3$ keV \protect\cite{ParticleDataGroup:2024cfk}, and $\Gamma_{\phi}$ is given in the table. The above derived constants are all combinations of $B_0$, $F$, $m_{\chi}$, the meson masses, $g_{5\chi}$, and the $\tilde{g}_q$. $\tilde{G}_{\phi\pi^+ K^+}=( \tilde{g}^2_{\phi\pi^+ K^+}(E_{\pi^+})G_FFV_{us}V_{ud}^*)\Bigg|_{E_{\pi^+}=\frac{m_{K^+}^2+m_{\pi^+}^2-m_{\phi}^2}{2m_{K^+}}}$, and $\tilde{g}_{\phi\pi^+ K^+}(E_{\pi^+})$ is defined in Eq.~\eqref{gphipiK}.  The $K^+\rightarrow\pi^+\chi\bar{\chi}$ decay rate is defined in Eq.~\eqref{eqn:Kpichichip}.}
    \centering
     \hspace*{-2 cm}  \renewcommand{\arraystretch}{1.5}
    \begin{tabular}{|c|c|c|c|}
            \hline
\multicolumn{4}{|c|}{\textbf{P$\otimes$P}} \\
\hline\rule{0pt}{0pt}
       \parbox[t][0.2cm][c]{2cm}{\textbf{Interaction}} & \parbox[t][0.2cm][c]{3cm}{ \textbf{$\frac{d\sigma}{d\Omega}$ or $\Gamma$ }}& \parbox[t][][t]{2.8cm}{ $\mathbf{\sigma}$ \textbf{ in Light Mediator Limit}}& \parbox[t][][t]{3.5cm}{ $\mathbf{\sigma}$ \textbf{ in Heavy Mediator Limit}} \\[12pt] 
        
        \hline
        
         \rule{0pt}{22pt}$\chi N\rightarrow\chi N $& 
         $\displaystyle\frac{ \tilde{h}_N^2g_{5\chi}^2 t^2}{64 \pi^2s\left. \left(t-m_{\phi}^2\right)^2 
         \right.         }$&$\displaystyle\frac{\tilde{h}_N^2g_{5\chi}^2}{16\pi (m_{\chi} + m_N)^2}$&$\displaystyle\frac{ \tilde{h}_N^2g_{5\chi}^2 m_{\chi}^4m_N^4v_{rel}^4}{3 \pi(m_{\chi}+m_N)^6m_{\phi}^4 }$\\[18pt] 
         
         \hline
         
         \rule{0pt}{27pt}$\chi \gamma\rightarrow\chi \gamma $&
         $\displaystyle\frac{- t^3}{256\pi^2 s }\left|\sum_j^{\pi^0/\eta/\eta'/\phi}\frac{\tilde{g}_{5j\chi} \tilde{g}_{ j\gamma}}{\left(t-m_{j}^2\right) 
         }\right|^2$
         &
         &$\displaystyle
             \left(\frac{ 2 E_{\gamma}^6}{5\pi m_{\chi}^2 }\left(\sum_j^{\pi^0/\eta/\eta'/\phi}\frac{\tilde{g}_{5j\chi} \tilde{g}_{ j\gamma}}{m_j^2}\right)^2\right)_{MT}
         $ \\[15pt]
         \hline
         \rule{0pt}{27pt}$\gamma\gamma\rightarrow\chi\bar{\chi}$&  $\displaystyle\frac{s^2}{256\pi^2 }\left|\sum_j^{\pi^0/\eta/\eta'/\phi}\frac{\tilde{g}_{5j\chi} \tilde{g}_{ j\gamma}}{\left(s-m_{j}^2\right) +im_{j}\Gamma_{j}}\right|^2\sqrt{\frac{s-4m_\chi^2}{s}} $ & \multicolumn{2}{|c|}{$\displaystyle\frac{E_{\gamma CM}^3\sqrt{E_{\gamma CM}^2-m_{\chi}^2}}{4\pi }\left|\sum_j^{\pi^0/\eta/\eta'/\phi}\frac{\tilde{g}_{5j\chi} \tilde{g}_{ j\gamma}}{\left(4E_{\gamma CM}^2-m_{j}^2\right) +im_{j}\Gamma_{j}}\right|^2$}
         \\[15pt]
         \hline
         
\rule{0pt}{27pt}$\chi\bar{\chi}\rightarrow\gamma\gamma$&$\displaystyle\frac{s^2}{512\pi^2 }\left|\sum_j^{\pi^0/\eta/\eta'/\phi}\frac{\tilde{g}_{5j\chi} \tilde{g}_{ j\gamma}}{\left(s-m_{j}^2\right) +im_{j}\Gamma_{j}}\right|^2\sqrt{\frac{s}{s-4m_\chi^2}}$&$\displaystyle\frac{ 1}{64\pi v_{rel} }\left(\sum_j^{\pi^0/\eta/\eta'/\phi}\tilde{g}_{5j\chi} \tilde{g}_{ j\gamma}\right)^2$&$\displaystyle\frac{ m_{\chi}^4}{4\pi v_{rel} }\left(\sum_j^{\pi^0/\eta/\eta'/\phi}\frac{\tilde{g}_{5j\chi} \tilde{g}_{ j\gamma}}{m_{j}^2}\right)^2$\\[15pt]
\hline
\rule{0pt}{30pt}$\pi^0/ \eta/\eta'\rightarrow\chi\bar\chi$& $\displaystyle\frac{\tilde{g}_{5(\pi^0/\eta/\eta')\chi}^2\sqrt{m_{(\pi^0/\eta/\eta')}^2-4m_{\chi}^2}}{8\pi}$&\multicolumn{2}{|c|}{}\\[5pt]
\hline
\rule{0pt}{25pt}$K^+\rightarrow\pi^+\phi$&$\displaystyle \frac{\tilde{G}_{\phi\pi^+ K^+}^2\sqrt{(m_{K^+}^2+m_{\pi^+}^2-m_{\phi}^2)^2-4m_{K^+}^2m_{\pi^+}^2}}{32\pi m_{K^+}^3}$&\multicolumn{2}{|c|}{}\\[10pt]
\hline
\rule{0pt}{15pt}$K^+\rightarrow\pi^+\chi\bar{\chi}$&See Eq.~\eqref{eqn:Kpichichip}&\multicolumn{2}{|c|}{}\\[5pt]\hline\rule{0pt}{50pt}$\phi\rightarrow$all&\multicolumn{3}{|c|}{$\begin{gathered}

\left(\frac{\tilde{g}_{5\phi\chi}^2\sqrt{m_{\phi}^2-4m_{\chi}^2}}{8\pi}\right)_{\chi\bar{\chi}}+\left(\frac{\tilde{g}_{\phi\gamma}^2m_{\phi}^3}{64\pi}\right)_{\gamma\gamma}+\left(\int^{\frac{s-3m_{\pi^0}^2}{2\sqrt{s}}}_{m_{\pi^0}} dE_{\pi^0} \frac{g_{\pi^0\pi^0\pi^0\phi}^2Y(E_{\pi^0},\sqrt{s},m_{\pi^0},m_{\pi^0})}{384\pi^3m_{\phi}}\right)_{\pi^0\pi^0\pi^0}\\
+\left(\int^{\frac{s+m_{\pi^0}^2-4m_{\pi^+}^2}{2\sqrt{s}}}_{m_{\pi^0}} dE_{\pi^0} \frac{Y(E_{\pi^0},\sqrt{s},m_{\pi^0},m_{\pi^+})(g_{\pi^+\pi^-\pi^0\phi}+2g_{\pi^+\pi^-\partial\pi^0\phi}(m_{\pi^0}^2-m_{\pi^+}^2+m_{\phi}^2-3E_{\pi^0}m_{\phi}))}{64\pi^3 m_{\phi}}\right)_{\pi^+\pi^-\pi^0} 
\end{gathered}$}\\[40pt]
\hline
    \end{tabular}

    \label{tab:pseudoscalar}
\end{table}

\begin{table}[]
    \caption{\textbf{As in Table \ref{tab:pseudoscalar}, but for DM interacting with quarks via a scalar mediator with $g_{\chi}\phi\chi\bar{\chi}$ and $\sum_q-i{g}_{5q}\phi q\gamma^5\bar{q}$ coupling terms}. $\tilde{g}_{j\chi}$ is defined in full in Eq.~\eqref{eq:pschicoup} and in the small angle approximation in Eq.~\eqref{eq:main-pseudocouplingslim}.   The $K^+\rightarrow\pi^+\chi\bar{\chi}$ decay rate is defined in Eq.~\eqref{eqn:Kpichichips}.}
    \label{tab:pseudoscalar+scalar}
    \centering
    \hspace*{-1 cm}  \renewcommand{\arraystretch}{1.5}
    \begin{tabular}{|c|c|c|c|}
            \hline
\multicolumn{4}{|c|}{\textbf{S$\otimes$P}} \\
\hline\rule{0pt}{0pt}
       \parbox[t][0.2cm][c]{2cm}{\textbf{Interaction}} & \parbox[t][0.2cm][c]{3cm}{ \textbf{$\frac{d\sigma}{d\Omega}$ or $\Gamma$ }}& \parbox[t][][t]{3cm}{ $\mathbf{\sigma}$ \textbf{ in Light Mediator Limit}}& \parbox[t][][t]{4cm}{ $\mathbf{\sigma}$ \textbf{ in Heavy Mediator Limit}} \\[12pt] 
        
        \hline
        
         \rule{0pt}{20pt}$\chi N\rightarrow\chi N $&$\displaystyle\frac{-g_\chi^2\tilde{h}_N^2}{64\pi^2}\frac{t(4m_{\chi}^2-t)}{s\left(t-m_{\phi}^2\right)^2
         }$ & $\displaystyle\left(\frac{g_\chi^2\tilde{h}_N^2}{8\pi m_N^2 v_{rel}^2}\right)_{MT}$&$\displaystyle\frac{g_\chi^2\tilde{h}_N^2}{2\pi}\frac{\mu^2v_{rel}^2m_{\chi}^2}{(m_{\chi}+m_{P})^2m_{\phi}^4}$\\[20pt]          
         \hline
         
         \rule{0pt}{25pt}$\chi \gamma\rightarrow\chi \gamma $&
        $\displaystyle\frac{t^2(4m_{\chi}^2-t)}{256\pi^2 s }\left|\sum_j^{\pi^0/\eta/\eta'/\phi}\frac{\tilde{g}_{ j\chi} \tilde{g}_{ j\gamma}}{\left(t-m_{j}^2\right)
        }\right|^2$
         &
         &$\displaystyle
            \left(\frac{E_{\gamma}^4}{2\pi }\left(\sum_j^{\pi^0/\eta/\eta'/\phi}\frac{\tilde{g}_{ j\chi} \tilde{g}_{ j\gamma}}{m_{j}^2}\right)^2\right)_{MT}$ \\[15pt]
         \hline
         \rule{0pt}{26pt}$\displaystyle\gamma\gamma\rightarrow\chi\bar{\chi}$&$\displaystyle\frac{s^{1/2}(s-4m_{\chi}^2)^{3/2}}{256\pi^2 }\left|\sum_j^{\pi^0/\eta/\eta'/\phi}\frac{\tilde{g}_{ j\chi} \tilde{g}_{ j\gamma}}{\left(s-m_{j}^2\right) +im_{j}\Gamma_{j}}\right|^2$&\multicolumn{2}{|c|}{$\displaystyle\frac{E_{\gamma \textrm{CM}}(E_{\gamma \textrm{CM}}^2-m_{\chi}^2)^{3/2}}{4\pi }\left|\sum_j^{\pi^0/\eta/\eta'/\phi}\frac{\tilde{g}_{ j\chi} \tilde{g}_{ j\gamma}}{\left(4E_{\gamma CM}^2-m_{j}^2\right) +im_{j}\Gamma_{j}}\right|^2$}
    \\[15pt]
         \hline
         
\rule{0pt}{27pt}$\chi\bar{\chi}\rightarrow\gamma\gamma$&$\displaystyle\frac{s^{3/2}(s-4m_{\chi}^2)^{1/2}}{512\pi^2 }\left|\sum_j^{\pi^0/\eta/\eta'/\phi}\frac{\tilde{g}_{ j\chi} \tilde{g}_{ j\gamma}}{\left(s-m_{j}^2\right) +im_{j}\Gamma_{j}}\right|^2$&$\displaystyle\frac{v_{rel}}{256\pi }\left(\sum_j^{\pi^0/
    \phi/\eta/\eta'}\tilde{g}_{ j\chi} \tilde{g}_{ j\gamma}\right)^2$&$\displaystyle\frac{m_{\chi}^4v_{rel}}{16\pi }\left(\sum_j^{\pi^0/
    \phi/\eta/
    \eta'}\frac{\tilde{g}_{ j\chi} \tilde{g}_{ j\gamma}}{m_{j}^2}\right)^2$\\[15pt]
\hline
\rule{0pt}{20pt}$\pi^0/ \eta/\eta'\rightarrow\chi\bar\chi$&$\displaystyle\frac{\tilde{g}_{(\pi^0/\eta/\eta')\chi}^2(m_{(\pi^0/\eta/\eta')}^2-4m_{\chi}^2)^{3/2}}{8\pi m_{(\pi^0/\eta/\eta')}^2}$&\multicolumn{2}{|c|}{}\\[10pt]
\hline
\rule{0pt}{25pt}$K^+\rightarrow\pi^+\phi$&$\displaystyle \frac{\tilde{G}_{\phi\pi^+ K^+}^2\sqrt{(m_{K^+}^2+m_{\pi^+}^2-m_{\phi}^2)^2-4m_{K^+}^2m_{\pi^+}^2}}{32\pi m_{K^+}^3}$&\multicolumn{2}{|c|}{}\\[10pt]
\hline
$K^+\rightarrow\pi^+\chi\bar{\chi}$&See Eq.~\eqref{eqn:Kpichichips}&\multicolumn{2}{|c|}{}\\\hline\rule{0pt}{50pt}$\phi\rightarrow$all&\multicolumn{3}{|c|}{$\begin{gathered}

\left(\frac{\tilde{g}_{\phi\chi}^2(m_{\phi}^2-4m_{\chi}^2)^{3/2}}{8\pi m_{\phi}^2}\right)_{\chi\bar{\chi}}+\left(\frac{\tilde{g}_{\phi\gamma}^2m_{\phi}^3}{64\pi}\right)_{\gamma\gamma}+\left(\int^{\frac{s-3m_{\pi^0}^2}{2\sqrt{s}}}_{m_{\pi^0}} dE_{\pi^0} \frac{g_{\pi^0\pi^0\pi^0\phi}^2Y(E_{\pi^0},\sqrt{s},m_{\pi^0},m_{\pi^0})}{384\pi^3m_{\phi}}\right)_{\pi^0\pi^0\pi^0}\\
+\left(\int^{\frac{s+m_{\pi^0}^2-4m_{\pi^+}^2}{2\sqrt{s}}}_{m_{\pi^0}} dE_{\pi^0} \frac{Y(E_{\pi^0},\sqrt{s},m_{\pi^0},m_{\pi^+})(g_{\pi^+\pi^-\pi^0\phi}+2g_{\pi^+\pi^-\partial\pi^0\phi}\left(m_{\pi^0}^2-m_{\pi^+}^2+m_{\phi}^2-3E_{\pi^0}m_{\phi})\right)^2}{64\pi^3 m_{\phi}}\right)_{\pi^+\pi^-\pi^0} 
\end{gathered}$}\\[40pt]
\hline

    \end{tabular}

\end{table}

In the limit that \begin{equation}
\label{eq:pseudolimit}
    |m_{\pi^0\phi}^2|\ll|m_{\phi}^2-m_{\pi^0}^2|\, \text{, }|m_{\eta\phi}^2|\ll|m_{\phi}^2-m_{\eta}^2|\, \text{, and } |m_{\eta' \phi}^2|\ll|m_{\phi}^2-m_{\eta'}^2|\,,
\end{equation} the mixing angles are small and the rotated couplings and masses can be approximated as
\begin{equation}
\label{eq:main-pseudocouplingslim}
    \begin{split}
\tilde{g}_{\phi\gamma}&\approx\sum_j-\tilde{g}_{j\gamma}\frac{m_{j\phi}^2}{(m_{j}^2-m_{\phi}^2)} \\
    \tilde{g}_{(5)j\chi}&\approx g_{(5)\chi}\frac{m_{j\phi}^2}{(m_{j}^2-m_{\phi}^2)}\\
    \tilde{g}_{(5)\phi\chi}&\approx g_{(5)\chi} \\
    m_{\phi}^2&\approx m_{\phi_b}^2
    \end{split}
\end{equation} where $m_{i\phi}^2$ for each meson are defined in Eq.~\eqref{eq:pseudoaxialmixings}.

We check the reliability of the small angle approximation in Fig.~\ref{fig:pseudoapprox}.  In these plots we show the fractional error on the rotated couplings, $1-g_{\text{approx}}/g_{\text{true}}$ as a function of the strength of the quark couplings $\tilde{g}_u$, $\tilde{g}_d$, and $\tilde{g}_s$ and as a function of $m_{\pi^0\phi}^2/(m_{\pi^0}^2-m_{\phi}^2)$.  For illustrative purposes we choose a quark coupling ratio of ($\tilde{g}_u$, $\tilde{g}_d$, $\tilde{g}_s)\propto$ (2:-1:-1) corresponding to the quark charges.  We do not show $\tilde{g}_{\eta'\chi}$ in these plots as it goes to 0 when ($\tilde{g}_u$, $\tilde{g}_d$, $\tilde{g}_s)\propto$ (2:-1:-1).   We take $m_{\phi_b}=0.9m_{\pi^0}$
in the left panel, and $m_{\phi_b}=(1-10^{-5})m_{\pi^0}$ in the right panel.  The left panel shows how the small angle approximation behaves when $m_{\phi_b}$ is far from any of the meson masses, while the right shows how the small angle approximation behaves near a mass resonance between a meson and the $\phi$.  The red regions show where the mixing becomes too large to remove and the theory becomes unstable. The vertical dashed line shows where the largest mixing angle between any of the mesons and $\phi$ equals 0.3, corresponding to where we stop using the approximation in our calculations. The region between the dashed line and the red block shows where we expect the small angle approximation to break down. The upper axis shows $m_{\pi^0\phi}^2 /|m_{\pi^0}^2-m_{\phi}^2|$ (the value of the $\pi^0-\phi$ mixing angle in the small angle approximation) for a given value of the quark coupling. These plots show that for most models the small angle approximation very accurately estimates the coupling parameters and $m_{\phi}$.  The parameter space between where the quark coupling is large enough to break the small angle approximation but small enough to not destabilize the entire model is rather narrow.  That said, when the $m_{\phi_b}$ is near the mass of one of the mesons, the small angle approximation breaks down over a much larger range of quark coupling strengths. So, while the small mixing approximation can be used in most situations, the masses and couplings should be calculated exactly for models with large quark couplings or an $m_{\phi_b}$ that is nearly resonant with one of the meson masses.

Using these couplings and the Feynman diagrams in Fig.~\ref{fig:pseudoscatter}, we calculate the cross section for  $\chi$-photon scattering, annihilation and production cross sections (see Fig.~\ref{fig:pseudoscatter:second}), and  the meson decay rates to $\chi\bar{\chi}$ (see Fig.~\ref{fig:pseudoscatter:third}). The $K^+$ decay rate to $\pi^+\chi\bar{\chi}$  and the $K^+$ decay rate to $\pi^+\phi$ are worked out in Appendix \ref{Ap:Kdecay}.  The resulting cross sections and decay rates can be found in Table \ref{tab:pseudoscalar} for the $P\otimes P$ case and in Table \ref{tab:pseudoscalar+scalar} in the $S\otimes P$ case.

Several cross sections depend on the decay widths of the neutral mesons or the $\phi$.  We take $\Gamma_{\pi^0}=7.7\times 10^{-3}$ keV, and $\Gamma_{\eta} = 1.3$ keV \cite{ParticleDataGroup:2024cfk} and show the calculated value of $\Gamma_{\phi}^{P\otimes P}$ in Table \ref{tab:pseudoscalar} and  $\Gamma_{\phi}^{S\otimes P}$ in Table \ref{tab:pseudoscalar+scalar}.

\subsection{Axial Vector Coupling to Quarks - ${A\otimes A}$ Case}
For a vector mediator that couples to quarks via $\sum_q \lambda_qq\gamma_{\mu}\gamma_5 \bar{q}A'^{\mu}$, we will only explore $A'$ that couple to the $\chi$ via $ \lambda_{\chi}\chi\gamma_{\mu}\gamma_5 \bar{\chi}A'^{\mu}$ because particles that couple to quarks via $\sum_q \lambda_qq\gamma_{\mu}\gamma_5 \bar{q}A'^{\mu}$ and $\chi$ via $ \lambda_{\chi}\chi\gamma_{\mu} \bar{\chi}A'^{\mu}$ do not produce new couplings to two photons, which is related to the Landau--Yang theorem.  The effective DM-photon scattering, production, and annihilation cross sections and the meson decays that result from this coupling to quarks are summarized in Table \ref{tab:axialvector}.  Details of how these cross sections and decay rates are calculated are provided below.
\subsubsection{DM-Nucleon Scattering}

If DM interacts with quarks via vector mediator with an axial vector coupling to quarks, the corresponding Lagrangian terms read:

\begin{equation}
    \mathcal{L}\supset \lambda_{\chi}\chi\gamma_{\nu}\gamma_5\bar{\chi}A'^{\nu}+\frac{1}{2}m^2_{A'} A'_{\mu}A'^{\mu}+\sum_q \lambda_qq\gamma_{\beta}\gamma_5 \bar{q}A'^{\beta}-m_{\chi}\chi\bar{\chi}\,.
\end{equation}
This results in $\chi$--nucleon interactions of the form
\begin{equation}
    \mathcal{L}\supset a_NN\gamma_{\mu}\gamma_5\bar{N}A'^{\mu}\,,
\end{equation}
where
\begin{equation}
\label{eqn:axialan}
    a_N = \sum_q \lambda_q\Delta_q^N\,.
\end{equation}
$\Delta_q^N$ are defined below as
\cite{Liang:2018pis} 
\begin{equation}
    \label{tab:axialnvcoup}
        \Delta_u^n=\Delta_d^p =-0.407 \,,\,\Delta_d^n=\Delta_u^p= 0.847\,,\, \Delta_s^n=\Delta_s^p=-0.035\,.
\end{equation}

The resulting $\chi$--N tree level cross section can be found using the Feynman diagram shown in Fig.~\ref{fig:main-axfeyn:f} and is shown
below.
\begin{equation}
\begin{split}   \left(\frac{d\sigma_{\chi N\rightarrow\chi N}}{d\Omega}\right)^{A\otimes A}=&\frac{\lambda_{\chi}^2 a_N^2}{32\pi^2m_{A'}^4s(m_{A'}^2-t)^2} \mathcal{G}_{A\otimes A}(s,t)\,,\\
\end{split}
\end{equation}
where
\begin{equation}
    \mathcal{G}_{A\otimes A}(s,t) =  2m_N^4 m_{A'}^4+4 m_N^2 (m_\chi^2 (5 m_{A'}^4 - 4 m_{A'}^2 t + 2 t^2) - m_{A'}^4 (s + t)) + 
 m_{A'}^4 (2 m_\chi^4 - 4 m_\chi^2 (s + t) + 2 s^2 + 2 s t + t^2)\,.
\end{equation}

For nonrelativistic scattering with a heavy mediator, we have
\begin{equation}
    \mathcal{G}_{A\otimes A}(s,t) \approx 
    24 m_{A'}^4m_\chi^2m_N^2\,.
\end{equation}

\begin{figure}
    \centering
    \begin{subfigure}{0.3\textwidth}
    \begin{fmffile}{f-diagram}
 \begin{fmfgraph*}(120,80)
   \fmfleft{i1,i2}
   \fmfright{o1,o2}
   \fmf{fermion,label={\large$N$},label.side=right}{i1,v1}
   \fmf{fermion,label={\large$N$}}{v1,o1}
   \fmf{fermion,label={\large$\chi$}}{i2,v2}
   \fmf{fermion,label={\large$\chi$}}{v2,o2}
   \fmf{photon,label={\large{$A'_{\mu}$}}}{v1,v2}
 \end{fmfgraph*}

\end{fmffile}
\caption{\label{fig:main-axfeyn:f}}
 \end{subfigure}
    \begin{subfigure}{0.3\textwidth}
    \begin{fmffile}{s-diagram}
 \begin{fmfgraph*}(120,80)
   \fmfleft{i1,i2}
   \fmfright{o1,o2}
   \fmf{photon,label={\large$\gamma$},label.side=right}{i1,v1}
   \fmf{photon,label={\large$\gamma$}}{v1,o1}
   \fmf{fermion,label={\large$\chi$}}{i2,v2}
   \fmf{fermion,label={\large$\chi$}}{v2,o2}
   \fmf{dashes,label={\large{$\pi_0/\eta/\eta'$}}}{v1,v2}
 \end{fmfgraph*}

\end{fmffile}
\caption{\label{fig:main-axfeyn:s}}
 \end{subfigure}
\begin{subfigure}{0.3\textwidth}
\begin{fmffile}{t-diagram}
 \begin{fmfgraph*}(120,80)
   \fmfleft{i1}
   \fmfright{o1,o2}
   \fmf{fermion,label={\large$\bar{\chi}$}}{o2,v1}
   \fmf{fermion,label={\large$\chi$},label.side=right}{v1,o1}
   \fmf{scalar,label={\large{$\pi_0/\eta/\eta'$}}}{i1,v1}
 \end{fmfgraph*}
\end{fmffile}
    
\caption{\label{fig:main-axfeyn:t}}
\end{subfigure}

    \caption{Feynman diagrams relevant to $\chi$--nucleon scattering (a), $\chi$--photon scattering (b), and meson decay to $\chi\bar{\chi}$ (c) for the $A\otimes A$ model.}\label{fig:main-axfeyn}
\end{figure}
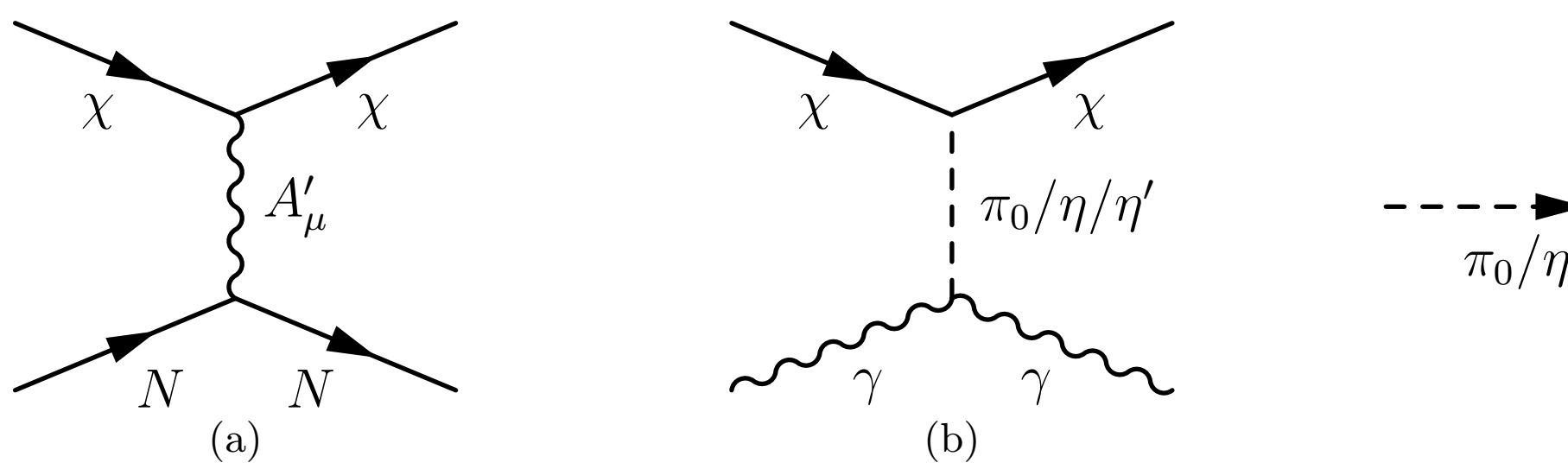
\subsubsection{DM-Photon Interactions and Meson Decays}
\label{sec:axial}
As in the pseudoscalar case, the axial vector coupling case will be complicated by mixing terms between the $A'_{\mu}$ and the light mesons. 
\begin{equation}
\label{axialmix}
    \mathcal{L}\supset -m_{\pi^0A'}A'_{\mu}\partial^{\mu}\pi^0_b-m_{\eta A'}A'_{\mu}\partial^{\mu}\eta_b-m_{\eta' A'}A'_{\mu}\partial^{\mu}\eta'_b\,,
\end{equation}
where $m_{\pi^0A'}$, $m_{\eta A'}$ and $m_{\eta' A'}$ are defined in Eq.~\eqref{eq:pseudoaxialmixings} and come from the $\mathcal{O}(p^2)$ terms and $L_4$ and $L_5$ terms of  $\mathcal{O}(p^4)$ of the ChEFT Lagrangian.

To calculate cross sections and decay rates involving these light mesons we shift to a basis without any mass or kinetic mixing terms.  As in the pseudoscalar coupling case, we will refer to the mesons in the un-shifted and unrotated basis (the one appearing in the Lagrangian) as ``bare'' mesons (marked with a ``b'' subscript), and assume that the rotated propagating basis is the one that corresponds to observations of the meson masses and couplings to photons.  To remove the mixing terms we shift $A'_{\mu}\rightarrow \tilde{A}_{\mu}'+\frac{m_{\pi^0A'}}{ m_{A'}^2}\partial_{\mu}\pi^0_b+\frac{m_{\eta A'}}{m_{A'}^2}\partial_{\mu}\eta_b+\frac{m_{\eta' A'}}{m_{A'}^2}\partial_{\mu}\eta'_b$.  This removes the $A'$-meson kinetic mixing terms, but introduces kinetic mixing between the $\pi^0$, $\eta$, and $\eta'$.  We can rotate the meson fields into each other to remove these kinetic mixing terms.  Doing so introduces a new mass mixing between the light mesons.  The fields must be rotated into each other one last time to remove the mass mixing. The rotation angles relevant for this must be calculated numerically; details on this can be found in Appendix \ref{app:axialvector}.  We limit ourselves to cases where $m_{A'}$ is larger than all of the mixing terms ($m_{j A'}$) in  Eq.~\eqref{axialmix}, as when $m_{jA'}\geq m_{A'}$ the kinetic mixing terms become too large to rotate away. 

In the rotated basis the following terms are relevant to interactions between the $\chi$ and photons
\begin{equation}
\begin{split}
\label{eq:rotatedax}
    \mathcal{L}\supset&\frac{1}{4}\tilde{g}_{\pi^0\gamma}\pi^0F^{\mu\nu}\tilde{F}_{\mu\nu}+ \frac{1}{4}\tilde{g}_{\eta\gamma}\eta F^{\mu\nu}\tilde{F}_{\mu\nu}+ \frac{1}{4}\tilde{g}_{\eta'\gamma}\eta' F^{\mu\nu}\tilde{F}_{\mu\nu}\\
    &+\lambda_{\pi^0\chi}\chi\gamma_{\beta}\gamma^5\bar{\chi}\partial^{\beta}\pi^0+\lambda_{\eta\chi}\chi\gamma_{\beta}\gamma^5\bar{\chi}\partial^{\beta}\eta+\lambda_{\eta'\chi}\chi\gamma_{\beta}\gamma^5\bar{\chi}\partial^{\beta}\eta'\,,
\end{split}
\end{equation}
where $F^{\mu\nu} = \partial^{\mu}A^{\nu}-\partial^{\nu}A^{\mu}$, $\tilde{F}_{\mu\nu} = \epsilon_{\sigma\rho\mu\nu}(\partial^{\sigma}A^{\rho}-\partial^{\rho}A^{\sigma})$, and $\tilde{g}_{j\gamma}$ are defined in Eq.~\eqref{eqn:pseudocouplings}.

The $\lambda_{j\chi}$'s are complicated expressions that emerge from the field rotations applied to remove the kinetic mixing terms. They depend on $\lambda_{\chi}$, the mixing parameters, $m_{iA'}$, and $m_{A'}$, and  are defined in Eq.~\eqref{eq:lambda}. Their derivation is described in more detail in Appendix \ref{app:axialvector}.  $\lambda_{j\chi}$ must be calculated numerically.  As with the pseudoscalar case, we provide a Python calculator to assist with the numerical calculations~\cite{avis_kozar_2026_24508201}.

In the limit that
\begin{equation}
\label{axlim}
    m_{jA'} \ll m_{A'},
\end{equation} 
one can estimate each of the $\lambda_{j\chi}$'s using the small angle approximation as:
\begin{equation}
    \begin{split}
        \lambda_{j\chi}&\approx\frac{\lambda_{\chi}m_{jA'}}{m_{A'}^2}\,.
    \end{split}
\end{equation}

\begin{figure}
    \centering
     \begin{subfigure}{0.47\textwidth}
\includegraphics[width=\linewidth]{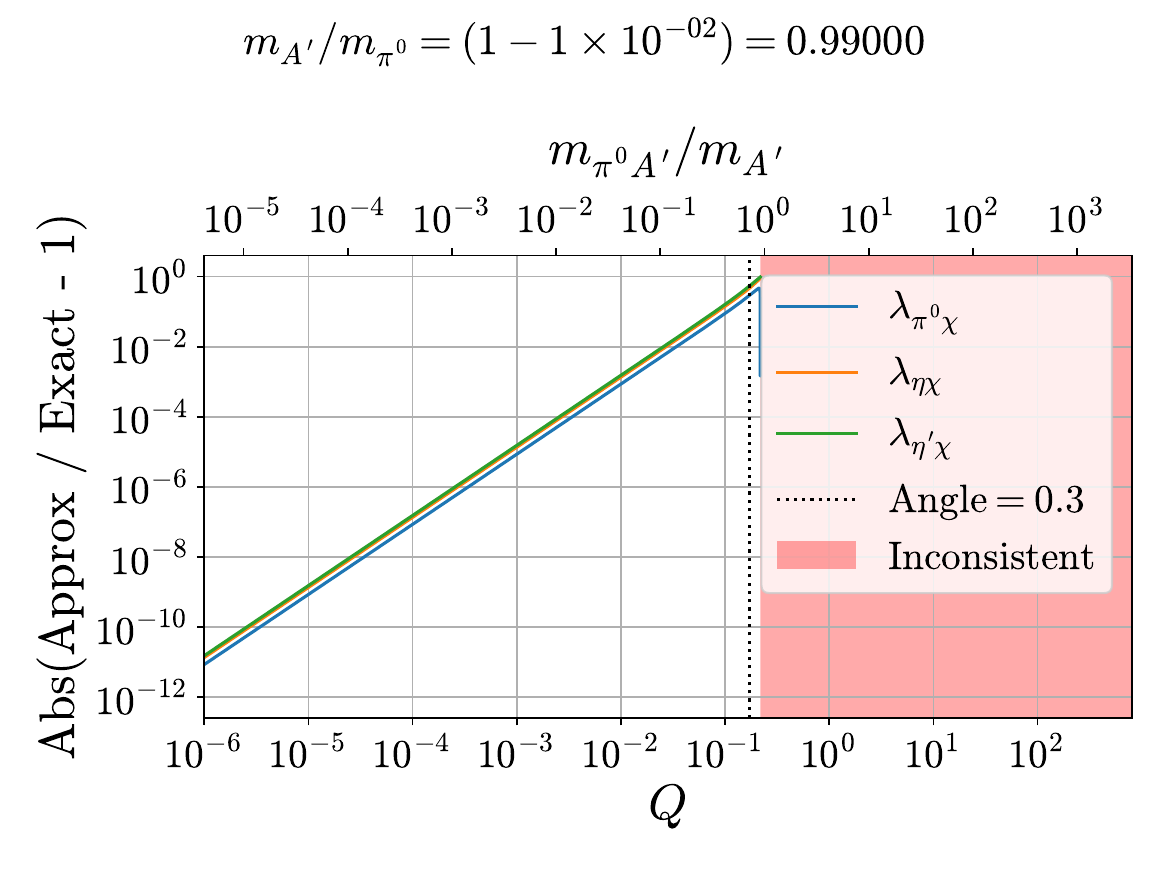}
   
    \end{subfigure}
    \begin{subfigure}{0.47\textwidth}
    \includegraphics[width=\linewidth]{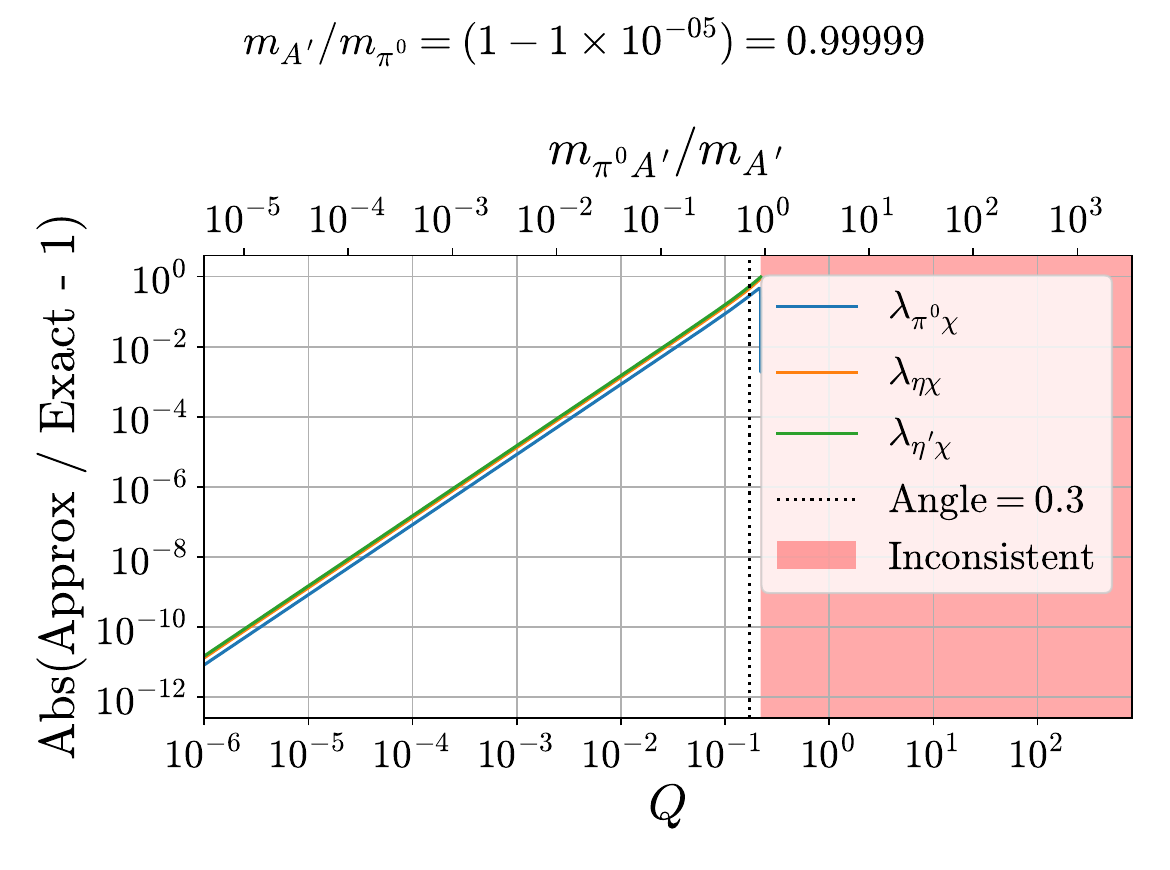}
   
    \end{subfigure}
   \caption{Plots showing the fractional error between the $\lambda_{j\chi}$'s found using the small angle approximation and their exact calculated values.  As an example we take the quark couplings to be proportional to their charge, setting  $\lambda_u =2Q$, $\lambda_d=-Q$, $\lambda_s=-Q$.  The red highlighted region shows where the large mixing causes the DM model to become unstable.  The vertical dotted line marks where the kinetic mixing angle $=0.3$.  The small mixing approximation should break down to the right of this line.  The left plot shows these ratios when $m_{A'} = (1-10^{-2}) m_{\pi^0}$, while the right plot shows these ratios when $m_{A'} = (1-10^{-5}) m_{\pi^0}$.  The upper axis shows  $m_{\pi^0 A'}^2/m_{A'}'^2$ for a given value of $Q$.  $m_{\pi^0 A'}^2/m_{A'}'^2$ approximates the mixing angle between the $\pi^0$ and the A'. \label{fig:axialapprox}}
\end{figure}

We check the validity of the small angle approximation in Fig.~\ref{fig:axialapprox}.  Here we show the fractional error between the exact values of the $\lambda_{j\chi}$'s and their values in the small angle approximation.  As an example, we take the quark couplings to be proportional to their charge.  $Q=(2\lambda_u,-\lambda_u,-\lambda_u$).  The red highlighted region shows where the large mixing causes the DM model to break down.  The vertical dotted line marks where the largest of the rotation angles $=0.3$.  The small angle approximation should break down to the right of this line.  In the left plot we set $m_{A'} = (1-10^{-2}) m_{\pi^0}$.  While the right takes $m_{A'} = (1-10^{-5}) m_{\pi^0}$.  The upper axis shows  values of $m_{\pi^0 A'}^2/m_{A'}'^2$ (which approximates the $\pi^0-A'$ mixing angle in the small angle limit) that correspond with each $Q$.   It is clear in these plots that the small angle approximation provides very accurate estimates of $\lambda_{j\chi}$ for most allowed quark coupling strengths.  The region between where the mixing angle becomes large and where the theory breaks down is very narrow, less than an order of magnitude wide, and does not grow as the mediator approaches a mass resonance with the mesons.  We can conclude that the small angle approximation for these calculations is generally reliable, except for a small set of strong quark couplings near their maximal allowed value.  When $m_{j A'}/m_{A'}$  exceeds 0.3, the exact values of the mixing angles and couplings should be calculated, however the parameter space where this can happen is fairly limited.

The $\lambda_{j\chi}$'s lead to a tree level interaction between the $\chi$ and photons and novel meson decays.  The resulting $\chi$-photon scattering cross section can be calculated from the Feynman diagram in Fig.~\ref{fig:main-axfeyn:s}, along with the production and annihilation to photon cross sections.  The meson decays to $\chi\bar{\chi}$ can be found using the diagram in Fig.~\ref{fig:main-axfeyn:t}.  The calculation of the $K^+\rightarrow\pi^+\chi\bar{\chi}$ decay rate and the $K^+\rightarrow \pi^+ A'$ decay rate is detailed in Appendix 
\ref{Ap:Kdecay}.  All of these cross sections and decay rates can be found in Table \ref{tab:axialvector}.

\begin{table}[h]
  \caption{\textbf{As in Tables \ref{tab:scalar}-\ref{tab:pseudoscalar+scalar} but for Fermionic DM interacting with quarks via a vector mediator with $A'^{\mu}\chi\gamma^5\gamma_{\mu}\bar{\chi}$ and $A'^{\mu} q\gamma^5\gamma_{\mu}\bar{q}$ coupling terms}. $a_N$ is defined in Eq.~\eqref{eqn:axialan}, $\lambda_{j\chi}$ are defined in Eq.~\eqref{eq:lambdasexact} and \eqref{eq:lambdasapprox} and $\tilde{g}_{j\gamma}$ are defined in Eq.~\eqref{eqn:pseudocouplings}.  The above derived constants are all combinations of $F$, the pion, kaon, and $A'$ masses $\lambda_{\chi}$ and the couplings between $A'$ and the light quarks $\lambda_q$. The $K^+\rightarrow\pi^+ A'_{\mu}$ decay rate is given in Eq.~\eqref{eq:K2piA}.  The $K^+\rightarrow\pi^+\chi\bar{\chi}$ decay rate is given in Eq.~\eqref{eqn:kpichichiA}.  $Y$ is defined in Eq.~\eqref{eq:Y}.}
    \label{tab:axialvector}
    
    \centering
    \hspace*{-1 cm}\renewcommand{\arraystretch}{1.5}
    \begin{tabular}{|c|c|c|c|}
            \hline
\multicolumn{4}{|c|}{\textbf{A$\otimes$A}} \\
\hline\rule{0pt}{0pt}
       \parbox[t][0.2cm][c]{2cm}{\textbf{Interaction}} & \parbox[t][0.2cm][c]   {3cm}{ \textbf{$\frac{d\sigma}{d\Omega}$ or $\Gamma$ }}& \parbox[t][][t]{3cm}{ $\mathbf{\sigma}$ \textbf{ in Light Mediator Limit}}& \parbox[t][][t]{4cm}{ $\mathbf{\sigma}$ \textbf{ in Heavy Mediator Limit}} \\[12pt] 
        
        \hline
        
         \rule{0pt}{20pt}$\chi N\rightarrow\chi N $& 
         $\displaystyle\frac{\lambda_{\chi}^2 a_N^2}{32\pi^2m_{A'}^4s(m_{A'}^2-t)^2} \mathcal{G}_{A\otimes A}(s,t)$ & $\displaystyle\frac{\lambda_{\chi}^2a_N^2}{\pi}\frac{m_N^2m_{\chi}^2}{m_{A'}^4(m_N+m_{\chi})^2}$&$\displaystyle\frac{3\lambda_{\chi}^2a_N^2\mu_{\chi n}^2}{\pi m_{A'}^4}$\\ [10pt]
         
         \hline
         
         \rule{0pt}{25pt}$\chi \gamma\rightarrow\chi \gamma $&
         $\displaystyle-\frac{m_{\chi}^2t^3}{64\pi^2s}\left(\sum_j^{\pi^0/\eta/\eta'}\frac{\lambda_{j\chi}\tilde{g}_{j\gamma}}{t-m_{j}^2}\right)^2$
         &
         &$\displaystyle\left(
            \frac{ 8 E_{\gamma}^6 }{5\pi}\left(\sum_j^{\pi^0/\eta/\eta'}\frac{\lambda_{j\chi}\tilde{g}_{j\gamma}}{m_{j}^2}\right)^2\right)_{MT}
         $ \\[15pt]
         \hline
         \rule{0pt}{25pt}$\gamma\gamma\rightarrow\chi\bar{\chi}$& $\displaystyle\frac{m_{\chi}^2s^{2}}{64\pi^2}\left|\sum_j^{\pi^0 /\eta/\eta'}\frac{\lambda_{j\chi}\tilde{g}_{j\gamma}}{(s-m_{j}^2)+im_j\Gamma_j}\right|^2\sqrt{\frac{s-4m_{\chi}^2}{s}}$&\multicolumn{2}{|c|}{$\displaystyle\frac{m_{\chi}^2E_{\gamma \textrm{CM}}^3}{\pi}\left|\sum_j^{\pi^0 /\eta/\eta'}\frac{\lambda_{j\chi}\tilde{g}_{j\gamma}}{(4E_{\gamma \textrm{CM}}^2-m_{j}^2)+im_j\Gamma_j}\right|^2\sqrt{E_{\gamma \textrm{CM}}^2-m_{\chi}^2}$}
         \\[15pt]
         \hline
         
\rule{0pt}{30pt}$\chi\bar{\chi}\rightarrow\gamma\gamma$&$\displaystyle\frac{m_{\chi}^2s^{2}}{128\pi^2}\left|\sum_j^{\pi^0 /\eta/\eta'}\frac{\lambda_{j\chi}\tilde{g}_{j\gamma}}{(s-m_{j}^2)+im_j\Gamma_j}\right|^2\sqrt{\frac{s}{s-4m_{\chi}^2}}$&$\displaystyle\frac{ m_{\chi}^2}{16\pi v_{rel}}\left(\sum_j^{\pi^0 /\eta/\eta'}\lambda_{j\chi}\tilde{g}_{j\gamma}\right)^2$&$\displaystyle\frac{m_{\chi}^6}{\pi v_{rel}}\left(\sum_j^{\pi^0 /\eta/\eta'}\frac{\lambda_{j\chi}\tilde{g}_{j\gamma}}{m_{j}^2}\right)^2$\\[15pt]
\hline
\rule{0pt}{25pt}$\pi^0/ \eta/\eta'\rightarrow\chi\bar\chi$&$\displaystyle\frac{\lambda_{j \chi}^2 m_{\chi}^2\sqrt{\frac{m_{j}^2}{4}-m_{\chi}^2}}{\pi}$&\multicolumn{2}{|c|}{}\\[7pt]
\hline
\rule{0pt}{10pt}$K^+\rightarrow\pi^+A'_{\mu}
$&See Eq.~\eqref{eq:K2piA}&\multicolumn{2}{|c|}{}\\
\hline
\rule{0pt}{10pt}$K^+\rightarrow\pi^+\chi\bar{\chi}$&See Eq.~\eqref{eqn:kpichichiA}&\multicolumn{2}{|c|}{}\\\hline
\rule{0pt}{25pt}$\tilde{A}'\rightarrow$ All&\multicolumn{3}{|c|}{{$\displaystyle\left(\frac{\lambda_{\chi}^2(m_{A'}^2-4m_{\chi}^2)^{3/2}}{12\pi m_{A'}^2}\right)_{\chi\bar{\chi}}+\left(\int_{m_{\pi^0}}^{\frac{m_{A'}^2+m_{\pi^0}^2-4m_{\pi^+}^2}{2m_{A'}}}dE_{\pi^0}Y(E_{\pi^0},m_{A'},m_{\pi^0},m_{\pi^+})\frac{(\lambda_u-\lambda_d)^2(E_{\pi^0}^2-m_{\pi^0}^2)}{8\pi^3F^2}\right)_{\pi^0\pi^+\pi^-}$}} \\[15pt]\hline

    \end{tabular}
  \end{table}
  
\section{Constraints}
\label{sec:limts}
Here we show limits on the DM-nucleon cross section that result from the effective interactions with photons and mesons.

\subsection{Relic Abundance}
\label{sec:relicabundance}
The indirect coupling of DM to photons leads to unavoidable production from freeze-in via $\gamma \gamma \rightarrow \bar \chi \chi$. Even if the DM starts out thermally decoupled, this production can rapidly saturate the relic abundance, leading to $\Omega_\chi h^2 \ge 0.12$. At higher DM mass, pion annihilation to $\chi \chi$ can dominate despite the Boltzmann suppression, and if the mediator is present in the bath, it too can contribute to freeze-in and freeze-out. The freeze-in abundance can na\"\i{}vely be computed by directly integrating the yield as in e.g. \cite{Hall:2009bx,Bernal:2017kxu,Cox:2025toz}. However, this can lead to a large overestimate of the relic abundance, as the tree-level annihilation to the mediator, $\bar \chi \chi \rightarrow \text{med} +\text{med}$ can start to contribute as the dark coupling increases \cite{Chu:2011be}. We therefore must solve the full Boltzmann equation for $\chi$:\footnote{Here we are ignoring effects due to the phase space distribution, see e.g. \cite{Du:2021jcj}.}
\begin{equation}
        \frac{1}{a^3} \frac{d(n_\chi a^3)}{dt} = \sum_X \left( \sv_{XX \rightarrow \chi \chi} n_{X}^2 - \sv_{\chi \chi \rightarrow XX} n_{\chi}^2 \right),
\end{equation}
where the sum is over bath species, the first term represents $\chi$ production from bath particles, while the second represents depletion.

Detailed balance states that the \textit{equilibrium} population of particles satisfies $\sv_{XX \rightarrow \chi \chi} n_{X,\mathrm{eq}}^2 = \sv_{\chi \chi \rightarrow XX} n_{\chi,\mathrm{eq}}^2$. Cross sections for $\chi$  annihilation to all species can be found in Appendix \ref{app:relic}.  We will assume a reheat temperature of $T_{RH} = 10$ MeV. This is a conservative choice, in that any higher initial temperature will result in more overproduction. At the same time, it remains high enough not to spoil BBN.\@ This means that the plasma temperature is always too cold for nucleons or free quarks to contribute. Light mesons are Boltzmann suppressed but remain in the plasma. For DM masses heavier than the pion mass, they will continue to compete with photons for freeze-in production as the latter's loop suppression can compensate for the Boltzmann factor. In all cases, we will assume that the initial population of DM and of the mediator is zero.\footnote{Starting with an already-coupled bath of DM or mediators will of course yield a larger relic abundance.}

Every contributing species is in equilibrium, so we may write:
\begin{equation}
        \frac{1}{a^3} \frac{d(n_\chi a^3)}{dt} =   \left( \sv_{\chi \chi\rightarrow \text{bath}  } n_{\chi,\mathrm{eq}}^2 - \sv_{\chi \chi \rightarrow \text{all}} n_{\chi}^2 \right),
        \label{eq:regularBE}
\end{equation}
 which by detailed balance is valid even if equilibrium is never actually realized for the $\chi$. Here, bath denotes a sum over all cross sections and decay rates (and inverse decay rates) to $ \{\gamma\gamma,\, \pi^+\pi^-,\,\pi^0\pi^0\}$,\,  and ``all'' to all kinematically allowed channels $\{\gamma\gamma$,  $\pi^+\pi^-$, $\pi^0\pi^0$, med med,  $\pi^0\pi^0$med,  $\pi^+\pi^-$med,  $\pi^0\pi^0\pi^0$, $\pi^+\pi^-\pi^0 \}$. We also include conversion: $\pi^\pm \phi \leftrightarrow \pi^\pm \pi^0$, and the Primakoff-like processes $e^\pm \phi \leftrightarrow e^\pm \gamma$, $\pi^\pm \phi \leftrightarrow \pi^\pm \gamma$.  At higher temperatures additional production and decay channels involving heavier mesons such as the $\eta$ and the $K$ may become important, but we do not include them. Defining the yield $Y = n_\chi/s$ in the usual way, where $s = 2\pi^2g_{\star s}T^3/45$ is the entropy density, $x = m_\chi/T$, and the dimensionless rate
 \begin{equation}
     \lambda_{a}(x) = \sqrt{\frac{\pi}{45}}M_{Pl} m_\chi\sv_{a}(x),
 \end{equation}
for process $a$, we can rewrite Eq.~\eqref{eq:regularBE} into the form:
\begin{equation}
\frac{dY}{dx} = \frac{1}{x^2}\,\xi \, \frac{g_{\star s}}{\sqrt{g_\star}}\Big[\lambda_{\rm bath}Y_{\rm eq}^2 - \lambda_{\rm all}Y^2\Big] \,
    \label{eq:soupedupBE}
\end{equation}
The factor
\begin{equation}
    \xi = \left(1+ \frac{1}{3}\frac{d\ln g_{\star s}}{d\ln T} \right)
\end{equation}
 accounts for the change in the number of degrees of freedom as a function of temperature, as described in e.g. \cite{srednicki1988calculations,Steigman:2012nb}. We solve Eq.~\eqref{eq:soupedupBE} numerically, integrating from $x = m_\chi/T_\mathrm{RH}$, when the Universe reheats. The relic abundance can be read out as usual as $\Omega_\chi h^2 = {m_\chi s_0 Y_\infty}/(\rho_c/h^2)$, with the present-day entropy density $s_0 = 2891.2$ cm$^{-3}$ and $\rho_c/h^2 = 1.054\times 10^{-5}$ GeV cm$^{-3}$ \cite{ParticleDataGroup:2024cfk}.  If $\chi$ ever achieves equilibrium, then the $\lambda$'s are identical and Eq.~\eqref{eq:soupedupBE} reduces to the standard freeze-out Boltzmann equation. 

As the couplings are raised, Eq.~\eqref{eq:soupedupBE} will yield a growing relic abundance, which will plateau and turn around as depletion outpaces production. This means that for a given DM and mediator mass, there are two solutions that yield the observed relic abundance: one from freeze-in, the other from freeze-out. To find these solutions, we fix the couplings between the mediator ($\phi/A'$) and the DM, sample the quark coupling to the mediator on a coarse grid, then converge on the values of the DM-mediator coupling that yield $\Omega_\chi h^2 = 0.12$ via numerical root-finding. For most cases illustrated here, the freeze-out solution only matters at the higher end of the $m_\chi$ range, as freeze-out for lower masses typically requires couplings larger than unity, violating the perturbativity requirement. The freeze-in and freeze-out branches will converge a little above the DM mass $m_\chi \simeq T_{RH}$, where DM production is always Boltzmann suppressed.

If the mediator mass $m_{\rm med}$ is low enough for the mediator to be present in the plasma (in practice, for $m_{\rm med} \le 10\,T_{RH}$), then the DM and mediator Boltzmann equations must be solved simultaneously. In this case Eq.~\eqref{eq:soupedupBE} becomes
\begin{align}
  \frac{dY_\chi}{dx} &=
     \frac{1}{x^2}\,\xi \, \frac{g_{\star s}}{\sqrt{g_\star}}
     \Big[\lambda_{\rm bath}\,Y_{\chi,\rm eq}^2 - \lambda_{\rm all}\,Y_\chi^2\Big]
   + \frac{2\,\xi}{x}\,\frac{\Gamma_{\text{med}\to\bar\chi\chi}}{H}\,
     \frac{K_1(m_{\rm med}/T)}{K_2(m_{\rm med}/T)}\big(Y_\text{med} - Y_{\text{med},\rm eq}\big),
     \label{eq:coupledBE_chi}\\
  \frac{dY_\text{med}}{dx} &=
   - \frac{\xi}{x}\,\frac{\Gamma_\text{med}}{H}\,
     \frac{K_1(m_{\rm med}/T)}{K_2(m_{\rm med}/T)}
     \big(Y_\text{med}-Y_{\text{med},\rm eq}\big),
     \label{eq:coupledBE_med}
\end{align}
where $\Gamma_\text{med} = \Gamma_{\text{med}\to\gamma\gamma} + \Gamma_{\text{med}\to\text{mesons}} + \Gamma_{\text{med}\to\bar\chi\chi}$ is the total mediator width, $H$ is the Hubble rate, the ratio of Bessel functions is the thermal average of the time-dilated decay rate, and $Y_{i,\rm eq} = \frac{45}{4\pi^4}\frac{g_i}{g_{\star s}}\left(\frac{m_i}{T}\right)^2 K_2\!\left(\frac{m_i}{T}\right)$. The thermally averaged cross section $\lambda_{\rm bath}$ includes the $s$-channel resonance, $\text{SM SM}\to\text{med}\to\bar\chi\chi$ with an on-shell mediator. Integrated over the Breit--Wigner peak, this is equivalent to $\chi$ production from decays of a mediator population at its equilibrium abundance. The second term in Eq.~\eqref{eq:coupledBE_chi} replaces that equilibrium abundance with the actual one, so that the resonant contribution is $\propto Y_\text{med}$ rather than $Y_{\text{med},\rm eq}$ (see e.g. \cite{Giudice:2003jh}). When $\Gamma_\text{med} \gg H$, $Y_\text{med}$ approaches $Y_{\text{med},\rm eq}$, the source term vanishes, and Eq.~\eqref{eq:soupedupBE} is recovered. The DM production cross sections for photons can be found in Tables \ref{tab:scalar} for the $S\otimes S$ case, \ref{tab:psuedo+scalar} for the $P\otimes S$ case, \ref{tab:pseudoscalar} for the $P\otimes P$ case, \ref{tab:pseudoscalar+scalar} for the $S\otimes P$ case, and \ref{tab:axialvector} for the $A\otimes A$ case.  The DM annihilation rates to all species, including mesons, can be found in Appendix \ref{app:relic}.

Finally, we comment on the relic abundance production for the axial vector $A\otimes A$ case. If $m_{A'} < 2m_\chi$, then the axial vector is in principle stable, and the $A'$ na\"\i{}vely becomes a dark matter component. However, such a new mediator coupled with quarks should also mix with the $Z$ boson, in a way that is ultimately dependent on the UV completion, leading to decays to $e^+e^-$ and $\bar \nu \nu$. The ChEFT-only contributions lead to lifetimes that can spoil BBN via photodisintegration, but the true widths are likely to be larger. This means that in a full theory the $A'$ will likely decay before BBN, but on long timescales with respect to $H(z)$. For our results, we will ignore mixing with $Z$, and simply assume the $A'$ does not contribute to the final relic abundance of DM.  

\subsection{Effective Number of Relativistic Degrees of Freedom $N_{eff}$} 
\label{sec:Neff}
If the mediator ($\phi/A'$) or the DM, $\chi$, is lighter than a few MeV, it can still carry entropy after neutrino decoupling, and that entropy can end up in two places. A species that is still in contact with the electromagnetic plasma when it becomes non-relativistic will ``dump'' its entropy into the photon sector. This raises $T_\gamma/T_\nu$ and gives $\Delta N_{\rm eff} < 0$. A species that loses contact with the photons while still relativistic keeps its entropy and free-streams as dark radiation, giving $\Delta N_{\rm eff} > 0$. In our model the second case happens only for mediators produced by $\chi$ (see below).

Computing $\Delta N_{\rm eff}$ exactly means integrating the coupled Boltzmann equations of the dark and Standard Model sectors through neutrino decoupling~\cite{Escudero:2018mvt,EscuderoAbenza:2020cmq,Burns:2023sgx}; see also \cite{Ho:2012ug,Nollett:2013pwa,Sabti:2019mhn,Depta:2019lbe,Giovanetti:2021izc}. That is too slow for a parameter scan, so we use a modified version of the approach of \cite{Boehm:2013jpa,Wilkinson:2016gsy}. We define $N_{\rm eff}$ in the usual way, as a neutrino-equivalent contribution:
\begin{equation}
  N_{\rm eff} \simeq N_{\rm eff}^{\rm SM}\left[1 + \frac{7}{22}\sum_i w_i\,\frac{g_i}{2}\,F_i(y_i)\right]^{-4/3},
  \qquad y_i = \frac{m_i}{T_{\nu,\rm dec}}, \label{eq:Neffdef}
\end{equation}
with $N_{\rm eff}^{\rm SM} = 3.044$ and
\begin{equation}
  F(y) = \frac{30}{7\pi^4}\int_y^\infty d\zeta \,
         \frac{(4\zeta^2-y^2)\sqrt{\zeta^2-y^2}}{e^\zeta \pm 1},
\end{equation}
with $-$ for bosons and $+$ for fermions. $F$ is the entropy of a species in units of one massless Weyl pair of fermions, so $F_{\rm F}(0) = 1$ and $F_{\rm B}(0) = 8/7$. We take $T_{\nu,\rm dec} = 2$~MeV. The sum runs over $\chi$ ($g_\chi = 4$, Dirac) and the mediator ($g_\phi = 1$ for $\phi$ and $\tilde\phi$, and $g_{A'} = 3$).

The standard form of Eq.~\eqref{eq:Neffdef}, with $w_i = 1$, assumes that every species is in equilibrium with the plasma at $T_{\nu,\rm dec}$ and stays in contact until it has given up its entropy. To approximate the effect of partial contact, we weight each species by
\begin{equation}
w_i = \left[1 - \exp\left\{-\int_{T_{\nu,\rm dec}}^{T_{\rm RH}}\frac{\Gamma_i}{H}\frac{dT}{T}\right\}\right]
      \left[1 - \exp\left\{-\int_{T_{\nu,\rm dec}/100}^{T_{\nu,\rm dec}} \frac{\Gamma_i}{H}\frac{dT}{T}\right\}\right], \label{eq:wdef}
\end{equation}
where $\Gamma_i$ is the rate linking species $i$ to the plasma and $T_{\rm RH}$ is the reheat temperature. The first factor measures how far the species has equilibrated with the plasma by neutrino decoupling. The second measures whether it stays in contact afterwards, so that its entropy goes to the photons rather than being carried away. 

The mediator couples through inverse decay $\gamma \gamma \rightarrow \mathrm{med}$, through pair production $\gamma \gamma \rightarrow \mathrm{med+med}$, as well as through the conversion and Primakoff processes listed in Sec. \ref{sec:relicabundance}. By detailed balance, and including the time dilation of the decay rate~\cite{KOLB1980224},
\begin{equation}
\Gamma_{\text{med}} = \Gamma_{\text{med}\to\gamma\gamma}\,\frac{K_1(m_\text{med}/T)}{K_2(m_\text{med}/T)} + \frac{2\,\mathcal{R}_{\gamma\gamma\to\text{med}+\text{med}}}{n_\text{med}^{\rm eq}} + \frac{\mathcal{R}_{\pi^\pm\gamma\to\pi^\pm\text{med}} + \mathcal{R}_{e^\pm\gamma\to e^\pm\text{med}} + \mathcal{R}_{e^+e^-\to\gamma \, \text{med}}}{n_\text{med}^{\rm eq}},
\end{equation}
where $K_\nu(x)$ are modified Bessel functions of the second kind and $\mathcal{R}$ is defined in Eq.~\eqref{eq:Rgg} below. The $\pi^\pm$ term applies to $S\otimes S$ and $P\otimes S$ with $m_\text{med} < 2m_\pi$, and the $e^\pm$ terms to all but the axial vector. For light mediators, where $K_1/K_2 \to m_\text{med}/2T$, these conversion terms dominate. DM couples to the photon bath through $\gamma\gamma\leftrightarrow\bar\chi\chi$, as well as through the mediator: $\bar\chi\chi\leftrightarrow\text{med}+\text{med}$ followed by $\text{med}\leftrightarrow\gamma\gamma$. That chain can go no faster than its slower step, so
\begin{equation}
\Gamma_\chi = \max\Big(\Gamma_{\gamma\gamma\to\bar\chi\chi},\;
  \min\big(\tfrac{1}{2}n_\chi^{\rm eq}\langle\sigma v\rangle_{\bar\chi\chi\to\text{med}+\text{med}},\,\Gamma_{\text{med}}\big)\Big),
\end{equation}
with $n_\chi^{\rm eq}$ (counting $\chi$ and $\bar\chi$) evaluated at $T_\gamma$ and $\Gamma_{\gamma\gamma\to\bar\chi\chi} = \mathcal{R}_{\gamma\gamma\to\bar\chi\chi}/n_\gamma$. Here $\mathcal{R}$ is the thermally averaged reaction density, i.e.\ the number of $\gamma\gamma\to\bar\chi\chi$ reactions per unit volume per unit time, computed with equilibrium photon distributions~\cite{Gondolo:1990dk}:
\begin{equation}
  \mathcal{R}_{\gamma\gamma\to\bar\chi\chi}(T)
  = \frac{1}{2}\,\big(n_\gamma^{\rm eq}\big)^2\,\langle\sigma v\rangle_{\gamma\gamma\to\bar\chi\chi}
  = \frac{T}{16\pi^4}\int_{4m_\chi^2}^{\infty} ds\;
    \sigma_{\gamma\gamma\to\bar\chi\chi}(s)\,s^{3/2}\,K_1\!\left(\frac{\sqrt{s}}{T}\right),
  \label{eq:Rgg}
\end{equation}
where $n_\gamma^{\rm eq} = 2\zeta(3)T^3/\pi^2$. $\mathcal{R}_{\gamma\gamma\to\text{med}+\text{med}}$ follows from the same expression with $\sigma_{\gamma\gamma\to\text{med}+\text{med}}$ and threshold $4m_\text{med}^2$.  $\Gamma_{\text{med}}$ and $\sigma_{\gamma\gamma\rightarrow\chi\bar{\chi}}$ can be found in the summary tables for each mediator: \ref{tab:scalar}, \ref{tab:psuedo+scalar}, \ref{tab:pseudoscalar}, \ref{tab:pseudoscalar+scalar}, and \ref{tab:axialvector}, while ${d\sigma_{\bar{\chi}\chi\rightarrow\text{med}+\text{med}}}/{d\Omega}$ can be found in Appendix \ref{app:relic}.
 
A mediator that is not held in equilibrium by the SM bath ($w_{\text{med}} < 1$) can still be produced by $\chi$ through $\bar\chi\chi\to\text{med}+\text{med}$, and through $\bar\chi\chi\to\text{med}$ when $m_{\text{med}} > 2m_\chi$. This can happen even if $\chi$ loses contact before $T_{\nu,\rm dec}$. The mediator produced this way free-streams as dark radiation. So does a mediator that equilibrates through a rate that shuts off before $T_{\nu,\rm dec}$ ($\pi^\pm\gamma\to\pi^\pm\text{med}$, $\gamma\gamma\to\text{med}+\text{med}$): it decouples with the neutrinos and enters at $T_\text{med} = T_\nu$. A mediator still held by the $e^\pm$ processes after $T_{\nu,\rm dec}$ is released as free radiation when they shut off, at the temperature where its optical depth, counted from recombination, reaches one. We follow its energy density $\rho_{\text{med}}$ with a relaxation form of the Boltzmann equation that reduces to freeze-in when $\rho_\text{med} \ll \rho_\text{med}^{\rm eq}$ and to equilibrium when production is fast. Its target is $(1-w_\text{med})\,\rho_\text{med}^{\rm eq}$, since the fraction $w_\text{med}$ is already counted in Eq.~\eqref{eq:Neffdef}. Its rate is capped by $\Gamma_\chi$, since the energy reaches the mediator from the photons through $\chi$. Under the assumption of a thermal spectrum, we remove the entropy that the mediator carries from the bath:
\begin{equation}
s_\text{med} = \frac{2\pi^2}{45}\,g_\text{med} T_\text{med}^3,\qquad T_\text{med} = \left(\frac{30\,\rho_\text{med}}{\pi^2 g_\text{med}}\right)^{1/4}.
\end{equation}
Imposing conservation of the combined photon-plus-mediator entropy from $T_{\nu,\rm dec}$ to recombination then fixes $T_\gamma/T_\nu$. Any mediator that decays to $\gamma\gamma$ or $\pi\pi$ 
before recombination returns its entropy to the photons, and we remove it from the free-streaming budget. Mediators heavy enough to decay to mesons ($m_\text{med} > 2m_\pi$) are not produced in any number at $T \le T_\text{RH}$, so in practice only $\gamma\gamma$ decays matter. The surviving mediator then adds its kinetic energy density at recombination to $N_{\rm eff}$. With $\rho_\text{med} = 0$ this procedure reproduces Eq.~\eqref{eq:Neffdef} exactly. Details are given in Appendix~\ref{app:Neff}.
 
We apply the Planck 2018 measurement at recombination, $N_{\rm eff} = 2.99 \pm 0.17$~\cite{Planck:2018vyg}, i.e.\ $\Delta N_{\rm eff} = -0.054 \pm 0.17$ relative to the Standard Model value $3.044$~\cite{Bennett:2020zkv}. We treat it as a two-sided bound and exclude parameter points more than $2\sigma$ from the central value. This means that in some plots (namely at low mediator mass), there will be a gap between excluded regions as $\Delta \Neff$ goes from negative to positive.

\subsection{Rare Meson Decays}

\label{sec:decays}
\label{sec:mesondecay}
If $\chi$ particles interact with quarks via a scalar mediator $\phi$  with a $\sum_q-i\tilde{g}_q\phi q\gamma^5\bar{q}$ quark coupling  or via a vector $A_{\mu}'$ with a $A_{\mu}'q\gamma^5\gamma^{\mu}\bar{q}$ quark coupling, then the light neutral mesons, the $\pi^0$, $\eta$, and $\eta'$ can decay to $\chi\bar{\chi}$ as long as $m_{\chi}< \frac{m_{j}}{2}$.  We use $m_j$ to denote the mass of the $j$th meson.  The neutral meson rest frame decay rate to $\chi\bar{\chi}$, $\Gamma_{(\pi^0/\eta/\eta')\rightarrow\chi\bar{\chi}}$, has been calculated in this work and can be found in Table \ref{tab:pseudoscalar} for the $P\otimes P$ case, Table \ref{tab:pseudoscalar+scalar} for the $S\otimes P$ case and Table \ref{tab:axialvector} for the $A\otimes A$ case.

We acquire limits on these DM models by requiring that for each meson $j$, $\Gamma_{j\rightarrow\chi\bar{\chi}}$ be less than the  limit on $\Gamma_{j\rightarrow\text{invisible}}$ as reported in Ref.\cite{ParticleDataGroup:2024cfk}:

    \begin{eqnarray}
        \Gamma_{\pi^0\rightarrow\text{invisible}} &<& 3.43\times10^{-17}\text{ GeV}, \nonumber \\
        \Gamma_{\eta\rightarrow\text{invisible}} &<& 3.14\times10^{-11}\text{ GeV},\\
        \Gamma_{\eta'\rightarrow\text{invisible}} &<& 3.95\times10^{-8}\text{ GeV}. \nonumber
    \end{eqnarray}

\label{sec:kaondecay}

DM coupling to quarks via a scalar, pseudoscalar or axial vector coupling can cause $K^+$ mesons to decay to $\pi^+\chi\bar{\chi}$ if $m_{\chi}\leq\frac{m_{K^+}-m_{\pi^+}}{2}$, or decay to $\pi^+ +$ mediator if $m_{\text{med}}<m_{K^+}-m_{\pi^+}$.  We place limits on a model by requiring that the predicted values of $\Gamma_{K^+\rightarrow\pi^+\chi\bar{\chi}}$ and $\Gamma_{K^+\rightarrow\pi^+ +\text{med}}$ not exceed the 90$\%$ upper confidence limit on the $K^+$ decay rate to $\pi^+\nu\bar{\nu}$,  $\Gamma_{K^+\rightarrow\pi^+\nu\bar{\nu}}$, as reported in Ref.~\cite{ParticleDataGroup:2024cfk}. The NA62 analysis that this limit is based on used a signal region that removed the pion resonance as well as the region above the $K \rightarrow 3\pi$ threshold, because of the large backgrounds in those regions \cite{NA62:2021zjw}. For this reason, we only apply this bound in the signal regions $s = [0, 0.010]$ GeV$^2$ and $s =  [0.026, 0.068]$ GeV$^2$. To avoid sharp binning effects, we smear the window with a Gaussian of width 0.001 GeV$^2$, corresponding to the approximate analysis resolution.

It is possible that after $K^+$ decays to a mediator, the mediator subsequently decays to photons, leading to a non-invisible event. To account for the possibility of the mediator decaying back into visibles before leaving the detector, we apply a smooth weight:
\begin{equation}
    \Gamma_{K^+\rightarrow\pi^+ +\text{med}}\,
    \frac{\Gamma_{\text{med}\rightarrow\chi\bar{\chi}}
          +\Gamma_{\text{med}\rightarrow\text{vis}}\,e^{-L/\ell}}
         {\Gamma_{\text{med}}^{\text{tot}}}
    \leq 6.06\times10^{-27}\text{ GeV}\,,
\end{equation}
where
\begin{equation}
    \ell=\frac{\sqrt{E_{\text{med}}^2-m_{\text{med}}^2}}{m_{\text{med}}}\,
    \frac{\hbar c}{\Gamma_{\text{med}}^{\text{tot}}}\,,
    \qquad
    \Gamma_{\text{med}\rightarrow\text{vis}}
    =\Gamma_{\text{med}}^{\text{tot}}-\Gamma_{\text{med}\rightarrow\chi\bar{\chi}}\,,
\end{equation}
where $L=100$~m and $E_{\text{med}}=50$~GeV are typical length and energy for NA62.

\subsection{Dark Matter Annihilation}

\subsubsection{Indirect Detection}
\label{sec:indirect}

The DM models we consider can annihilate to a photon line at loop level, allowing us to recast constraints from hard X-ray or low-energy gamma-ray searches for annihilating DM.\@ Specifically, we use the constraints derived in Ref.~\cite{Laha:2020ivk} using data from INTEGRAL.\@ In the S$\otimes$P and S$\otimes$S models, the annihilation cross section is $p$-wave, meaning that the constraints derived in the above reference cannot be used directly. In those cases, we rescale the results using the J-factor computations presented in Ref.~\cite{Boddy:2018ike}.

\subsubsection{CMB Bounds on Annihilation}
\label{sec:CMBlims}

If DM annihilates at a large enough rate around the time of recombination, the resulting energy injection can affect recombination and thus leave detectable imprints on the CMB.\@ Ref.~\cite{Slatyer:2015jla} presents limits on DM annihilation to photons from the CMB, which can be recast using, as above, the fact that the DM we consider annihilates to photons at loop level.

\subsection{Cosmological Bounds on Dark Matter-Photon Scattering}
\label{sec:cmbscatter}
If DM scatters with photons---or other relativistic particles---the resulting energy transfer can wash out density perturbations and thus affect structure formation~\cite{Crumrine:2024sdn}. This scattering can also produce spectral distortions in the CMB, as discussed in Ref.~\cite{Ali-Haimoud:2021lka}. We have derived limits by recasting the constraints presented in those references, but have confirmed that they are always much weaker than the other constraints we show, so we opt to exclude them from the figures presented here.

\subsection{Constraints on the Mediator}

The DM models explored in this work are not merely constrained by their effective cross sections with photons and the meson decays they induce.  The mass of the mediator and the strength of its couplings, to both the $\chi$ and the SM are limited by fifth force experiments, the Bullet Cluster, SN1987a, and neutron star cooling limits.  These limits are relevant when, for the model chosen, a given value of $\sigma_{\chi N}$ can only be reached when the mediator mass and couplings take on combinations of values that are already excluded.  We also use these constraints to limit the range of mediator masses and couplings we explore in our multidimensional scans of DM parameter space shown in Figs \ref{fig:mx_sigma_6d},  \ref{fig:mx_sigma_mphi10}, and \ref{fig:mx_sigma_mphi10_mfv}.   These limits are particularly relevant for low mediator masses, and often set a lower limit on the mediator masses that can be considered for a given cross section. Below we describe the constraints that apply to specific combinations of mediator mass and coupling  that must be considered when setting effective constraints.

\subsubsection{Fifth Force Constraints}
\label{sec:fifthforce}

\begin{figure}[t]
  \centering
  \includegraphics[width=0.65\textwidth]{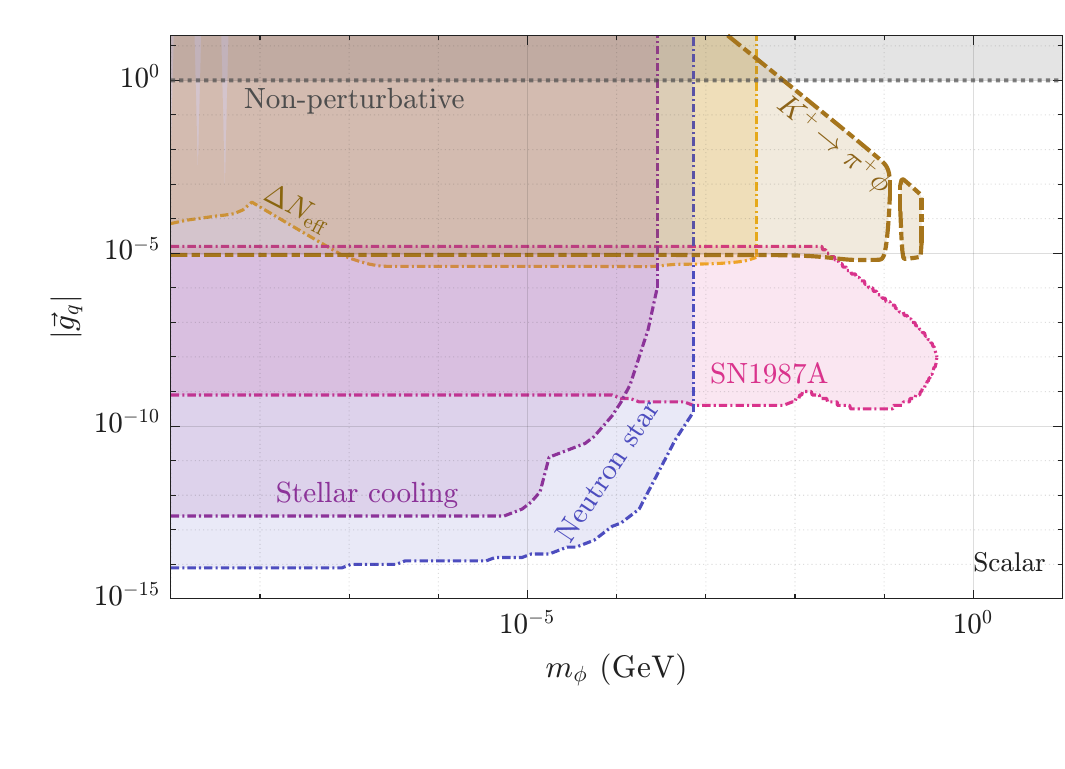}\\[1pt]
  \includegraphics[width=0.45\textwidth]{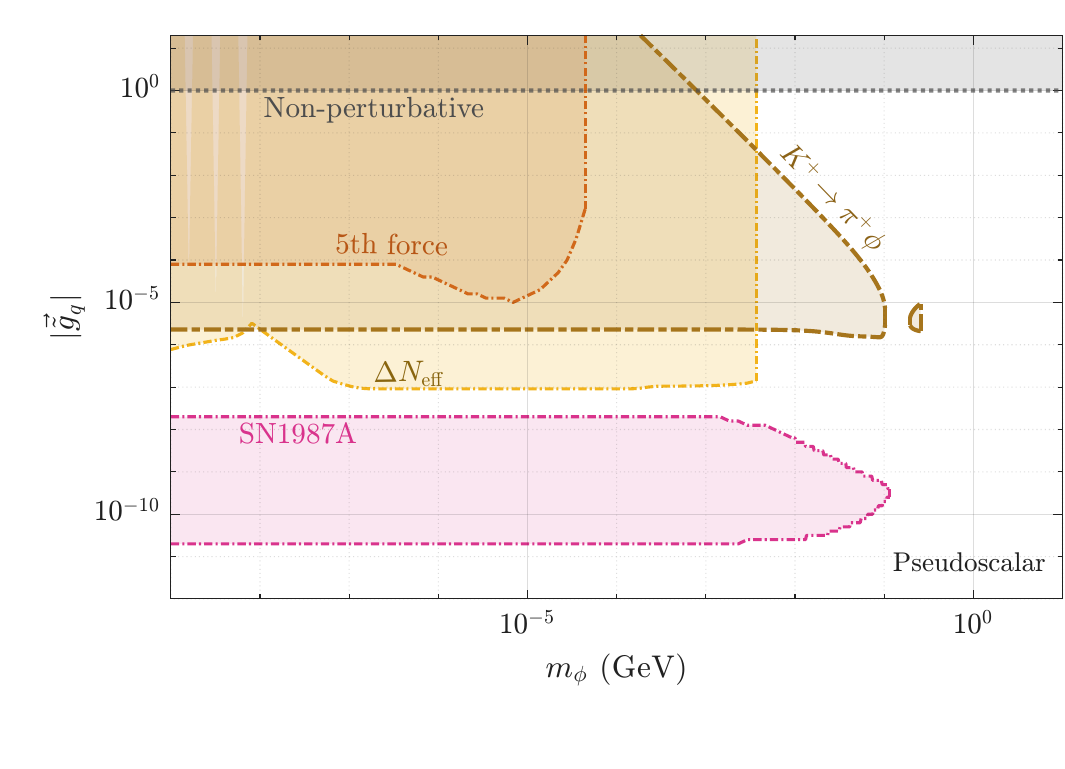} 
  \includegraphics[width=0.45\textwidth]{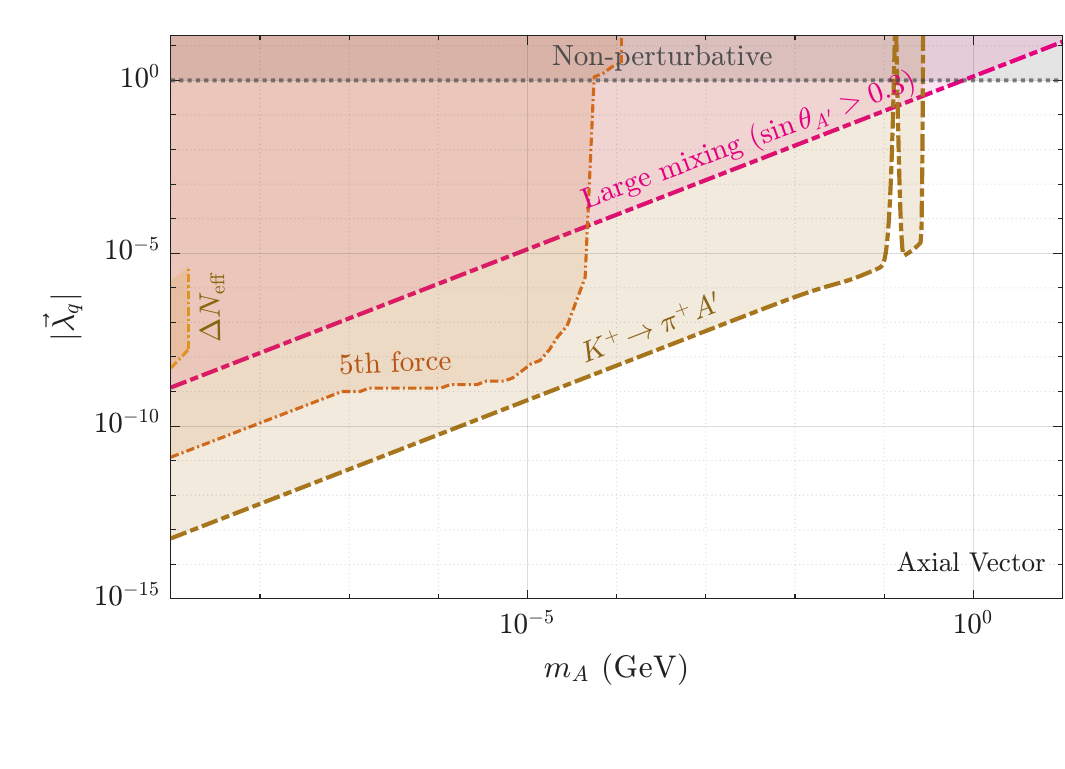}
  \caption{Constraints on the magnitude of the couplings between the mediator only and the quarks for scalar (top), pseudoscalar (bottom-left) and axial vector mediators (bottom-right). Here, the quark coupling is proportional to charge $[2, -1, -1]$. The gap in the $K$ decay constraint is due to the window used by NA62, which excludes the pion resonance region. Bounds shown: $\Delta N_{\rm eff}$~\cite{Planck:2018vyg}, $K^+\!\to\pi^+ +$ mediator~\cite{ParticleDataGroup:2024cfk,NA62:2021zjw}, fifth force~\cite{Cong_2025}, neutron star cooling~\cite{Fiorillo:2025zzx}, stellar cooling~\cite{Hardy:2016kme}, and SN1987A~\cite{Hardy:2024gwy,Ramani:2019jam}.}
  \label{fig:mediatorlims}
\end{figure}

Fifth force limits focus on new short distance forces between SM particles that result from the existence of a light mediator coupled to SM particles.  These constraints restrict how large the mediator coupling to the SM can be for a given mediator mass.  We apply combined limits from fifth force experiments presented in Ref. \cite{Cong_2025}.  Limits relevant to a scalar coupling between $\phi$ and quarks ($S\otimes S$ and $P\otimes S$) can be found in Figure 14 (c) of Ref. \cite{Cong_2025}.  Here $\lambda = 1/m_{\phi}$ and $g^N_s = h_N$.  Limits relevant to a pseudoscalar coupling between $\phi$ and quarks ($S\otimes P$ and $P\otimes P$) are taken from Figure 13 of Ref. \cite{Cong_2025} where $\lambda = 1/m_{\phi_b}$ and $g^N_p = \tilde{h}_{N}$. Axial vector limits ($A\otimes A$) are from Figure 10 (b) of Ref. \cite{Cong_2025} where $\lambda = 1/m_{A'}$ and $g^N_A = a_N$.

\subsubsection{Stellar Limits}
\label{sec:stellarlimits}
If the mediator is light, it may be produced from plasma processes in stars, supernovae and stellar remnants.  Ref.~\cite{Fiorillo:2025zzx} places the most stringent current limits on $h_N$ for $\phi$ in the eV to MeV mass range based on excess emission of $\phi$ particles altering neutron star cooling.  These are the leading limits for scalar couplings to quarks $S\otimes S$ and $P\otimes S $ cases.  We use the limits presented in Figure 1 of Ref.~\cite{Fiorillo:2025zzx}, where $g_N=h_N$. We further include scalar emission from stellar cooling bounds on the tip of the red giant branch and horizontal branch stars, from Ref. \cite{Hardy:2016kme}. 

We include limits on both $h_N$ and $\tilde{h}_N$, the scalar and pseudoscalar couplings to nuclei, from observations of SN1987A.  Excess production of light mediators in the proto-neutron star core of SN1987A cools the SN too quickly, reducing the number of neutrinos emitted by the star below observed levels.  We take limits on $h_N$ from Figure 6 of Ref. \cite{Hardy:2024gwy} where $y_N$ corresponds to $h_N$.  Although they consider particles that can also decay to a dark sector (and therefore not be reabsorbed), these assume instantaneous decay. We thus take the more conservative limits that do not assume mediator decay. The SN1987A and stellar limits are extended to masses below those presented in the original works, where the emission rate is independent of the mediator mass.
For $\tilde{h}_N$  we use the constraints found in Table 1 of Ref. \cite{Ramani:2019jam}. Here $g_N$ corresponds to $\tilde{h}_{N}$ and $m_{A'}$ corresponds to $m_{\phi_b}$.

\subsubsection{Bullet Cluster Self-Interaction Limits}
\label{sec:bulletcluster}

To get limits on the strength of the mediator coupling to the $\chi$ as a function of mediator mass we turn to Bullet Cluster limits on DM self-interactions~\cite{Markevitch_2004}.  We follow the approach described in Ref.~\cite{Ramani:2019jam}, who use the momentum-transfer cross section in the Born approximation, $ \sigma_T^{\rm Born}$, requiring ${\sigma_T^{\rm Born}}/{m_{\chi}}\leq1$ cm$^2/$g.  For each mediator-DM coupling $\sigma_T^{\rm Born}$ is
\begin{equation}
    \begin{split}
        S\otimes S \,  (S\otimes P)&: \sigma_T^{\rm Born} \sim\frac{\accentset{(\sim)}{g}_{\chi} {}^4}{2\pi m_{\chi}^2v_{rel}^4}\left(\text{log}(1+R^2)-R^2/(1+R^2)\right)\\
        P\otimes S \, (P\otimes P)&: \sigma_T^{\rm Born} \sim\frac{\accentset{(\sim)}{g}_{5\chi} {}^4}{32 m_{\chi}^2}\frac{\left(6(R^2+1)\text{log}(1+R^2)+(R^4-3R^2-6)R^2\right)}{R^4(R^2+1)}\\
        A\otimes A&:\sigma_T^{\rm Born} \sim\frac{6\lambda_{\chi}^4}{\pi m_{\chi}^2v_{rel}^4}\left(\text{log}(1+R^2)-R^2/(1+R^2)\right),\\
    \end{split}
\end{equation}
where $R = \frac{m_{\chi}v_{rel}}{m_\text{med}}$ and the relative velocity between $\chi$'s is taken to be $v_{rel} \sim 0.01$.

\begin{figure}[htbp]
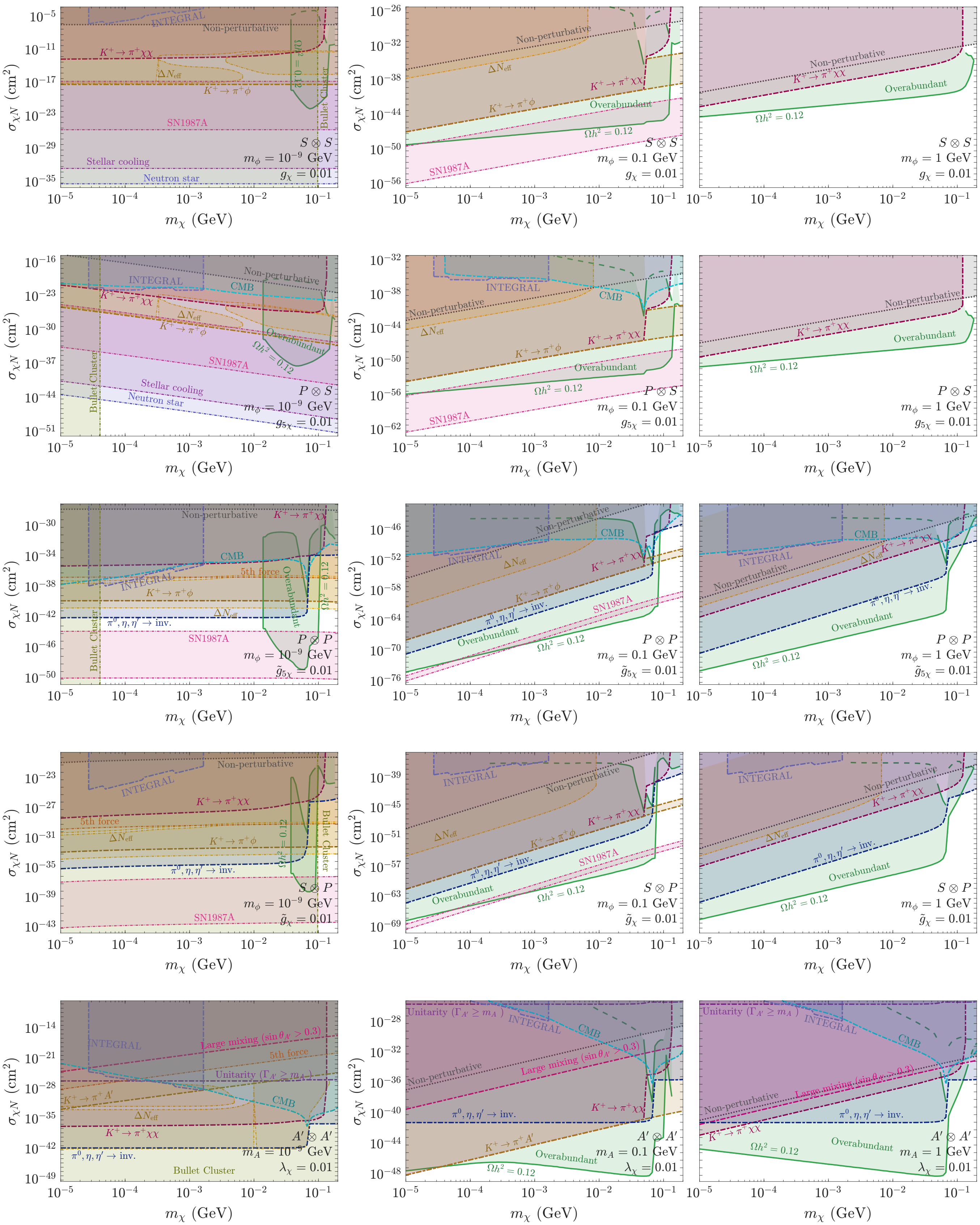

  \centering
  \panelrow{figures/figures_g/compare_limits_g_lightScalar_mphi1e-09_gx0.01_sigma}{figures/figures_g/compare_limits_g_heavyScalar_mphi0.1_gx0.01_sigma}{figures/figures_g/compare_limits_g_heavyScalar_mphi1_gx0.01_sigma}\\
\panelrow{figures/figures_g/compare_limits_g_lightScalarPS_mphi1e-09_gx0.01_sigma}{figures/figures_g/compare_limits_g_heavyScalarPS_mphi0.1_gx0.01_sigma}{figures/figures_g/compare_limits_g_heavyScalarPS_mphi1_gx0.01_sigma}\\
\panelrow{figures/figures_g/compare_limits_g_lightScalarPP_mphi1e-09_gx0.01_sigma}{figures/figures_g/compare_limits_g_heavyScalarPP_mphi0.1_gx0.01_sigma}{figures/figures_g/compare_limits_g_heavyScalarPP_mphi1_gx0.01_sigma}\\
\panelrow{figures/figures_g/compare_limits_g_lightScalarSP_mphi1e-09_gx0.01_sigma}{figures/figures_g/compare_limits_g_heavyScalarSP_mphi0.1_gx0.01_sigma}{figures/figures_g/compare_limits_g_heavyScalarSP_mphi1_gx0.01_sigma}\\
\panelrow{figures/figures_g/compare_limits_g_axialVector_mphi1e-09_gx0.01_sigma}{figures/figures_g/compare_limits_g_axialVector_mphi0.1_gx0.01_sigma}{figures/figures_g/compare_limits_g_axialVector_mphi1_gx0.01_sigma}\\
  \caption{Bounds on each of the models considered here, for fixed values of the mediator mass, and fixed ratios between the mediator-quark couplings $(g_u, g_d, g_s)\propto (2, -1, -1)$, with the mediator-DM coupling $g_\chi = 0.01$. $m_\text{med} = 10^{-9},\,0.1,\,1$~GeV (left to right); the right-hand pair shares one $y$ axis. See text for details on all constraints. $S\otimes P$ denotes scalar coupling to DM, pseudoscalar coupling to quarks, etc.} 
  \label{fig:slices_gx0p01}
\end{figure}

\begin{figure}[htbp]
  \centering
  \panelrow{figures/figures_g_mfv/compare_limits_g_lightScalar_mphi1e-09_gx0.01_mfv_sigma}{figures/figures_g_mfv/compare_limits_g_heavyScalar_mphi0.1_gx0.01_mfv_sigma}{figures/figures_g_mfv/compare_limits_g_heavyScalar_mphi1_gx0.01_mfv_sigma}\\
\panelrow{figures/figures_g_mfv/compare_limits_g_lightScalarPS_mphi1e-09_gx0.01_mfv_sigma}{figures/figures_g_mfv/compare_limits_g_heavyScalarPS_mphi0.1_gx0.01_mfv_sigma}{figures/figures_g_mfv/compare_limits_g_heavyScalarPS_mphi1_gx0.01_mfv_sigma}\\
  \caption{As in Figure \ref{fig:slices_gx0p01}, but with quark couplings fixed along the Yukawa direction $\hat{g} \propto [m_u, m_d, m_s]$, $g_\chi = 0.01$. $m_\phi = 10^{-9},\,0.1,\,1$~GeV (left to right); the right-hand pair shares one $y$ axis.}
  \label{fig:slices_mfv}
\end{figure}

\renewcommand{\panelw}{0.5\textwidth}

\begin{figure}[p]
  \centering
  \includegraphics[width=.8\textwidth]{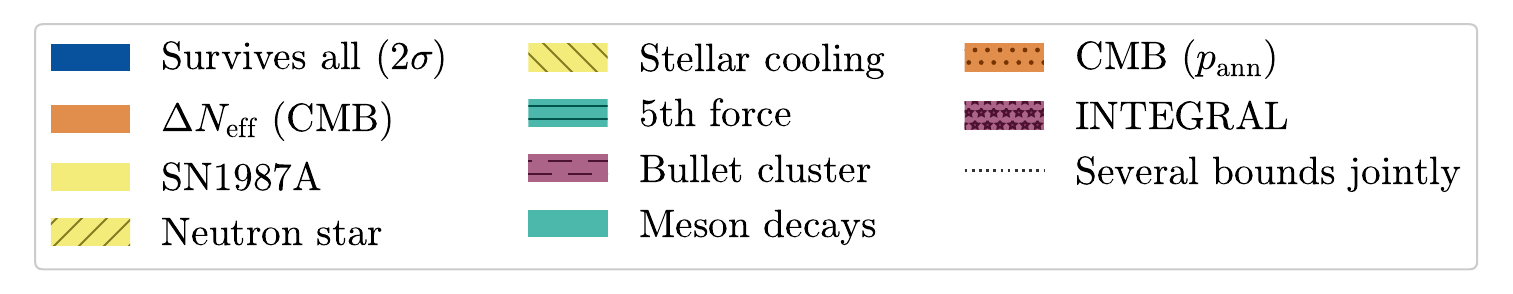}\\
  \includegraphics[width=\panelw]{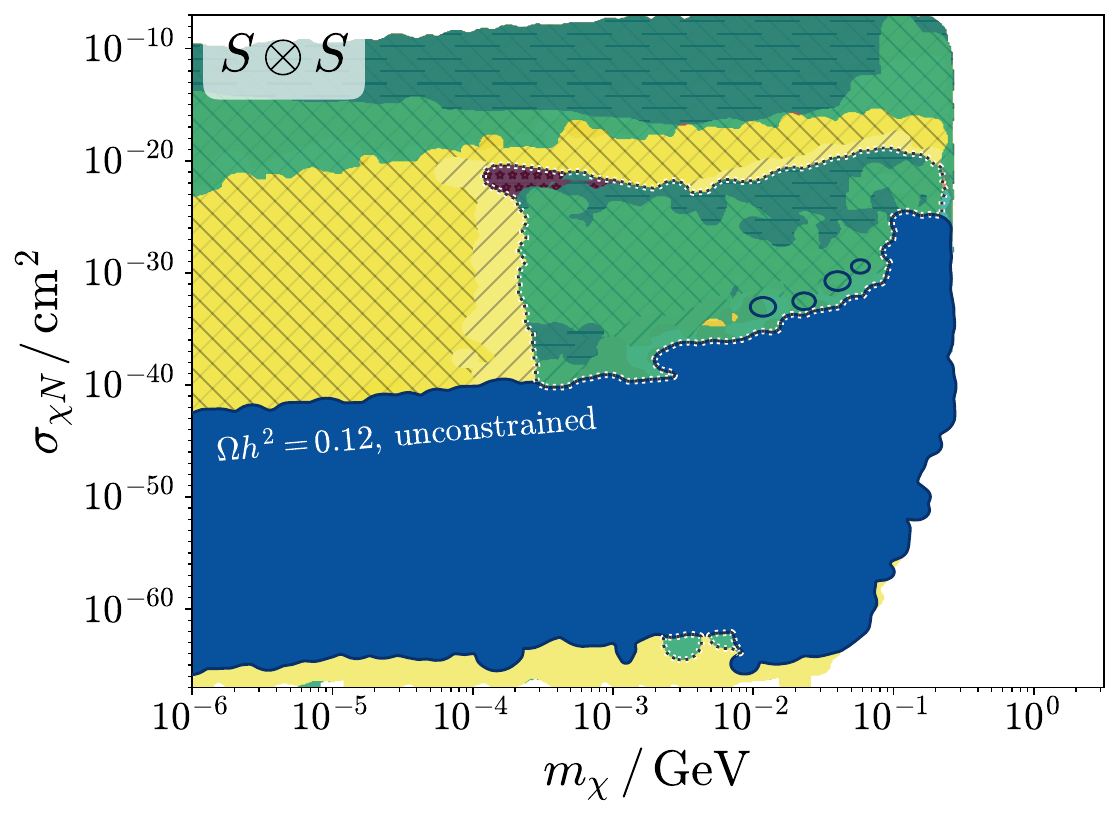}\hfill
  \includegraphics[width=\panelw]{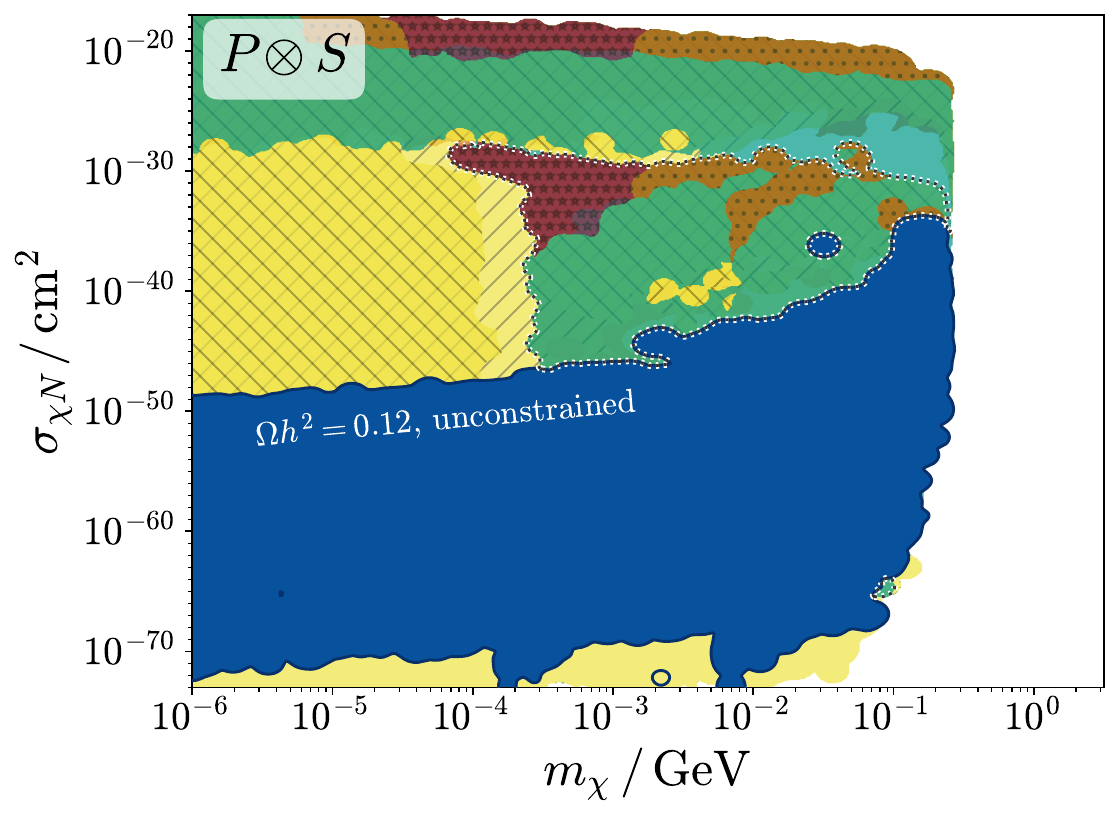}\\[\medskipamount]
  \includegraphics[width=\panelw]{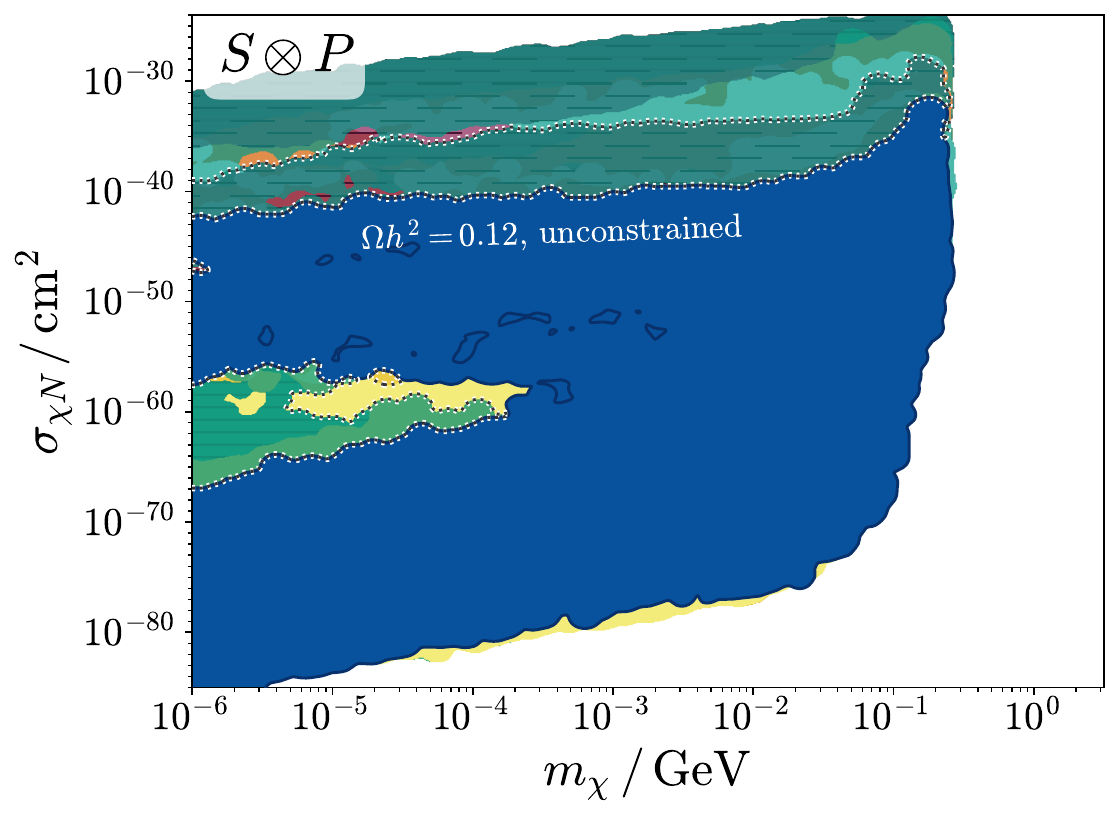}\hfill
  \includegraphics[width=\panelw]{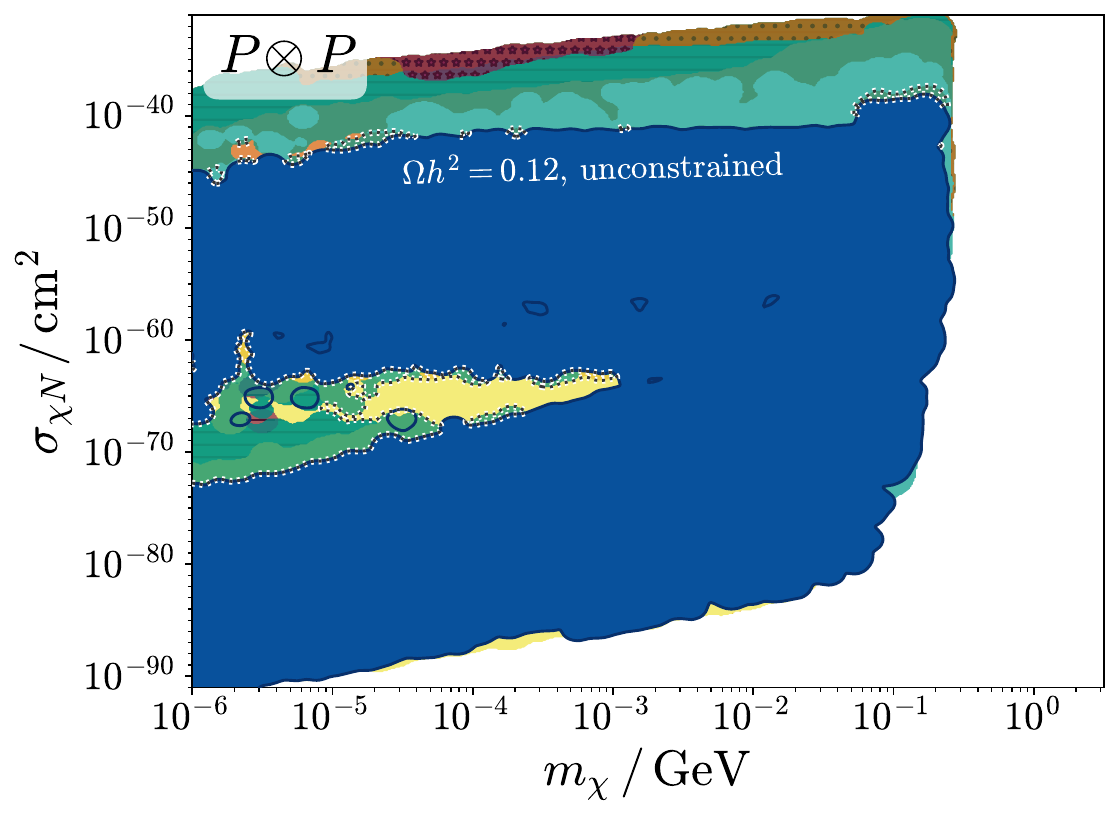}\\[\medskipamount]
  \includegraphics[width=\panelw]{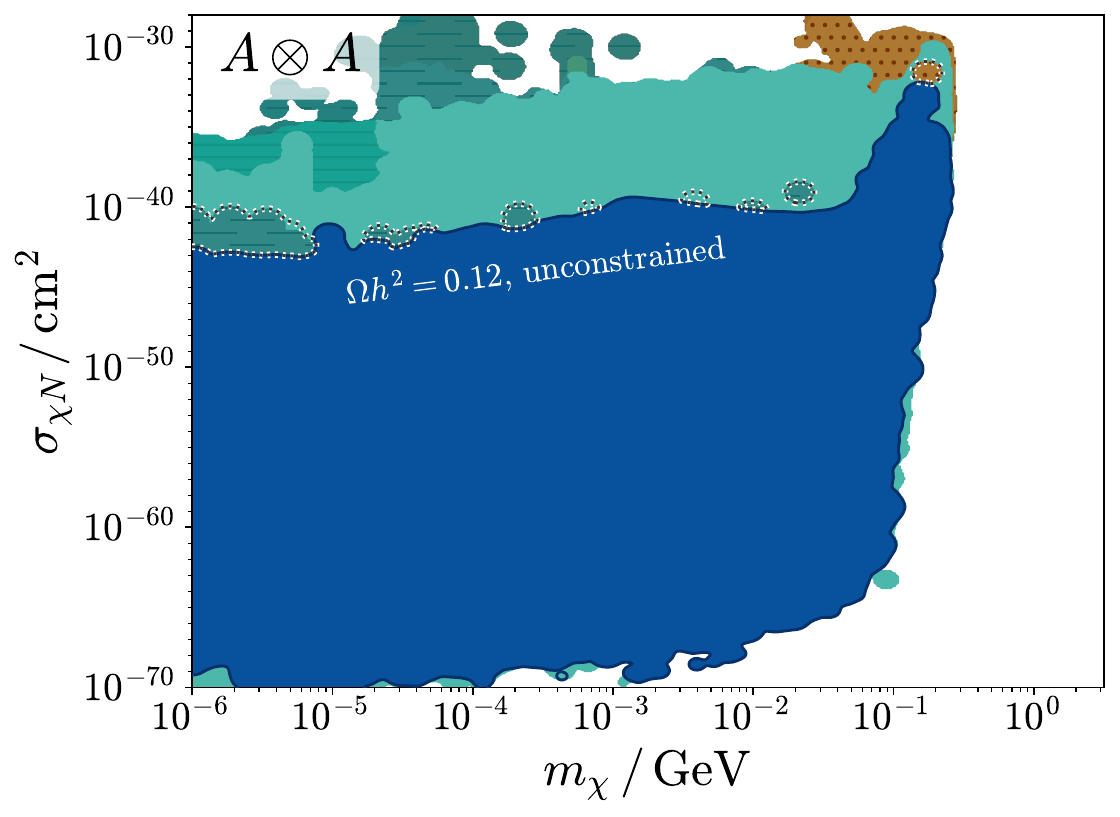}\\[\medskipamount]

  \caption{Parameter space in the $m_\chi$--$\sigma_{\chi N}$ plane that can yield the observed DM relic abundance through thermal processes such as $\chi \chi \leftrightarrow \gamma \gamma$, for $T_{RH} = 10$ MeV, allowing all 6 parameters to vary: \{$m_\chi,m_\text{med}, g_\chi, g_u,g_d,g_s$\}. The non-white regions can yield $\Omega_\chi h^2 = 0.12$, while the blue region denotes space that can survive all other constraints. The watercolor regions indicate which additional constraints rule out each of those points. $S\otimes P$ denotes scalar coupling to DM, pseudoscalar coupling to quarks, etc.}
  \label{fig:mx_sigma_6d}
\end{figure}

\begin{figure}[p]
  \centering
  \includegraphics[width=\textwidth]{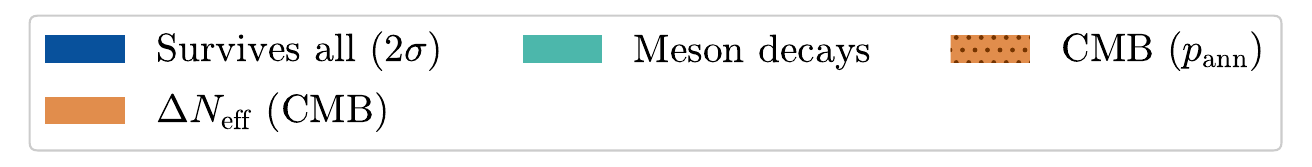}\\
  \includegraphics[width=\panelw]{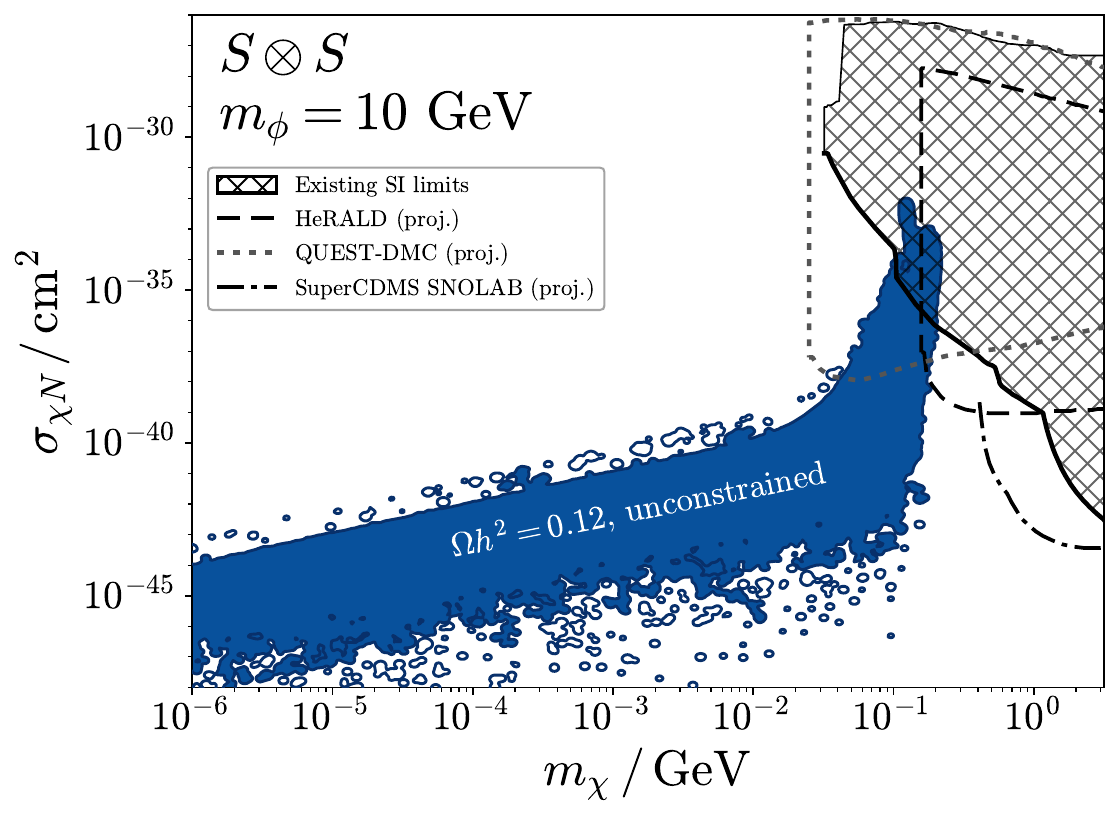}\hfill
  \includegraphics[width=\panelw]{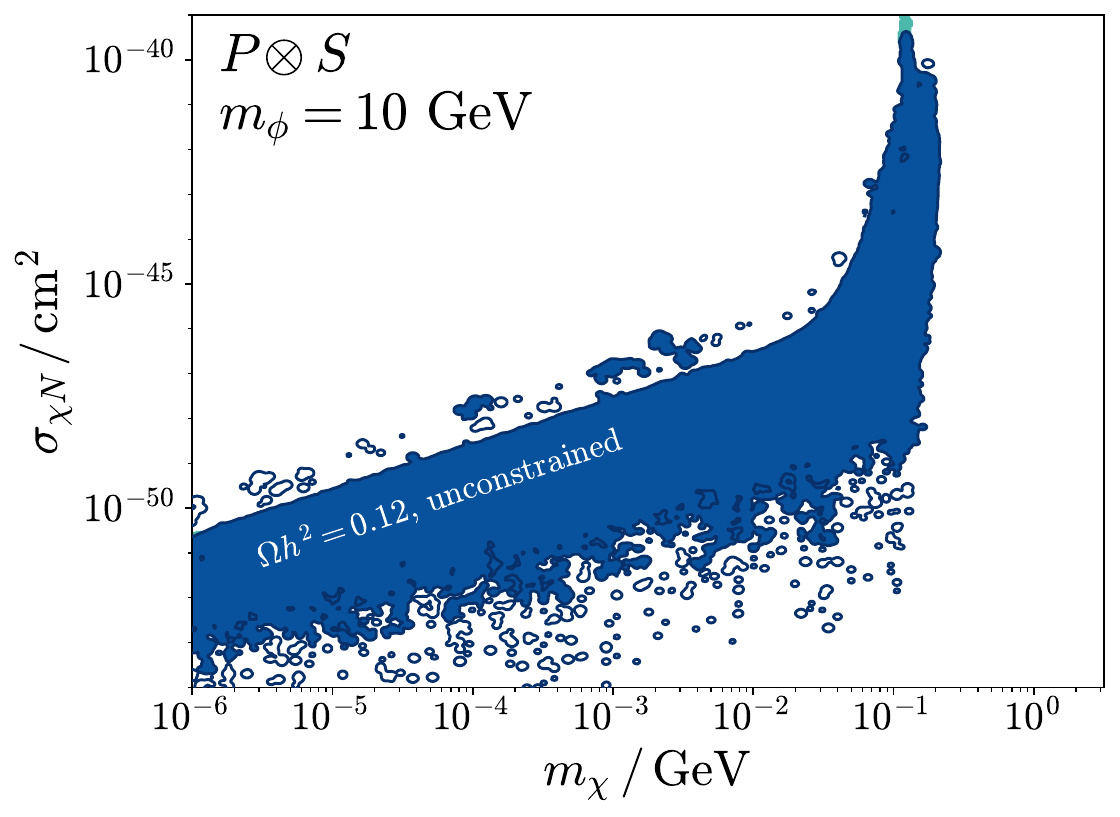}\\[\medskipamount]
  \includegraphics[width=\panelw]{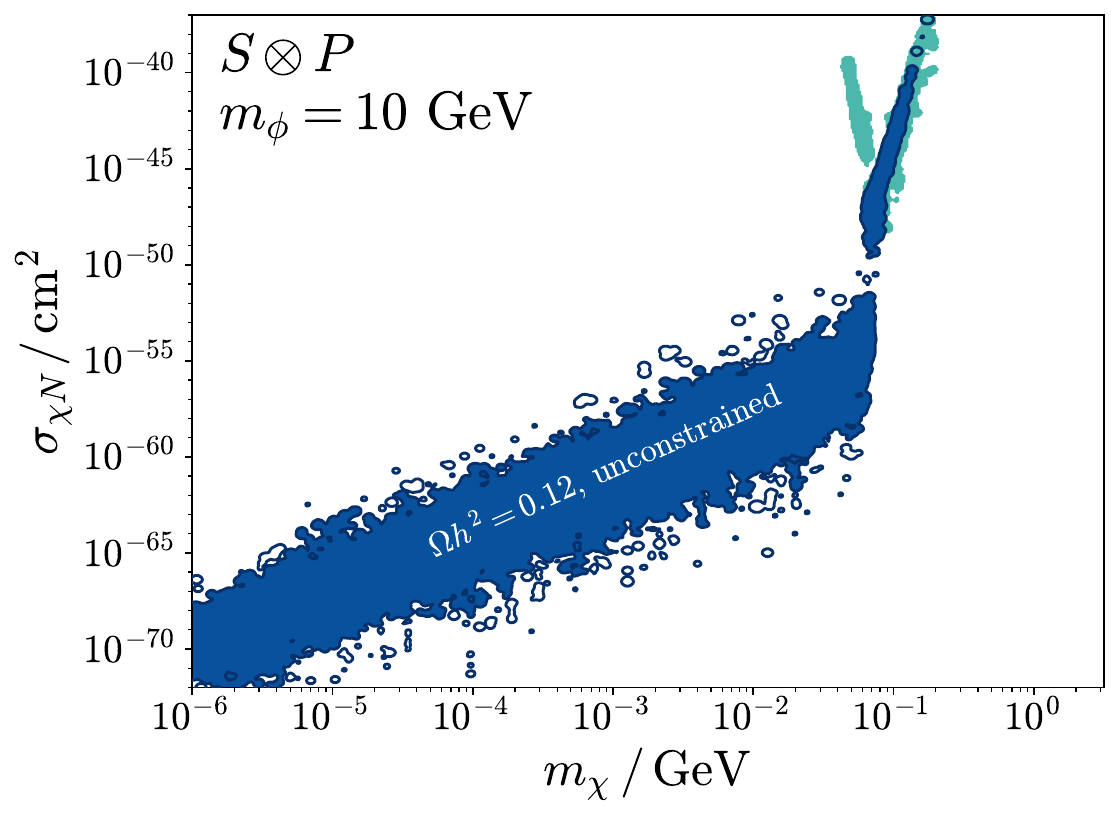}\hfill
  \includegraphics[width=\panelw]{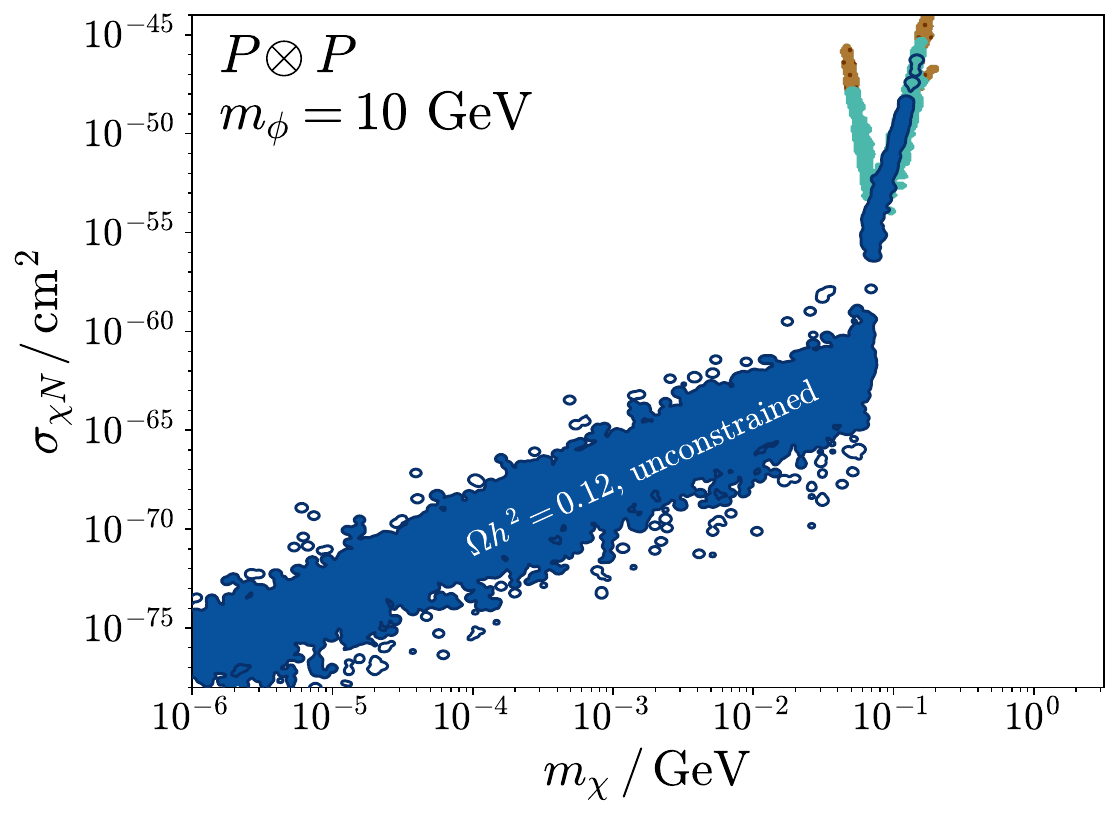}\\[\medskipamount]
  \includegraphics[width=\panelw]{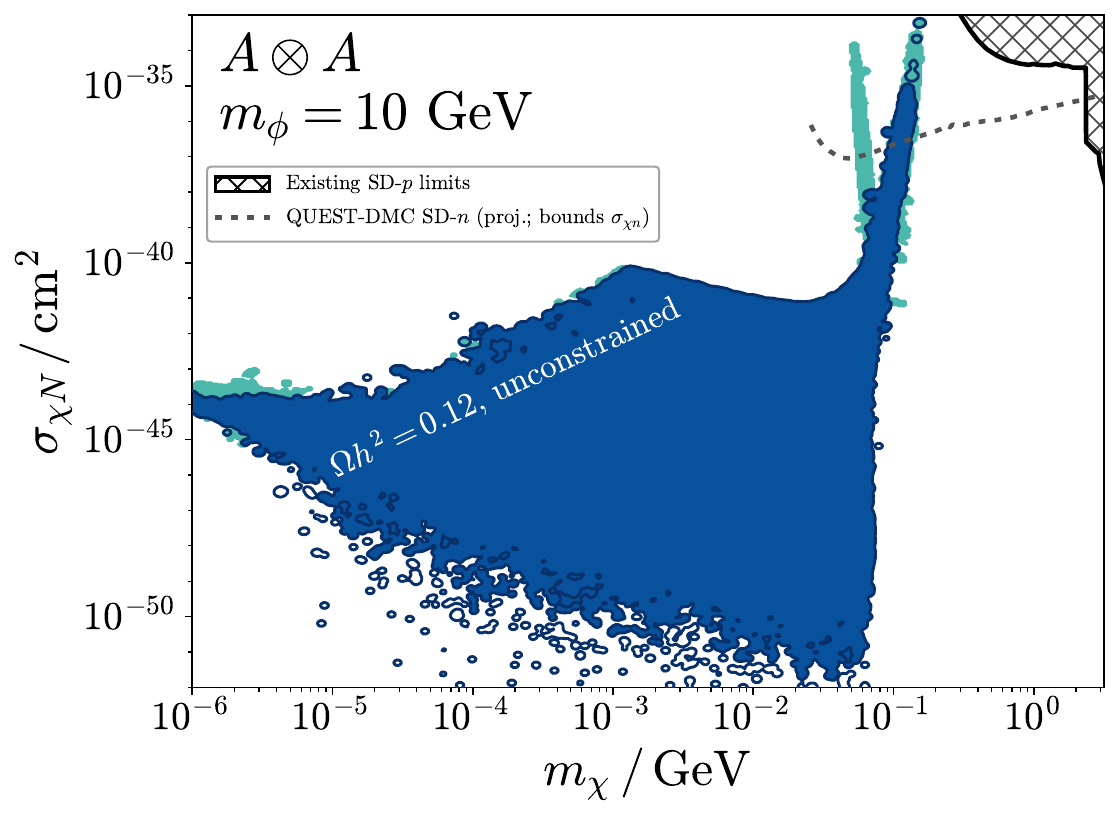}\\[\medskipamount]

  \caption{As Fig.~\ref{fig:mx_sigma_6d}, but for fixed mediator mass $m_{\text{med}} = 10$~GeV. Also shown for $S\otimes S$ are existing direct detection constraints~\cite{CRESST:2017ues,LUX:2018akb,CRESST:2019jnq,CDEX:2019hzn,XENON:2019zpr,CDEX:2021cll,2022PhRvD.106f2004A}  on spin-independent DM-nucleon cross section, and projected sensitivities from SuperCDMS \cite{SuperCDMS:2016wui}, QUEST-DMC \cite{QUEST-DMC:2025miz} and HeRALD \cite{Hertel:2018aal}. Spin-dependent ($A \otimes A$) limits are from \cite{Behnke:2016lsk,SuperCDMS:2017nns,PICO:2017tgi,CRESST:2022dtl}. We note that the QUEST-DMC spin-dependent sensitivity is for $\chi$-neutron scattering, whereas the axis plotted here is for proton scattering. }
  \label{fig:mx_sigma_mphi10}
\end{figure}

\begin{figure}[ht]
  \centering
  
  \includegraphics[width=\textwidth]{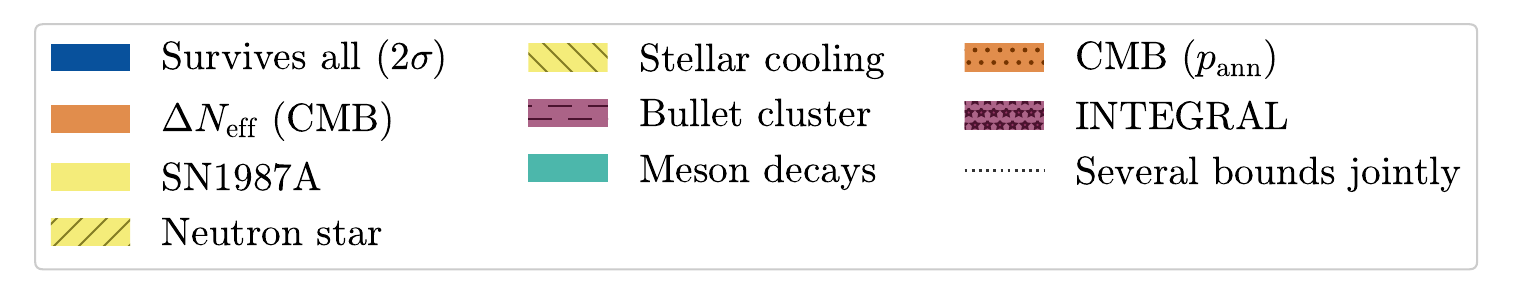}\\
  \includegraphics[width=\panelw]{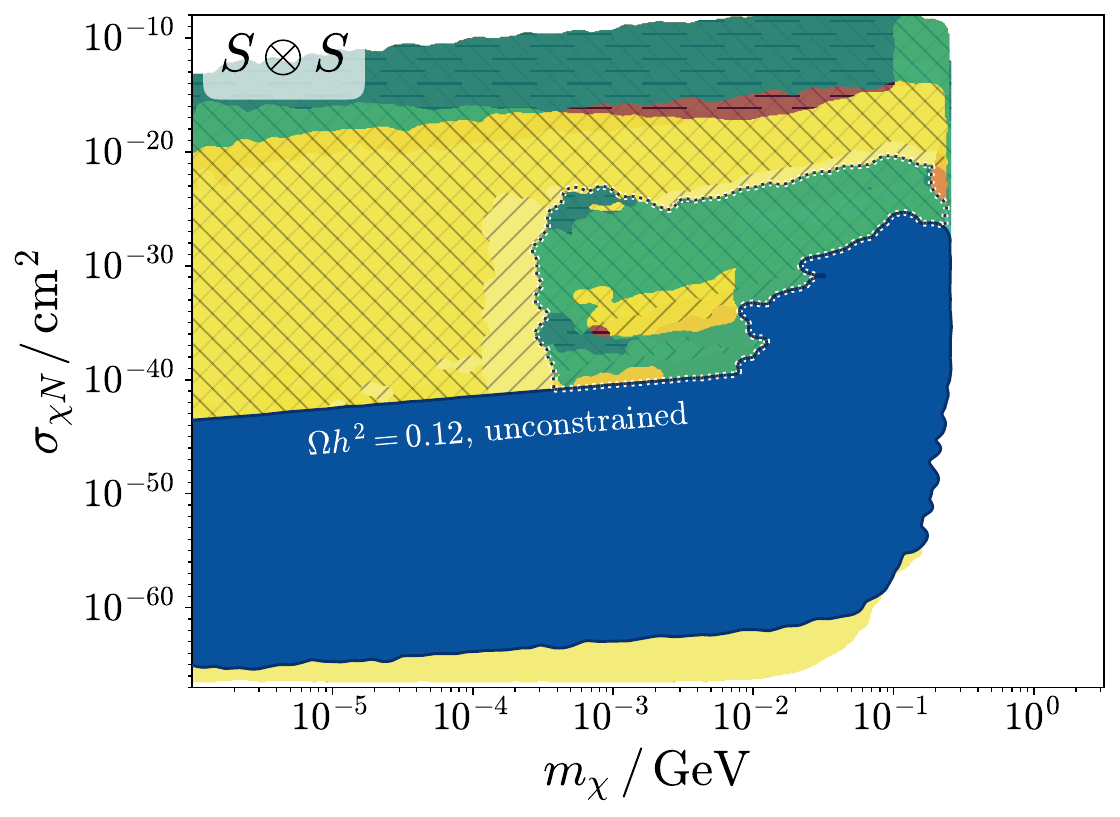}\hfill
  \includegraphics[width=\panelw]{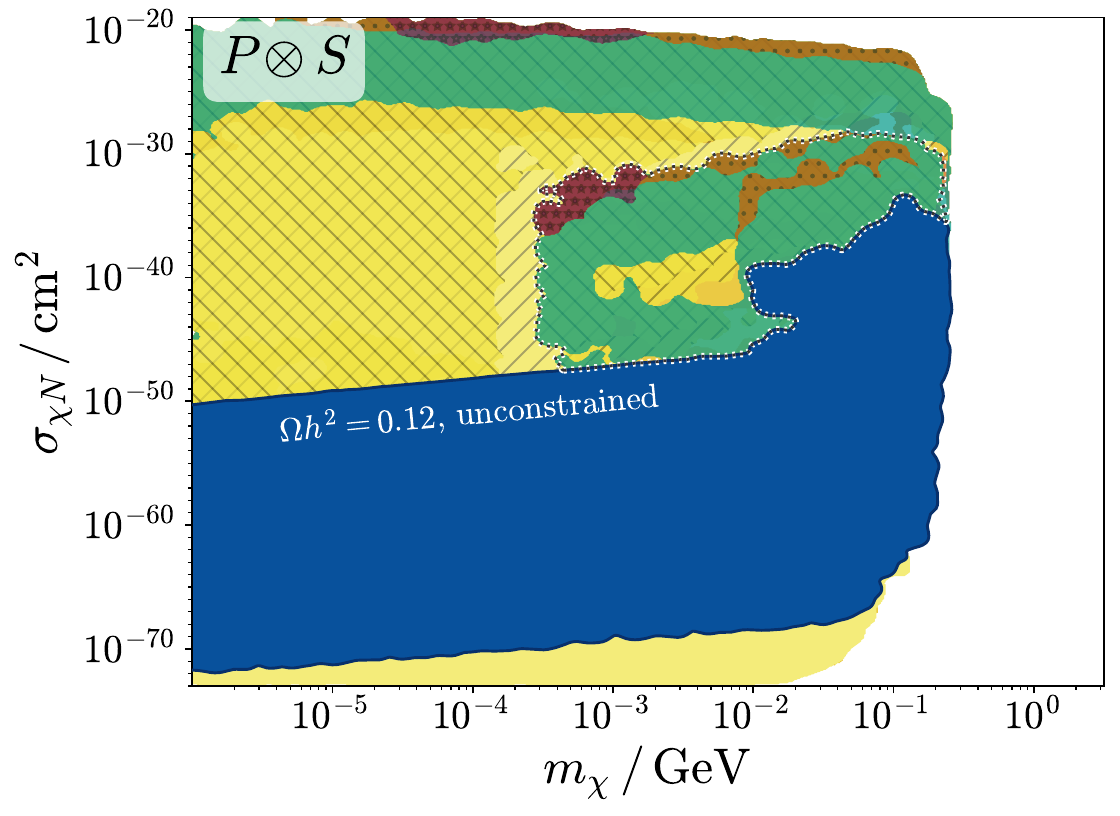}\\[\medskipamount]
  \includegraphics[width=\panelw]{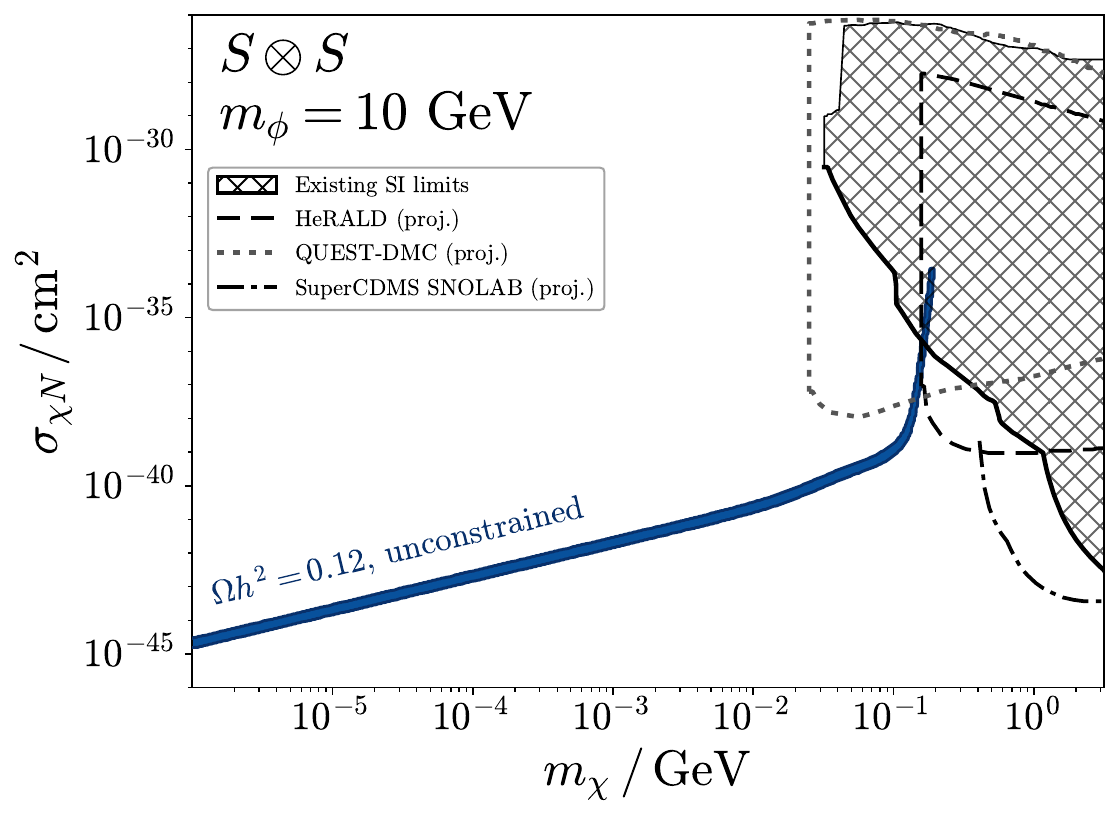}\hfill
  \includegraphics[width=\panelw]{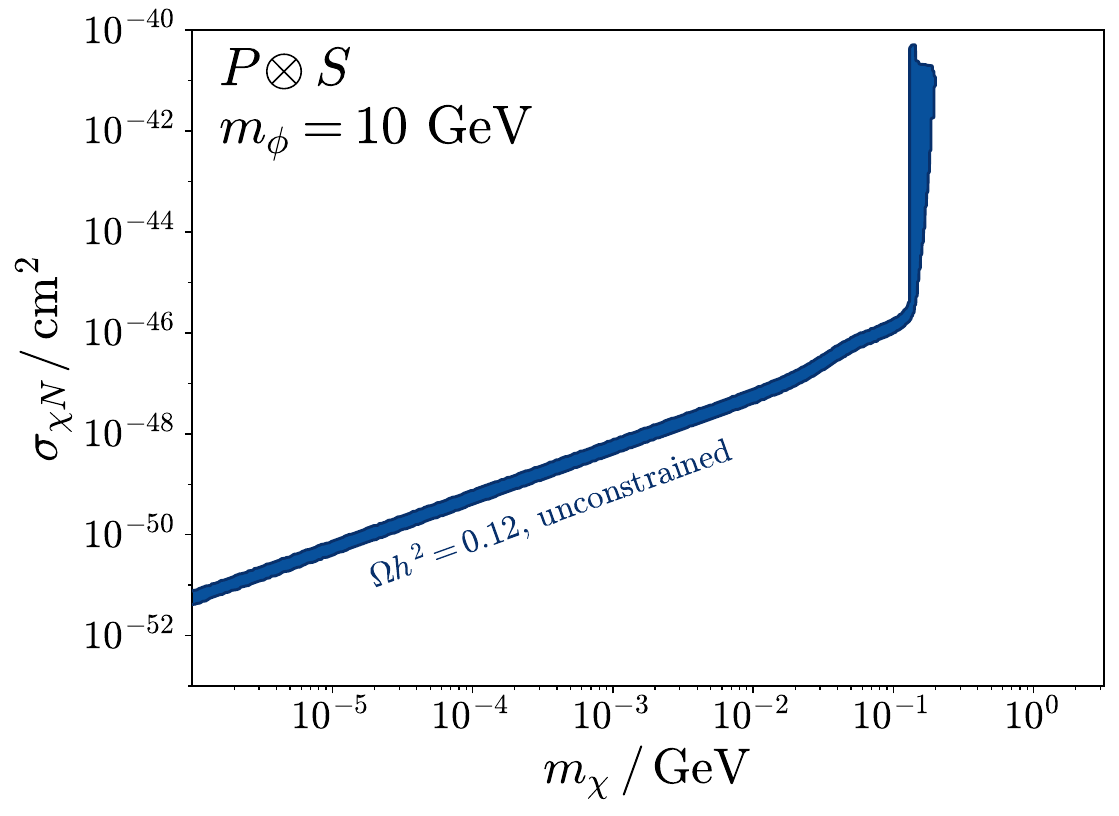}
  \caption{As Fig.~\ref{fig:mx_sigma_6d}, but for mediators with a minimally flavor-violating scalar coupling to quarks, i.e. $g_q \propto$ Yukawa, and varying the normalization, along with the DM coupling $g_{(5)\chi}$ and masses. Top: allowing $1$ eV $\le m_\phi \le 10$ GeV; bottom: fixing $m_\phi = 10$ GeV.}
  \label{fig:mx_sigma_mphi10_mfv}
\end{figure}

\section{Results}
\label{sec:results}
Using these effective interactions and decays we recast several existing limits on DM-photon couplings and DM-meson couplings into limits on  $\sigma_{\chi N}$. All the limits are placed by directly evaluating respective rates from the models themselves; the DM-nucleon scattering cross sections plotted here are derived quantities, defined in Tables \ref{tab:scalar}--\ref{tab:axialvector}. For velocity-dependent cross sections we set $v_{rel} = 220$ km/s. This means that some of the cross sections will be very small ($\ll 10^{-70}$ cm$^{2}$ in some cases). As these are velocity-dependent, they will vary depending on the context in which they are observed, we present them simply to reflect the ability of a direct detection experiment to probe them. 

We considered limits from overproduction of DM by photons in the early universe (Sec.~\ref{sec:relicabundance}), altering $N_{eff}$ (\ref{sec:Neff}), anomalous meson decays (\ref{sec:decays}),  DM annihilation to photons (observed either by INTEGRAL, Sec. \ref{sec:indirect} or through its effect on the CMB, \ref{sec:CMBlims}), DM scattering with CMB photons (\ref{sec:cmbscatter}) as well as fifth force (\ref{sec:fifthforce}), stellar cooling (\ref{sec:stellarlimits}) and  Bullet Cluster limits on the mediators (\ref{sec:bulletcluster}).

We first show how these constraints look when fixing mediator masses and couplings. Consistent with prior results, we will find the parameter space largely ruled out. We will then allow all parameters to vary, resulting not only in reopened parameter space, but also large volumes for which the observed relic abundance is obtained while satisfying all the constraints considered here.\footnote{A public version of the software used to produce these constraints is available at \url{https://github.com/aaronvincent/loophole}.} 

\subsection{Constraints for Fixed Mediator Mass and Couplings}
\label{sec:slices}
We apply these limits to $\sigma_{\chi N}$ in Figs.~\ref{fig:slices_gx0p01}, and \ref{fig:slices_mfv} for specific choices of mediator masses, values of the coupling between the mediator and $\chi$ and ratios between the mediator couplings to the quarks. For the relic density computation, we fix $T_{RH} = 10$ MeV as was done by Ref. \cite{Cox:2025toz}. This is high enough to allow BBN to occur unimpeded, while being as low as possible such that the overproduction is minimized.
Moving from left to right, each grid of plots shows the limits on the cross section for a low mass mediator (1 eV), an intermediate mass mediator (100 MeV), and a high mass mediator (1 GeV).  Moving from top to bottom, each grid shows limits for the $S\otimes S$ model, the $P\otimes S$ model, the $P\otimes P$ model, the $S\otimes P$ model and the $A\otimes A$ model.   In Fig.~\ref{fig:slices_gx0p01} we assume the coupling to the mediator is 0.01 and that the quark couplings are proportional to their electric charges ($g_u:g_d:g_s\propto2:-1:-1$), and in Fig.~\ref{fig:slices_mfv} we assume a mediator coupling of 0.01 with Minimal Flavour Violation (MFV)-inspired couplings, i.e. with couplings proportional to the quark masses, similar to a Higgs portal interaction.  In all of these plots, models in the colored regions are excluded due to one of the limits described above. Only the INTEGRAL and CMB ($p_\text{ann}$) bounds rely on the candidate making up the entire DM mass budget; these limits are always subdominant to others.  In these plots, we vary $\sigma_{\chi N}$ by varying the coupling between the mediator and the quarks, while holding all other variables constant.  This means that the Bullet Cluster limits appearing in these plots apply to arbitrarily small cross sections, because these limits do not depend on the coupling between the mediator and the quarks.

The green line marks where, via freeze-in or freeze-out from the photon bath, the cross section produces the observed relic abundance of DM.\@ The highlighted green region between these lines shows where DM would be overabundant.  Notably, the region that can produce the correct amount of or an excess of DM is much smaller in the light mediator case.  This is because the DM largely annihilates away to the light mediator outside of the green region.  Outside of the green region we could expect DM to be underabundant unless an additional production mechanism outside of freeze-in and freeze-out was introduced for these models. Limits on the mediator that come from production in neutron stars (indigo dash-dotted line),  SN1987a (pink region enclosed by dash-dotted lines), and fifth force experiments (orange dash-dotted line) as well as limits from rare kaon decays (maroon dash-dotted line for $K\rightarrow \pi \chi\bar{\chi}$ and brown dash-dotted line for $K\rightarrow\pi+$med) are independent of the $\chi$ abundance. The $N_{eff}$ bound, marked by an orange dashed line, comes from the contribution of both $\chi$ and the mediator, and self-consistently includes the thermal production in the Early Universe, so does not rely on the DM saturating the relic abundance. Many of the subplots share spike features around  $m_{\chi}=m_{\pi^+}/2$, and around $m_{\chi} = m_{\rm med}/2$, due to resonances at those masses.  

We also mark regions on these plots where the DM model needed to achieve the given cross section is not self-consistent.  This could happen for a number of reasons.  The couplings between the DM and the mediator or the mediator and the quarks needed to reach a large cross section could be large enough to make the model non-perturbative (marked by a gray dotted line).  Large couplings can also cause the mediator's decay rate to become larger than its mass (marked by a dash-dotted purple line), making our description of the mediator in the calculations incorrect.  Neither region should be considered a realistic cross section for the model shown.  
Finally, we draw a dark pink dash-dotted line to show where the small mixing approximation breaks down, corresponding to where any one of the mixing angles reaches a value of 0.3.   Calculations of cross sections and decay rates done in the small mixing approximation will not be reliable above this line. That said, at a given $m_{\chi}$ points located above  the small angle limit line never result in cross sections or decay rates that are smaller than those along the line. 
 This means that bounds that apply along this line also apply to all points above it. 
 We also note that couplings only slightly larger than those that cause the small angle approximation to break down (see Appendices \ref{ap:pseudoscalar} and \ref{app:axialvector} for details), very quickly lead to pathological behavior of the meson masses or DM models, and should be avoided.

Generally speaking, models with light mediators are strongly constrained by the combination of constraints considered here. At low mediator masses, the constraints are mainly on the mediator, from stellar and neutron star cooling and DM self-interaction constraints. Of course, if $g_\chi$ were lower, the Bullet Cluster constraints would relax. These constraints weaken for increasing $m_\phi$, leading to dominant constraints from freeze-in overproduction. The axial vector mediator, on the other hand, remains strongly constrained for all three values of $m_{A'}$. This is \textit{not} because freeze-in production via the $A'$ is more efficient---in fact it is less so than through the scalar and pseudoscalar. It is instead because we are plotting the DM-nucleon scattering cross section, which is even more highly suppressed.

Fig. \ref{fig:slices_mfv} shows the equivalent parameter space under the MFV assumptions, with $g_q \propto$ Yukawas.  The qualitative features are similar to the charge-proportional cases in Fig. \ref{fig:slices_gx0p01}. We note that this is not exactly equivalent to a Higgs-portal model, since we are not including tree-level couplings to leptons. 

Taken together, these plots depict a landscape that is extremely constrained for sub-GeV fermionic DM, especially so in the case of a light mediator.  However it is also clear that the constraints remain very sensitive to model choice, and fixing even one parameter does not tell the whole story.  For example, varying the mediator mass for a given coupling model can significantly change what parts of parameter space are constrained. 

\subsection{Limits on Full Parameter Space Scans}
\label{sec:globalfit}
So far we have presented the constrained sub-GeV DM in slices, fixing the mediator mass and the ratios between the up, down and strange quark couplings. As we shall see, however, freeing these parameters \textit{vastly} opens up the available parameter space. These results are shown in Figs.~\ref{fig:mx_sigma_6d}--\ref{fig:mx_sigma_mphi10_mfv}.

We perform six-dimensional scans over all of the free parameters of each mediator model we consider. We use GAMBIT-light\footnote{GAMBIT-light is a lightweight version of GAMBIT \cite{GAMBIT:2017yxo}, available at \url{https://github.com/GambitBSM/gambit_light_1.0}.} with a combination of the Diver \cite{Martinez:2017lzg} and \texttt{paraprof} \cite{paraprof2025} scanners. Diver is a differential evolution algorithm that excels at finding small features in large parameter volumes, while \texttt{paraprof} is tuned to explore parameter spaces more widely than normal minimizers. We first allow the DM and mediator masses to vary, along with the couplings $g_\chi, g_u, g_d$ and $g_s$. We scan $m_\chi\in[10^{-6},1]$\,GeV, $m_{\rm med}\in[10^{-9},10]$\,GeV (or fixed $m_{\rm med}=10$\,GeV), and $g_\chi\in[10^{-13},1]$ ($[10^{-16},1]$ for P$\otimes$P), all with log-flat priors; the quark couplings $g_{u,d,s}\in[-1,1]$ are scanned independently with a signed-log prior that is log-flat in $|g_q|$ on either side of zero and flat across $|g_q|<10^{-11}$ ($10^{-16}$ for P$\otimes$P). For A$\otimes$A the same ranges and priors apply to the axial-vector mediator mass $m_{A'}$, the dark coupling $\lambda_\chi$ and the axial quark couplings $\lambda_{u,d,s}$.
 
Our scan targets a likelihood function that only includes the relic abundance, so that the points that we keep are all capable of producing $\Omega_\chi h^2 = 0.12$ within two standard deviations. This step amounts to a two-dimensional profile-likelihood of the 6D space. We then check whether these combinations of parameter are ruled out by any of the limits described above, and project our constraints down to the $\sigma_{\chi N}-m_\chi$ plane.  If all combinations of parameters that could produce a given value of $\sigma_{\chi N}$ are ruled out, we consider that model ruled out, and shade that area in with the appropriate constraint. The orange region is excluded by $\Delta N_{\rm eff}$ and the yellow region by SN1987A; within the yellow, forward and backward hatching mark the neutron-star and stellar-cooling bounds, respectively. Fifth-force searches exclude the dark green, horizontally striped region, and the Bullet Cluster the pink region with dashed stripes. Anomalous meson decays rule out the turquoise region. DM annihilation is constrained by CMB distortions (orange, dotted) and by INTEGRAL X-ray observations (pink, dotted).  A dotted line outlines regions where no single constraint alone rules the entire viable model space out, but rather several constraints working together rule out the entire viable model space.  Outside of the dotted outline, the available model space is ruled out by the single limit marked on the plots. If any combination of parameters for a given value of $\sigma_{\chi N}$ is not ruled out and leads to the correct relic abundance of DM, we treat that value of $\sigma_{\chi N}$ as not ruled out; those areas are shown as a shaded blue region, labeled ``$\Omega_\chi h^2 = 0.12$, unconstrained''. Anything in the white outer region cannot yield the observed relic abundance from thermal processes alone, at the reheat temperature we consider. The blue region should not be taken to mean that all models that produce these cross sections are allowed; rather, it means that there are at least some parameter values that yield the correct relic abundance and have that DM mass and cross section, and do not run afoul of other constraints. This means that the blue regions represent viable targets for DM searches.

As in Figs.~\ref{fig:slices_gx0p01}, \ref{fig:slices_mfv}, we limit ourselves to healthy theories: models that are perturbative, with mixings between the mediator and the mesons that are small enough to be rotated away, and where the mediator mass is larger than its decay rate. Fig.~\ref{fig:mx_sigma_6d} shows 6D scan results for all five coupling models considered in this work. The available parameter space where some model exists that produces the correct cross section and relic abundance without being constrained by any limits considered in this work is surprisingly large, and much more open than would appear from considering the slice plots that focus on more specific choices of parameters. We note that most small-scale uneven features seen here are simply due to the sampling and binning method. We describe these results below; Appendix \ref{ap:relic} shows detailed figures of which channels lead to the majority of the $\chi$ abundance in each case. 

The majority of the allowed regions are produced from freeze-in. For scalar and pseudoscalar couplings to quarks, the region is allowed to go very low in coupling thanks to resonant production of the mediator from inverse-decay, $\gamma \gamma \rightarrow \phi$ and $\pi^+ \pi^- \rightarrow \phi$, followed by decay into $\bar \chi \chi$. When couplings are large enough, direct annihilation of SM particles into $\chi$ pairs begins to dominate as the mediator lifetime is suppressed. 

For scalar quark couplings (top panels), reannihilation, where the full Boltzmann equation is required, starts to be relevant in the very top of the blue region, with a few valid points produced by freeze-out alone in the high mass, high cross section region. 

There are two distinct branches in the pseudoscalar-coupled case (middle panels). The bottom branch is almost entirely produced from freeze-in, with very little reannihilation. Again, the top of that band is regular off-resonance freeze-in, while the bottom 10 decades can still yield sufficient DM thanks to resonant production of the mediator, followed by decay. The top branch, in contrast, mainly populates $\phi$ from the Primakoff-like process $e^\pm \gamma \rightarrow e^\pm \phi$, followed by $\phi \phi \rightarrow \bar \chi \chi$ ($e^+e^- \rightarrow \phi \gamma$ also occurs, but is subdominant by two orders of magnitude). At the very top, freeze-out from $\bar \chi \chi \rightarrow \phi \phi$ is the main production mechanism.

DM production in the case of an axial vector portal is almost entirely freeze-in, with the $A'$ first brought into equilibrium with $\pi^\pm A' \leftrightarrow \pi^\pm \pi^0$, followed by $A'$ decay to DM pairs. At the very top, direct annihilation of photons into $\bar \chi \chi$ via the neutral mesons that mix with the $A'$ dominates production. 

Overlaid over the relic abundance are the various external constraints that limit these scenarios. Astrophysical bounds (SN1987A, neutron star and stellar cooling) shown in yellow, bracket the parameter space, and are especially strong for large cross sections generated by low mediator masses. A combination of meson decays, stellar cooling and neutron star cooling limit the high mass region. Very large cross sections are only attainable with mediator masses and couplings that violate fifth force experiments (scalar mediator-$\chi$ couplings) or X-ray and CMB bounds (pseudoscalar mediator-$\chi$ couplings, thanks to $s$-channel annihilation). The axial vector model is mainly bounded from above by meson decay. 

Fig.~\ref{fig:mx_sigma_mphi10} shows the equivalent results, but in the limit of a heavy mediator, with fixed $m_\phi = 10$ GeV. In this case, available parameter space is much more restricted, but still includes viable DM candidates from $\sim$ keV to a few hundred MeV in mass. Most of the parameter space that can yield the observed relic density survives other bounds, since the most constraining ones in Fig.~\ref{fig:mx_sigma_6d} were mainly due to restrictions on light mediators (or e.g. in the case of the Bullet Cluster, large interaction rates due to the small mediator mass). Since these models are produced by freeze-in, they do not contribute meaningfully to the relativistic plasma at neutrino decoupling, so for most of the parameter space are not constrained by $N_{eff}$ either. At large mediator masses, most of the parameter space is produced from pion annihilation for scalar quark couplings, except at larger couplings where photonic freeze-in is dominant. For pseudoscalar and axial quark couplings, the relic abundance at high mediator mass is entirely generated by $\gamma \gamma \rightarrow \bar \chi \chi$. The V-shape at the top of these figures surrounds the $2 m_\chi = m_{\pi^0}$ resonance. Invisible pion decay bounds constrain the left half of the V. The right half remains partially constrained instead by the bound on $\eta \rightarrow$ invisible.

Next, we restrict the coupling ratios, focusing again on MFV Yukawa-like scalar couplings to quarks, shown in Fig.~\ref{fig:mx_sigma_mphi10_mfv}.  These four-dimensional scans (over DM and mediator masses, $g_\chi$ and normalization of the quark couplings) yield qualitatively very similar results to the cases where the quark couplings were free to vary. The slightly smoother boundaries are not a physical effect, they are just due to the lower-dimensionality of the scan allowing for better sampling. When $m_\phi$ is also fixed, the relic-abundance region is fully specified, leading to a thin band in the bottom panels of Fig. \ref{fig:mx_sigma_mphi10_mfv}. The allowed region lies on the plane 
\begin{equation}
  g_\chi \, g_q \simeq 2\times10^{-4}
  \left(\frac{m_\phi}{10\,\mathrm{GeV}}\right)^{2}
  \left(\frac{m_\chi}{\mathrm{GeV}}\right)^{-1/2}
  \left(\frac{\Omega h^2}{0.12}\right)^{1/2}
  \qquad (T_\mathrm{RH} = 10\,\mathrm{MeV},\ m_\chi \lesssim 10\,\mathrm{MeV}),
\end{equation}
as expected from UV-dominated freeze-in through the contact limit
$\sigma_{\gamma\gamma\to\chi\chi}\propto g_\chi^2\,\mathcal{D}\,E^4/m_\phi^4$ with
$\mathcal{D}\propto g_q^2$, where $g_q$ is the scaling factor of the quark couplings. Above $m_\chi \sim 10$\,MeV the band bends upward, as Boltzmann suppression at $T_\mathrm{RH}$ sets in, partly offset by meson-pair production.

\section{Conclusions}
\label{sec:conclusion}
This work extends the study of the effective and loop level interactions that result from hadrophilic dark matter models carried out in Refs.~\cite{Diamond:2023fsm, Cox:2024rew, Cox:2025toz}.   Ref.~\cite{Diamond:2023fsm} considered how hadrophilic DM that interacted with quarks via a vector coupling.  In this case the DM's strongest effective interaction is with electrons making these models observable at electron recoil detectors. Ref.~\cite{Cox:2024rew} used an EFT formalism to explore the many phenomenological consequences of hadrophilic DM that interacts with quarks via a scalar coupling involving a heavy mediator.  Ref.~\cite{Cox:2025toz} extended this to include vector, pseudoscalar and axial vector couplings via a heavy mediator. In contrast, we have allowed the mediator mass to vary below the UV cutoff, yielding rich new phenomenology. Allowing for light mediators with distinct couplings to DM versus quarks greatly impacts the predicted DM phenomenology. Decay rates and effective interactions with photons and mesons calculated for this work are presented in Tables \ref{tab:scalar}, \ref{tab:psuedo+scalar}
, \ref{tab:pseudoscalar}, \ref{tab:pseudoscalar+scalar}, and \ref{tab:axialvector}  in a model-independent manner, such that the reader may calculate the effective interactions and resulting limits for hadrophilic DM models of their choosing.  

These effective couplings can dramatically change the phenomenology of seemingly simple hadrophilic DM.\@ Through its effective coupling to photons, hadrophilic DM can be abundantly produced by photon annihilation in the early universe, as well as other processes which remain relevant even for the reheat temperature of 10 MeV that we have chosen here to preserve BBN.  Fixing the mediator mass and the ratio of couplings, it is easy to find that many hadrophilic DM models are ruled due to overproduction from this freeze-in process. New effective interactions also generate many new constraints on hadrophilic DM. The effective coupling to photons can cause hadrophilic DM to modify $N_{eff}$, annihilate to photons (which can produce excess observable X-rays and also distort the CMB), or wash out CMB perturbations though scattering with photons.  The coupling to light mesons produces several new mesons decay channels which are constrained by measured meson decay rates.  Together, these photon and meson observables place many new limits on hadrophilic DM models, particularly in the sub-GeV mass range. This can be seen in Figs. \ref{fig:slices_gx0p01} and \ref{fig:slices_mfv}.

However, by freeing these model parameters, we have shown that vast areas of the parameter space are not only allowed, but can also produce the observed DM relic abundance while evading other constraints. The 6D free parameter scans in Fig \ref{fig:mx_sigma_6d}, as well as the minimal flavor violating scans in Fig \ref{fig:mx_sigma_mphi10} show where viable DM models remain and can provide a new set of benchmark targets for DM experiments.


Moving forward, we intend for this study to be a useful tool for those developing and testing hadrophilic DM models, and a guide to help experimental searches find viable DM targets.  This work, along with Refs.~\cite{Diamond:2023fsm, Cox:2024rew,Cox:2025toz} demonstrates how even minimal hadrophilic DM models naturally have interactions with other Standard Model particles which cannot be ignored when studying their phenomenology.  Future work could consider the effective interactions of hadrophilic DM at high energies, or how effective and loop-level interactions can produce novel phenomenology for other types of DM, such as scalar DM or neutrinophilic DM.

  \acknowledgements
We thank Joachim Brod, James Cline, Simon Caron-Huot, Adrian Liu, and Brooks Thomas for helpful discussions relevant to this work. MDD was supported by the Arthur B. McDonald Canadian Astroparticle Physics Research Institute and the Canada Foundation for Innovation.  MDD was supported in part by the Natural Sciences and Engineering Research Council of Canada (NSERC). CVC was generously supported by Washington University in St. Louis through the Edwin Thompson Jaynes Postdoctoral Fellowship. ACV and NAK were supported by NSERC.\@ Computations were performed at the Queen's Centre for Advanced Computing.  The work made use of SymPy~\cite{10.7717/peerj-cs.103}  SciPy~\cite{2020SciPy-NMeth}, Julia \cite{bezanson2017julia} and Matplotlib~\cite{Hunter:2007}. We used Claude Opus to code up our computed rates into likelihood functions, interface with GAMBIT, produce plotting scripts, and check for errors in equations and text. The authors are responsible for the content of the work. 

\bibliography{main.bib}
\appendix
\section{Meson Mixing, Rotation Angles, and Rotated Parameters for the  Pseudoscalar Quark Coupling}
\label{ap:pseudoscalar}

When an external scalar couples to quarks via
\begin{equation}
    \mathcal{L} \supset\text{ } \sum_q -i\tilde{g}_q\phi q\gamma^5\bar{q}\,.
\end{equation}
The interaction can be described in the low energy limit, described by the chiral effective Lagrangian
\begin{equation}
    \mathcal{L}_{ChEFT}\supset\frac{F^2}{4}\text{Tr}\left(\zeta^{\dagger}U+\zeta U^{\dagger}\right)+L_6\text{Tr}\left(U^{\dagger}\zeta+\zeta^{\dagger}U\right)^2+L_7\text{Tr}\left(U^{\dagger}\zeta-\zeta^{\dagger}U\right)^2
\end{equation}
where $U= e^{\frac{i}{F}\bf{\pi}}$ contains the mesons, $F = 92.4$ MeV \cite{holstein_1990} is the pion decay constant, $\zeta=2B_0 (\mathcal{M}+\textbf{g}_q\phi+i\tilde{\textbf{g}}_q\phi)$, $\mathcal{M}=$diag$(m_u,m_d,m_s)$, $\textbf{g}_q$ = diag($g_u$,$g_d$,$g_s$), $\tilde{\textbf{g}}_q$ = diag($\tilde{g}_u$,
$\tilde{g}_d$,$\tilde{g}_s$), $L_6=1.46\times10^{-3}$, $L_7=-0.39\times10^{-3}$ \cite{Bijnens_2014}, 
 $B_0 = \frac{m_{\pi^+}^2}{m_u+m_d}\simeq2.8$ GeV, where  $m_{\pi^+}$ is the charged pion mass, and $m_q$ represent the quark masses. 
\begin{equation}
\label{pmatrix}
    \pi= 
    \begin{pmatrix}
    \pi^0+\frac{\eta_8}{\sqrt{3}}+\frac{\sqrt{2}}{\sqrt{3}}\eta_0&\sqrt{2}\pi^+ &\sqrt{2}K^+ \\ \sqrt{2}\pi^-&-\pi^0+\frac{\eta_8}{\sqrt{3}}+\frac{\sqrt{2}}{\sqrt{3}}\eta_0 &\sqrt{2}K^0\\
    \sqrt{2}K^-&\sqrt{2}\bar{K}^0 & -\frac{2}{\sqrt{3}}\eta_8+\frac{\sqrt{2}}{\sqrt{3}}\eta_0
    \end{pmatrix}\,.
\end{equation}\,
We note that the $\eta_0$ is not usually included with the octet of mesons presented above, however it is relevant when a mediator couples to quarks via a pseudoscalar or axial vector coupling, so we include it here.  Expanding out the relevant interaction terms in the Lagrangian we can see that the $\phi$ picks up mass mixing terms with the $\pi^0$, $\eta$, and $\eta'$ of the form
\begin{equation}
    \mathcal{L}\supset m_{\pi^0\phi}^2\phi_{b}\pi^0_b+ m_{\eta \phi}^2\phi_{b}\eta_b+m^2_{\eta'\phi}\phi\eta'_b
\end{equation}
where 
\begin{equation}
\label{mjphi}
    \begin{split}
        m_{\pi^0\phi}^2&=(\tilde{g}_u -\tilde{g}_d)\left(B_0 F+\frac{32B_0^2L_6}{F}\sum_i^{u,s,d}m_i\right)+\frac{32B_0^2L_7}{F}(\tilde{g}_u+\tilde{g}_d+\tilde{g}_s)(m_u-m_d)\\
m_{\eta\phi}^2&=\frac{(\tilde{g}_u +\tilde{g}_d-2\tilde{g}_s)}{\sqrt{3}}\left(B_0 F+\frac{32B_0^2L_6}{F}\sum_i^{u,s,d}m_i\right)+\frac{32B_0^2L_7}{\sqrt{3}F}(\tilde{g}_u+\tilde{g}_d+\tilde{g}_s)(m_u+m_d-2m_s)\\
m^2_{\eta'\phi}&=\frac{\sqrt{6}(\tilde{g}_u+\tilde{g}_d+\tilde{g}_s)}{3}\left(B_{0}F+(L_6+L_7)\frac{32B_0^2}{   F}\sum_i^{u,s,d}m_i\right)\,.\\
    \end{split}
\end{equation}
For these calculations we treat the $\eta\simeq \eta_8$ and $\eta'\simeq\eta_0$.  

We must remove the above mixing terms to proceed with calculations. We can do this by rotating to a basis that does not contain these terms.  We mark the particles in the mixed basis with a ``b'' to denote that these are the bare particles.  The rotated basis without mixing terms corresponds to the propagating basis.  We assume the particles in this rotated basis correspond to those observed in experiments.  We apply the rotation matrix $R_{\phi}$ to shift between the starting or ``bare'' basis and the propagating basis with the mixing terms removed
\begin{equation}
    \begin{pmatrix}
   \phi \\ \pi^0 \\
    \eta \\ \eta'
    \end{pmatrix}= R_{\phi}
    \begin{pmatrix}
   \phi_{b}\\ \pi^0_b \\
    \eta_b\\\eta'_b
    \end{pmatrix}\,.
\end{equation}
Here
\begin{equation}
\label{eq:rotationmatrix}
    R_{\phi}={}^4R_{12}(\alpha){}^4R_{13}(\beta){}^4R_{14}(\gamma){}^4R_{23}(\delta){}^4R_{24}(\zeta){}^4R_{34}(\kappa)\,,
\end{equation}
where for example
\begin{equation}
\label{eq:rotationangles}
\begin{split}
    {}^4R_{12}(\theta) &= 
    \begin{pmatrix}
        \cos(\theta) &-\sin(\theta)&0&0\\
        \sin(\theta)& \cos(\theta)&0&0\\0&0&1&0\\
        0&0&0&1
    \end{pmatrix}\,.\\
\end{split}
\end{equation}
The lower indices denote the positions in the matrix that mix, i.e. 12 indicates the mixing between the first and second particles.  The upper index describes the number of rows and columns in the matrix.  All of the other rotation matrices follow the pattern of the one shown above.  
As we assume the propagating basis is the observable basis we take the observed masses of the $\pi^0$, $\eta$, and $\eta'$ as the correct meson masses in the rotated basis, while the masses of the unrotated  states are left as a free parameter.  The mass of the mediator $\phi_{b}$ as it appears in the Lagrangian is its mass in the unrotated basis as calculations of the $\phi_{b}$ couplings to nucleons based on its couplings to quarks in \cite{Dienes:2013xya} do not consider mixing with light mesons in their coupling estimates.  The rotation angles $\alpha$, $\beta$, and $\gamma$ are found by solving
\begin{equation}
    \begin{pmatrix}
       m_{\phi}^2&0&0&0 \\
        0& m_{\pi^0}^2&0&0\\
        0&0&m_{\eta}^2&0\\0&0&0&m_{\eta'}^2
    \end{pmatrix}
    =R_{\phi}\begin{pmatrix}
       m_{\phi_{b}}^2 &\frac{m_{\pi^0 \phi}^2}{2}&\frac{m_{\eta \phi}^2}{2}&\frac{m_{\eta' \phi}^2}{2} \\
        \frac{m_{\pi^0 \phi}^2}{2}& m_{\pi^0b}^2 &0&0\\
        \frac{m_{\eta \phi}^2}{2}&0&m_{\eta b}^2&0\\
        \frac{m_{\eta' \phi}^2}{2}&0&0&m_{\eta' b}^2
    \end{pmatrix}R_{\phi}^T\,.
\end{equation}
In this $\alpha$, $\beta$, $\gamma$, $\delta$, $\zeta$, $\kappa$, $m_{\pi^0b}$, $m_{\eta b}$, $m_{\eta'b}$, and $m_{\phi}$ are unknown.  There is not a tractable closed form analytic solution for these masses and mixing angles.  Instead they must be calculated numerically based on specific values of $m_{\phi_{b}}$, $\tilde{g}_u$, $\tilde{g}_d$, and $\tilde{g}_s$.  We have assembled a numerical solver to use for this purpose~\cite{avis_kozar_2026_24508201}. The values of $\alpha$, $\beta$, $\gamma$, $\delta$, $\zeta$, $\kappa$ will be important for calculating decay rates and cross sections. The values of $m_{\pi^0b}$, $m_{\eta b}$, $m_{\eta' b}$, and $m_{\phi}$ do not play a direct role in estimating cross sections, but do provide an important check to the reliability of the calculation.  $m_{\pi^0b}$, $m_{\eta b}$, $m_{\eta' b}$, and $m_{\phi}$ should all be positive and real.  If they are not, then the large mixing has made the theory unstable.  

In the limit that $\mathbf{|m_{\pi^0 \phi}^2|\ll|m_{\phi}^2-m_{\pi^0}^2|}$ \textbf{and} $\mathbf{|m_{\eta \phi}^2|\ll|m_{\phi}^2-m_{\eta}^2|}$  \textbf{and} $\mathbf{|m_{\eta' \phi}^2|\ll|m_{\phi}^2-m_{\eta'}^2|}$ we can approximate the values of the mixing angles without doing the full numeric calculation.  In the small mixing limit, the mixing angles are
\begin{equation}
    \begin{split}
        \alpha&\approx\frac{m_{\pi^0 \phi}^2}{(m_{\pi^0}^2-m_{\phi}^2)}\\
        \beta&\approx \frac{m_{\eta \phi}^2}{(m_{\eta}^2-m_{\phi}^2)} \\
        \gamma&\approx \frac{m_{\eta' \phi}^2}{(m_{\eta'}^2-m_{\phi}^2)}\\
        \delta&\approx \zeta\approx \kappa\approx 0\,.\\
    \end{split}
\end{equation}

With the rotation angles worked out, the $\phi$ mass in the propagating frame can be found to be

\begin{equation}
\label{eq:mphiexact}
     m_{\phi}^2 = \frac{m_{\phi_b}^2-(m_{\pi^0}^2 \left(\cos(\beta)\cos(\gamma)\sin(\alpha)\right)^2+m_{\eta}^2 \left(\cos(\gamma ) \sin(\beta
   )\right)^2+m_{\eta'}^2\sin(\gamma)^2)}{\left(\cos(\alpha)\cos(\beta)\cos(\gamma)\right)^2}\,.
\end{equation}

The couplings between $\phi$, mesons and $\chi$ can be found using--

\begin{equation}
\begin{split}
     \begin{pmatrix}
\tilde{g}_{(5)\phi\chi}&\tilde{g}_{(5)\pi^0\chi} &
        \tilde{g}_{(5)\eta\chi}&\tilde{g}_{(5)\eta'\chi}
    \end{pmatrix}&= 
    \begin{pmatrix}
        \tilde{g}_{(5)\phi_{b}\chi}&0&0&
        0
    \end{pmatrix}R_{\phi}^T\,.\\
\end{split}
\end{equation}

Expanding this out gives
\begin{equation}
\label{eq:pschicoup}
    \begin{split}
\tilde{g}_{(5)\pi^0\chi}&=\tilde{g}_{(5)\chi}\cos(\gamma)\cos(\beta)\sin(\alpha)\\
    \tilde{g}_{(5)\phi\chi}&=\tilde{g}_{(5)\chi} \cos(\alpha)\cos(\gamma)\cos(\beta)\\
    \tilde{g}_{(5)\eta\chi}&=\tilde{g}_{(5)\chi} \cos(\gamma)\sin(\beta)\\
     \tilde{g}_{(5)\eta'\chi}&=\tilde{g}_{(5)\chi}\sin(\gamma)\,.\\
    \end{split}
\end{equation}

The meson couplings to photons, $\tilde{g}_{j\gamma}$, are measured values, they can be found in Eq.\eqref{eqn:pseudocouplings}.  We take the measured values to be true values in the rotated basis.  The $\phi$ while not coupled to photons in the unrotated basis, picks up a coupling in the rotated basis.  It can be found by solving 
\begin{eqnarray}
    \begin{pmatrix}
\tilde{g}_{\phi\gamma}&\tilde{g}_{\pi^0\gamma} &
        \tilde{g}_{\eta\gamma}&\tilde{g}_{\eta'\gamma}
    \end{pmatrix}R_{\phi}&= 
    \begin{pmatrix}
        0&\tilde{g}_{\pi^0_b\gamma} &
        \tilde{g}_{\eta_b\gamma}&\tilde{g}_{\eta'_b\gamma}
    \end{pmatrix}
\end{eqnarray}
which gives
\begin{equation}
\label{eq:phiphoton}
\begin{split}
    \tilde{g}_{\phi\gamma}&=\frac{-\tilde{g}_{\pi^0\gamma}\left( c_{\zeta}(c_{\delta}s_{\alpha}+c_{\alpha}s_{\beta}s_{\delta})+c_{\alpha}c_{\beta}s_{\gamma}s_{\zeta}\right)+\tilde{g}_{\eta\gamma}\left(c_{\kappa}(s_{\alpha}s_{\delta}-c_{\alpha}c_{\delta}s_{\beta})+s_{\kappa}((c_{\gamma}s_{\alpha}+c_{\alpha}s_{\beta}s_{\delta})s_{\zeta}-c_{\alpha}c_{\beta}c_{\zeta}s_{\gamma})\right)}{c_{\alpha}c_{\beta}c_{\gamma}}\\
   &+ \frac{\tilde{g}_{\eta'\gamma}\left(((c_{\alpha}c_{\delta}s_{\beta}-s_{\alpha}s_{\delta})s_{\kappa})+c_{\kappa}((c_{\gamma}s_{\alpha}+c_{\alpha}s_{\beta}s_{\delta})s_{\zeta}-c_{\alpha}c_{\beta}c_{\zeta}s_{\gamma})\right)}{c_{\alpha}c_{\beta}c_{\gamma}}
   \end{split}\,,
\end{equation}
where $c_x = \cos(x)$ and $s_x=\sin(x)$.

In the small mixing angle limit all of the above derived masses and couplings can be approximated as
\begin{equation}
\label{eq:app-pseudocouplingslim}
    \begin{split}
\tilde{g}_{\phi\gamma}&\approx\sum_j-\tilde{g}_{j\gamma}\frac{m_{j\phi}^2}{(m_{j}^2-m_{\phi}^2)} \\
    \tilde{g}_{(5)j\chi}&\approx g_{(5)\chi}\frac{m_{j\phi}^2}{(m_{j}^2-m_{\phi}^2)}\\
    \tilde{g}_{(5)\phi\chi}&\approx g_{(5)\chi} \\
    m_{\phi}^2&\approx m_{\phi_b}^2
    \end{split}
\end{equation}
where the $m_{j\phi}$'s are defined in Eq.~\eqref{mjphi}.

\section{Meson Mixing, Rotation Angles, and Rotated Parameters for the  Axial Vector Quark Coupling}
\label{app:axialvector}
When DM interacts with quarks through a vector mediator  with couplings of the form
\begin{equation}
    \mathcal{L}\supset \lambda_{\chi}\chi\gamma_{\mu}\gamma_5\bar{\chi}A'^{\mu}+\frac{1}{2}m^2_{A'} A'_{\mu}A'^{\mu}+\sum_q \lambda_qq\gamma_{\mu}\gamma_5 \bar{q}A'^{\mu}
\end{equation}
the interaction can be described in the low energy limit by expanding the relevant terms of the chiral effective Lagrangian.  For the axial interaction those terms are 
\begin{equation}
    \mathcal{L}_{ChEFT}\supset  \frac{F^2}{4}\left(\text{Tr}\left(D_{\mu}U D^{\mu}U^{\dagger}\right) \right)
     +L_4\text{Tr}\left(D_{\mu}U D^{\mu}U^{\dagger}\right)\text{Tr}\left(\zeta^{\dagger}U+\zeta U^{\dagger}\right)+L_5\text{Tr}\left(D_{\mu}U D^{\mu}U^{\dagger}\left(\zeta^{\dagger}U+\zeta U^{\dagger}\right)\right)
\end{equation}
where Tr() denotes the trace, $U= e^{\frac{i}{F}\bf{\pi}}$, and $\pi$ is defined in Eq.~\eqref{pmatrix}.  $F = 92.4$ MeV \cite{holstein_1990} is the pion decay constant.  Interactions with external scalars and pseudoscalars are captured in the $\zeta$ term where $\zeta=2B_0 (\mathcal{M}+\textbf{g}_q\phi+i\tilde{\textbf{g}}_q\phi)$. To focus on the axial component of the interactions we will take $\textbf{g}_q$ = 0, $\tilde{\textbf{g}}_q$ = 0. $B_0 = \frac{m_{\pi^+}^2}{m_u+m_d}\simeq2.8$ where  $m_{\pi^+}$ is the charged pion mass. $\mathcal{M}=$diag$(m_u,m_d,m_s)$ contains the quark masses.  Contributions from the axial current are captured in the derivative terms
\begin{equation}
\begin{split}
    &D_{\mu}U =\partial_{\mu}U-i\textbf{A'}_{\mu} U-iU\textbf{A'}_{\mu}\,,\\
    &D_{\mu}U^{\dagger}=\partial_{\mu}U^{\dagger}+iU^{\dagger}\textbf{A'}_{\mu}+i\textbf{A'}_{\mu} U^{\dagger}\,,\\
\end{split}
\end{equation}
where the matrix
$\textbf{A'}_{\mu}=  A'_{\mu}\text{diag}(\lambda_u,\lambda_d,\lambda_s)$, captures the axial couplings to the quarks. The $L_4$ and $L_5$ terms in Eq.~\eqref{eqn:ChEFT Lagrangian} are only relevant for the axial vector interactions.  $L_4 = 1.87\times 10^{-3}$ and $L_5 = 1.22\times10^{-3}$, $L_6=1.46\times10^{-3}$, $L_7=-0.39\times10^{-3}$ \cite{Bijnens_2014}.
 Expanding these axial couplings results in kinetic mixing terms between the $A'_{\mu}$ and the $\pi^0$, $\eta$, and $\eta'$ of the form
\begin{equation}
    \mathcal{L}\supset -m_{\pi^0A'}A'_{\mu}\partial^{\mu}\pi^0_b-m_{\eta A'}A'_{\mu}\partial^{\mu}\eta_b-m_{\eta^{\prime} A}A'_{\mu}\partial^{\mu}\eta^{\prime}_{b}\,,
\end{equation}
where
\begin{equation}
\begin{split}
\label{mjAs}
    m_{\pi^0A'}&= F\left((\lambda_d-\lambda_u)\left(1+\frac{16 B_0 L_4}{F^2}\sum_i^{u,s,d}m_i\right)+\frac{16 B_0 L_5(\lambda_um_u-\lambda_dm_d)}{F^2}\right)\\
    m_{\eta A'} &= \frac{\sqrt{3} F}{3}\left((\lambda_u+\lambda_d-2\lambda_s)\left(1+\frac{16 B_0 L_4}{F^2}\sum_i^{u,s,d}m_i\right)+\frac{16 B_0 L_5(\lambda_um_u+\lambda_dm_d-2\lambda_sm_s)}{F^2}\right)\\
    m_{\eta' A'} &= \frac{\sqrt{6} F}{3}\left((\lambda_u+\lambda_d+\lambda_s)\left(1+\frac{16 B_0 L_4}{F^2}\sum_i^{u,s,d}m_i\right)+\frac{16 B_0 L_5(\lambda_um_u+\lambda_dm_d+\lambda_sm_s)}{F^2}\right)\,.
\end{split}
\end{equation}
To account for the mixing terms, we shift 
\begin{equation}
    A'_{{\mu}}\rightarrow \tilde{A}_{\mu}'+\frac{m_{\pi^0A'}}{ m_{A'}^2}\partial_{\mu}\pi^0_b+\frac{m_{\eta A'}}{m_{A'}^2}\partial_{\mu}\eta_b+\frac{m_{\eta^{\prime} A'}}{m_{A'}^2}\partial_{\mu}\eta^{\prime}_{b}\,.
\end{equation}  This absorbs the mixing between the $A'$ and the mesons but introduces kinetic mixing terms between the $\pi^0_b$, $\eta_b$ and $\eta^{\prime}_{b}$ which we must rotate away for calculations.  This can be achieved with
\begin{equation}
    \begin{pmatrix}
        \pi^0\\
        \eta \\
        \eta^{\prime}
    \end{pmatrix}=R_{A'} R_{Kin}^{-1}\begin{pmatrix}
        \pi^0_b\sqrt{1-\frac{m_{\pi^0A'}^2}{m_{A'}^2}}\\
        \eta_b\sqrt{1 -\frac{m_{\eta A'}^2}{ m_{A'}^2}} \\
        \eta^{\prime}_{b}\sqrt{1 -\frac{m_{\eta^{\prime} A'}^2}{ m_{A'}^2}}
    \end{pmatrix}
\end{equation}
where the square rooted factors are added to normalize the kinetic terms in the Lagrangian, and
\begin{equation}
    \begin{split}
        R_{Kin}& = \begin{pmatrix}
            1&-\tan(u)&(\tan(u)\sin(w)-\sin(v)/\cos(u))/D\\
            0&1/\cos(u)&(\tan(u)\sin(v)-\sin(w)/\cos(u))/D\\
            0&0&\cos(u)/D
        \end{pmatrix}\\
    \end{split}
\end{equation}
where $D =\sqrt{1-\sin(u)^2-\sin(v)^2-\sin(w)^2+2\sin(u)\sin(v)\sin(w)}$. 
The mixing angles \( u \), \( v \), and \( w \) are
\begin{equation}
    \begin{split}
    \sin(u)=& \frac{-m_{\pi^0 A'}m_{\eta A'}}{2m_{A'}^2\sqrt{\left(1-\frac{m_{\pi^0 A'}^2}{m_{A''}^2}\right)\left(1-\frac{m_{\eta A'}^2}{ m_{A'}^2}\right)}}\\
        \sin(v)=& \frac{-m_{\pi^0 A'}m_{\eta^{\prime} A}}{2m_{A'}^2\sqrt{\left(1-\frac{m_{\pi^0 A}^2}{m_{A'}^2}\right)\left(1-\frac{m_{\eta^{\prime} A}^2}{ m_{A'}^2}\right)}}\\
        \sin(w)=& \frac{-m_{\eta^{\prime} A}m_{\eta A}}{2m_{A'}^2\sqrt{\left(1-\frac{m_{\eta^{\prime} A}^2}{m_{A'}^2}\right)\left(1-\frac{m_{\eta A}^2}{ m_{A'}^2}\right)}}\,.\\
    \end{split}
\end{equation}
The rotation matrix $R_{A'}$ is 
\begin{equation}
    R_{A'}= {}^3R_{12}(\alpha){}^3R_{13}(\beta){}^3R_{23}(\gamma)\,.
\end{equation}The rotation matrices ${}^nR_{ij}$ are defined in Eq.~\eqref{eq:rotationangles}.  The rotation angles $\alpha$, $\beta$, and $\gamma$ must be found uniquely for the axial vector case by solving
\begin{equation}
\begin{pmatrix}
           m_{\pi^0}^2&0&0 \\
        0&m_{\eta}^2&0\\
        0&0&m_{\eta'}^2
    \end{pmatrix}
    =R_{A'}R_{Kin}^{T}\begin{pmatrix}
        m_{\pi^0b}^2&0&0 \\
        0&m_{\eta_{b}}^2&0\\
        0&0&m_{\eta' b}^2
    \end{pmatrix}R_{Kin}R_{A'}^T\,.
\end{equation}
The mixing angles must be solved for numerically.  One can do so using the solver~\cite{avis_kozar_2026_24508201}.
As before, we assume the fields in the rotated basis are the propagating fields relevant to observation and measurement.  We mark the unrotated fields with a ``b'' to indicate that they are in their ``bare'' state.  In the rotated or propagating basis the following terms in the Lagrangian are relevant to interactions between $\chi$ and photons 
\begin{equation}
\label{eq:lambda}
\begin{split}
    \mathcal{L}\supset&\frac{\tilde{g}_{\pi^0\gamma}}{4}\pi^0F^{\mu\nu}\tilde{F}_{\mu\nu}+ \frac{\tilde{g}_{\eta\gamma}}{4}\eta F^{\mu\nu}\tilde{F}_{\mu\nu}+ \frac{\tilde{g}_{\eta'\gamma}}{4}\eta' F^{\mu\nu}\tilde{F}_{\mu\nu}\\
    &+\lambda_{\pi^0\chi}\chi\gamma^\beta\gamma^5\bar{\chi}\partial_\beta\pi^0+\lambda_{\eta\chi}\chi\gamma^\beta\gamma^5\bar{\chi}\partial_\beta\eta+\lambda_{\eta^{\prime}\chi\chi}\chi\gamma^\beta\gamma^5\bar{\chi}\partial_\beta\eta^{\prime}
\end{split}
\end{equation}
where
\begin{equation}
    \begin{split}
        \lambda_{\pi^0\chi}=&\frac{\lambda_{\chi}}{m_{A'}^2} \frac{m_{\pi^0 A'}}{\left(1-\frac{m_{\pi^0A'}^2}{m_{A'}^2}\right)^{1/2}}\left(s_{\alpha}t_{u}s_{\beta}c_{\gamma}+c_{\alpha}t_{u}s_{\gamma}+c_{\beta}c_{\gamma}+\frac{(s_{\alpha}s_{\gamma}-c_{\alpha}s_{\beta}c_{\gamma})(t_{u} s_{w}-s_{v}/c_{u})}{D}\right)\\
        +&\frac{\lambda_{\chi}}{m_{A'}^2} \frac{m_{\eta A'}}{\left(1-\frac{m_{\eta A'}^2}{ m_{A'}^2}\right)^{1/2}}\left(\frac{-s_{\alpha}s_{\beta}c_{\gamma}-c_{\alpha}s_{\gamma}}{c_{u}}+\frac{(s_{\alpha}s_{\gamma}-c_{\alpha}s_{\beta}c_{\gamma})(t_{u} s_{v}-s_{w}/c_{u})}{D}\right)\\
        +&\frac{\lambda_{\chi}}{m_{A'}^2}\frac{m_{\eta^{\prime} A'}}{\left(1-\frac{m_{\eta^{\prime} A'}^2}{ m_{A'}^2}\right)^{1/2}}\left(\frac{s_{\alpha}c_{u}s_{\gamma}-c_{u}c_{\alpha}s_{\beta}c_{\gamma}}{D}\right)\\
        \lambda_{\eta\chi}=&\frac{\lambda_{\chi} }{m_{A'}^2}\frac{m_{\pi^0 A'}}{\left(1-\frac{m_{\pi^0A'}^2}{m_{A'}^2}\right)^{1/2}}\left(s_{\alpha}t_{u} s_{\beta}s_{\gamma}-c_{\alpha}t_{u} c_{\gamma}+c_{\beta}s_{\gamma}-\frac{(c_{\alpha}s_{\beta}s_{\gamma}+s_{\alpha}c_{\gamma})(t_{u}s_{w}-s_{v}/c_{u})}{D}\right)\\
        +&\frac{\lambda_{\chi} }{m_{A'}^2}\frac{m_{\eta A'}}{\left(1-\frac{m_{\eta A'}^2}{m_{A'}^2}\right)^{1/2}}\left(\frac{c_{\alpha}c_{\gamma}-s_{\alpha}s_{\beta}s_{\gamma}}{c_{u}}-\frac{(c_{\alpha}s_{\beta}s_{\gamma}+s_{\alpha}c_{\gamma})(t_{u}s_{v}-s_{w}/c_{u})}{D}\right)\\
        -&\frac{\lambda_{\chi} }{m_{A'}^2}\frac{m_{\eta^{\prime} A'}}{\left(1-\frac{m_{\eta^{\prime}A'}^2}{m_{A'}^2}\right)^{1/2}}\left(\frac{c_{u}c_{\alpha}s_{\beta}s_{\gamma}+s_{\alpha}c_{u}c_{\gamma}}{D}\right).\\
        \lambda_{\eta^{\prime}\chi}=& \frac{\lambda_{\chi} }{m_{A'}^2}\frac{m_{\pi^0 A'}}{\left(1-\frac{m_{\pi^0A'}^2}{m_{A'}^2}\right)^{1/2}}\left(s_{\beta}-s_{\alpha}t_{u}c_{\beta}+\frac{c_{\alpha}c_{\beta}(t_{u}s_{w}-s_{v}/c_{u})}{D}\right)\\
        +&\frac{\lambda_{\chi} }{m_{A'}^2}\frac{m_{\eta A'}}{\left(1-\frac{m_{\eta A'}^2}{m_{A'}^2}\right)^{1/2}}\left(s_{\alpha}c_{\beta}/c_{u}+\frac{c_{\alpha}c_{\beta}(t_{u}s_{v}-s_{w}/c_{u})}{D}\right)\\
        +&\frac{\lambda_{\chi} }{m_{A'}^2}\frac{m_{\eta^{\prime} A'}}{\left(1-\frac{m_{\eta^{\prime} A'}^2}{m_{A'}^2}\right)^{1/2}}\frac{c_{u}c_{\alpha}c_{\beta}}{D}\,.
    \end{split}
    \label{eq:lambdasexact}
\end{equation}
We represent $\sin(x)=s_{x}$, $\cos(x)=c_{x}$ and $\tan(x)=t_{x}$. 
In the limit that all $m_{jA'}$ satisfy $m_{jA'}\ll m_{A'}'$, the above mixed coupling terms simplify to
\begin{equation}
    \begin{split}
        \label{eq:lambdasapprox}\lambda_{j\chi}&\approx\frac{\lambda_{\chi}m_{jA'}}{m_{A'}^2}\,.\\
    \end{split}
\end{equation}
The $\tilde{g}_{j\gamma}$'s are known and measured values.  They can be found in Eq.~\eqref{eqn:pseudocouplings}.  The $m_{jA'}$'s are defined in Eq.~\eqref{mjAs}.

\section{Calculations for \texorpdfstring{$K^+\rightarrow\pi^++\chi+\bar{\chi}$}{K+ -> pi+ chi chibar}}
DM-quark couplings can also cause the $K^+$ to decay to $\pi^+\chi\bar{\chi}$, $\pi^+\phi$, $\pi^+\phi$, or  $\pi^+A'$.  These decays arise due to terms in the chiral Lagrangian that describe weak interactions between the light mesons.  The relevant terms are
\begin{equation}
    \mathcal{L}\supset \frac{i G_FF^4 } {\sqrt{2}}V_{us}V_{ud}^*\left(g_8\text{Tr}(\lambda (U^{\dagger}D_{\mu}U)(U^{\dagger}D^{\mu}U))+g_8^s(U^{\dagger}D_{\mu}U)_{3,2}\text{Tr}(U^{\dagger}D^{\mu}U)\right)
\end{equation}
where $\lambda=(\lambda_6-i\lambda_7)/2$, $\lambda_{6,7}$ are the Gell-Mann matrices, $G_F = 1.17\times 10^{-5}$ GeV$^{-2}$, $V_{us}^* = 0.222$, $V_{ud} = 0.974$, $g_8=3.62$ and $g_8^s=-0.38 g_8$ \cite{Gerard:2005yk}.  The (3,2) subscript denotes the 3,2 term in the matrix.  Expanding this out, the relevant terms for $K^+$ decay are
\begin{equation}
\label{eq:KLagrangianapp}
    \begin{split}
        \mathcal{L}\supset&\frac{ G_FF^2 } {\sqrt{2}}V_{us}V_{ud}^*\Bigg[-2g_8(\partial_{\mu}\pi^-)(\partial^{\mu}K^+)+\frac{ig_8}{F}\partial_{\mu}K^+(\pi^0\partial^{\mu}\pi^--\pi^-\partial^{\mu}\pi^0)\\
        &+\frac{ig_8\sqrt{3}}{3F}(-3\eta\partial_{\mu}K^+\partial^{\mu}\pi^-+2K^+\partial_{\mu}\eta\partial^{\mu}\pi^-+\pi^-\partial_{\mu}K^+\partial^{\mu}\eta)\\
        &+\frac{i\sqrt{6}}{3F}\partial^{\mu}\eta'_b(\partial_{\mu}K^+\pi^--\partial_{\mu}\pi^-K^+)(2g_8+3g_8^s)\\
        &+2iA'^{\mu} \left(-\partial_{\mu}K^+\pi^-(g_8(\lambda_s+\lambda_u)+g_8^s(\lambda_u+\lambda_s+\lambda_d))+\partial_{\mu}\pi^-K^+(g_8(\lambda_u+\lambda_d)+g_8^s(\lambda_u+\lambda_s+\lambda_d))\right)\Bigg]\,.
    \end{split}
\end{equation}
The first term in the brackets above produces a small kinetic mixing between the $\pi^+$ and the $K^+$.  This combined with the $\phi$ coupling to $\pi^+$ and $K^+$ in Eq.~\eqref{eq:expandedChEFTL} leads to a small three-point interaction between the $\phi$, $\pi^-$ and $K^+$ that allows $K^+$ to decay into $\pi^+\chi\bar{\chi}$.  The remaining terms in Eq.~\eqref{eq:KLagrangianapp} are for three-point interactions involving the $K^+$, the $\pi^-$ and either the neutral mesons ($\pi^0$, $\eta$ and $\eta'$) or the external vector mediator $A'^{\mu}$.  The mixing between the neutral mesons and the $\phi$ described above would cause the $\phi$ to pick up all of the interactions of the neutral mesons, and the mesons to pick up interactions with $\chi\bar{\chi}$, leading all four particles to potentially  mediate $K^+$ decay to $\pi^+\chi\bar{\chi}$.  The mixing between the $A_{\mu}'$ and the neutral mesons causes all of the neutral mesons to couple to $\chi\bar{\chi}$, so all of the three-point couplings terms in Eq.~\eqref{eq:KLagrangian} will be relevant to $K^+\rightarrow\pi^+\chi\bar{\chi}$ in the axial vector coupling case. 

\label{Ap:Kdecay}
\subsection{Scalar coupling to quarks}
\begin{figure}
    \centering
\begin{subfigure}{0.3\textwidth}
\begin{fmffile}{se-diagram}
 \begin{fmfgraph*}(120,80)
   \fmfleft{i1}
   
   \fmfright{o1,o2,o3}
\fmfforce{(0w,0.9h)}{i1}
\fmfforce{(1w,0.9h)}{o3}
\fmfforce{(1w,0.1h)}{o1}
\fmfforce{(1w,0.6h)}{o2}
   \fmfforce{(0.4w,0.9h)}{v1}
   \fmfforce{(0.7w,0.35h)}{v2}
   \fmf{fermion,label={\large$\bar{\chi}$},label.side=right}{o2,v2}
   \fmf{fermion,label={\large$\chi$},label.side=right}{v2,o1}
   \fmf{scalar,label={\large{$K^+$}},label.side=left}{i1,v1}
   \fmf{scalar,label={\large{$\pi^+$}},label.side=left}{v1,o3}
    \fmf{scalar,label={\large{$\phi$}}}{v1,v2}
 \end{fmfgraph*}
\end{fmffile}
    
\caption{}
\end{subfigure}
\begin{subfigure}{0.3\textwidth}
\begin{fmffile}{fifth-diagram}
\begin{fmfgraph*}(120,80)
   \fmfleft{i1}
   \fmfright{o1,o2}
   \fmf{scalar,label={\large$\pi^+$}}{v1,o2}
   \fmf{scalar,label={\large$\phi$},label.side=right}{v1,o1}
   \fmf{scalar,label={\large{$K^+$}}}{i1,v1}
 \end{fmfgraph*}
\end{fmffile}
    
\caption{}
\end{subfigure}

    \caption{Feynman diagrams relevant to  $K^+$ decay to $\pi^+\chi\bar{\chi}$ (a), and $K^+$ decay to $\pi^+\phi$ (b) for the $S\otimes S$ and $P\otimes S$ models.}
    \label{fig:scalardecays}
\end{figure}
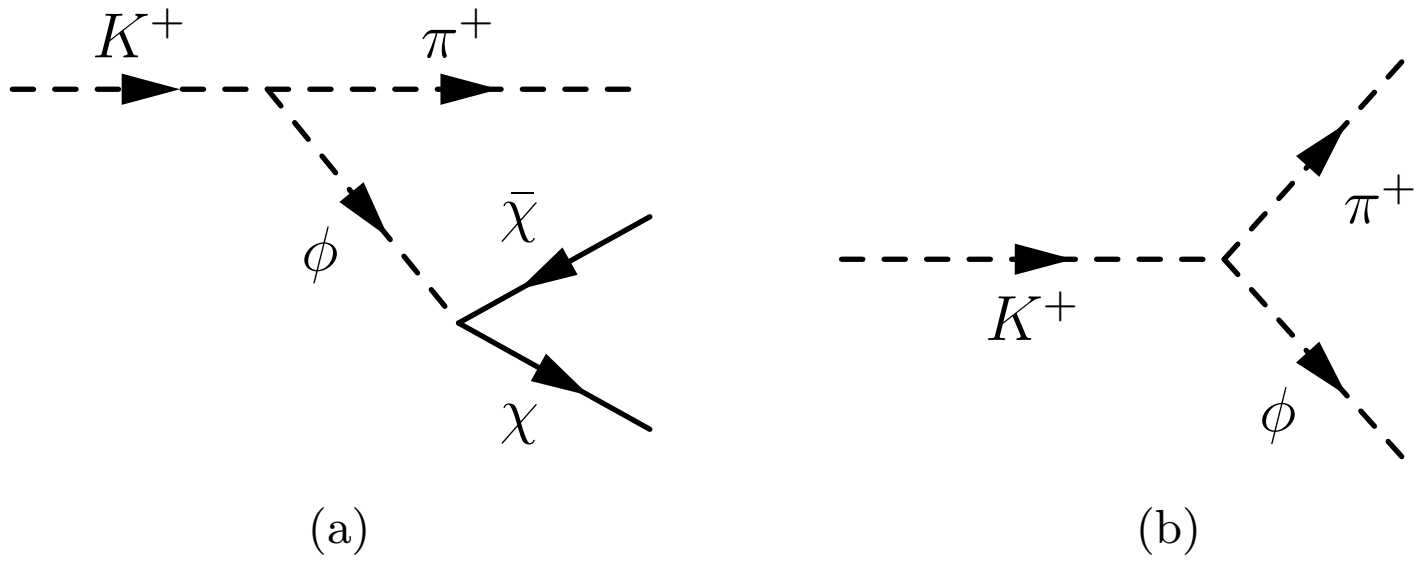
For the scalar quark coupling cases, the term in Eq.~\eqref{eq:KLagrangianapp}
\begin{equation}
    \mathcal{L}_{\Delta S=1}\supset-\frac{ G_FF^2 } {\sqrt{2}}V_{us}V_{ud}^*2g_8(\partial_{\mu}\pi^-)(\partial^{\mu}K^+)
\end{equation}
contains the terms that control both $K^+$ decay into  $\pi^+\chi\bar{\chi}$ mediated by $\phi$ and  $K^+$ decay to $\pi^+\phi$.  The above kinetic mixing term between the kaon and pion can be rotated away to produces a three-point interaction between the $K^+$, the $\pi^+$ and the $\phi$ of the form
\begin{equation}
\label{eqn:gpiKphi}
\begin{split}
     c_{\pi^+ K^+\phi}=&B_0\frac{-\sin(2\theta)((g_d+g_u)(2\delta^2-1)+(g_s+g_u))+2(g_d+g_u)\delta\sqrt{1-\delta^2}\cos(2\theta)}{2(\delta^2-1)}\\
     \approx& \delta B_0\left(\frac{(g_s+g_u) m_{\pi^+}^2-(g_u+g_d) m_{K^+}^2}{m_{K^+}^2-m_{\pi^+}^2}\right)
\end{split}
\end{equation}
where
    \begin{equation}
    \begin{split}
    \delta=& -\frac{G_F}{\sqrt{2}}F^2 V_{us}^*V_{ud}g_8\\
        \theta =& \frac{1}{2}\arctan\left(\frac{2\delta\sqrt{1-\delta^2}m_{\pi^+}^2}{m_{K^+}^2-(1-2\delta^2)m_{\pi^+}^2}\right)
    \end{split}
\end{equation}
where we take $G_F = 1.17\times 10^{-5}$ GeV$^{-2}$, $V_{us}^* = 0.222$, $V_{ud} = 0.974$, and $g_8=3.62$.  $\delta$ is very small, making $\theta$ very small.  One can reasonably use the approximate coupling found by taking the small angle limit for $\theta$, shown above.
The decay rate for the $S\otimes S$ case is then 
\begin{equation}
    \label{eq:scalardecay}\Gamma_{K^+\rightarrow\pi^+\chi\bar{\chi}}=\frac{g_{\chi}^2c_{\pi^+ K^+\phi}^2}{32\pi^3m_{K^+}}\mathcal{F}_{S\otimes S}(m_{\chi},m_{\phi})
\end{equation}
where
\begin{equation}
\label{eqn:phasespaces}    \mathcal{F}_{S\otimes S}(m_{\chi},m_{\phi})=\int_{m_{\pi^+}}^{E_{\pi \text{Max}}}dE_{\pi^+}\int_{E_{\chi\text{ Min}}}^{E_{\chi\text{Max}}}dE_{\chi}\frac{(m_{K^+}^2+m_{\pi^+}^2-2m_{K^+}E_{\pi^+}-4m_{\chi}^2)}{(m_{K^+}^2+m_{\pi^+}^2-2E_{\pi^+}m_{K^+}-m_{\phi}^2)^2}\,.
\end{equation}
where $E_{\chi \rm{Max}}$ and $E_{\chi \rm{Min}}$ are defined respectively as the larger and smaller values of
\begin{equation}
\label{Echimax}
   E_{\chi\text{Min/Max}}= \frac{1}{2} \left(m_{K^+}-E_{\pi^+}\pm\sqrt{\frac{\left(E_{\pi^+}^2-m_{\pi^+}^2\right)  \left(m_{K^+}^2+m_{\pi^+}^2-2 E_{\pi^+}m_{K^+}-4 m_{\chi}^2\right)}{m_{K^+}^2+m_{\pi^+}^2-2 E_{\pi^+}m_{K^+}}
   }\right)
\end{equation}
and
\begin{equation}
\label{Epimax}
    E_{\pi^+\text{Max}} = \frac{m_{K^+}^2+m_{\pi^+}^2-4m_{\chi}^2}{2m_{K^+}}\,.
\end{equation}
Because there is no $E_{\chi}$ dependence in the integrand above, the $E_{\chi}$ integral gives 
\begin{equation}
    \mathcal{F}_{S\otimes S}(m_{\chi},m_{\phi})=\int_{m_{\pi^+}}^{E_{\pi \text{Max}}}dE_{\pi^+}\frac{(m_{K^+}^2+m_{\pi^+}^2-2m_{K^+}E_{\pi^+}-4m_{\chi}^2)}{(m_{K^+}^2+m_{\pi^+}^2-2m_{K^+}E_{\pi^+}-m_{\phi}^2)^2}Y(E_{\pi^+},m_{K^+},m_{\pi^+},m_{\chi})
\end{equation}
where 
\begin{equation}
    \label{eq:Y}Y(E_{\pi^+},m_{K^+},m_{\pi^+},m_{\chi})=\sqrt{\frac{\left(E_{\pi^+}^2-m_{\pi^+}^2\right)  \left(m_{K^+}^2+m_{\pi^+}^2-2 E_{\pi^+}m_{K^+}-4 m_{\chi}^2\right)}{m_{K^+}^2+m_{\pi^+}^2-2 E_{\pi^+}m_{K^+}}}\,.
\end{equation}
The above expression must be calculated numerically for specific values of $m_{\chi}$ and $m_{\phi}$.
For the $P\otimes S $ case, where $\phi$ couples to the $\chi$ via $\phi\chi\gamma^5\bar{\chi}$, $\Gamma_{K^+\rightarrow\pi^+\chi\bar{\chi}}^{P\otimes S}$ can be found by taking the expression in Eq.\,\eqref{eq:scalardecay} and replacing $\mathcal{F}_{S\otimes S}$ with $\mathcal{F}_{P\otimes S}$ where
\begin{equation}
\label{eqn:phasespacePS}
    \mathcal{F}_{P\otimes S}=\int_{m_{\pi^+}}^{E_{\pi^+ \text{Max}}}dE_{\pi^+}\frac{(m_{K^+}^2+m_{\pi^+}^2-2m_{K^+}E_{\pi^+})}{(m_{K^+}^2+m_{\pi^+}^2-2E_{\pi^+}m_{K^+}-m_{\phi}^2)^2}Y(E_{\pi^+},m_{K^+},m_{\pi^+},m_{\chi})\,.
\end{equation}
$E_{\chi\text{(Min/Max)}}$ is defined in Eq. \eqref{Echimax} and $E_{\pi^+\text{Max}}$ is defined in Eq. \eqref{Epimax}

The decay rate to $K^+\rightarrow\pi^+\phi$ is
the same for both the $S\otimes S$ case and the $P\otimes S$ case:
\begin{equation}
    \Gamma_{K^+\rightarrow\pi^+\phi} = \frac{c_{\pi^+ K^+\phi}^2\sqrt{(m_{K^+}^2+m_{\pi^+}^2-m_{\phi}^2)^2-4m_{K^+}^2m_{\pi^+}^2}}{16\pi m_{K^+}^3}\,.
\end{equation}
\subsection{Pseudoscalar coupling to quarks}

 The following Lagrangian terms are relevant to $K^+$ decay to $\pi^+ +\chi+\bar{\chi}$.  There is mixing between the $K^+$ and the $\pi^+$ as in the scalar case.  We neglect it in this calculation because the mixing angle is quite small, and is less relevant to this calculation. 
 
\begin{equation}
\begin{split}
    \mathcal{L}\supset &\frac{i G_FF } {\sqrt{2}}V_{us}V_{ud}^* \Bigg( g_8\partial_{\mu}K^-\left( \pi^0_b\partial^{\mu}\pi^+ -\pi^+\partial^{\mu}\pi^0_b\right) +\frac{ g_8}{\sqrt{3}}\left(-3\eta_b\partial^{\mu}\pi^+\partial_{\mu}K^-+2K^-\partial^{\mu}\eta_b\partial_{\mu}\pi^++\pi^+\partial_{\mu}\eta_b\partial^{\mu}K^-
    \right)\\
    &+\frac{ \sqrt{6}\partial^{\mu}\eta'}{3}(2g_8+3g_8^s)\left(\partial_{\mu}\pi^+K^--\pi^+\partial_{\mu}K^-\right) \Bigg)\\
\end{split}
\end{equation}
where $G_F = 1.17\times 10^{-5}$ GeV$^{-2}$, $V_{us}^* = 0.222$, $V_{ud} = 0.974$, and $g_8=3.62$. The kaon decay amplitude is then
\begin{equation}
    iM=\frac{i G_FF} {\sqrt{2}}V_{us}V_{ud}^*\left(\sum_j^{\pi^0,\phi,\eta,\eta'}  \frac{\tilde{g}_{j\chi}\tilde{g}_{j\pi^+ K^+}(E_{\pi^+})}{(m_{K^+}^2+m_{\pi^+}^2-2m_{K^+}E_{\pi^+}-m_{j}^2)}\right)\chi\gamma^5\bar{\chi}\,,
\end{equation}

where
\begin{equation}
\label{gphipiK}
    \begin{split}
    \tilde{g}_{\phi\pi^+ K^+}(E_{\pi^+})&=g_8(2m_{K^+}E_{\pi^+}-m_{K^+}^2)s_{\alpha}c_{\beta}c_{\gamma}+\frac{g_8(m_{K^+}^2-6m_{K^+}E_{\pi^+}+2m_{\pi^+}^2)}{\sqrt{3}}s_{\beta}c_{\gamma}+\frac{\sqrt{6}(3g_8^s+2g_8)(m_{\pi^+}^2-m_{K^+}^2)}{3}s_{\gamma}\\
    \tilde{g}_{\pi^0\pi^+ K^+}(E_{\pi^+})&=g_8(2m_{K^+}E_{\pi^+}-m_{K^+}^2)\Big(c_{\zeta}(c_{\alpha}c_{\delta}-s_{\alpha}s_{\beta}s_{\delta})-c_{\beta}s_{\alpha}s_{\gamma}s_{\zeta}\Big)\\&+\frac{g_8(m_{K^+}^2-6m_{K^+}E_{\pi^+}+2m_{\pi^+}^2)}{\sqrt{3}}\Big(c_{\beta}c_{\zeta}s_{\delta}-s_{\beta}s_{\gamma}s_{\zeta}\Big)+\frac{\sqrt{6}(3g_8^s+2g_8)(m_{\pi^+}^2-m_{K^+}^2)}{3}c_{\gamma}s_{\zeta}\\
    \tilde{g}_{\eta\pi^+ K^+}(E_{\pi^+})&=g_8(2m_{K^+}E_{\pi^+}-m_{K^+}^2)\Big(c_{\kappa}(-c_{\delta}s_{\alpha}s_{\beta}-c_{\alpha}s_{\delta})+(-c_{\beta}c_{\zeta}s_{\alpha}s_{\gamma}-(c_{\alpha}c_{\delta}-s_{\alpha}s_{\beta}s_{\delta})s_{\zeta})s_{\kappa}\Big)\\&+\frac{g_8(m_{K^+}^2-6m_{K^+}E_{\pi^+}+2m_{\pi^+}^2)}{\sqrt{3}}\Big(c_{\beta}c_{\delta}c_{\kappa}+(-c_{\zeta}s_{\beta}s_{\gamma}-c_{\beta}s_{\delta}s_{\zeta})s_{\kappa}\Big)+\frac{\sqrt{6}(3g_8^s+2g_8)(m_{\pi^+}^2-m_{K^+}^2)}{3}c_{\gamma}c_{\zeta}s_{\kappa}\\
    \tilde{g}_{\eta'\pi^+ K^+}(E_{\pi^+})&=g_8(2m_{K^+}E_{\pi^+}-m_{K^+}^2)\Big(c_{\kappa}(-c_{\beta}c_{\zeta}s_{\alpha}s_{\gamma}-(c_{\alpha}c_{\delta}-s_{\alpha}s_{\beta}s_{\delta})s_{\zeta})-(-c_{\delta}s_{\alpha}s_{\beta}-c_{\alpha}s_{\delta})s_{\kappa}\Big)\\&+\frac{g_8(m_{K^+}^2-6m_{K^+}E_{\pi^+}+2m_{\pi^+}^2)}{\sqrt{3}}\Big(c_{\kappa}(-c_{\zeta}s_{\beta}s_{\gamma}-c_{\beta}s_{\delta}s_{\zeta})-c_{\beta}c_{\delta}s_{\kappa}\Big)+ \frac{\sqrt{6}(3g_8^s+2g_8)(m_{\pi^+}^2-m_{K^+}^2)}{3}c_{\gamma}c_{\zeta}c_{\kappa}\,.
\end{split}
\end{equation}

$s_\theta=\sin(\theta)$ and $c_\theta=\cos(\theta)$.  The rotation angles, $\alpha$, $\beta$, and $\gamma$ are defined in Appendix \ref{ap:pseudoscalar}.  In the limit that $|m_{\phi\pi^0}^2|\ll|m_{\phi}^2-m_{\pi^0}^2|$,  $|m_{\phi\eta}^2|\ll|m_{\phi}^2-m_{\eta}^2|$, and $|m_{\phi\eta'}^2|\ll|m_{\phi}^2-m_{\eta'}^2|$,
\begin{equation}
\label{gphipiKsmall}
    \begin{split}
    \tilde{g}_{\phi\pi^+ K^+}(E_{\pi^+})&\approx g_8(2m_{K^+}E_{\pi^+}-m_{K^+}^2)\frac{m_{\phi\pi^0}^2}{2(m_{\pi^0}^2-m_{\phi}^2)}+\frac{g_8(m_{K^+}^2-6m_{K^+}E_{\pi^+}+2m_{\pi^+}^2)}{\sqrt{3}}\frac{m_{\phi\eta}^2}{2(m_{\eta}^2-m_{\phi}^2)}\\&+\frac{\sqrt{6}(3g_8^s+2g_8)(m_{\pi^+}^2-m_{K^+}^2)}{3}\frac{m_{\phi\eta'}^2}{2(m_{\eta'}^2-m_{\phi}^2)}\\
    \tilde{g}_{\pi^0\pi^+ K^+}(E_{\pi^+})&\approx g_8(2m_{K^+}E_{\pi^+}-m_{K^+}^2)\\
    \tilde{g}_{\eta\pi^+ K^+}(E_{\pi^+})&\approx\frac{g_8(m_{K^+}^2-6m_{K^+}E_{\pi^+}+2m_{\pi^+}^2)}{\sqrt{3}}\\
    \tilde{g}_{\eta'\pi^+ K^+}(E_{\pi^+})&\approx\frac{\sqrt{6}(3g_8^s+2g_8)(m_{\pi^+}^2-m_{K^+}^2)}{3}\,.
\end{split}
\end{equation}
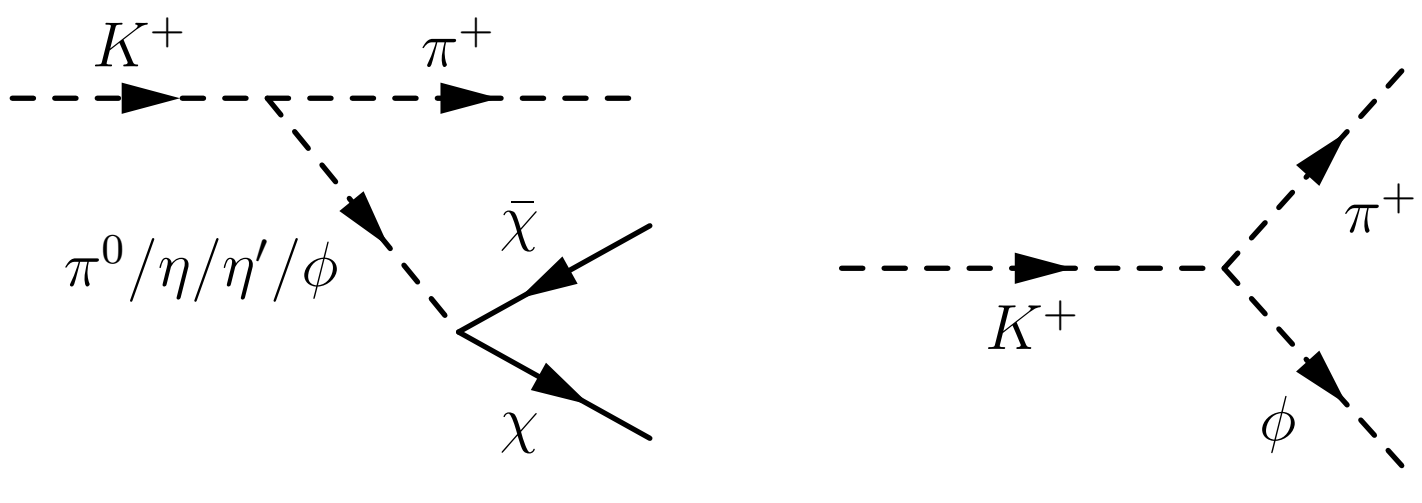
\begin{figure}
     \centering
\begin{subfigure}{0.3\textwidth}
\begin{fmffile}{fourth-diagram}
 \begin{fmfgraph*}(120,80)
   \fmfleft{i1}
   
   \fmfright{o1,o2,o3}
\fmfforce{(0w,0.9h)}{i1}
\fmfforce{(1w,0.9h)}{o3}
\fmfforce{(1w,0.6h)}{o2}
\fmfforce{(1w,0.1h)}{o1}
   \fmfforce{(0.4w,0.9h)}{v1}
   \fmfforce{(0.7w,0.35h)}{v2}
   \fmf{fermion,label={\large$\bar{\chi}$},label.side=right}{o2,v2}
   \fmf{fermion,label={\large$\chi$},label.side=right}{v2,o1}
   \fmf{scalar,label={\large{$K^+$}},label.side=left}{i1,v1}
   \fmf{scalar,label={\large{$\pi^+$}},label.side=left}{v1,o3}
    \fmf{scalar,label={\large{$\pi^0/\eta/\eta'/\phi$}}}{v1,v2}
 \end{fmfgraph*}
\end{fmffile}
    
\caption{}
\end{subfigure}
\begin{subfigure}{0.3\textwidth}
\begin{fmffile}{fifthf-diagram}
\begin{fmfgraph*}(120,80)
   \fmfleft{i1}
   \fmfright{o1,o2}
   \fmf{scalar,label={\large$\pi^+$}}{v1,o2}
   \fmf{scalar,label={\large$\phi$},label.side=right}{v1,o1}
   \fmf{scalar,label={\large{$K^+$}}}{i1,v1}
 \end{fmfgraph*}
\end{fmffile}
    
\caption{}
\end{subfigure}

    \caption{Feynman diagrams relevant to $K^+$ decay to $\pi^+\chi\bar{\chi}$ (a) and $K^+$ decay to $\pi^+ \phi$ (b) for the $P\otimes P$ and $S\otimes P$ models}
    \label{fig:pseudoK}
\end{figure}
We note that $m_{\phi}$ here is its rotated mass, and not the mass that appears in the Lagrangian.  It is defined exactly in Eq.~\eqref{eq:mphiexact} and in  the small mixing limit in Eq.~\eqref{eq:app-pseudocouplingslim}.
The decay rate is then
\begin{equation}
\label{eqn:Kpichichip}
\begin{split}
    \Gamma_{K^+\rightarrow\pi^+\chi\bar{\chi}} = \frac{(G_FF V_{us}V_{ud}^*)^2}{64\pi^3m_{K^+}}\int_{m_{\pi^+}}^{E_{\pi ^+\text{Max}}}&dE_{\pi^+}(m_{K^+}^2+m_{\pi^+}^2-2m_{K^+}E_{\pi^+})Y(E_{\pi^+},m_{K^+},m_{\pi^+},m_{\chi})\\&\times\left(\sum_j^{\pi^0,\phi,\eta,\eta'}  \frac{\tilde{g}_{j\chi}\tilde{g}_{j\pi K}(E_{\pi^+})}{(m_{K^+}^2+m_{\pi^+}^2-2E_{\pi^+}m_{K^+}-m_{j}^2)}\right)^2\,.
\end{split}
\end{equation}$E_{\chi\text{(Min/Max)}}$ is defined in Eq. \eqref{Echimax} and $E_{\pi^+\text{Max}}$ is defined in Eq. \eqref{Epimax}.
In the case that $\phi$ couples to $\chi$ as $g_\chi \phi\chi\chi$ ($S\otimes P$) the decay rate becomes
\begin{equation}
\label{eqn:Kpichichips}
\begin{split}
    \Gamma_{K^+\rightarrow\pi^+\chi\bar{\chi}} = \frac{(G_FF V_{us}V_{ud}^*)^2}{64\pi^3m_{K^+}}\int_{m_{\pi^+}}^{E_{\pi \text{Max}}}&dE_{\pi^+}(m_{K^+}^2+m_{\pi^+}^2-2m_{K^+}E_{\pi^+}-4m_{\chi}^2)Y(E_{\pi^+},m_{K^+},m_{\pi^+},m_{\chi})\\&\times\left(\sum_j^{\pi^0,\phi,\eta,\eta'}  \frac{\tilde{g}_{j\chi}\tilde{g}_{j\pi K}(E_{\pi^+})}{(m_{K^+}^2+m_{\pi^+}^2-2m_{K^+}E_{\pi^+}-m_j^2)}\right)^2\,.
\end{split}
\end{equation}
$E_{\chi\text{(Min/Max)}}$ is defined in Eq.~\eqref{Echimax} and $E_{\pi\text{Max}}$ is defined in Eq.~\eqref{Epimax}.

The decay rate to $K^+\rightarrow\pi^+\phi$ is
\begin{equation}
    \Gamma_{K^+\rightarrow\pi^+\phi} = \tilde{g}^2_{\phi\pi^+ K^+}(E_{\pi^+})\Bigg|_{E_{\pi^+}=\frac{m_{K^+}^2+m_{\pi^+}^2-m_{\phi}^2}{2m_{K^+}}}\frac{(G_FFV_{us}V_{ud}^*)^2}{32\pi m_{K^+}^3}\sqrt{(m_{K^+}^2+m_{\pi^+}^2-m_{\phi}^2)^2-4m_{K^+}^2m_{\pi^+}^2}\,,
\end{equation}
where $\tilde{g}_{\phi\pi^+ K^+}(E_{\pi^+})$ can be found in Eq.~\eqref{gphipiK} or in Eq.~\eqref{gphipiKsmall} in the small angle limit. $\tilde{g}_{j\chi}$ can be found in Eq.\eqref{eq:pschicoup} and in the small angle approximation in Eq.~\eqref{eq:app-pseudocouplingslim}. $m_{j\phi}^2$ are defined in Eq.~\eqref{eq:pseudoaxialmixings}.

\subsection{Axial Vector Coupling to Quarks}

The relevant interaction terms in Eq.~\eqref{eq:KLagrangianapp} for $K^+$ decays involving $A'^{\mu}$ are
\begin{equation}
  \begin{split}  \mathcal{L}\supset 
 &\frac{i G_FF } {\sqrt{2}}V_{us}V_{ud}^* \Bigg( g_8\partial_{\mu}K^-\left( \pi^0_b\partial^{\mu}\pi^+ -\pi^+\partial^{\mu}\pi^0_b\right) +\frac{ g_8}{\sqrt{3}}\left(-3\eta_b\partial^{\mu}\pi^+\partial_{\mu}K^-+2K^-\partial^{\mu}\eta_b\partial_{\mu}\pi^++\pi^+\partial_{\mu}\eta_b\partial^{\mu}K^-
    \right)\\
    &+\frac{ \sqrt{6}\partial^{\mu}\eta'}{3}(2g_8+3g_8^s)\left(\partial_{\mu}\pi^+K^--\pi^+\partial_{\mu}K^-\right) \Bigg)\\
   &+\frac{2i G_FF^2 }{\sqrt{2}}V_{us}V_{ud}^* A'^{\mu}\left(g_8((\lambda_s+\lambda_u)\pi^+\partial_{\mu}K^--(\lambda_u+\lambda_d)\partial_{\mu}\pi^+K^-)-g_8^s(\lambda_u+\lambda_s+\lambda_d)(K^-\partial_{\mu}\pi^+-\partial_{\mu}K^-\pi^+)\right)\\
\end{split}
\end{equation}
\begin{figure}
    \centering

\begin{subfigure}{0.3\textwidth}
\begin{fmffile}{four-diagram}
 \begin{fmfgraph*}(120,80)
   \fmfleft{i1}
   
   \fmfright{o1,o2,o3}
\fmfforce{(0w,0.9h)}{i1}
\fmfforce{(1w,0.9h)}{o3}
\fmfforce{(1w,0.6h)}{o2}
\fmfforce{(1w,0.1h)}{o1}
   \fmfforce{(0.4w,0.9h)}{v1}
   \fmfforce{(0.7w,0.35h)}{v2}
   \fmf{fermion,label={\large$\bar{\chi}$},label.side=right}{o2,v2}
   \fmf{fermion,label={\large$\chi$},label.side=right}{v2,o1}
   \fmf{scalar,label={\large{$K^+$}},label.side=left}{i1,v1}
   \fmf{scalar,label={\large{$\pi^+$}},label.side=left}{v1,o3}
    \fmf{scalar,label={\large{$\pi^0/\eta/\eta'$}}}{v1,v2}
 \end{fmfgraph*}
\end{fmffile}
    
\caption{}
\end{subfigure}
\begin{subfigure}{0.3\textwidth}
\begin{fmffile}{four-diagram2}
 \begin{fmfgraph*}(120,80)
   \fmfleft{i1}
   
   \fmfright{o1,o2,o3}
\fmfforce{(0w,0.9h)}{i1}
\fmfforce{(1w,0.9h)}{o3}
\fmfforce{(1w,0.6h)}{o2}
\fmfforce{(1w,0.1h)}{o1}
   \fmfforce{(0.4w,0.9h)}{v1}
   \fmfforce{(0.7w,0.35h)}{v2}
   \fmf{fermion,label={\large$\bar{\chi}$},label.side=right}{o2,v2}
   \fmf{fermion,label={\large$\chi$},label.side=right}{v2,o1}
   \fmf{scalar,label={\large{$K^+$}},label.side=left}{i1,v1}
   \fmf{scalar,label={\large{$\pi^+$}},label.side=left}{v1,o3}
    \fmf{photon,label={\large{$A'_{\mu}$}}}{v1,v2}
 \end{fmfgraph*}
\end{fmffile}
    
\caption{}
\end{subfigure}
\begin{subfigure}{0.3\textwidth}
\begin{fmffile}{six-diagram}
 \begin{fmfgraph*}(120,80)
   \fmfleft{i1}
   \fmfright{o1,o2}
   \fmf{scalar,label={\large$\pi^+$}}{v1,o2}
   \fmf{photon,label={\large$A'_{\mu}$},label.side=right}{v1,o1}
   \fmf{scalar,label={\large{$K^+$}}}{i1,v1}
 \end{fmfgraph*}
\end{fmffile}
    
\caption{}
\end{subfigure}

    \caption{ Feynman diagrams shown in (a) and (b) both contribute to $K^+$ decay to $\pi^+\chi\bar{\chi}$ due to the $A'^{\mu}$ coupling to quarks. The diagram in panel (c) describes the decay $K^+\rightarrow \pi^+ A'_{\mu}$}\label{fig:app-axfeyn}
\end{figure}
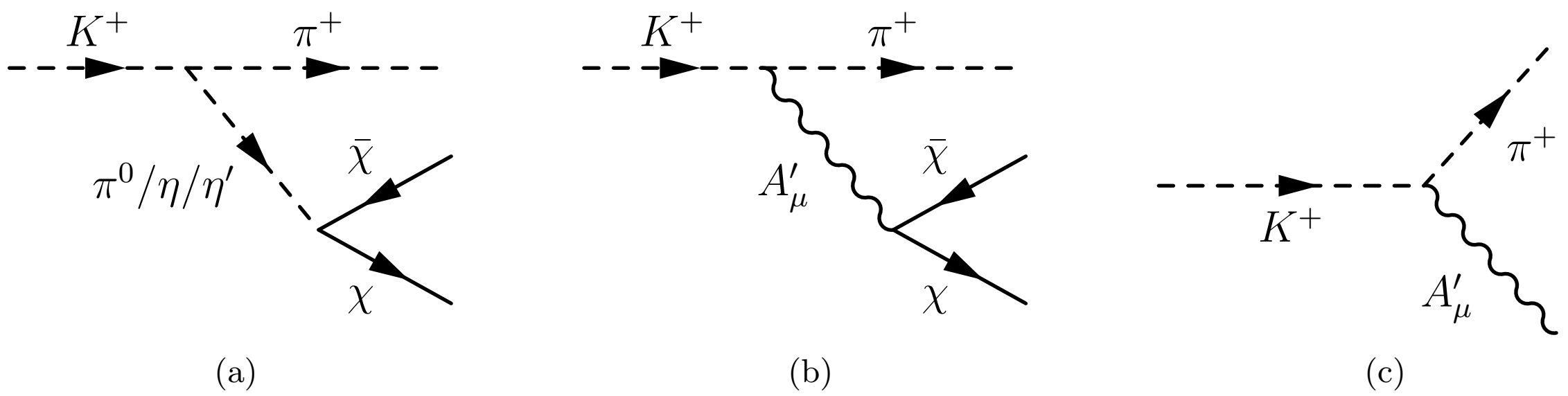
where $G_F = 1.17\times 10^{-5}$ GeV$^{-2}$, $V_{us}^* = 0.222$, $V_{ud} = 0.974$, $g_8=3.62$, and $g_8^s=-0.38 g_8$ \cite{Gerard:2005yk}. 
Shifting $ A'_{{\mu}'}\rightarrow \tilde{A}_{\mu}'+\frac{m_{\pi^0A'}}{ m_{A'}^2}\partial_{\mu}\pi^0_b+\frac{m_{\eta A'}}{m_{A'}^2}\partial_{\mu}\eta_b+\frac{m_{\eta^{\prime} A'}}{m_{A'}^2}\partial_{\mu}\eta^{\prime}_{b}$ to remove mixing terms between $A'$ and the mesons will introduce mixing between the mesons which must be rotated away. In the new shifted and rotated basis (see Appendix \ref{app:axialvector} for details on the shifting and rotating) the amplitude for decay is
\begin{equation}
\begin{split}
    iM_{K^+\rightarrow\pi^+\chi\bar{\chi}} = &\frac{ G_FF^4 } {\sqrt{2}}V_{us}V_{ud}^*\left(\lambda_{\chi}\lambda_{A'\pi K}(p_K^{\nu},p_{\pi^+}^{\nu})\frac{\left(g^{\mu\nu}-\frac{(p_K-p_{\pi^+})^{\mu}(p_K-p_{\pi^+})^{\nu}}{m_{A'}^2}\right)}{(p_K-p_{\pi^+})^2-m_{A'}^2}\right.\\
    &\left.-\left[\sum_j^{\pi^0,\eta,\eta'}\frac{\lambda_{j\chi}\lambda_{j\pi K}(E_{\pi^+})}{(p_K-p_{\pi^+})^2-m_j^2}\right](p_K-p_{\pi^+})^{\mu}\right)\bar{u}(p_{\chi})\gamma^5\gamma_{\mu}\bar{v}(p_{\bar{\chi}})\,.
    \end{split}
\end{equation}

This gives a total decay rate of 
\begin{equation}
\label{eqn:kpichichiA}
\begin{split}
    \Gamma_{K^+\rightarrow\pi^+\chi\bar{\chi}} =& \frac{ G_F^2F^8 } {128\pi^3m_{K^+}}|V_{us}|^2|V_{ud}|^{2} \int_{m_{\pi^+}}^{E_{\pi \text{Max}}}dE_{\pi^+}\int_{E_{\chi\text{ Min}}}^{E_{\chi\text{Max}}}dE_{\chi}\\
    &\times\Bigg\{8m_{\chi}^2(m_{K^+}^2+m_{\pi^+}^2-2E_{\pi^+}m_{K^+})\left|\sum_j^{\pi^0/\eta/\eta^{\prime}}\frac{\lambda_{j\chi}\lambda_{j \pi K}}{(m_{K^+}^2+m_{\pi^+}^2-2E_{\pi^+}m_{K^+})-m_j^2+im_j\Gamma_j}
    \right|^2 \\
    -&\frac{32 \lambda_{\chi}  m_{\chi}^2 \left(2 E_{\pi^+}
   m_{K^+}+m_{A'}^2-m_{K^+}^2-m_{\pi^+}^2\right) \left(E_{\pi^+} m_{K^+} (\lambda_{A'K}-\lambda_{A'\pi})-\lambda_{A' K} m_{K^+}^2+\lambda_{A'\pi} m_{\pi^+}^2\right)
   }{F^2 m_{A'}^2 \left(\left(2 E_{\pi^+} m_{K^+}+m_{A'}^2-m_{K^+}^2-m_{\pi^+}^2\right)^2+ \Gamma _{A'}^2
   m_{A'}^2\right) }\\
\times&\sum_j^{\pi^0/\eta/\eta^{\prime}}\left(\frac{|\lambda_{ j \pi K}| \lambda_{ j \chi}\left(\left(2 E_{\pi^+} m_{K^+}+m_{A'}^2-m_{K^+}^2-m_{\pi^+}^2\right) \left(2 E_{\pi^+}
   m_{K^+}-m_{K^+}^2-m_{\pi^+}^2+m_j^2\right)+\Gamma _{A'} \Gamma _{j} m_{A'}
   m_j\right)}{\left(\left(2 E_{\pi^+} m_{K^+}-m_{K^+}^2-m_{\pi
   }^2+m_j^2\right)^2+ \Gamma _{j}^2 m_j^2\right) }\right)\\
    -&\frac{8 \lambda_{\chi}^2}{F^4 m_{A'}^4 \left(\left(-2 E_{\pi^+}
   m_{K^+}-m_{A'}^2+m_{K^+}^2+m_{\pi^+}^2\right)^2+m_{A'}^2\Gamma_{A'}^2\right)}\Bigg(8 E_{\pi^+}^3 m_{K^+}^3 m_{\chi}^2
   (\lambda_{A'K}-\lambda_{A'\pi})^2 \\
   +&4 E_{\pi^+}^2
   m_{K^+}^2 m_{\chi}^2 (\lambda_{A'K}-\lambda_{A'\pi}) \Big(2
   m_{A'}^2 (\lambda_{A'K}-\lambda_{A'\pi})+m_{K^+}^2
   (\lambda_{A'\pi}-5 \lambda_{A'K})-m_{\pi^+}^2
   (\lambda_{A'K}-5 \lambda_{A'\pi})\Big)\\-&2 E_{\pi^+}
   m_{K^+} \Bigg[m_{A'}^4 \left(m_{K^+}^2 (\lambda_{A'K}+\lambda_{A'\pi})^2+4 \lambda_{A'K} \lambda_{A'\pi}
   m_{\chi}^2\right)+8 m_{A'}^2 m_{\chi}^2 (\lambda_{A'K}-\lambda_{A'\pi}) \left(\lambda_{A'K}
   m_{K^+}^2-\lambda_{A'\pi} m_{\pi^+}^2\right)\\-&4 m_{\chi}^2
   \left(\lambda_{A'K} m_{K^+}^4 (2 \lambda_{A'K}-\lambda_{A'\pi})+m_{K^+}^2 m_{\pi^+}^2
   \left(\lambda_{A'K}^2-4 \lambda_{A'K} \lambda_{A'\pi}+\lambda_{A'\pi}^2\right)+\lambda_{A'\pi} m_{\pi^+}^4
   (2 \lambda_{A'\pi}-\lambda_{A'K})\right)\Bigg]\\+&4 E_{\chi}^2
   m_{A'}^4 m_{K^+}^2 (\lambda_{A'K}+\lambda_{A'\pi})^2+4
   E_{\chi} m_{A'}^4 m_{K^+}^2 (E_{\pi^+}-m_{K^+}) (\lambda_{A'K}+\lambda_{A'\pi})^2\\
   +&m_{A'}^4 \Big(m_{K^+}^4 (\lambda_{A'K}+\lambda_{A'\pi})^2+m_{K^+}^2 \left(m_{\pi^+}^2 (\lambda_{A'K}+\lambda_{A'\pi})^2-4 \lambda_{A'K}^2 m_{\chi}^2\right)-4
   \lambda_{A'\pi}^2 m_{\pi^+}^2 m_{\chi}^2\Big)\\+&8 m_{A'}^2
   m_{\chi}^2 \left(\lambda_{A'K} m_{K^+}^2-\lambda_{A'\pi}
   m_{\pi^+}^2\right)^2-4 m_{\chi}^2 \left(m_{K^+}^2+m_{\pi^+}^2\right) \left(\lambda_{A'K} m_{K^+}^2-\lambda_{A'\pi}
   m_{\pi^+}^2\right)^2\Bigg) \Bigg\},
\end{split}
\end{equation}
where $E_{\chi\text{(Min/Max)}}$ is defined in Eq.~\eqref{Echimax} and $E_{\pi^+\text{Max}}$ is defined in Eq.~\eqref{Epimax}.  The term in the first line comes from the mesons acting as mediators, the terms in the second and third lines are mixing terms between the amplitudes of the axial vector and meson mediators and the terms in the fourth row and below come from decays mediated by the axial vector.  
\begin{equation}
\begin{split}   
\lambda_{A'K}=&g_8(\lambda_s+\lambda_u)+g_8^s(\lambda_u+\lambda_d+\lambda_s)\\
\lambda_{A'\pi} =&g_8(\lambda_u+\lambda_d)+g_8^s(\lambda_u+\lambda_d+\lambda_s)\\
\lambda_{A'\pi K}(p_K^{\nu},p_{\pi^+}^{\nu})=&\frac{2i}{F^2}\left(p_K^{\nu}\lambda_{A'K}+p_{\pi^+}^{\nu}\lambda_{A'\pi}\right)\\
    \lambda_{\pi^0\pi K}=&\frac{\lambda_{\pi^0_b\pi K}}{\left(1-\frac{m_{\pi^0A'}^2}{m_{A'}^2}\right)^{1/2}} \left(s_{\alpha}t_{u}s_{\beta}c_{\gamma}+c_{\alpha}t_us_{\gamma}+c_{\beta}c_{\gamma}+\frac{(s_{\alpha}s_{\gamma}-c_{\alpha}s_{\beta}c_{\gamma})(t_u s_w-s_v/c_u)}{D}\right)\\
    +&\frac{\lambda_{\eta_b\pi K}}{\left(1-\frac{m_{\eta A'}^2}{ m_{A'}^2}\right)^{1/2}} \left(\frac{-s_{\alpha}s_{\beta}c_{\gamma}-c_{\alpha}s_{\gamma}}{c_u}+\frac{(s_{\alpha}s_{\gamma}-c_{\alpha}s_{\beta}c_{\gamma})(t_u s_v-s_w/c_u)}{D}\right)\\
        +&\frac{\lambda_{\eta^{\prime}_{b}\pi K}}{\left(1-\frac{m_{\eta^{\prime} A'}^2}{ m_{A'}^2}\right)^{1/2}}\left(\frac{s_{\alpha}c_us_{\gamma}-c_uc_{\alpha}s_{\beta}c_{\gamma}}{D}\right)\\             \lambda_{\eta\pi K}=&\frac{\lambda_{\pi^0_b\pi K}}{\left(1-\frac{m_{\pi^0A'}^2}{m_{A'}^2}\right)^{1/2}}\left(s_{\alpha}t_us_{\beta}s_{\gamma}-c_{\alpha}t_uc_{\gamma}+c_{\beta}s_{\gamma}-\frac{(c_{\alpha}s_{\beta}s_{\gamma}+s_{\alpha}c_{\gamma})(t_us_w-s_v/c_u)}{D}\right)\\      +&\frac{\lambda_{\eta_b\pi K}}{\left(1-\frac{m_{\eta A'}^2}{m_{A'}^2}\right)^{1/2}}\left(\frac{c_{\alpha}c_{\gamma}-s_{\alpha}s_{\beta}s_{\gamma}}{c_u}-\frac{(c_{\alpha}s_{\beta}s_{\gamma}+s_{\alpha}c_{\gamma})(t_us_v-s_w/c_u)}{D}\right)\\
        -&\frac{\lambda_{\eta^{\prime}_{b}\pi K}}{\left(1-\frac{m_{\eta^{\prime}A'}^2}{m_{A'}^2}\right)^{1/2}}\left(\frac{c_uc_{\alpha}s_{\beta}s_{\gamma}+s_{\alpha}c_uc_{\gamma}}{D}\right)\\        
        \lambda_{\eta^{\prime}\pi K}=& \frac{\lambda_{\pi^0_b\pi K} }{\left(1-\frac{m_{\pi^0A'}^2}{m_{A'}^2}\right)^{1/2}}\left(s_{\beta}-s_{\alpha}t_uc_{\beta}+\frac{c_{\alpha}c_{\beta}(t_us_w-s_v/c_u)}{D}\right)\\       +&\frac{\lambda_{\eta_b\pi K}}{\left(1-\frac{m_{\eta A'}^2}{m_{A'}^2}\right)^{1/2}}\left(s_{\alpha}c_{\beta}/c_u+\frac{c_{\alpha}c_{\beta}(t_us_v-s_w/c_u)}{D}\right)\\+&\frac{\lambda_{\eta^{\prime}_{b}\pi K}}{\left(1-\frac{m_{\eta^{\prime} A'}^2}{m_{A'}^2}\right)^{1/2}}\frac{c_uc_{\alpha}c_{\beta}}{D}\,.
\end{split}
\end{equation}
In the limit of small mixing angles this is
\begin{equation}
\begin{split}   
    \lambda_{j\pi K}\approx&\lambda_{j_b\pi K} \\
\end{split}
\end{equation}
and
\begin{align}
    \lambda_{\pi^0_b\pi K}(E_{\pi^+})&=\frac{ig_8}{F^3}(2E_{\pi^+}m_{K^+}-m_{K^+}^2)+\lambda_{A'\pi K\pi^0_b}(E_{\pi^+})\\
     \lambda_{\eta_b\pi K}(E_{\pi^+})&=\frac{\sqrt{3}ig_8}{3F^3}(m_{K^+}^2-6m_{K^+}E_{\pi^+}+2m_{\pi^+}^2)+\lambda_{A'\pi K\eta_b}(E_{\pi^+})\\
      \lambda_{\eta^{\prime}_{b}\pi K}(E_{\pi^+})&=\frac{i\sqrt{6}}{3F^3}(2g_8+3g_8^s)(m_{\pi^+}^2-m_{K^+}^2)+\lambda_{A'\pi K\eta'_b}(E_{\pi^+})\\
      \lambda_{A'\pi K j}(E_{\pi^+})&= -\frac{2im_{jA'} }{F^2m_{A'}^2}\left(g_8\left((\lambda_d+\lambda_u)(E_{\pi^+}m_{K^+}-m_{\pi^+}^2)-(\lambda_s+\lambda_u)(E_{\pi^+}m_{K^+}-m_{K^+}^2)\right)+g_8^s(m_{K^+}^2-m_{\pi^+}^2)(\lambda_u+\lambda_d+\lambda_s)\right)\,.
\end{align}
In the small mixing limit we can take $\lambda_{A'\pi K j}\approx0$.

The decay rate for $K^+\rightarrow\pi^+A'$ is
\begin{equation}
\label{eq:K2piA}
\begin{split}
    \Gamma_{K\rightarrow\pi A'}=&  G_F^2F^4 |V_{us}|^2|V_{ud}|^{2}\frac{\left(\left(-m_{A'}^2+m_{K^+}^2+m_{\pi^+}^2\right)^2-4 m_{K^+}^2 m_{\pi^+}^2\right)^{3/2}}{128 \pi  m_{A'}^2 m_{K^+}^7}\\ 
    \times&\Bigg(\lambda_{A'\pi}^2
   \left(m_{A'}^4+2 m_{A'}^2
   \left(m_{K^+}^2-m_{\pi^+}^2\right)+\left(m_{K^+}^2+m_{\pi^+}^2\right)^2\right)+4 \lambda_{A'\pi} \lambda_{A'K} m_{K^+}^2
   \left(-m_{A'}^2+m_{K^+}^2+m_{\pi^+}^2\right)+4 \lambda_{A'K}^2
   m_{K^+}^4\Bigg).
   \end{split}
\end{equation}.

The $\lambda_{j\chi}$'s are defined exactly in Eq.~\eqref{eq:lambdasexact} and in the small angle approximation in Eq.~\eqref{eq:lambdasapprox}.
\section{Cross Sections for Relic Abundance Calculations}
\label{app:relic}
We note that this section only includes annihilation cross sections relevant to $\chi$'s with $m_{\chi}\leq 300$ MeV, as this represents the heaviest $\chi$'s we can produce at the correct DM abundance with a reheat temperature of 10 MeV.  Additional annihilation channels not included in this work will open up for higher mass $\chi$'s or for higher reheat temperatures.  Some of the following cross sections will depend on three- and four- meson interaction terms which were not included in Eq.~\eqref{eq:expandedChEFTL}.  These additional interaction terms can be found by expanding the ChEFT Lagrangian, Eq.~\eqref{eqn:ChEFT Lagrangian} out to 3rd or 4th order in 1/F. We neglect the $L_i$ terms, and interaction terms that do not contribute to the $\chi$ annihilation at the values of $m_{\chi}$ and reheat temperature considered in this work.  Doing so gives:
\begin{equation}
\label{eq:34mesons}
    \begin{split}
    \mathcal{L}\supset& \frac{F^2}{4}\times\Bigg[(\mathcal{O}(1, F^{-1}, F^{-2}) \text{ terms}) +\frac{B_0}{6F^3}(\tilde{g}_d-\tilde{g}_u)\phi\left((\pi^0)^3+2\pi^0\pi^+\pi^-\right)-\frac{B_0(\tilde{g}_u+\tilde{g}_d)}{2\sqrt{3}F^3}(\eta+\sqrt{2}\eta')((\pi^0)^2+2\pi^+\pi^-)\\&+\frac{B_0}{24F^4}(m_u+m_d)\left((\pi^0)^4+2(\pi^0)^2\pi^+\pi^-\right)-\frac{B_0(m_d-m_u)}{6\sqrt{3}F^4}(\eta+\sqrt{2}\eta')\left((\pi^0)^3+2\pi^0\pi^+\pi^-\right) \\&+\frac{1}{3 F^4}\left(\pi^0d\pi^0\left(\pi^+d\pi^-+\pi^-d\pi^+\right)-(d\pi^0)^2\pi^+\pi^--(\pi^0)^2d\pi^+d \pi^-\right)\\
    &-\frac{(\lambda_d-\lambda_u)^2}{F^2}A_{\mu}'A'^{\mu}\pi^+\pi^-+\frac{2}{3F^3}A'_{\mu}(\lambda_d-\lambda_u)\left(\pi^0\left(\pi^+\partial^{\mu}\pi^-+\pi^-\partial^{\mu}\pi^+\right)-2\pi^+\pi^-\partial^{\mu}\pi^0\right)\Bigg]\,.
    \end{split}
\end{equation}

The expressions in the first three lines will be relevant for $\chi$ annihilation when $\phi$ has a pseudoscalar coupling to quarks.  The last three lines are relevant when $A'^{\mu}$ has an axial vector coupling to quarks. 
\subsection{Scalar Couplings to Quarks}

The $\chi$ annihilation cross sections are as follows:

In the $S\otimes S$ case:
\begin{equation}
    \sigma^{S\otimes S}_{\chi\bar{\chi}\rightarrow\text{all}}=\sigma^{S\otimes S}_{\chi \chi\rightarrow\phi\phi}+\sigma^{S\otimes S}_{\chi\bar{\chi}\rightarrow\gamma\gamma}+\sigma^{S\otimes S}_{\chi\bar{\chi}\rightarrow\pi^+\pi^-}+\sigma^{S\otimes S}_{\chi\bar{\chi}\rightarrow \pi^0\pi^0}\,,
\end{equation}
where
\begin{equation}
\label{eq:anss}
    \begin{split}
        \sigma^{S\otimes S}_{\chi \chi\rightarrow\phi\phi}=&\frac{g_{\chi}^4 \sqrt{\left(s-4 m_{\chi}^2\right) \left(s-4 m_{\phi}^2\right)}}{32 \pi  s \left(m_{\chi}^2 \left(s-4 m_{\phi
   }^2\right)+m_{\phi}^4\right)}\times\left(\frac{16
   m_{\chi}^4+2 m_{\chi}^2 \left(s-8 m_{\phi}^2\right)+3 m_{\phi}^4}{4
   m_{\chi}^2-s}\vphantom{\frac{\left(\frac{\sqrt{\left(s-4 m_{\chi}^2\right) \left(4 m_{\phi
   }^2-s\right)}}{2 m_{\phi}^2-s}\right)}{\left(s-4 m_{\chi}^2\right)^{3/2}
   \left(2 m_{\phi}^2-s\right) \sqrt{4 m_{\phi}^2-s}}}\right.\\
   &+\left.\frac{2 \left(m_{\chi}^2 \left(s-4 m_{\phi}^2\right)+m_{\phi
   }^4\right) \left(-32 m_{\chi}^4+16 m_{\chi}^2 \left(s-m_{\phi}^2\right)+6
   m_{\phi}^4-4 m_{\phi}^2 s+s^2\right) \tan
   ^{-1}\left(\frac{\sqrt{\left(s-4 m_{\chi}^2\right) \left(4 m_{\phi
   }^2-s\right)}}{2 m_{\phi}^2-s}\right)}{\left(s-4 m_{\chi}^2\right)^{3/2}
   \left(2 m_{\phi}^2-s\right) \sqrt{4 m_{\phi}^2-s}}\right)\\
   \end{split}
   \end{equation}
   \begin{equation}
   \sigma^{S\otimes S}_{\chi\bar{\chi}\rightarrow\gamma\gamma}=\frac{g_\chi^2\alpha^2 \mathcal{D}}{18432 \pi^3} \frac{s \left(s-4m_\chi^2\right)}{\left(\left(s-m_\phi^2\right)^2+\left(m_{\phi}\Gamma_{\phi}^{S\otimes S}\right)^2\right)} \sqrt{\frac{s}{s-4m_\chi^2}}
   \end{equation}
   \begin{equation}
   \sigma^{S\otimes S}_{\chi\bar{\chi}\rightarrow \pi^+\pi^-}=\frac{B_0^2(g_u+g_d)^2g_{\chi}^2}{32\pi s}\frac{\sqrt{s-4m_{\pi^+}^2}\sqrt{s-4m_{\chi}^2}}{\left(\left(s-m_\phi^2\right)^2+\left(m_{\phi}\Gamma_{\phi}^{S\otimes S}\right)^2\right)}
      \end{equation}
   \begin{equation}
   \sigma^{S\otimes S}_{\chi\bar{\chi}\rightarrow \pi^0\pi^0}=\frac{B_0^2(g_u+g_d)^2g_{\chi}^2}{64\pi s}\frac{\sqrt{s-4m_{\pi^0}^2}\sqrt{s-4m_{\chi}^2}}{\left(\left(s-m_\phi^2\right)^2+\left(m_{\phi}\Gamma_{\phi}^{S\otimes S}\right)^2\right)}\,.
\end{equation}

For the $P\otimes S$ case the total annihilation rate is
\begin{equation}
    \sigma^{P\otimes S}_{\chi\bar{\chi}\rightarrow\text{all}}=\sigma^{P\otimes S}_{\chi \chi\rightarrow\phi\phi}+\sigma^{P\otimes S}_{\chi\bar{\chi}\rightarrow\gamma\gamma}+\sigma^{P\otimes S}_{\chi\bar{\chi}\rightarrow\pi^+\pi^-}+\sigma^{P\otimes S}_{\chi\bar{\chi}\rightarrow \pi^0\pi^0}\,,
\end{equation}
where
\begin{equation}
        \sigma^{P\otimes S}_{\chi\bar{\chi}\rightarrow\phi\phi}=\frac{g_{5\chi}^4}{16\pi s} \left( -\sqrt{\frac{s-4m_{\phi}^2}{s-4m_{\chi}^2}}\frac{2m_{\chi}^2(s-4m_{\sigma}^2)+3m_{\phi}^4}{2(m_{\chi}^2(s-4m_{\phi}^2)+m_{\phi}^4)}+\frac{6m_{\phi}^4-4m_{\phi}^2s+s^2}{(s-4m_{\chi}^2)(s-2m_{\phi}^2)}\tanh^{-1}\left(\frac{\sqrt{(s-4m_{\chi}^2)(s-4m_{\phi}^2)}}{s-2m_{\phi}^2}\right)\right)
    \end{equation}
   \begin{equation}
        \sigma^{P\otimes S}_{\chi\bar{\chi}\rightarrow\gamma\gamma}=\displaystyle\frac{g_{5\chi}^2 \alpha^2\mathcal{D}}{18432 \pi^3} \frac{s^2}{ {\left(s-m_\phi^2\right)}^2}\sqrt{\frac{s}{s-4m_\chi^2}}
           \end{equation}
   \begin{equation}
  \sigma^{P\otimes S}_{\chi\bar{\chi}\rightarrow\pi^+\pi^-}=\frac{B_0^2(g_u+g_d)^2g_{5\chi}^2}{32\pi(s-m_{\phi}^2)^2}\frac{\sqrt{s-4m_{\pi^+}^2}}{\sqrt{s-4m_{\chi}^2}}
     \end{equation}
   \begin{equation}
  \sigma^{P\otimes S}_{\chi\bar{\chi}\rightarrow \pi^0\pi^0}=\frac{B_0^2(g_u+g_d)^2g_{5\chi}^2}{64\pi(s-m_{\phi}^2)^2}\frac{\sqrt{s-4m_{\pi^0}^2}}{\sqrt{s-4m_{\chi}^2}}\,.
    \label{eq:annps}
\end{equation}

\subsection{Pseudoscalar Couplings to Quarks}
At a reheat temperature of 10 MeV the annihilation channels available to the $\chi$ are:  
\begin{equation}
    \begin{split}
    \label{eq:annpp}
        \sigma^{P\otimes P}_{\chi\bar{\chi}\rightarrow\text{all}} =\sigma^{P\otimes P}_{\chi\bar{\chi}\rightarrow \gamma\gamma}+\sum_j^{\phi/\pi^0/
    \eta/\eta'}\sigma^{P\otimes P}_{\chi\bar{\chi}\rightarrow jj}+\sigma^{P\otimes P}_{\chi\bar{\chi}\rightarrow \pi^0\pi^0\pi^0}+\sigma^{P\otimes P}_{\chi\bar{\chi}\rightarrow \pi^0\pi^+\pi^-}+\sigma^{P\otimes P}_{\chi\bar{\chi}\rightarrow \pi^0\pi^0\phi}+\sigma^{P\otimes P}_{\chi\bar{\chi}\rightarrow \pi^+\pi^-\phi}\,.
    \end{split}
    \end{equation}
All of the above annihilation channels can be calculated at tree-level using the mixing between the mesons and $\phi$ and the three and four meson interaction terms of the ChEFT Lagrangian shown in Eq.~\eqref{eq:34mesons}.  
We put the interaction terms drawn from this Lagrangian and needed for the annihilation calculations in Table \ref{Tab:3pis}. Note these interaction terms are all given in the small mixing limit (see Eq.~\eqref{eq:pseudolimit}).  The resulting annihilation cross sections are:
\begin{table}[h]

    \centering
    \begin{tabular}{|c|c |c| c| c|}
\hline
    &$\pi^0$&$\eta$&$\eta'$&$\phi$\\
    \hline
    $\pi^0\pi^0\pi^0$&$\displaystyle \frac{m_{\pi^+}^2}{F^2}$ &$\displaystyle -\frac{B_0(m_d-m_u)}{\sqrt{3}F^2}$ & $\displaystyle-\frac{2B_0(m_d-m_u)}{\sqrt{6}F^2}$& $\displaystyle\frac{B_0(\tilde{g}_d-\tilde{g}_u)}{F}-\sum_j^{\pi^0/\eta/\eta'}\frac{\tilde{g}_{\pi^0\pi^0\pi^0 j}m_{j \phi}^2}{(m_{\phi}^2-m_{j}^2)}$\\
    \hline
    $\pi^+\pi^-\pi^0$&$\displaystyle \frac{ 2B_0(m_u+m_d)}{3F^2}$&$\displaystyle -\frac{ B_0(m_d-m_u)}{3\sqrt{3}F^2}$&$\displaystyle -\frac{2 B_0(m_d-m_u)}{3\sqrt{6}F^2}$ &$\displaystyle \frac{B_0(\tilde{g}_d-\tilde{g}_u)}{3F}-\sum_j^{\pi^0/\eta/\eta'}\frac{\tilde{g}_{\pi^+\pi^-\pi^0j}m_{ j\phi}^2}{m_{\phi}^2-m_j^2}$\\
    \hline
    $\pi^+\pi^-\partial\pi^0$ & $\displaystyle\frac{1}{3F^2}$ & & & $\displaystyle \frac{-m_{\pi^0\phi}^2}{3F^2\left(m_{\phi}^2-m_{\pi^0}^2\right)}$ \\
    \hline
    $\pi^0\pi^0\phi$&$\displaystyle\frac{B_0(\tilde{g}_d-\tilde{g}_u)}{F}-\sum_j^{\pi^0/\eta/\eta'}\frac{\tilde{g}_{\pi^0\pi^0\pi^0 j}m_{j \phi}^2}{m_{\phi}^2-m_j^2}$&$\displaystyle -\frac{B_0(\tilde{g}_d+\tilde{g}_u)}{\sqrt{3}F}$&$\displaystyle -\frac{2B_0(\tilde{g}_d+\tilde{g}_u)}{\sqrt{6}F}$&$\displaystyle \sum_j^{\pi^0/\eta/\eta'}-\frac{m_{j \phi}^2}{\tilde{g}_{\pi^0\pi^0\phi j}(m_{\phi}^2-m_j^2)}$\\
    \hline
    $\pi^+\pi^-\phi$&$\displaystyle \frac{B_0(\tilde{g}_d-\tilde{g}_u)}{3F}-\sum_j^{\pi^0/\eta/\eta'}\frac{\tilde{g}_{\pi^+\pi^-\pi^0j}m_{ j\phi}^2}{m_{\phi}^2-m_j^2}$&$\displaystyle -\frac{2B_0(\tilde{g}_d+\tilde{g}_u)}{\sqrt{3}F}$&$\displaystyle -\frac{2B_0(\tilde{g}_d+\tilde{g}_u)}{\sqrt{6}F}$&$\displaystyle\sum_j^{\pi^0/\eta/\eta'}-\frac{\tilde{g}_{\pi^+\pi^-\phi j}m_{j \phi}^2}{m_{\phi}^2-m_j^2}$\\\hline
      
\end{tabular}
    \caption{Effective couplings relevant to $\chi$ annihilation for the $S\otimes P$ and $P\otimes P$ cases. These terms are displayed in the form $\tilde{g}_{ab}$ where the row headers= $a$, and the column headers = $b$.}
    \label{Tab:3pis}  $m_{j\phi}^2$ are defined in Eq.~\eqref{eq:pseudoaxialmixings}.
\end{table}

    \begin{equation}
   \sigma^{P\otimes P}_{\chi\bar{\chi}\rightarrow \gamma\gamma}= \frac{s^2}{128\pi }\left|\sum_j^{\pi^0/\eta/\eta'/\phi}\frac{\tilde{g}_{ j\chi} \tilde{g}_{ j\gamma}}{\left(s-m_{j}^2\right) +im_{j}\Gamma_{j}}\right|^2\sqrt{\frac{s}{s-4m_\chi^2}}
   \end{equation}
   \begin{equation}
        \sigma^{P\otimes P}_{\chi\bar{\chi}\rightarrow jj}=-\frac{\tilde{g}_{j\chi}^4}{16\pi s}\sqrt{\frac{s-4m_{j}^2}{s-4m_{\chi}^2}}\left(\frac{2m_{\chi}^2(s-4m_{j}^2)+3m_{j}^4}{2(m_{\chi}^2(s-4m_{j}^2)+m_{j}^4)}+\frac{6m_{j}^4-4m_{j}^2s+s^2}{(s-4m_{\chi}^2)(s-2m_{j}^2)}\tanh^{-1}\left(\frac{\sqrt{(s-4m_{\chi}^2)(s-4m_{j}^2)}}{s-2m_{j}^2}\right)\right)
        \end{equation}
        \begin{equation}
        \sigma^{P\otimes P}_{\chi\bar{\chi}\rightarrow \pi^0\pi^0\pi^0}=\frac{1}{768\pi^3\sqrt{1-\frac{4m_{\chi}^2}{s}}}\left|\sum_j^{\pi^0/\eta/\eta'/\phi}\frac{\tilde{g}_{j\chi} \tilde{g}_{ \pi^0\pi^0\pi^0j}}{\left(s-m_{j}^2\right) +im_{j}\Gamma_{j}}\right|^2\int^{\frac{s-3m_{\pi^0}^2}{2\sqrt{s}}}_{m_{\pi^0}} dE_{\pi^0} Y(E_{\pi^0},\sqrt{s},m_{\pi^0},m_{\pi^0})
        \end{equation}
        
In the small angle approximation,
        \begin{equation}
        \begin{split}
        \sigma_{\chi\bar{\chi}\rightarrow\pi^+\pi^-\pi^0}^{P\otimes P}&=\frac{1}{128 \pi^3\sqrt{1-\frac{4m_{\chi}^2}{s}}}\left|\sum_j^{\pi^0/\eta/\eta'/\phi}\frac{\tilde{g}_{j\chi} \tilde{g}_{\pi^+\pi^-\pi^0 j}}{\left(s-m_{j}^2\right) +im_{j}\Gamma_{j}}\right|^2\int^{\frac{s+m_{\pi^0}^2-4m_{\pi^+}^2}{2\sqrt{s}}}_{m_{\pi^0}}dE_{\pi^0}Y(E_{\pi^0},\sqrt{s},m_{\pi^0},m_{\pi^+})\\
        &-\frac{1}{1024\pi^3(s-m_{\pi^0}^2)\sqrt{1-\frac{4m_{\chi}^2}{s}}}\left(\sum_k^{(\phi/\pi^0)}\sum_j^{(\pi^0/\eta/\eta'/\phi)}\frac{ \tilde{g}_{j \chi}\tilde{g}_{\pi^0\pi^0\pi^0 j}\tilde{g}_{k\chi}\tilde{g}_{\pi^+\pi^-\partial\pi^0 k}\left((s-m_j^2)(s-m_{k}^2)+m_{j}\Gamma_{j}m_{k}\Gamma_{k}\right)  } {\left(\left(s-m_{j}^2\right)^2 +m_j^2\Gamma_j^2\right)\left(\left(s-m_{k}^2\right)^2 +m_{k}^2\Gamma_{k}^2\right)}\right)\\
        &\times\int^{\frac{s+m_{\pi^0}^2-4m_{\pi^+}^2}{2\sqrt{s}}}_{m_{\pi^0}}dE_{\pi^0}\left[\left(21 s-44 E_{\pi^0} \sqrt{s}-32 m_{\pi^+}^2+32 m_{\pi^0}^2\right)
   Y(E_{\pi^0},\sqrt{s},m_{\pi^0},m_{\pi^+})  \right]\\
   &+\frac{1}{98304 \pi ^3 
\sqrt{1-\frac{4m_{\chi}^2}{s}}}\left|\sum_k^{\pi^0/\phi}\frac{\tilde{g}_{k\chi} \tilde{g}_{\pi^+\pi^-\partial\pi^0 k}}{\left(s-m_{k}^2+i m_{k}\Gamma_{k}\right)}\right|^2\int^{\frac{s+m_{\pi^0}^2-4m_{\pi^+}^2}{2\sqrt{s}}}_{m_{\pi^0}}dE_{\pi^0} \Bigg[Y(E_{\pi^0},\sqrt{s},m_{\pi^0},m_{\pi^+})\\&\times\left(16 s Y(E_{\pi^0},\sqrt{s},m_{\pi^0},m_{\pi^+})^2+3\left(44 E_{\pi^0} \sqrt{s}+32 m_{\pi^+}^2-32 m_{\pi^0}^2-21
   s\right)^2\right)\Bigg]\\
   \end{split}
   \end{equation}
   
   \begin{equation}
   \sigma_{\chi\bar{\chi}\rightarrow\pi^0\pi^0\phi}^{P\otimes P}=\frac{1}{256\pi^3\sqrt{1-\frac{4m_{\chi}^2}{s}}}\left|\sum_j^{\pi^0/\eta/\eta'/\phi}\frac{\tilde{g}_{j\chi} \tilde{g}_{\pi^0\pi^0 \phi j}}{\left(s-m_{j}^2\right) +im_{j}\Gamma_{j}}\right|^2\int^{\frac{s+m_{\phi}^2-4m_{\pi^0}^2}{2\sqrt{s}}}_{m_{\phi}}dE_{\phi}Y(E_\phi,\sqrt{s},m_{\phi},m_{\pi^0})
   \end{equation}
   
   \begin{equation}
       \sigma_{\chi\bar{\chi}\rightarrow\pi^+\pi^-\phi}^{P\otimes P} = \frac{1}{128\pi^3\sqrt{1-\frac{4m_{\chi}^2}{s}}}\left|\sum_j^{\pi^0/\eta/\eta'/\phi}\frac{\tilde{g}_{j\chi} \tilde{g}_{\pi^+\pi^-\phi j}}{\left(s-m_{j}^2\right) +im_{j}\Gamma_{j}}\right|^2\int^{\frac{s+m_{\phi}^2-4m_{\pi^+}^2}{2\sqrt{s}}}_{m_{\phi}}dE_{\phi}Y(E_{\phi},\sqrt{s},m_{\phi},m_{\pi^+})
   \end{equation}
   For the $S\otimes P$ case we have
\begin{equation}
  \label{eq:annsp}
    \sigma^{S\otimes P}_{\chi\bar{\chi}\rightarrow\text{all}} =\sigma^{S\otimes P}_{\chi\bar{\chi}\rightarrow \gamma\gamma}+\sum_j^{\phi/\pi^0/
    \eta/\eta'}\sigma^{S\otimes P}_{\chi\bar{\chi}\rightarrow jj}+\sigma^{S\otimes P}_{\chi\bar{\chi}\rightarrow \pi^0\pi^0\pi^0}+\sigma^{S\otimes P}_{\chi\bar{\chi}\rightarrow \pi^0\pi^+\pi^-}+\sigma^{S\otimes P}_{\chi\bar{\chi}\rightarrow \pi^0\pi^0\phi}+\sigma^{S\otimes P}_{\chi\bar{\chi}\rightarrow \pi^+\pi^-\phi}
\end{equation}
   
        \begin{equation}
\sigma^{S\otimes P}_{\chi\bar{\chi}\rightarrow\gamma\gamma} =\frac{s^{3/2}(s-4m_{\chi}^2)^{1/2}}{128\pi }\left|\sum_j^{\pi^0/\eta/\eta'/\phi}\frac{\tilde{g}_{ j\chi} \tilde{g}_{ j\gamma}}{\left(s-m_{j}^2\right) +im_{j}\Gamma_{j}}\right|^2
    \end{equation}
    \begin{equation}
    \begin{split}
   \sigma^{S\otimes P}_{\chi\bar{\chi}\rightarrow jj}= &\frac{\tilde{g}_{j\chi}^4}{32\pi s(s-4m_{\chi}^2)}\Bigg[-\frac{\sqrt{(s-4m_{\chi}^2)(s-4m_{j}^2)}(16m_{\chi}^4+2m_{\chi}^2(s-8m_{j}^2)+3m_{j}^4)}{m_{\chi}^2(s-4m_{j}^2)+m_{j}^4}\\&+\frac{s^2-32m_{\chi}^4+16m_{\chi}^2(s-m_{j}^2)+6m_{j}^4-4m_{j}^2s}{s-2m_{j}^2}\tan^{-1}\left(\frac{\sqrt{(s-4m_{\chi}^2)(s-4m_{j}^2)}}{s-2m_{j}^2}\right)\Bigg]\,.
    \end{split}
\end{equation}\,.
For all of the three particle annihilations calculated here
\begin{equation}
    \sigma^{S\otimes P}_{\chi\chi\rightarrow\text{3 particles}} = \frac{g_{\chi}^2\left(1-\frac{4m_{\chi}^2}{s}\right)}{g_{5\chi}^2} \sigma^{P\otimes P}_{\chi\chi\rightarrow\text{3 particles}}\,.
\end{equation}
The $\tilde{g}_{j\chi}$ and  $\tilde{g}_{\phi\gamma}$ can be found in the exact form in Eq.~\eqref{eq:pschicoup}  and Eq.~\eqref{eq:app-pseudocouplingslim} respectively, and in their small angle approximation form in Eq.~\eqref{eq:phiphoton}. The remaining $\tilde{g}_{j\gamma}$'s can be found in Eq.~\eqref{eqn:pseudocouplings}.   The meson widths are taken to be $\Gamma_{\pi^0}=7.7\times 10^{-3}$ keV and $\Gamma_{\eta} = 1.3$ keV \protect\cite{ParticleDataGroup:2024cfk}, and the $\phi$ decay widths can be found in Table \ref{tab:pseudoscalar} for the $P\otimes P$  case and Table \ref{tab:pseudoscalar+scalar} for the $S \otimes P$ case. $Y(a,b,c,d)$ is defined in Eq.~\eqref{eq:Y}.

\subsection{Axial Vector Couplings to Quarks}
For reheat temperatures of 10 MeV, the relevant terms contributing to the $\chi$ annihilation cross section are.
\begin{equation}
    \sigma^{A\otimes A}_{\chi\bar{\chi}\rightarrow\text{all}} = \sigma^{A\otimes A}_{\chi\bar{\chi}\rightarrow\gamma\gamma}+\sigma^{A\otimes A}_{\chi\bar{\chi}\rightarrow A'A'}+\sum_{j}^{\pi^0/\eta/\eta'}\sigma^{A\otimes A}_{\chi\bar{\chi}\rightarrow jj}+\sigma^{A\otimes A}_{\chi\bar{\chi}\rightarrow\pi^0\pi^0\pi^0}+\sigma^{A\otimes A}_{\chi\bar{\chi}\rightarrow\pi^+\pi^-\pi^0}+\sigma^{A\otimes A}_{\chi\bar{\chi}\rightarrow\pi^+\pi^- A'}\,.
\end{equation}
These cross sections may all be found at tree level using the meson mixing with the $A'$, and 3 meson and 4 particle interaction terms shown in Eq.~\eqref{eq:34mesons}.  From Eq.~\eqref{eq:34mesons} we can extract the coupling coefficients relevant to $A'_{\mu}$ meson interactions.  In the small angle approximation (see Eq.~\eqref{axlim} for details) they are
\begin{equation}
\label{4lambda}
    \lambda_{A'A'\pi^+\pi^-} = - 2(\lambda_d-\lambda_u)^2\,\quad\text{ and } \quad \lambda_{\pi^+\pi^-\pi^0 A'} =\frac{2(\lambda_d-\lambda_u)}{3F}\,.
\end{equation}
The remaining relevant three and four meson coupling terms can be found in Table \ref{Tab:3pis}.  The $\chi$ annihlation cross sections are as follows:
\begin{equation}
       \sigma^{A\otimes A}_{\chi\bar{\chi}\rightarrow\gamma\gamma} =\frac{m_{\chi}^2s^{2}}{32\pi}\left(\sum_j^{\pi^0 /\eta/\eta'}\frac{\lambda_{j\chi}\tilde{g}_{j\gamma}}{s-m_{j}^2}\right)^2\sqrt{\frac{s}{s-4m_{\chi}^2}}
       \end{equation}

\begin{equation}
\begin{split}    
       \sigma^{A\otimes A}_{\chi\bar{\chi}\rightarrow A'A'}&=\frac{\lambda_{\chi}^4 \sqrt{\frac{s-4 m_{A'}^2}{s-4 m_{\chi}^2}}}{32 \pi  m_{A'}^4 s \left(m_{A'}^4-4 m_{A'}^2
   m_{\chi}^2+m_{\chi}^2 s\right)} \left[-8 m_{A'}^8+8
   m_{A'}^6 \left(2 m_{\chi}^2+s\right)+m_{A'}^4 \left(64 m_{\chi}^4-28
   m_{\chi}^2 s-3 s^2\right)\vphantom{\frac{\sqrt{\left(s-4 m_{j}^2\right)
   \left(s-4 m_{\chi}^2\right)}}{2 m_{j}^2-s}}\right.\\
   &+4 m_{A'}^2 m_{\chi}^2 s \left(3 s-16
   m_{\chi}^2\right)-2 m_{\chi}^2 s^2 \left(s-4
   m_{\chi}^2\right)\\
   &+2 \frac{\left(16 m_{A'}^8-16 m_{A'}^6 \left(4 m_{\chi}^2+s\right)+2
   m_{A'}^4 s \left(8 m_{\chi}^2+5 s\right)-4 m_{A'}^2 s \left(-16 m_{\chi}^4-2
   m_{\chi}^2 s+s^2\right)-16 m_{\chi}^4 s^2+s^4\right) }{\left(2 m_{A'}^2-s\right) \sqrt{\left(4
   m_{A'}^2-s\right) \left(s-4 m_{\chi}^2\right)}}\\
   &\left.\times \left(m_{A'}^4-4 m_{A'}^2 m_{\chi}^2+m_{\chi}^2
   s\right) \tan
   ^{-1}\left(\frac{\sqrt{\left(4 m_{A'}^2-s\right) \left(s-4 m_{\chi}^2\right)}}{2
   m_{A'}^2-s}\right)\right] 
   \end{split}
   \end{equation}
   
   \begin{equation}
       \begin{split}
   \sigma^{A\otimes A}_{\chi\bar{\chi}\rightarrow jj}=&\frac{1}{32 \pi  s}\lambda_{j\chi}^4 \left[\vphantom{\frac{\frac{\sqrt{\left(s-4 m_{j}^2\right)
   \left(s-4 m_{\chi}^2\right)}}{2 m_{j}^2-s}}{2 m_j^2-s}}\frac{\sqrt{\left(s-4 m_{j}^2\right) \left(s-4
   m_{\chi}^2\right)} \left(m_{\chi}^2 \left(4 m_{j}^4+8 m_{j}^2
   s-2 s^2\right)-3 m_{j}^4 s\right)}{m_{j}^4+m_{\chi}^2 \left(s-4
   m_{j}^2\right)}\right.\nonumber \\
   &+\left.\frac{2 \left(s \left(6 m_{j}^4-4 m_{j}^2
   s+s^2\right)+4 m_{\chi}^2 \left(2 m_{j}^4-4 m_{j}^2
   s+s^2\right)\right) \tanh ^{-1}\left(\frac{\sqrt{\left(s-4 m_{j}^2\right)
   \left(s-4 m_{\chi}^2\right)}}{2 m_{j}^2-s}\right)}{2 m_j^2-s}\right]
    \end{split}
    \label{eq:annaa}
\end{equation}
\begin{equation}
    \sigma_{\chi\bar{\chi}\rightarrow\pi^0\pi^0\pi^0}^{A\otimes A} = \frac{m_{\chi}^2 }{192\pi^3\sqrt{1-\frac{4m_{\chi}^2}{s}}}\left|\sum_j^{\pi^0/\eta/\eta'}\frac{\tilde{g}_{\pi^0\pi^0\pi^0 j}\lambda_{j\chi}}{s-m_j^2+i m_j\Gamma_j}\right|^2\int^{\frac{s-3m_{\pi^0}^2}{2\sqrt{s}}}_{m_{\pi^0}}dE_{\pi^0}Y(E_{\pi^0},\sqrt{s},m_{\pi^0},m_{\pi^0}) 
\end{equation}
\begin{equation}
\begin{split}
     \sigma_{\chi\bar{\chi}\rightarrow\pi^-\pi^+A'}^{A\otimes A} &= \frac{\lambda_{\chi}^2\lambda_{A'A'\pi^+\pi^-}^2}{1536\pi^3m_{A'}^6\left((s-m_{A'}^2)^2+m_{A'}^2\Gamma_{A'}^2\right)s\sqrt{1-\frac{4m_{\chi}^2}{s}}}\int^{\frac{s+m_{A'}^2-4m_{\pi^+}^2}{2\sqrt{s}}}_{m_{A'}}dE_{A'}\Bigg[Y(E_{A'},\sqrt{s},m_{A'},m_{\pi^+})\\
     &\times\left(12 m_{\chi}^2 s^2 \left(E_{A'}^2-4 m_{A'}^2\right)+m_{A'}^2 s \left(4m_{A'}^2-E_{A'}^2\right) \left(5 m_{A'}^2+24 m_{\chi}^2\right)+16 m_{A'}^4 m_{\chi}^2\left(E_{A'}^2-7 m_{A'}^2\right)\right)\Bigg]\\
   &+\frac{\lambda_{\pi^+\pi^-\pi^0 A'}^2 \lambda_{\pi^0\chi}^2 m_{\chi}^2 \sqrt{s}}{512 \pi ^3 m_{A'}^2 \left((s-m_{\pi^0}^2)^2+m_{\pi^0}^2\Gamma_{\pi^0}^2\right) \sqrt{1-\frac{4 m_{\chi}^2}{s}}}\int^{\frac{s+m_{A'}^2-4m_{\pi^+}^2}{2\sqrt{s}}}_{m_{A'}}dE_{A'}\Bigg[Y(E_{A'},\sqrt{s},m_{A'},m_{\pi^+})\\
   &\times \left(\sqrt{s} \left(16 E_{A'}^2-53 m_{A'}^2\right)-4 E_{A'}m_{A'}^2\right) \Bigg]\\
   &+\frac{\lambda_{\chi}\lambda_{A'A'\pi^+\pi^-} \lambda_{\pi^+\pi^-\pi^0A'} \lambda_{\pi^0\chi}m_{\chi}^2  \left(s-m_{A'}^2\right)\left(\left(s-m_{A'}^2\right) \left(s-m_{\pi^0}^2\right)+m_{A'}\Gamma_{A'} m_{\pi^0}\Gamma_{\pi^0}\right)}{32 \pi ^3 m_{A'}^4 \sqrt{1-\frac{4m_{\chi}^2}{s} }\left(\left(s-m_{A'}^2\right)^2+m_{A'}^2\Gamma_{A'}^2\right) \left(\left(s-m_{\pi^0}^2\right)^2+m_{\pi^0}^2\Gamma_{\pi^0}^2\right)}
   \\&\times\int^{\frac{s+m_{A'}^2-4m_{\pi^+}^2}{2\sqrt{s}}}_{m_{A'}}dE_{A'}Y(E_{A'},\sqrt{s},m_{A'},m_{\pi^+})\left(E_{A'}^2-4 m_{A'}^2\right)
\end{split}
\end{equation}
\begin{equation}
    \begin{split}
        \sigma_{\chi\bar{\chi}\rightarrow\pi^+\pi^-\pi^0}^{A\otimes A}=&\frac{\lambda_{\chi}^2 \lambda_{\pi^+\pi^-\pi^0 A'}^2 }{2048 \pi ^3 m_{A'}^4
   \left((s-m_{A'}^2)^2+m_{A'}^2\Gamma_{A'}^2\right)s\sqrt{1-\frac{4m_{\chi}^2}{s}}}\int_{m_{\pi^0}}^{\frac{s+m_{\pi^0}^2-4m_{\pi^+}^2}{2\sqrt{s}}}dE_{\pi^0}\Bigg[Y(E_{\pi^0},\sqrt{s},m_{\pi^0},m_{\pi^+}) \\
   &\times\Bigg(12 E_{\pi^0}^2 \left(4
   m_{\chi}^2 \left(4 m_{A'}^4-6 m_{A'}^2 s+3 s^2\right)-5 m_{A'}^4 s\right)-4 E_{\pi^0} \sqrt{s}
   \left(m_{\chi}^2 \left(44 m_{A'}^4-96 m_{A'}^2 s
    +48 s^2\right)+m_{A'}^4 s\right)\\
    &+48
   m_{\pi^0}^2 m_{A'}^4 \left(8 m_{\chi}^2-s\right)+s \left(4 m_{\chi}^2 \left(5 m_{A'}^4-32
   m_{A'}^2 s+16 s^2\right)+11 m_{A'}^4 s\right)\Bigg)\Bigg]\\
   &+\frac{\lambda_{\chi} \lambda_{\pi^0\chi }\lambda_{\pi^+\pi^-\pi^0 A'} \tilde{g}_{\pi^+\pi^-\partial\pi^0\pi^0}m_{\chi}^2 (s-m_{A'}^2)\left((s-m_{\pi^0}^2)(s-m_{A'}^2)+m_{\pi^0}\Gamma_{\pi^0}m_{A'}\Gamma_{A'}\right) }{1024 \pi ^3 m_{A'}^2 \left((s-m_{\pi^0}^2)^2+m_{\pi^0}^2\Gamma_{\pi^0}^2\right)\left((s-m_{A'}^2)^2+m_{A'}^2\Gamma_{A'}^2\right)\sqrt{s-4m_{\chi}^2}  }\\
   &\times \int_{m_{\pi^0}}^{\frac{s+m_{\pi^0}^2-4m_{\pi^+}^2}{2\sqrt{s}}}dE_{\pi^0}Y(E_{\pi^0},\sqrt{s},m_{\pi^0},m_{\pi^+})\left(3 E_{\pi^0}-2 \sqrt{s}\right) \Bigg[\left(76E_{\pi^0} \sqrt{s}+32 m_{\pi^+}^2-48 m_{\pi^0}^2-37 s\right) \Bigg]   \\
   &+\frac{\lambda_{\pi^0\chi}^2 \tilde{g}_{\pi^+\pi^-\partial\pi^0\pi^0}^2 m_{\chi}^2}{786432 \pi ^3\left((s-m_{\pi^0}^2)^2+m_{\pi^0}^2\Gamma_{\pi^0}^2\right) \sqrt{1-\frac{4m_{\chi}^2}{s}}}\int_{m_{\pi^0}}^{\frac{s+m_{\pi^0}^2-4m_{\pi^+}^2}{2\sqrt{s}}}dE_{\pi^0}\Bigg[8 Y(E_{\pi^0},\sqrt{s},m_{\pi^0},m_{\pi^+})\\&\times\left(16 s Y(E_{\pi^0},\sqrt{s},m_{\pi^0},m_{\pi^+})^2+3\left(76 E_{\pi^0}\sqrt{s}+32m_{\pi^+}^2-48m_{\pi^0}^2-38 s\right)^2\right)\Bigg]\\
   &+\frac{m_{\chi}^2}{32\pi^3\sqrt{1-\frac{4m_{\chi}^2}{s}}}\left|\sum_{j}^{\pi^0/\eta/\eta'}\frac{\lambda_{j\chi}\tilde{g}_{\pi^+\pi^-\pi^0 j}}{s-m_j^2+im_j\Gamma_j}\right|^2\int_{m_{\pi^0}}^{\frac{s+m_{\pi^0}^2-4m_{\pi^+}^2}{2\sqrt{s}}}dE_{\pi^0}Y(E_{\pi^0},\sqrt{s},m_{\pi^0},m_{\pi^+})\\
   &+\frac{\lambda_{\chi}\lambda_{\pi^+\pi^-\pi^0 A'} m_{\chi}^2(s-m_{A'}^2)}{32\pi^3m_{A'}^2\left((s-m_{A'}^2)^2+m_{A'}^2\Gamma_{A'}^2\right)\sqrt{s-4m_{\chi}^2}}\left(\sum_j^{\pi^0/\eta/\eta'}\frac{\lambda_{j\chi}\tilde{g}_{\pi^+\pi^-\pi^0 j} \left((s-m_j^2)(s-m_{A'}^2)+m_jm_{A'}\Gamma_j\Gamma_{A'}\right)}{\left((s-m_j^2)^2+m_j^2\Gamma_j^2\right)}\right)\\
   &\times\int_{m_{\pi^0}}^{\frac{s+m_{\pi^0}^2-4m_{\pi^+}^2}{2\sqrt{s}}}dE_{\pi^0}(3E_{\pi^0}-2\sqrt{s})Y(E_{\pi^0},\sqrt{s},m_{\pi^0},m_{\pi^+})\\
   &+\frac{\lambda_{\pi^0\chi} \tilde{g}_{\pi^+\pi^-\partial\pi^0\pi^0}}{512\pi^3\sqrt{1-\frac{4m_{\chi}^2}{s}}\left((s-m_{\pi^0}^2)^2+m_{\pi^0}^2\Gamma_{\pi^0}^2\right)}\left(\sum_j^{\pi^0/\eta/\eta'}\frac{\lambda_{j\chi} \tilde{g}_{\pi^+\pi^-\pi^0 j}\left((s-m_{j}^2)(s-m_{\pi^0}^2)+m_{\pi^0}\Gamma_{\pi^0}m_j\Gamma_j\right)}{\left((s-m_j^2)^2+m_j^2\Gamma_j^2\right)}\right)\\
   &\times\int_{m_{\pi^0}}^{\frac{s+m_{\pi^0}^2-4m_{\pi^+}^2}{2\sqrt{s}}}dE_{\pi^0}(76 E_{\pi^0}\sqrt{s}+32m_{\pi^+}^2-48m_{\pi^0}^2-37s)Y(E_{\pi^0},\sqrt{s},m_{\pi^0},m_{\pi^+})
   \end{split}\,.
\end{equation}

The $\lambda_{j\chi}$'s can be found in its exact form Eq.~\eqref{eq:lambdasexact} and in the small angle approximation in Eq.~\eqref{eq:lambdasapprox}.  The $\tilde{g}_{j\gamma}$'s can be found in Eq.~\eqref{eqn:pseudocouplings}.   The meson widths are taken to be $\Gamma_{\pi^0}=7.7\times 10^{-3}$ keV and $\Gamma_{\eta} = 1.3$ keV \protect\cite{ParticleDataGroup:2024cfk}, and the $A'$ decay width can be found in Table \ref{tab:axialvector}.

\section{$N_{eff}$ from $\chi$ Production of Light Mediators}
\label{app:Neff}

In this appendix we describe how the contribution to $N_{\rm eff}$ from mediators produced by DM annihilation is computed. We write med for any of $\phi$ and $A'$. The neutrino temperature $T_\nu$ scales as $1/a$ and equals the photon temperature $T_\gamma$ above $T_{\nu,\rm dec}$.

A full treatment requires solving the coupled Boltzmann equations for the momentum distributions of $\chi$, the mediator and the Standard Model species~\cite{Escudero:2018mvt,EscuderoAbenza:2020cmq}. We instead make the following approximations: 1) while it produces med, $\chi$ is in kinetic and chemical equilibrium at $T_\gamma$, and the mediator is produced towards equilibrium at the same temperature, so that a dark sector with its own temperature is captured only through a bottleneck on the production rate, introduced below; 2) the mediators carry a thermal spectral shape, which fixes both their entropy and the kinetic fraction of their energy at recombination; 3) partial contact with the plasma, $0 < w_i < 1$, corresponds to a partial transfer of entropy; 4) the inverse processes are included through a relaxation factor that is exact in the freeze-in and equilibrium limits and interpolates between them.

The energy density of the mediator evolves according to
\begin{equation}
  \dot\rho_\text{med} + 4H\rho_\text{med}
  = \Big[\frac{1}{4}\big(n_\chi^{\rm eq}\big)^2\langle\sigma v E\rangle_{\bar\chi\chi\to\text{med}+\text{med}}
       + \Gamma_{\text{med}\to\bar\chi\chi}\,m_\text{med}\,n_\text{med}^{\rm eq}\Big]
    \left(1 - \frac{\rho_\text{med}}{\rho_\text{med}^{\rm eq}}\right),
  \label{eq:rhophi}
\end{equation}
where $n_\chi^{\rm eq}$ counts $\bar \chi$ and $\chi$, $\langle\sigma v E\rangle$ denotes the thermal average of the cross section weighted by the total energy of the annihilating pair, and $n_\text{med}^{\rm eq}$ and $\rho_\text{med}^{\rm eq}$ are the equilibrium number and energy densities of the mediator at $T_\gamma$. The term in square brackets is the rate at which $\chi$ deposits energy into an empty mediator population. The first contribution comes from $\bar\chi\chi\to\text{med}+\text{med}$ and the second from inverse decays $\bar\chi\chi\to\text{med}$, which are open for $m_\text{med} > 2m_\chi$; by detailed balance the latter inject energy at the rate $\Gamma_{\text{med}\to\bar\chi\chi}\,m_\text{med}\,n_\text{med}^{\rm eq}$, the time-dilation factor of the decay rate cancelling against the energy of the produced mediator. The final factor accounts for the inverse processes. It is exact in the freeze-in limit $\rho_\text{med} \ll \rho_\text{med}^{\rm eq}$, halts production once the mediator reaches equilibrium, and interpolates between the two regimes.

Two further ingredients are needed. First, the energy must reach the mediator from the photon bath through $\chi$, so the transfer can proceed no faster than $\chi$'s own coupling to the plasma, $\Gamma_\chi$. Second, the fraction $w_\text{med}$ of the mediator population that is held in equilibrium by the plasma is already included in Eq.~\eqref{eq:Neffdef}, and must not be counted again. Using $\rho_\text{med}/T_\nu^4$, which is constant for free-streaming radiation, Eq.~\eqref{eq:rhophi} then becomes
\begin{equation}
  \frac{d}{d\ln(1/T_\gamma)}\left(\frac{\rho_\text{med}}{T_\nu^4}\right)
  = \frac{\Gamma_{\bar\chi\chi\to\text{med}}}{H}
    \left[(1-w_\text{med})\,\frac{\rho_\text{med}^{\rm eq}}{T_\nu^4} - \frac{\rho_\text{med}}{T_\nu^4}\right],
  \label{eq:Qrelax}
\end{equation}
with the effective production rate
\begin{equation}
  \Gamma_{\bar\chi\chi\to\text{med}} = \min\left(
    \frac{\frac{1}{4}\big(n_\chi^{\rm eq}\big)^2\langle\sigma v E\rangle_{\bar\chi\chi\to\text{med}+\text{med}}
       + \Gamma_{\text{med}\to\bar\chi\chi}\,m_\text{med}\,n_\text{med}^{\rm eq}}{\rho_\text{med}^{\rm eq}},\;
    \Gamma_\chi\right).
\end{equation}
We integrate Eq.~\eqref{eq:Qrelax} from $\rho_\text{med} = 0$ at $T_{\rm RH}$ until $\chi$ has annihilated away, after which $n_\chi^{\rm eq}$ is negligible and $\rho_\text{med}/T_\nu^4$ is frozen. Note that the mediator is produced whenever $\chi$ equilibrates at some point above $T_{\nu,\rm dec}$, even if $\chi$ loses contact with the plasma before neutrino decoupling.

The entropy that the mediator carries away is not transferred to the photons, and Eq.~\eqref{eq:Neffdef} must be modified accordingly. After neutrino decoupling the total entropy of the electromagnetic bath and of the mediator is conserved, and production of the mediator by $\chi$ merely transfers entropy from the former to the latter. The bath consists of photons, $e^\pm$, and the dark species in contact with them, and has
\begin{equation}
  g_{\star s}^{\rm EM}(T_\gamma) = 2 + \frac72 F_{\rm F}\!\left(\frac{m_e}{T_\gamma}\right)
    + \frac78\sum_i w_i\, g_i\, F_i\!\left(\frac{m_i}{T_\gamma}\right)
\end{equation}
entropic degrees of freedom. Comoving entropy conservation below $T_{\nu,\rm dec}$ then reads
\begin{equation}
  \frac{2\pi^2}{45}\,g_{\star s}^{\rm EM}(T_\gamma)\left(\frac{T_\gamma}{T_\nu}\right)^3
  + \frac{s_\text{med}}{T_\nu^3}
  = \frac{2\pi^2}{45}\left[\frac{11}{2} + \frac78\sum_i w_i\,g_i\,F_i(y_i)\right]
  + \frac{s_\text{med}}{T_\nu^3}\bigg|_{T_{\nu,\rm dec}},
  \label{eq:Sbalance}
\end{equation}
where $s_\text{med}$ is the entropy density defined in Sec.~\ref{sec:Neff}, and the first term on the right-hand side is the entropy of the bath at $T_{\nu,\rm dec}$. Eq.~\eqref{eq:Sbalance} determines $T_\gamma/T_\nu$, which in turn enters $\rho_\text{med}^{\rm eq}/T_\nu^4$ in Eq.~\eqref{eq:Qrelax}; the two equations are therefore solved simultaneously.
In two further cases the mediator becomes free radiation without being produced by $\chi$. In the first, it is in equilibrium with the plasma by $T_{\nu,\rm dec}$, but only through processes that switch off before then ($\pi^\pm\gamma\to\pi^\pm\text{med}$, $\gamma\gamma\to\text{med}+\text{med}$). It decouples together with the neutrinos, as a thermal population at $T_\nu$. Its share is $w_\text{free}\,\rho_\text{med}^{\rm eq}$, where $w_\text{free}$ is the fraction that has equilibrated by $T_{\nu,\rm dec}$ (the first factor of Eq.~\eqref{eq:wdef}) but is no longer in contact afterwards (one minus the second factor). At $T_{\nu,\rm dec}$ we therefore set $\rho_\text{med}$ to the larger of $w_\text{free}\,\rho_\text{med}^{\rm eq}$ and what $\chi$ has produced. In the second, it is still held in the bath by the $e^\pm$ processes after $T_{\nu,\rm dec}$. These switch off as the $e^\pm$ annihilate, and it leaves the bath where its optical depth, counted from recombination, reaches unity. At that temperature its equilibrium energy density $w_\text{med}\,\rho_\text{med}^{\rm eq}$ is added to $\rho_\text{med}$, and its entropy moves from the bath to $s_\text{med}$: we set $w_\text{med} = 0$ in $g_{\star s}^{\rm EM}$ and in Eq.~\eqref{eq:Qrelax}.
Mediators that decay to Standard Model states before recombination return their entropy to the photons. We account for this by multiplying the frozen $\rho_\text{med}/T_\nu^4$ by the surviving fraction
\begin{equation}
  \exp\left[-\int_{T_{\rm rec}}^{T_{\nu,\rm dec}}
    \frac{\Gamma_{\text{med}\to\rm SM}}{H}\,
    \frac{K_1(m_\text{med}/T_\gamma)}{K_2(m_\text{med}/T_\gamma)}\,\frac{dT_\gamma}{T_\gamma}\right],
\end{equation}
where $\Gamma_{\text{med}\to\rm SM}$ is the sum of the widths to $\gamma\gamma$, and $\pi\pi$ and then evaluating Eq.~\eqref{eq:Sbalance} at recombination, where $g_{\star s}^{\rm EM} = 2$, with the correspondingly reduced $s_\text{med}$. The entropy of the decayed mediators is thereby deposited in the photons.

The effective number of neutrinos at recombination receives two contributions: the neutrinos, through the modified ratio $T_\gamma/T_\nu$, and the surviving mediators, through their kinetic energy,
\begin{equation}
  N_{\rm eff} = N_{\rm eff}^{\rm SM}\left[\frac{4}{11}\left(\frac{T_\gamma}{T_\nu}\right)^3_{\rm rec}\right]^{-4/3}
  + \frac{\rho_\text{med}}{\rho_{1\nu}}\,
    \frac{\int_0^\infty d\zeta \, \zeta^2\big(\sqrt{\zeta^2+y^2}-y\big)/(e^\zeta-1)}
         {\int_0^\infty d\zeta \, \zeta^3/(e^\zeta-1)},
  \qquad y = \frac{m_\text{med}}{T_\text{med}}\bigg|_{\rm rec}.
  \label{eq:Neffphi}
\end{equation}
Here $\rho_\text{med}$ is the surviving energy density redshifted as radiation to recombination, $T_\text{med}$ is obtained from it as in Sec.~\ref{sec:Neff}, and $\rho_{1\nu} = \frac78\frac{\pi^2}{15}\big[(4/11)^{1/3}T_\gamma\big]^4$ is the energy density of a single Standard Model neutrino species. The ratio of integrals is the kinetic fraction of a relativistically decoupled Bose--Einstein distribution at $T_\text{med}$, and tends to unity for $m_\text{med} \ll T_\text{med}$. In the absence of mediator production, $\rho_\text{med} = 0$, Eq.~\eqref{eq:Sbalance} gives $2\,(T_\gamma/T_\nu)^3_{\rm rec} = \frac{11}{2} + \frac78\sum_i w_i g_i F_i(y_i)$, and Eq.~\eqref{eq:Neffphi} reduces to Eq.~\eqref{eq:Neffdef}.

\subsection{Additional Cross Sections Relevant to Mediator Production}
In addition to $\chi$ annihilation to the mediator, several additional processes may contribute to the mediator abundance needed to estimate modifications to $N_{eff}$.  The following cross sections not described in the main body of this work are relevant to correctly estimating the mediator abundance and its impact on $N_{eff}$.  We limit ourselves to interactions that could be relevant at a reheat temperature of $10$ MeV.  For higher reheat temperatures, more mediator production processes may become relevant.  

\subsubsection{Scalar Coupling to Quarks}
When $\phi$ has a scalar coupling to quarks (the $S\otimes S$ and $P\otimes S$ models) it can be produced at one loop by photon annihilation and at tree level by pion annihilation.  The one-loop $\phi$ coupling to photons also allows electrons to annihilate to the $\phi$ and a photon and  for photons to convert into $\phi$ by scattering with an electron or charged pion.  The relevant cross sections are shown below, and the derivation of the photon annihilation cross section is shown in more detail.

\begin{figure}[h]
    \centering
    \begin{subfigure}{0.3\textwidth}
    \begin{fmffile}{scalar2photon1}
 \begin{fmfgraph*}(120,80)
   \fmfleft{i1,i2}
   \fmfright{o1,o2}
   \fmf{scalar,label={\large$\phi$},label.side=left}{i2,v1}
   \fmf{photon,label={\large$\gamma$},label.side=left}{v2,o2}
   \fmf{scalar,label={\large$\pi^+/K^+$},label.side=left}{v1,v2}
   \fmf{scalar,label={\normalsize$\pi^+/K^+$},label.side=left}{v2,v3}
   \fmf{scalar,label={\normalsize$\pi^+/K^+$},label.side=left}{v3,v4}
   \fmf{scalar,label={\normalsize$\pi^+/K^+$},label.side=left}{v4,v1}
   \fmf{scalar,label={\large{$\phi$}},label.side=right}{i1,v4}
\fmf{photon,label={\large{$\gamma$}},label.side=left}{o1,v3}
 \end{fmfgraph*}
\end{fmffile}
 \end{subfigure}
 \begin{subfigure}{0.3\textwidth}
    \begin{fmffile}{scalar2photon2}
 \begin{fmfgraph*}(120,80)
   \fmfleft{i1,i2}
   \fmfright{o1,o2}
   \fmfforce{(0.35w,0.7h)}{v1}
   \fmfforce{(0.65w,0.7h)}{v2}
   \fmfforce{(0.65w,0.3h)}{v3}
   \fmfforce{(0.35w,0.3h)}{v4}
   \fmf{scalar,label={\large$\phi$},label.side=left}{i2,v1}
   \fmf{photon,label={\large$\gamma$},label.side=right}{v3,o2}
   \fmf{scalar,label={\large$\pi^+/K^+$},label.side=left}{v1,v2}
   \fmf{scalar}{v2,v3}
   \fmf{scalar,label={\normalsize$\pi^+/K^+$},label.side=left}{v3,v4}
   \fmf{scalar,label={\normalsize$\pi^+/K^+$},label.side=left}{v4,v1}
   \fmf{scalar,label={\large{$\phi$}},label.side=right}{i1,v4}
\fmf{photon,label={\large{$\gamma$}},label.side=right}{o1,v2}
 \end{fmfgraph*}
\end{fmffile}
 \end{subfigure}\\[30pt]
  \begin{subfigure}{0.3\textwidth}
    \begin{fmffile}{scalar2photon3}
 \begin{fmfgraph*}(120,80)
   \fmfleft{i1,i2}
   \fmfright{o1,o2}
   \fmfforce{(0.35w,0.7h)}{v1}
   \fmfforce{(0.65w,0.7h)}{v2}
   \fmfforce{(0.65w,0.3h)}{v3}
   \fmfforce{(0.35w,0.3h)}{v4}
   \fmf{scalar,label={\large$\phi$},label.side=right}{i2,v4}
   \fmf{photon,label={\large$\gamma$},label.side=right}{v3,o2}
   \fmf{scalar,label={\large$\pi^+/K^+$},label.side=left}{v1,v2}
   \fmf{scalar}{v2,v3}
   \fmf{scalar,label={\normalsize$\pi^+/K^+$},label.side=left}{v3,v4}
   \fmf{scalar}{v4,v1}
   \fmf{scalar,label={\large{$\phi$}},label.side=left}{i1,v1}
\fmf{photon,label={\large{$\gamma$}},label.side=right}{o1,v2}
 \end{fmfgraph*}
\end{fmffile}
 \end{subfigure}
 \begin{subfigure}{0.3\textwidth}
    \begin{fmffile}{scalar2photon4}
 \begin{fmfgraph*}(120,80)
   \fmfleft{i1,i2}
   \fmfright{o1,o2}
     \fmfforce{(0.35w,0.7h)}{v1}
   \fmfforce{(0.65w,0.7h)}{v2}
   \fmfforce{(0.65w,0.3h)}{v3}
   \fmfforce{(0.35w,0.3h)}{v4}
   \fmf{scalar,label={\large$\phi$},label.side=right}{i2,v4}
   \fmf{photon,label={\large$\gamma$},label.side=left}{v2,o2}
   \fmf{scalar,label={\large$\pi^+/K^+$},label.side=left}{v1,v2}
   \fmf{scalar,label={\normalsize$\pi^+/K^+$},label.side=left}{v2,v3}
   \fmf{scalar,label={\normalsize$\pi^+/K^+$},label.side=left}{v3,v4}
   \fmf{scalar}{v4,v1}
   \fmf{scalar,label={\large{$\phi$}},label.side=left}{i1,v1}
\fmf{photon,label={\large{$\gamma$}},label.side=left}{o1,v3}
 \end{fmfgraph*}
\end{fmffile}
 \end{subfigure}
    \caption{Feynman diagrams relevant for $\phi$ annihilation to photons for the $S\otimes S$ and $P\otimes S$ models and involving a meson loop.}
    \label{scalar2photon}
\end{figure}
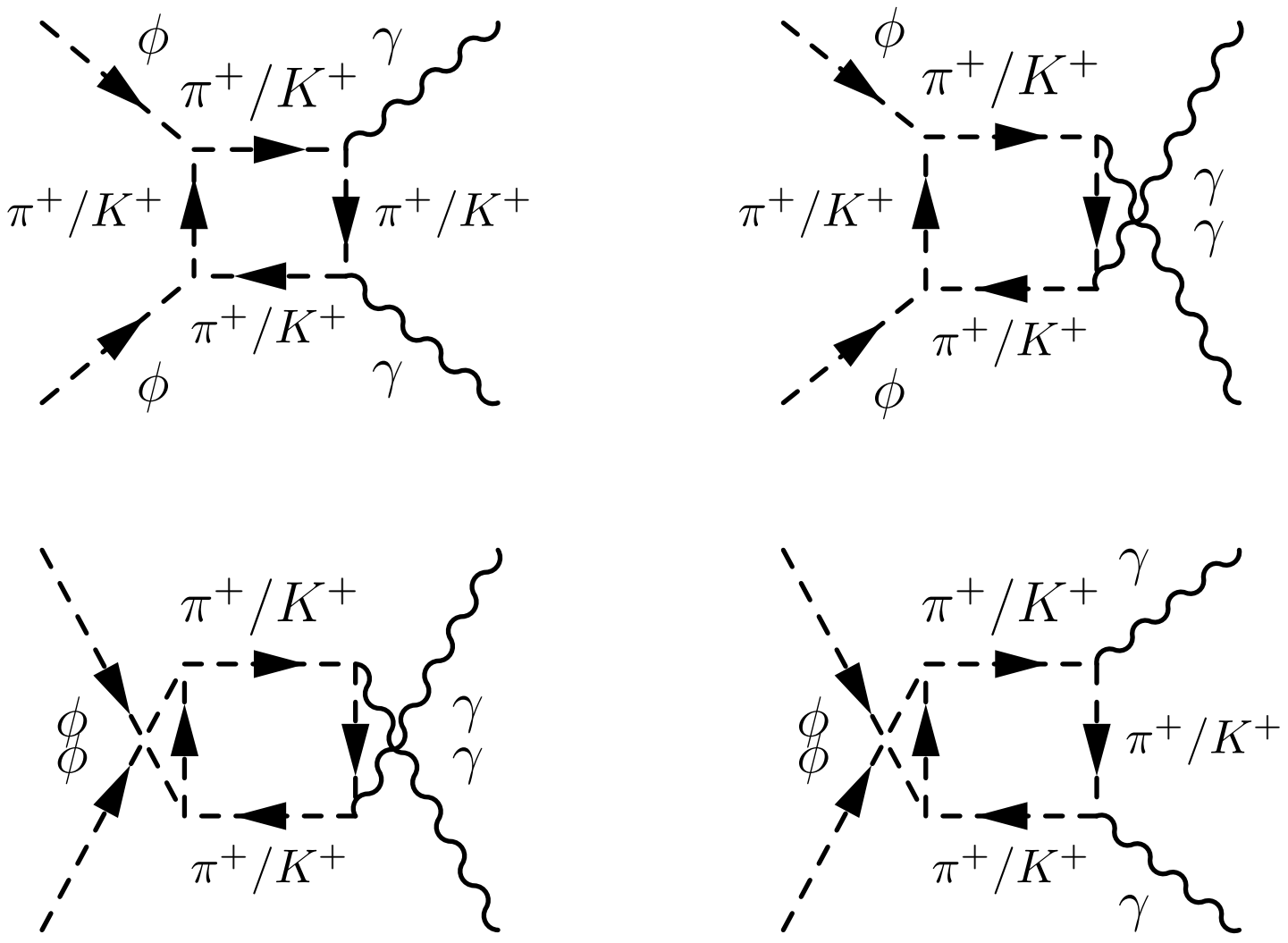
In the limit that $m_{\phi}\ll m_{\pi^+}$ and $s\ll m_{\pi^+}^2$ the amplitude of $\phi$ annihilation to photons is 
\begin{equation}
    M^{\mu\nu} = \frac{ie^2B_0^2g^{\mu\nu}}{24\pi^2}\left(\frac{(g_u+g_d)^2}{m_{\pi^+}^2}+\frac{(g_u+g_s)^2}{m_{K^+}^2}\right)\,.
\end{equation}
Which gives an annihilation cross section of
\begin{equation}
      \sigma_{\gamma\gamma\rightarrow\phi\phi}^{S/P \otimes S}=\frac{\alpha^2B_0^4}{1152\pi^3s}\left(\frac{(g_u+g_d)^2}{m_{\pi^+}^2}+\frac{(g_u+g_s)^2}{m_{K^+}^2}\right)^2\frac{\sqrt{s-4m_{\phi}^2}}{\sqrt{s}}\,,
\end{equation}
  where $s = 4E_{\gamma CM}^2$.

  $\phi$'s may also be produced by pion or electron annihilation.

  \begin{equation}
    \sigma_{\pi^+\pi^-\rightarrow\phi\phi}^{S/P\otimes S} = \frac{B_0^4(g_u+g_d)^4}{32\pi s(s-m_{\pi^+}^2)^2}\frac{\sqrt{s-4m_{\phi}^2}}{\sqrt{s-4m_{\pi^+}^2}}
\end{equation}

\begin{equation}
    \sigma_{\pi^0\pi^0\rightarrow\phi\phi}^{S/P\otimes S} = \frac{B_0^4}{32\pi s}\frac{\sqrt{s-4m_{\phi}^2}}{\sqrt{s-4m_{\pi^0}^2}}\left(\frac{(g_u+g_d)^2}{(s-m_{\pi^0}^2)}+\frac{2(g_u-g_d)^2}{3(s-m_{\eta'}^2)}+\frac{(g_u-g_d)^2}{3(s-m_{\eta}^2)}\right)^2
\end{equation}
\begin{equation}
    \frac{d\sigma_{e^+e^-\rightarrow \gamma\phi}^{P/S \otimes S}}{d\Omega} = \frac{g_{\phi\gamma}^2 \alpha  (s-m_{\phi}^2)  \left(2 \left(m_e^4+m_e^2 \left(m_{\phi}^2-2 t\right)+t
   \left(-m_{\phi}^2+s+t\right)\right)+s^2\right)}{64 \pi  s^{5/2} \sqrt{s-4 m_e^2}}\,.
\end{equation}

Finally, we include contributions to $\phi$ production from photon conversions to $\phi$ via scattering with a charged pion or electron.

\begin{equation}
\begin{split}
    \frac{d\sigma_{\pi^{\pm}\gamma\rightarrow \pi^{\pm}\phi}^{P/S\otimes S}}{d\Omega}=&\frac{B_0^2(g_u+g_d)^2 \alpha\sqrt{m_{\phi}^4-2 m_{\phi}^2 \left(m_{\pi^{\pm}}^2+s\right)+\left(m_{\pi^{\pm}}^2-s\right)^2}}{32 \pi s \left(s-m_{\pi^{\pm}}^2\right)^3 \left(m_{\pi^{\pm}}^2-t\right)^2}\\
   &\times\left(m_{\phi}^2 \left(m_{\pi^{\pm}}^2-s\right) \left(m_{\pi^{\pm}}^2-t\right)+6 m_{\pi^{\pm}}^6-2 m_{\pi^{\pm}}^4 (s+2 t)+m_{\pi^{\pm}}^2 t (t-7 s)+3
   s t (s+t)\right)
\end{split}
\end{equation}
\begin{equation}
    \frac{d\sigma_{e\gamma\rightarrow e\phi}^{P/S \otimes S}}{d\Omega} = \frac{\alpha g_{\phi\gamma}^2 \left(2 \left(s-m_e^2\right) \left(s-m_e^2-m_{\phi}^2\right)+2 s t+t^2\right) \sqrt{m_{e}^4-2 m_{e}^2
   \left(m_{\phi}^2+s\right)+\left(m_{\phi}^2-s\right)^2}}{64 \pi s t \left(s-m_{e}^2\right)}
\end{equation}
where $g_{\phi\gamma}=\frac{\alpha}{12 \pi}B_0 \left( \frac{g_u+g_d}{m_{\pi^+}^2}+\frac{g_u+g_s}{m_{K^+}^2}\right)$.
  \subsubsection{Pseudoscalar coupling to quarks}
   For $\phi$ with a pseudoscalar coupling  to quarks (the $P\otimes P$ and $S\otimes P$ models), $\phi$ may be produced at tree level via photon annihilation,  pion annihilation, and neutral pion conversion to $\phi$ via scattering with a charged pion.  The effective $\phi$ coupling to photons also allows the $\phi$ to be produced via electron annihilation and photon conversion by scattering off of an electron.  
 
\begin{equation}
  \sigma_{\phi \phi \rightarrow \gamma \gamma}^{S/P\otimes P} =    \frac{\tilde{g}_{\phi\gamma}^4 \left(\left(2 m_{\phi}^2-s\right) \sqrt{s \left(s-4
   m_{\phi}^2\right)} \left(6 m_{\phi}^4-4 m_{\phi}^2 s+5
   s^2\right)-8 m_{\phi}^4 \left(3 m_{\phi}^4-4 m_{\phi}^2
   s+s^2\right) \tanh ^{-1}\left(\frac{\sqrt{s \left(s-4 m_{\phi}^2\right)}}{2
   m_{\phi}^2-s}\right)\right)}{128 \pi  s^2 \left(2 m_{\phi}
   ^2-s\right)}.
\end{equation}
\begin{equation}
    \sigma_{\pi^0\pi^0\rightarrow\phi\phi}^{S/P\otimes P} = \frac{\tilde{g}_{\pi^0\pi^0\phi\phi}^2}{32\pi s}\frac{\sqrt{s-4m_{\phi}^2}}{\sqrt{s-4m_{\pi^0}^2}}
\end{equation}
\begin{equation}
    \sigma_{\pi^+\pi^-\rightarrow\pi^0\phi}^{S/P\otimes P} = \frac{\sqrt{m_{\phi}^4-2 m_{\phi}^2 \left(m_{\pi^0}^2+s\right)+ \left(m_{\pi^0}^2-s\right)^2} \left(\tilde{g}_{\pi^+\pi^-\pi^0\phi}-\tilde{g}_{\pi^+\pi^-\partial\pi^0\phi}
   \left(m_{\phi}^2+2 m_{\pi^+}^2+m_{\pi^0}^2-3 s\right)\right)^2}{16 \pi  s^{3/2} \sqrt{s-4 m_{\pi^+}^2}}
\end{equation}
\begin{equation}
    \frac{d\sigma_{e^+e^-\rightarrow \gamma\phi}^{P/S \otimes P}}{d\Omega} =-\frac{\alpha\tilde{g}_{\phi\gamma}^2 ( s-m_{\phi}^2)  \left(2 m_{e}^4 s+2 m_{e}^2 \left(m_{\phi}^4-m_{\phi}^2
   s-2 s t\right)+s \left(m_{\phi}^4-2 m_{\phi}^2 (s+t)+s^2+2 s t+2 t^2\right)\right)}{64 \pi s^{3/2} t^2 \sqrt{s-4 m_{e}^2}}
\end{equation}
\begin{equation}
    \sigma_{\pi^0\pi^0\rightarrow\pi^0\phi}^{S/P\otimes P} = \frac{\tilde{g}_{\pi^0\pi^0\pi^0\phi}^2}{16\pi s}\frac{\sqrt{m_{\phi}^4-2 m_{\phi}^2 m_{\pi^0}^2-2 m_{\phi}^2 s+m_{\pi^0}^4-2 m_{\pi^0}^2 s+s^2}}{\sqrt{s}\sqrt{s-4m_{\pi^0}^2}}
\end{equation}

\begin{equation}
    \frac{d\sigma_{\pi^{\pm}\pi^0\rightarrow\pi^{\pm}\phi}^{S/P\otimes P}}{d\Omega} = \frac{\sqrt{m_{\phi}^4-2 m_{\phi}^2 \left(m_{\pi^+}^2+s\right)+\left(m_{\pi^+}^2-s\right)^2} \left(\tilde{g}_{\pi^+\pi^-\pi^0\phi}-\tilde{g}_{\pi^+\pi^-\partial\pi^0\phi}
   \left(m_{\phi}^2+2 m_{\pi^+}^2+m_{\pi^0}^2-3 t\right)\right)^2}{64 \pi ^2 s \sqrt{m_{\pi^+}^4-2 m_{\pi^+}^2
   \left(m_{\pi^0}^2+s\right)+\left(m_{\pi^0}^2-s\right)^2}}
\end{equation}
\begin{equation}
    \frac{d\sigma_{e\gamma\rightarrow e\phi}^{P/S \otimes P}}{d\Omega} = \frac{\tilde{g}_{\phi\gamma}^2 \alpha \left(s-m_{e}^2\right) \sqrt{m_{e}^4-2 m_{e}^2
   \left(m_{\phi}^2+s\right)+\left(m_{\phi}^2-s\right)^2}}{16 \pi s t}
\end{equation}
The four particle couplings $\tilde{g}_{\pi^0\pi^0\phi\phi}$, $\tilde{g}_{\pi^+\pi^-\pi^0\phi}$, $\tilde{g}_{\pi^+\pi^-\partial\pi^0\phi}$, and $\tilde{g}_{\pi^0\pi^0\pi^0\phi}$ come from expanding the ChEFT Lagrangian out to  third order in 1/F, and  are defined in Table \ref{Tab:3pis}.
\subsubsection{Axial Vector coupling to quarks}
The $A'^{\mu}$ mediator, in addition to being produced by $\chi$ annihilation, can be produced by pion annihilation or neutral pion conversion via scattering with a charged pion.  We display the relevant production cross sections below.

\begin{equation}
    \sigma_{\pi^{+}\pi^-\rightarrow A'A'}^{A\otimes A} = 
    \frac{\lambda_{A'A'\pi^+\pi^- }^2(12m_{A'}^4-4m_{A'}^2s+s^2)}{128\pi m_{A'}^4s}\frac{\sqrt{s-4m_{A'}^2}}{\sqrt{s-4m_{\pi^+}^2}}
\end{equation}
\begin{equation}
    \sigma_{\pi^{+}\pi^-\rightarrow \pi^0A'}^{A\otimes A} = \frac{9 \lambda_{\pi^+\pi^-\pi^0A'}^2 \left(\frac{m_{A'}^4-2 m_{A'}^2 \left(m_{\pi^0}^2+s\right)+\left(m_{\pi^0}^2-s\right)^2}{s}\right)^{3/2}}{256
   \pi ^2 m_{A'}^2 \sqrt{s-4 m_{\pi^+}^2}}
\end{equation}

\begin{equation}
\begin{split}
     \frac{d\sigma_{\pi^{+}\pi^0\rightarrow \pi^+A'}^{A\otimes A}}{d\Omega} = &\frac{9 \lambda_{\pi^+\pi^-\pi^0A'}^2 \sqrt{m_{A'}^4-2 m_{A'}^2 \left(m_{\pi^+}^2+s\right)+\left(m_{\pi^+}^2-s\right)^2}}{256 \pi ^2 m_{A'}^2 s \sqrt{m_{\pi^+}^4-2 m_{\pi^+}^2
   \left(m_{\pi^0}^2+s\right)+\left(m_{\pi^0}^2-s\right)^2}} 
   \times\left(m_{A'}^4-2
   m_{A'}^2 \left(m_{\pi^0}^2+t\right)+\left(m_{\pi^0}^2-t\right)^2\right)
\end{split} 
\end{equation}
\begin{equation}
    \sigma_{A'A'\rightarrow\gamma\gamma}^{A\otimes A} = \frac{ s^{3/2} \left(12 m_{A'}^4-4 m_{A'}^2 s+s^2\right)}{1024 \pi  m_{A'}^4 \sqrt{s-4 m_{A'}^2}}\left(\frac{\alpha}{12\pi}\left(\frac{(\lambda_d-\lambda_u)^2}{m_{\pi^{\pm}}^2}+\frac{(\lambda_s-\lambda_u)^2}{m_{K^{\pm}}^2}\right)\right)^2
\end{equation}
$\lambda_{A'A'\pi^+\pi^-}$ and $\lambda_{\pi^+\pi^-\pi^0A'}$ are defined in Eq.~\eqref{4lambda}.


\section{Relevant Production Channels for Relic Abundance}
\label{ap:relic}
\begin{figure}[p]
  \centering
  \relicwide{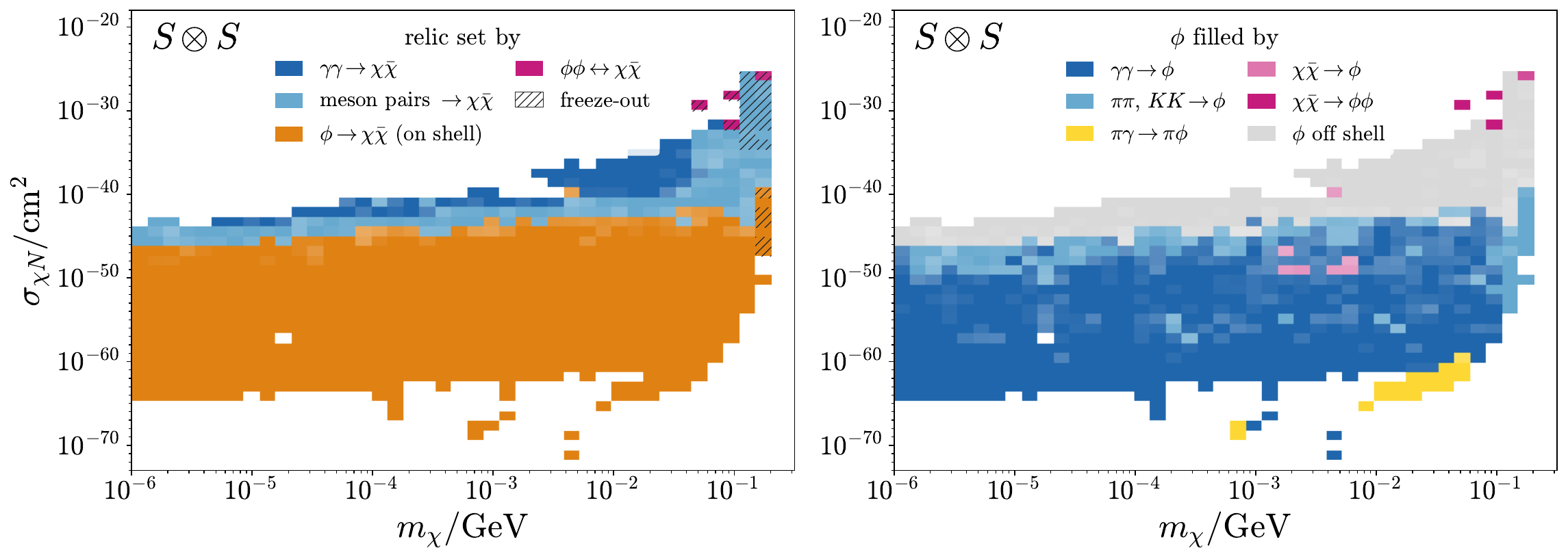}\\[1ex]
  \relicwide{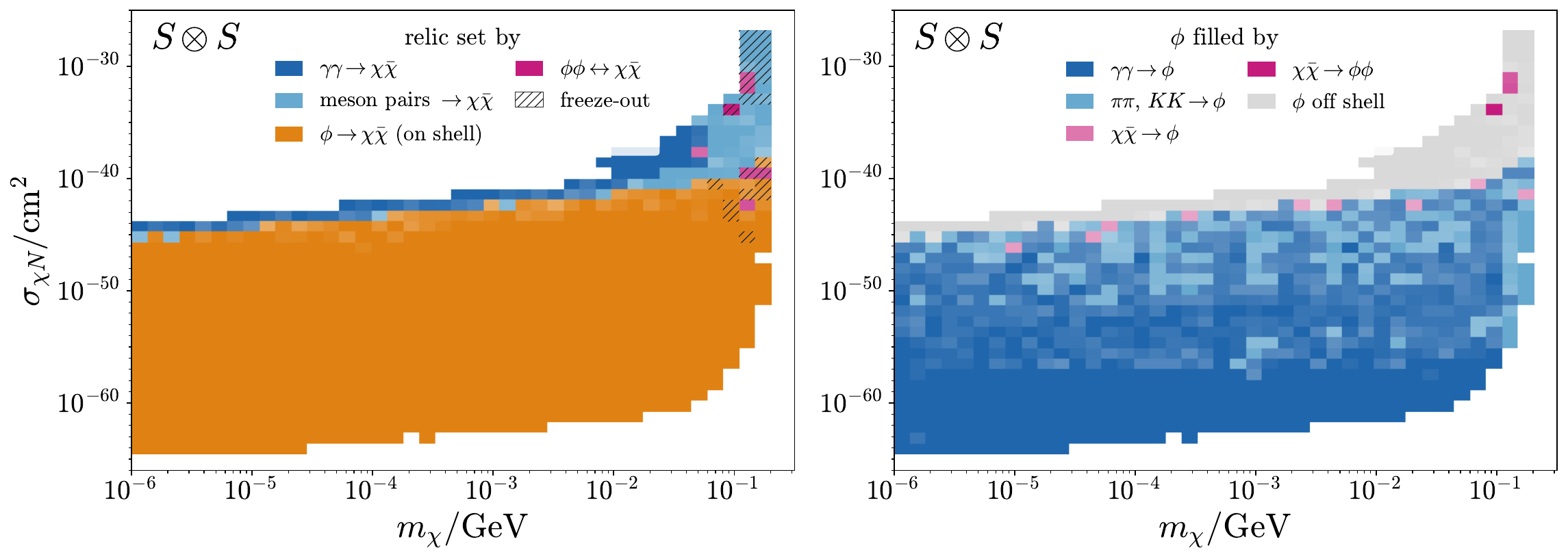}\\[1ex]
  \relicsingle{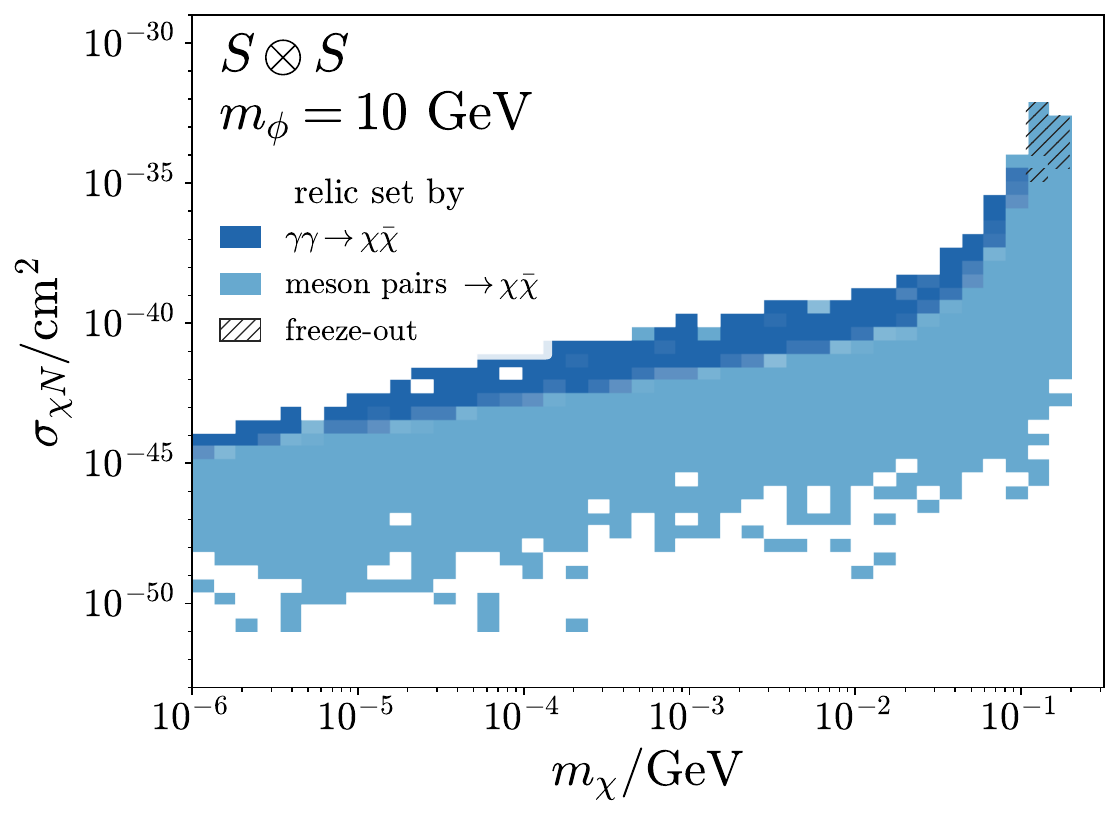}\hfill
  \relicsingle{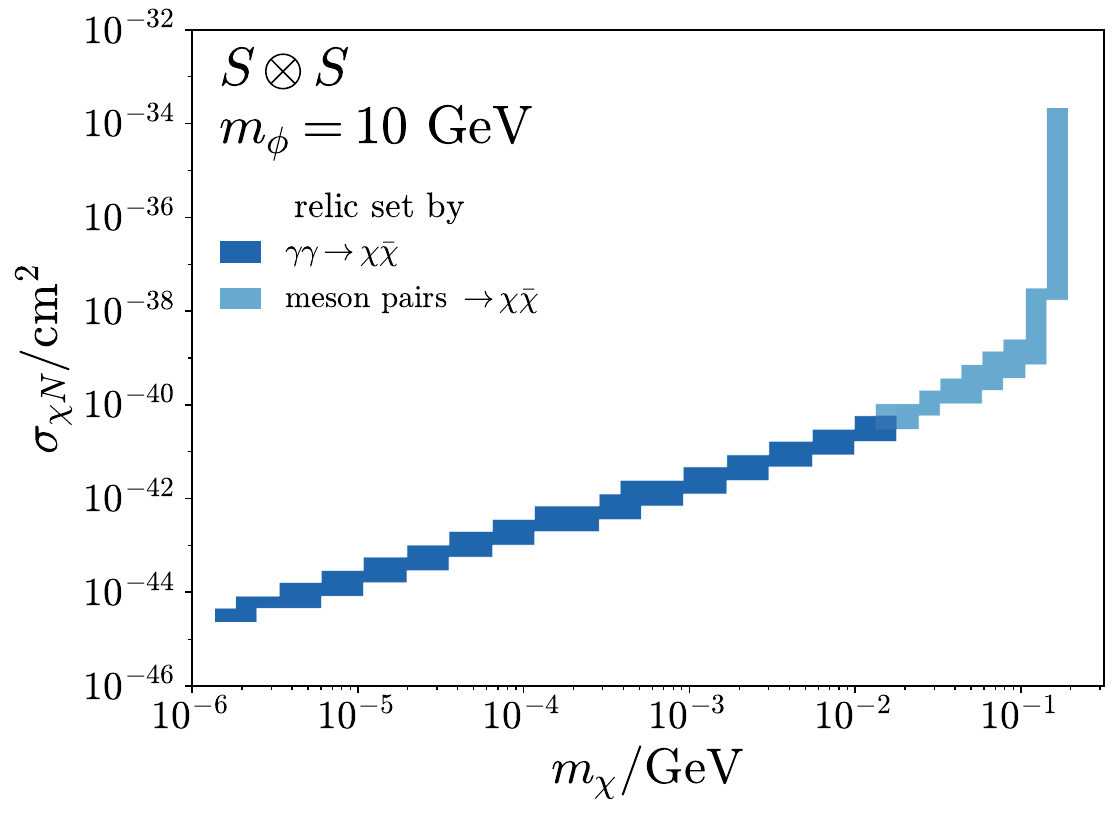}
  \caption{$S\otimes S$: the channel that set $\Omega h^2$ (left; freeze-out
    cells hatched) and what filled the mediator (right; grey where $\phi$ was
    exchanged only off shell), over the surviving region in
    $(m_\chi, \sigma_{\chi N})$.  Each cell shows its majority label, paler where
    the majority is weaker.  Top: full 6d scan; middle: MFV;
    bottom: $m_\phi = 10$~GeV, free coupling ratios (left) and MFV (right).}
  \label{fig:relic-source-SS}
\end{figure}

\begin{figure}[p]
  \centering
  \relicwide{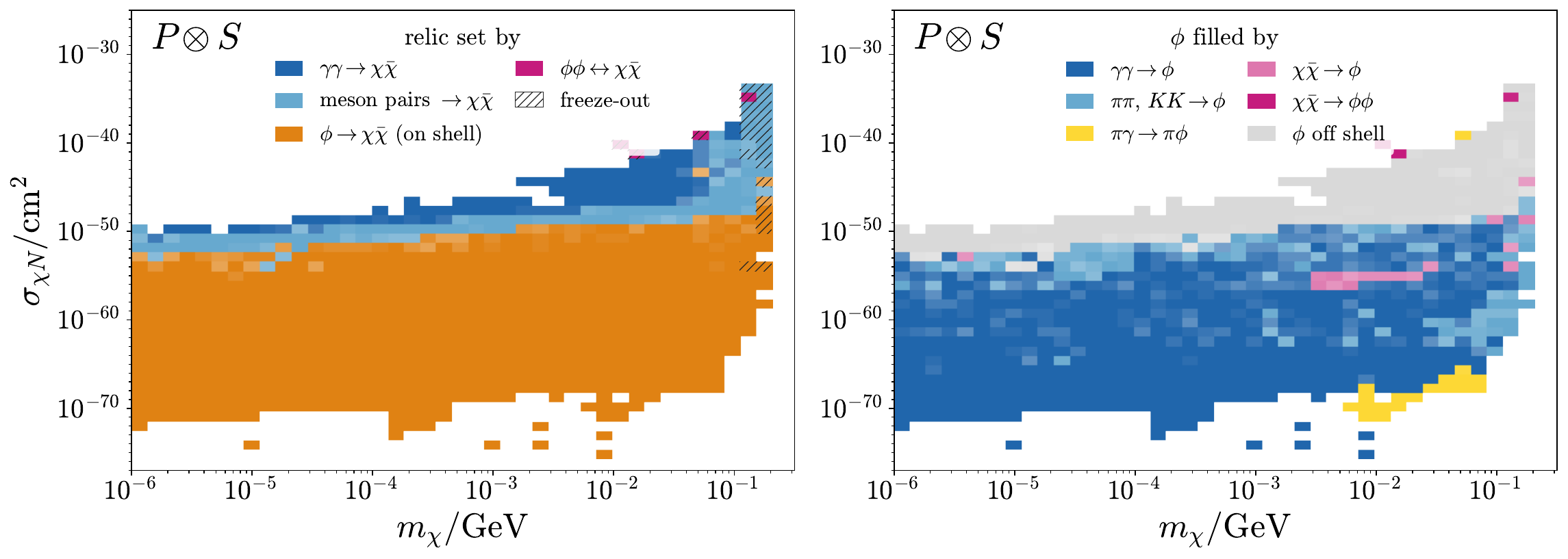}\\[1ex]
  \relicwide{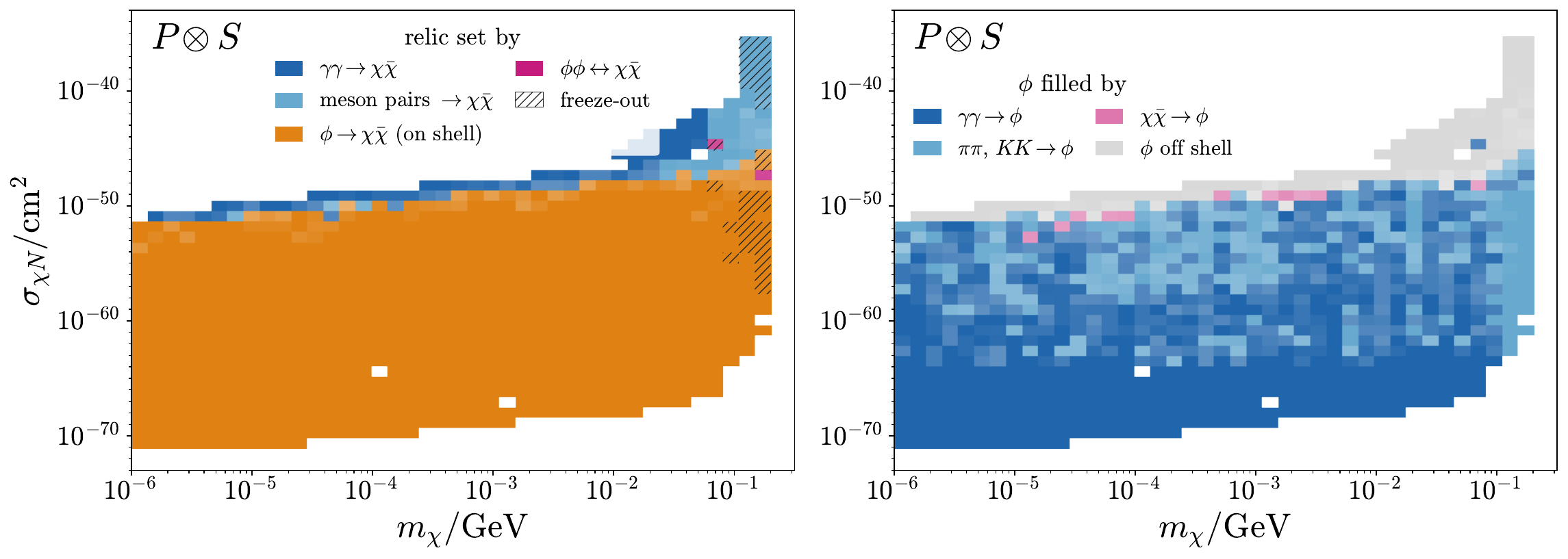}\\[1ex]
  \relicsingle{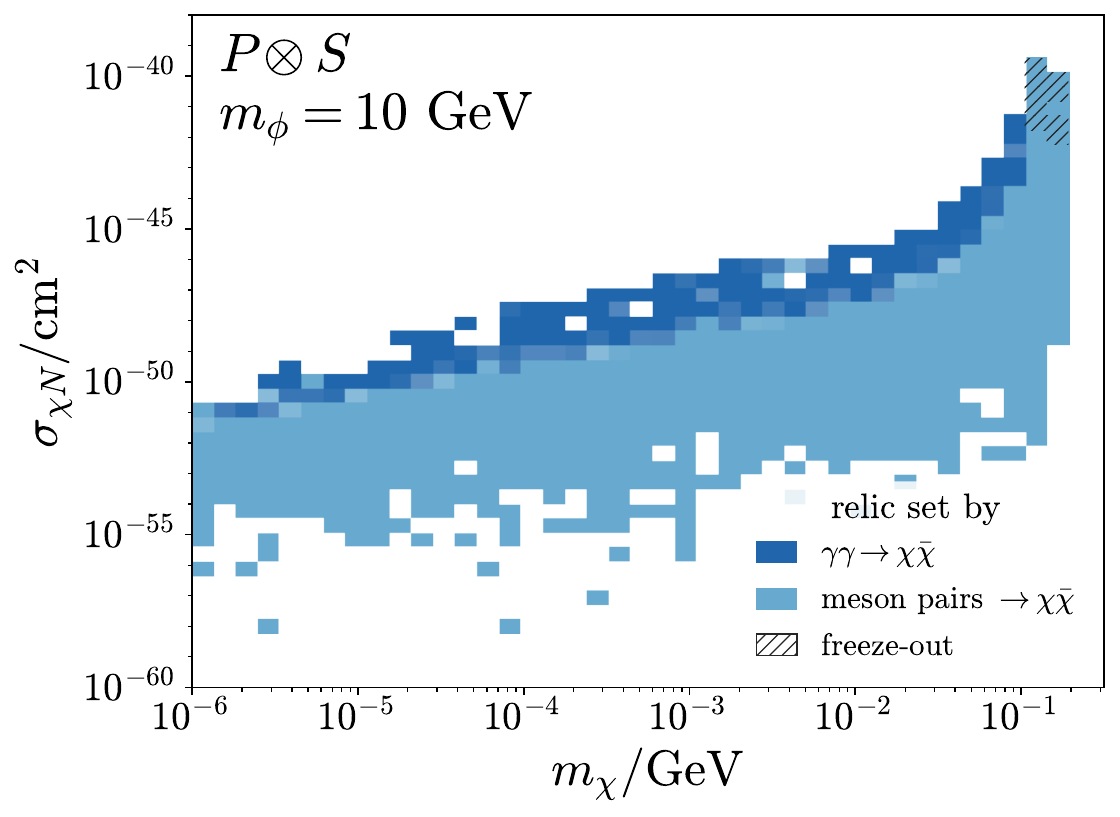}\hfill
  \relicsingle{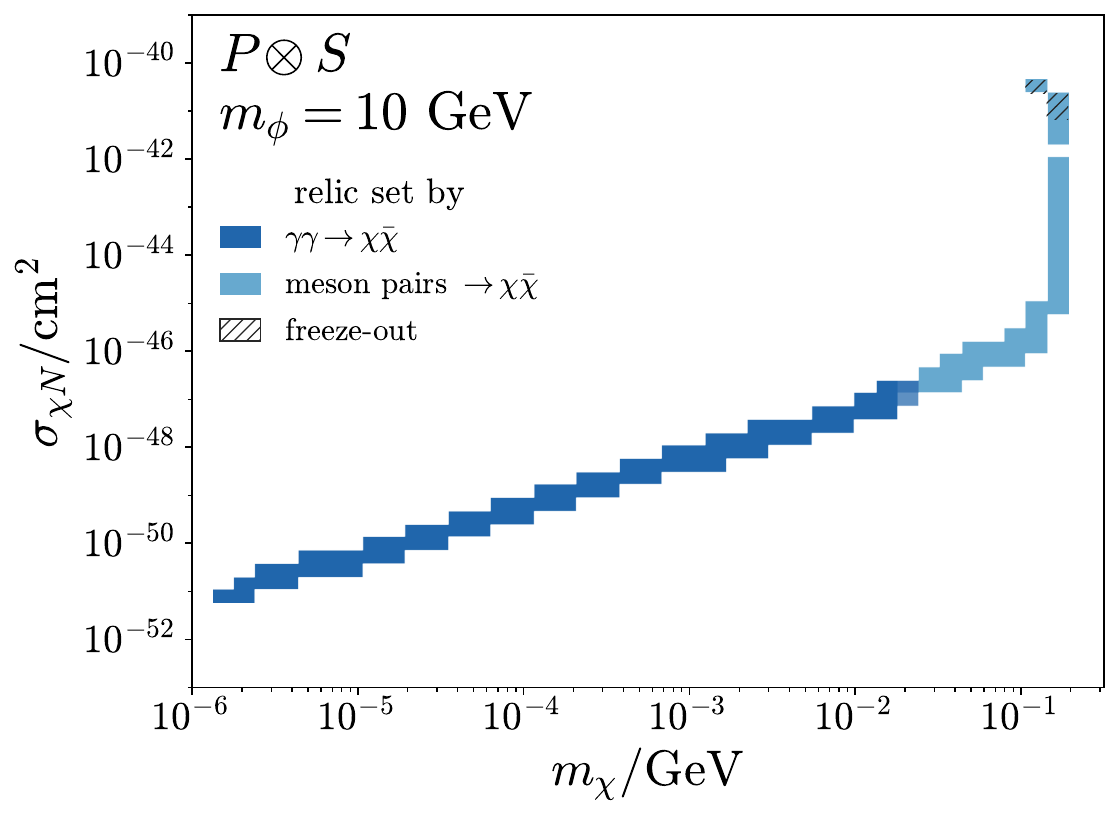}
  \caption{As Fig.~\ref{fig:relic-source-SS}, for $P\otimes S$.  Top:
    6d scans; middle: MFV; bottom: $m_\phi = 10$~GeV, free quark coupling ratios (left) and
    MFV (right).}
  \label{fig:relic-source-PS}
\end{figure}

\begin{figure}[p]
  \centering
  \relicwide{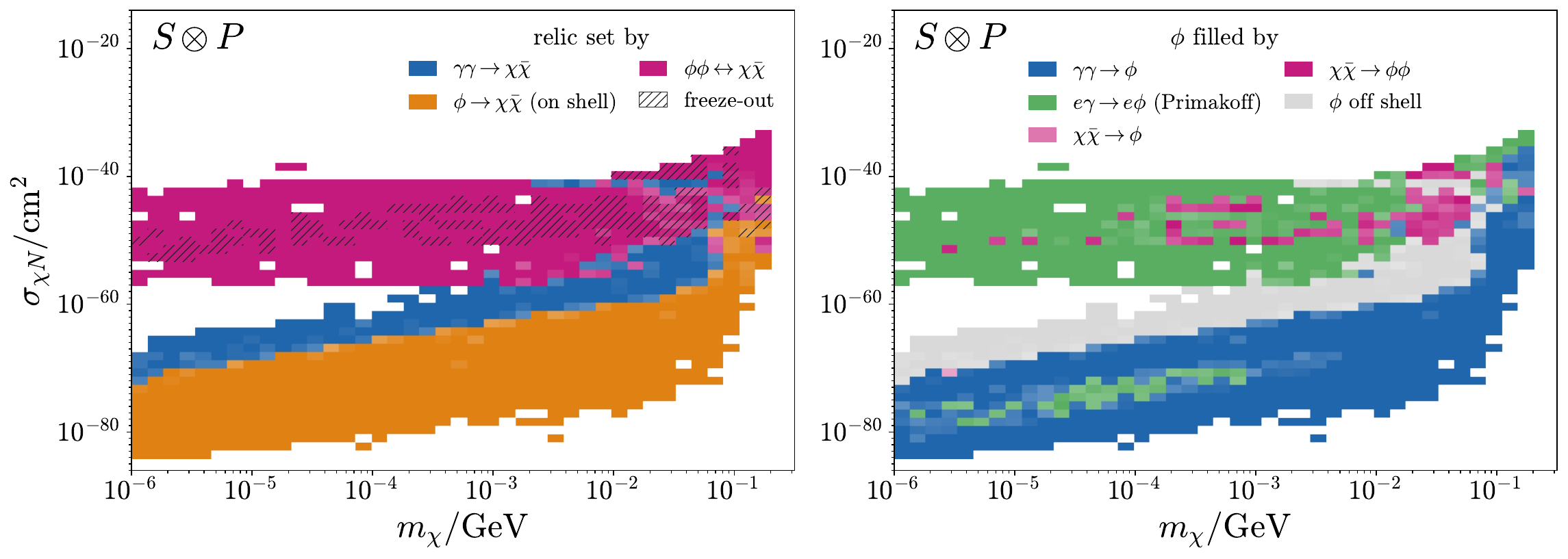}\\[1ex]
  \relicwide{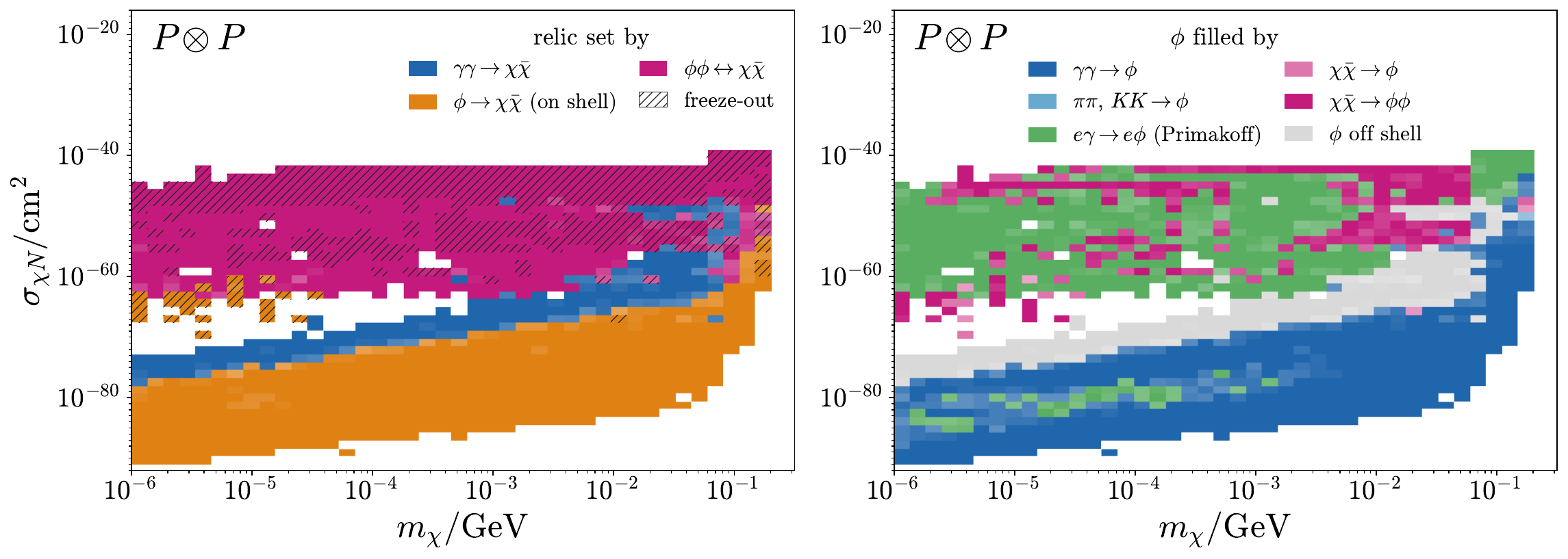}\\[1ex]
  \relicwide{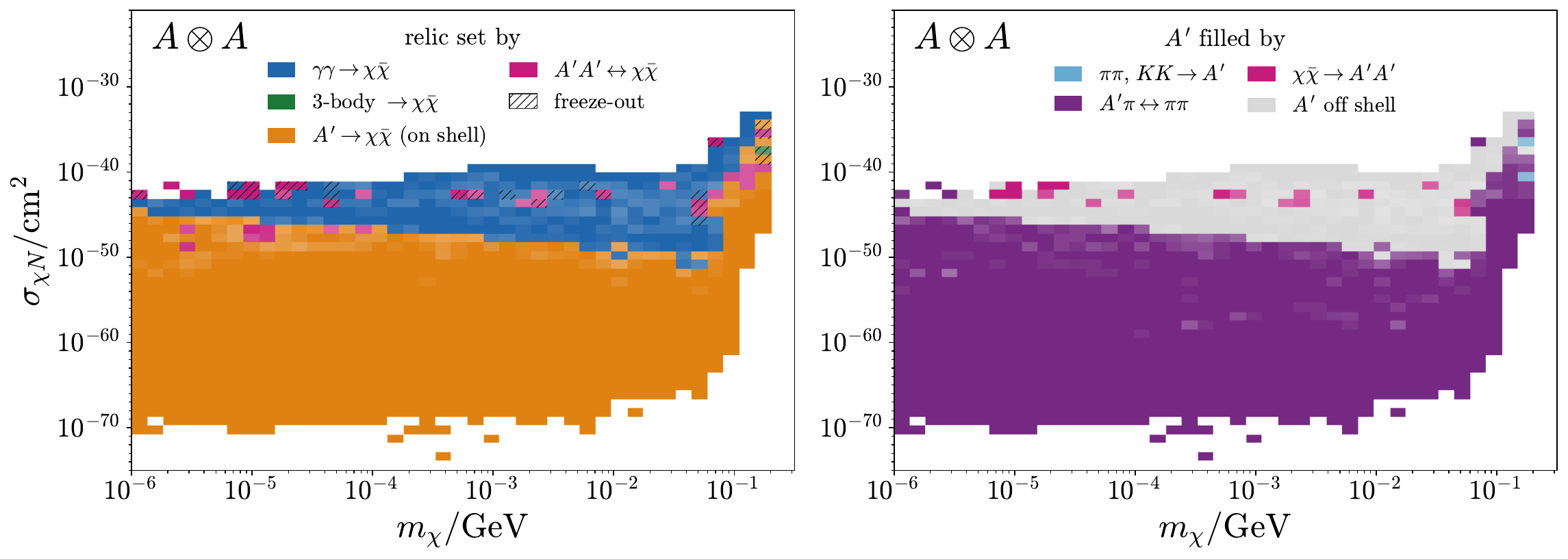}
  \caption{As Fig.~\ref{fig:relic-source-SS}, for the 6d scans of
    $S\otimes P$ (top), $P\otimes P$ (middle) and $A'\otimes A'$ (bottom).}
  \label{fig:relic-source-SP-PP-AA}
\end{figure}

In this Appendix, we show which are the dominant production channels involved in the (mostly) freeze-in relic abundance calculations. For each model, we show 1) which channel dominantly populates the $\chi$ DM abundance, and 2) when it is a two-step process via the mediator, which channel dominantly populates the mediator. Since the reheat temperature is set to 10 MeV, the heavy mediator case never populates the bath with $\phi$ or $A'$. See main text for full description. Fig. \ref{fig:relic-source-SS} shows the $S\otimes S$ model, Fig. \ref{fig:relic-source-PS} shows $P\otimes S$, and Fig. \ref{fig:relic-source-SP-PP-AA} shows the remaining models. For pseudoscalar and axial quark couplings with fixed $m_\phi = m_{A'} = 10$~GeV, $\gamma\gamma\to\chi\bar\chi$ sets $\Omega_\chi h^2$ over essentially the whole surviving region in all three, so we do not show those cases.

\end{document}